\documentclass[12pt,vi,twoside,singlespace]{mitthesis}
\usepackage{epsfig}
\usepackage[usenames,dvips]{color}
\usepackage{bibcontents}
\usepackage{graphicx}
\usepackage{amsmath}
\usepackage{natbib}
\usepackage{amssymb}
\usepackage{amstext}
\usepackage{deluxetable}
\usepackage{aastex_hack}
\usepackage[margin=20pt,font=small,labelfont=bf]{caption}
\usepackage{pstricks}
\usepackage{pstricks}
\usepackage{pst-node}
\usepackage{pst-tree}
\begin{document}

\title{Radiation Transport Around Kerr Black Holes}

\author{Jeremy David Schnittman}
\prevdegrees{B.\ A.\ Physics \\Harvard University (1999)}
\department{Department of Physics}
% If the thesis is for two degrees simultaneously, list them both
% separated by \and like this:
% \degree{Doctor of Philosophy \and Master of Science}
\degree{Doctor of Philosophy}
\degreemonth{February}
\degreeyear{2005}
\thesisdate{January 20, 2005}

%% By default, the thesis will be copyrighted to MIT.  If you need to copyright
%% the thesis to yourself, just specify the `vi' documentclass option.  If for
%% some reason you want to exactly specify the copyright notice text, you can
%% use the \copyrightnoticetext command.  
%\copyrightnoticetext{\copyright IBM, 1990.  Do not open till Xmas.}

% If there is more than one supervisor, use the \supervisor command
% once for each.
\supervisor{Edmund Bertschinger}{Professor of Physics}

% This is the department committee chairman, not the thesis committee
% chairman.  You should replace this with your Department's Committee
% Chairman.
\chairman{Thomas J. Greytak}{Professor of Physics\\Associate Department Head for Education}

% Make the titlepage based on the above information.  If you need
% something special and can't use the standard form, you can specify
% the exact text of the titlepage yourself.  Put it in a titlepage
% environment and leave blank lines where you want vertical space.
% The spaces will be adjusted to fill the entire page.  The dotted
% lines for the signatures are made with the \signature command.
\maketitle

% The abstractpage environment sets up everything on the page except
% the text itself.  The title and other header material are put at the
% top of the page, and the supervisors are listed at the bottom.  A
% new page is begun both before and after.  Of course, an abstract may
% be more than one page itself.  If you need more control over the
% format of the page, you can use the abstract environment, which puts
% the word "Abstract" at the beginning and single spaces its text.

%% You can either \input (*not* \include) your abstract file, or you can put
%% the text of the abstract directly between the \begin{abstractpage} and
%% \end{abstractpage} commands.

% First copy: start a new page, and save the page number.
\cleardoublepage
% Uncomment the next line if you do NOT want a page number on your
% abstract and acknowledgments pages.
% \pagestyle{empty}
\setcounter{savepage}{\thepage}
\begin{abstractpage}
%% The text of your abstract and nothing else (other than comments) goes here.
%% It will be single-spaced and the rest of the text that is supposed to go on
%% the abstract page will be generated by the abstractpage environment.  This
%% file should be \input (not \include 'd) from cover.tex.

This Thesis describes the basic framework of a relativistic ray-tracing
code for analyzing accretion processes around Kerr black
holes. We begin in Chapter 1 with a brief historical summary of the
major advances in black hole astrophysics over the past few
decades and outline some of the important questions still open today. In
Chapter 2 we present a detailed
description of the ray-tracing code, which integrates the geodesic
equations of motion for massless particles,
tabulating the position and momentum along each photon
trajectory. Coupled with an independent model for the emission and
absorption at each point in spacetime, time-dependent images and
spectra can be produced by integrating the radiative transfer
equation along these geodesic photon paths. This approach can be used
to calculate the transfer function between the plane of the
accretion disk and the detector plane, an important tool for
modeling relativistically broadened emission lines. 

Observations from the \textit{Rossi X-Ray Timing Explorer} have shown
the existence of high frequency quasi-periodic oscillations (HFQPOs)
in a growing number of black hole binary systems.
In Chapter 3, we employ a simple ``hot spot'' model
to explain the position and amplitude of these HFQPO peaks. Using the
exact geodesic equations of motion for the Kerr metric, we calculate the
trajectories of massive test particles, which are treated as
isotropic, monochromatic emitters in their rest frames, imaged with
the ray-tracing code described above. The power spectrum of the
periodic X-ray light curve consists of multiple peaks located at
integral combinations of the black hole coordinate frequencies. 
Additionally, we model the effects of
shearing the hot spot in the disk, producing an arc of emission that
also follows a geodesic orbit. By including non-planar
orbits that experience Lense-Thirring precession, we investigate the
possible connection between high and low frequency QPOs.

In Chapter 4, we introduce additional
features to the hot spot model to explain the broadening of
the QPO peaks as well as the damping of higher frequency harmonics in
the power spectrum. We present a number of analytic results that
agree closely with more detailed numerical calculations. Three
primary pieces are developed: the addition of multiple hot spots with
random phases, a finite width in the distribution of geodesic orbits,
and the scattering of
photons from the hot spot through a corona around the
black hole. The complete model is used to fit the observed power
spectra of both type A and type B QPOs seen in XTE J1550--564, giving
confidence limits on each of the model parameters. We also include a
discussion of higher-order statistics and the use of the bicoherence
to distinguish between competing QPO models.

To gain more insight into the overall accretion geometry, in
Chapter 5 we follow the formulation of \citet{novik73} for describing the
structure of a relativistic $\alpha$-disk around a Kerr black hole. The
resulting equations of vertical structure can be integrated at each
radius to give the complete density and temperature profile of the
steady-state disk. Inside of the ISCO, the gas is
propagated along a plunging geodesic trajectory, evolving according to
one-dimensional classical hydrodynamics in the local inertial frame of
the fluid. Given the surface temperature of the disk everywhere outside
of the horizon, the observed spectrum is calculated using the transfer
function mentioned above. The features of this modified
thermal spectrum may be used to infer the physical properties of the
accretion disk and the central black hole.

As an extension of the simple scattering model presented in Chapter 4,
in Chapter 6 we develop a Monte Carlo code to calculate the detailed
propagation of photons from a hot spot emitter scattering through a
high-temperature, low-density corona surrounding the black hole. Each
photon is followed until it is either captured by the black hole or is
``detected'' by a distant observer. The coronal scattering has two major
observable effects: the inverse-Compton process alters the photon
spectrum by adding a high energy power-law tail,
and the random scattering of each photon effectively damps out the
highest frequency
modulations in the X-ray light curve. We present simulated
photon spectra and light curves and compare with \textit{RXTE} data,
concluding with the implications for the hot spot model of HFQPOs.

\end{abstractpage}

% Additional copy: start a new page, and reset the page number.  This way,
% the second copy of the abstract is not counted as separate pages.
% Uncomment the next 6 lines if you need two copies of the abstract
% page.
% \setcounter{page}{\thesavepage}
% \begin{abstractpage}
% \input{abstract}
% \end{abstractpage}

\cleardoublepage

\section*{Acknowledgments}
\begin{flushright}
{\it
Gravity can not be held responsible for people falling in love.\\
\medskip

No, this trick won't work...How on earth are you ever going\\
to explain in terms of chemistry and physics so important\\
a biological phenomenon as first love? \\
\medskip
}
-Albert Einstein
\end{flushright}
\vspace{0.5cm}

\begin{center}
\it{
For Nomi\\

The love of my life, my closest friend, my foundation, my joy, my
$\alpha$ and my $\Omega$.\\
And sometimes even my $\upsilon$.}\\
\end{center}

This work is the cumulative result of so many people and so much
effort that I am almost certain to omit important names, places, and
events. My apologies in advance. 

In roughly chronological order, I would like to thank the following:
The Almighty, the Creator, for making a world so beautiful and full of
wonder. Albert Einstein, my childhood hero, who only gained in stature
and esteem as I grew older and was fortunate enough to learn more
about his unequaled accomplishments. He was a Giant standing on
the shoulders of giants, seeing farther than anyone else has before or
since. My parents, Michael and Suzanne Schnittman, for giving me a
lifetime full of love and encouragement, and always reminding me to
get back to work. Thank you dad for teaching me the value of a sense
of humor. Thank you mom for teaching me the value of an education,
whether formal or informal. My brother Aaron for keeping me in my
place and always giving me someone to look up to. My superb teachers
and classmates at Wilson Magnet High School, for igniting my passion
for science and mathematics. Steve Craxton, my
first research advisor, who introduced me to scientific
computing, proper writing style (may he forgive me for the pages to
come), ray-tracing and radiation transport, and the world of
professional physics research. 

Rabbis Blau, Brovender, Ebner, Kilimnick, Schrader, and Walk, for
teaching me about what is really important in this life. And my MIT
office-mates, for distracting me with what is really unimportant in
life. From the early days in 37-644, Tesla Jeltema, Josh Winn, and Jon
Miller; from
the mezzanine clubhouse, Jamie Portsmouth, Nick Morgan, Josh Faber,
and John Fregeau; from 37-638, Adrienne Juett, Ed Boyce, Alex
Shirokov, Kristin Burgess, Dave Pooley, and Adam Bolton; from the
silly putty foosball madhouse, Matt Muterspaugh, Jake Hartman, Eric
Pfahl, and my replacements Allyn Dullighan and Will Farr; from the
antechamber, Justin Kasper, Miriam Krauss, Judd Bowman, Molly Swanson,
and Bobby
Cohanim. Thanks to the D.\ Samuel Gottesman Library of the Albert
Einstein College of Medicine, for being my office away from home. 

Thank you to Paul Schechter, who took me under his wing my
first day on the job; Fred Rasio, who introduced me to black holes
and general relativity; Jack Wisdom, who trained me in the art of
chaotic dynamics; Scott Hughes, whose office was always wide open
for me to come in and bounce ideas off the white board; Jackie Hewitt,
for offering an outsider's point of view and fitting me into her crazy
schedule. Arlyn Hertz for fitting me into Jackie's schedule and
helping make the defense run so smoothly. Martha Bezzat for answering
all my random questions and requests for help with the mail, fax
machine, and MIT bureaucratic regulations, always promptly and with a
smile. And of course, many many thanks to
my advisor, Ed Bertschinger, who provided the inspiration and
direction for much of this work. His careful, critical eye kept me in
line, and his encouragement this past year brought the final product
to completion. Much of this work was supported by his NASA ATP grant
NAG5-13306.

Again, the biggest accolades are for my wife Nomi, for putting up with
me and supporting me, especially these past few months. I couldn't
have done it without you.

\begin{figure}[ht]
\begin{center}
\scalebox{1.1}{\includegraphics*[120,500][490,750]{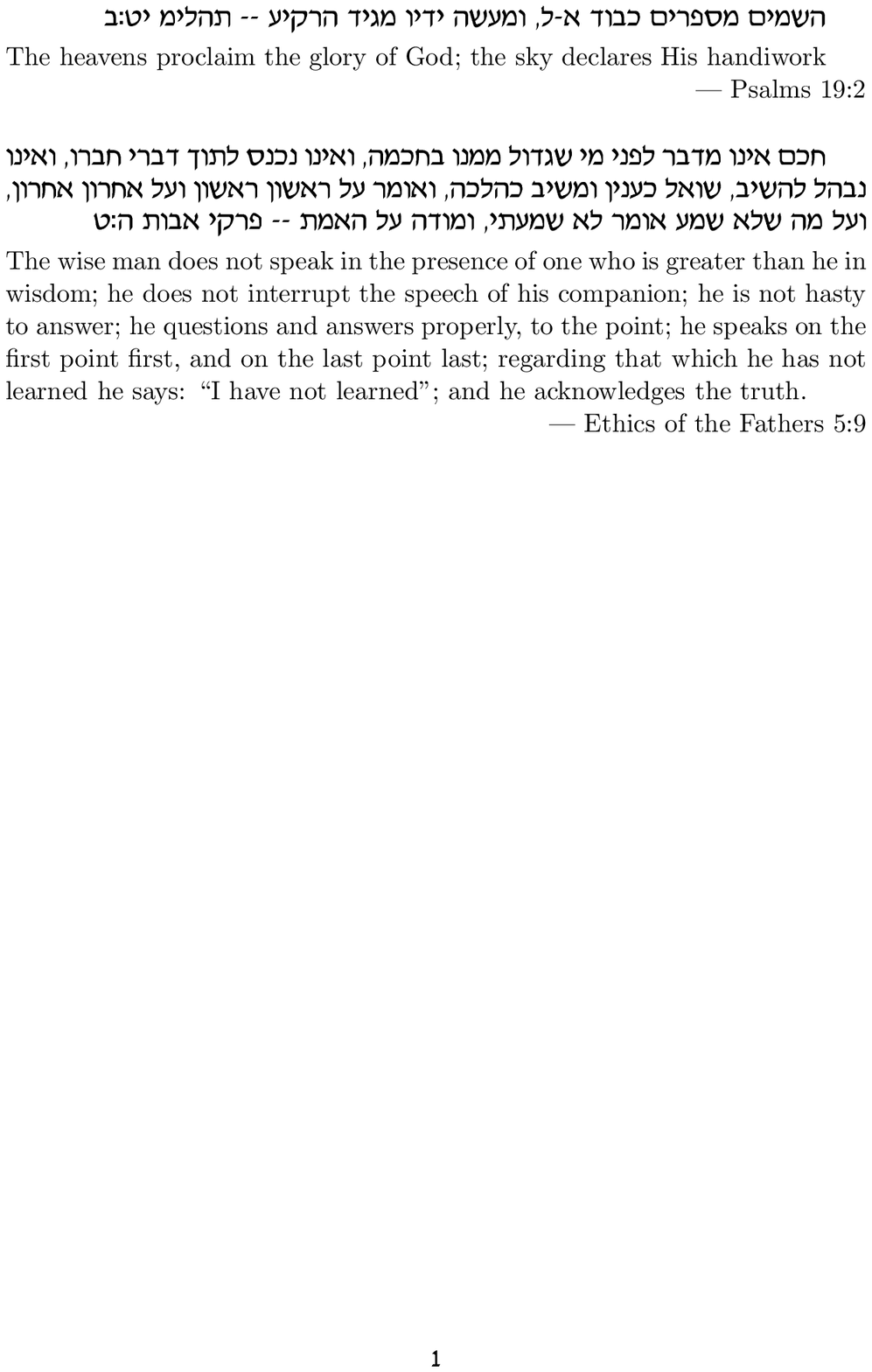}}
\end{center}
\end{figure}

\pagestyle{headings}
%% This file simply contains the commands that actually generate the table of
%% contents and lists of figures and tables.  You can omit any or all of
%% these files by simply taking out the appropriate command.  For more
%% information on these files, see appendix C.3.3 of the LaTeX manual. 
\tableofcontents
\newpage
\listoffigures
\newpage
\listoftables

\chapter{Introduction and Outline}
\begin{flushright}
{\it
I want to know God's thoughts; the rest are details.\\
\medskip

My religion consists of a humble admiration of the illimitable \\
superior Spirit who reveals Himself in the slight details \\
we are able to perceive with our frail and feeble mind.\\
\medskip
}
-Albert Einstein
\end{flushright}
\vspace{1cm}

For nearly a century now, Einstein's special and general theories of
relativity (SR/GR) have helped form our understanding of the
universe. Yet still many of Einstein's most fundamental predictions
have not been proven (or disproved, to be fair). Central among these
predictions are the laws governing matter and radiation in the strong
gravitational fields around neutron stars and black holes. As our
observational and theoretical capabilities continue to advance, the
ability to probe these strong field regions steadily
improves. Some of the most recent of these impressive advances are the
spatial and spectral
resolution attainable with the \textit{Chandra} and \textit{XMM-Newton}
observatories and the timing resolution of the \textit{Rossi X-ray
Timing Explorer} (\textit{RXTE}). The 
equally important developments in large-scale parallel processing have
made cost-effective computing widely available for accurate,
high-resolution astrophysical modeling.

However, the technological marvels of space observatories and the
impressive simulations of massive supercomputers are limited in their
ability to independently explain the laws of physics around black
holes. Unfortunately, many of the data analysis tools used today to fit
observations are lacking important fundamental physics. At the same
time, many theoretical models for accretion disks do not include
emission mechanisms that would allow us to compare them directly to
observations. It is within this context that we present a framework
that works towards bridging the gap between theory and observations of
black hole accretion disks.

The central foundation of this Thesis is a ray-tracing code for the
Kerr metric, the basis of a more generalized analysis
tool for relativistic accretion disk models. Often referred to as a
``post-processor,'' this tool could be used to analyze the raw output data
from an independent simulation that models the dynamic behavior of
accreting gas around the black hole. The ultimate goal for
this tool is to allow direct comparison of a variety of simulations
with spectral and timing observations of black hole binaries as well
as active galactic nuclei (AGN). In the following chapters, we will
describe the physics packages included in this analysis code and apply
it to a number of simple models. In doing so, we hope to gain some
insight into a number of observational features, such as the broad
iron emission line, quasi-periodic oscillations (QPOs) in the X-ray light
curves, and the shape of the continuum photon energy spectra from black hole
systems. 

\section{Motivation}\label{motivation}
In the past decade, observations of X-ray emission from accreting
neutron stars and black holes have introduced new
possibilities for astrophysical tests of fundamental physics. Recent
discoveries made by satellites such as \textit{ASCA},
\textit{BeppoSAX}, \textit{RXTE}, \textit{Chandra}, and
\textit{XMM-Newton} provide direct evidence
for strong-field gravitational effects in compact binary systems and
AGN. These results include Doppler-broadened
iron K$\alpha$ fluorescent emission from microquasars
and millisecond variability of the X-ray flux from black 
holes in low-mass X-ray binaries
[see \citet{mccli04} for an excellent review]. These measurements 
give the exciting prospect for determining a black hole's mass and
spin, as well as tests of general relativity in the strong-field
regime.

The strong gravitational fields near a black hole introduce
significant deviations from Newtonian physics, including the existence
of an inner-most stable circular orbit (ISCO), a feature absent in the
classical Kepler problem. Since accreting gas can efficiently lose
energy and angular momentum only outside of the ISCO, the hydrodynamic
and radiative behavior of the inner accretion disk should be strongly
dependent on the structure of the spacetime metric near the
ISCO. The famous ``no hair'' theorem states that the only observable
features of a stationary, electrically neutral black hole are
functions of its mass $M$ and specific angular momentum $a\equiv
J/M$. 

Yet the spin is a much more 
difficult quantity to measure, since the leading order curvature terms
scale as $\sim M/r^3$ for the mass and $\sim a/r^4$ for the spin
contributions [see, e.g.\ \citet{barde72}], analogous to the
relative scaling of monopole and dipole fields in 
electromagnetism.  Thus any observable that is sensitive to the spin
parameter will presumably originate from the regions closest to the
black hole. As with most observations in astrophysics, the most
difficult ones are also the most rewarding. While mass is also the
most important ingredient in Newtonian gravity, the spin is a
fundamentally relativistic feature, so is one of the ideal means of
probing strong gravity. By understanding the behavior of matter
near the ISCO, we can determine the mass and angular momentum, and
thus completely describe the black hole. 

While all of the current observations of GR in the weak field regime
are consistent with Einstein's theory [see \citet{will01} for a
review], it is still conceivable that GR may break down in the
strong gravity limit. Alternatively, the black hole spacetime may not
be strictly described by the Kerr metric, but possibly a perturbed
version of the stationary ``no-hair'' case \citep{colli04}.
If this is the case, the proof of such a
deviation would require precision measurements of the behavior of
space time around black holes \citep{dedeo04,psalt04a}. We are actually nearing
that level of
precision with the present state of X-ray observations, but are still
severely lacking in our theoretical understanding of the accretion and
radiation physics that produce the X-ray emission. Until the theory is
advanced significantly further, these phenomenal observations will not
be able to either prove or disprove general relativity.

\section{Historical Background}
Much of the early history presented in this section is based
on the excellent narrative in the book \textit{Black Holes and Time
Warps} by Kip Thorne (1994), a thoroughly satisfying
read for both astrophysicists and non-scientists alike.

\subsection{Theory}\label{history_theory}

Unlike many of the other fundamental physics developments of the
20$^{\rm th}$ century (e.g.\ quantum theory; standard model of
particle physics), black holes were conceived of in theory long before
any observational evidence pointed us in their direction. The original
idea of a black hole was in fact the natural consequence for 
Newtonian gravity and a finite speed of light. First proposed by John
Michell in 1783, a ``dark star'' was one for which the escape velocity
from the surface was equal to or greater than the speed of light. In
fact, for a given mass, this classical critical surface is identical
to the event horizon for a Schwarzschild black hole. With the advent
of the wave theory of light, Michell's thought experiment based on
light corpuscles was not quite so compelling. 

Only with the formulation of Einstein's general theory of relativity
as a geometric theory (in which even massless light waves would be
affected by gravity's pull) did the question of dark stars resurface. Mere
months after Einstein's original publication on GR, Karl Schwarzschild
successfully derived the complete spacetime metric for the inside and
outside of a spherically symmetric star, including the prediction of
an event horizon from which even light could not escape
\citep{schwa16a,schwa16b}. Despite the elegance of the Schwarzschild
solution, most physicists, including Einstein, resisted the idea of a
black hole for many years, based largely on an aesthetic distaste for
spacetime singularities where all known laws of physics would
fail. Not until the 1960s was the theoretical astrophysics community
more or less in consensus that dark stars were in fact physically
possible and perhaps even ``compulsory.'' Then in 1967, John Wheeler
first coined the term \textit{black hole}
\citep{wheel68}, and they passed from the realm of curious oddity to
scientific reality.

Thorne refers to the decade extending roughly from the mid-sixties to
the mid-seventies as the
\textit{Golden Age} of black hole research, when black holes evolved
from a scientific reality to an entire field of intense theoretical
and observational research. Some of the most important results to come
out of this golden age were the ``no hair'' conjecture \citep{ginzb64,
israe67, price72}, the Kerr metric for spinning black holes
\citep{kerr63, carte66, boyer67}, the Penrose process \citep{penro69},
black hole thermodynamics \citep{hawki73, beken73, hawki74, hawki75,
beken75}, and black hole perturbation theory \citep{teuko72}. During the same
period, the seminal papers were written on accretion theory for
compact objects \citep{lynde69, pring72, shaku73, novik73,
page74}. And finally, it was during this period that the GR Bible,
often referred to simply as MTW, was ``canonized'' \citep{mtw73}.

For the purposes of this Thesis, we are interested primarily in the
astrophysically observable characteristics of black holes, which
generally are the product of the accretion of hot gas and its emission
of radiation. Many authors have approached the problem of accretion in
compact binaries with a variety of different methods, both analytic
and computational. One of the earliest is that of \citet{bondi44}, who
consider spherically symmetric accretion. Surely the most popular
theoretical paper on the physics of accretion disks
is \citet{shaku73}, which derives the basic structure and
observational appearance of a steady-state thin accretion
disk. \citet{novik73} promptly extended this model to include full
relativistic effects in the Kerr metric, which we will explore in
greater detail in Chapter 5. \citet{shapi76} brought these results
closer in line with observations of hard photon spectra from Cygnus
X-1 by including a hot corona around
the disk, which is now a widely accepted feature of the accretion
geometry \citep{paczy78,haard93}. An early theory for the coupling of
magnetic fields between the disk and the black hole was proposed by
\citet{bland77} and is still one of the leading explanations for the
formation of relativistic jets.
The application of accretion disk theory to neutron stars was spear-headed
by \citet{ghosh78} with the addition of magnetic field effects in the
inner disk. 

In addition to the classical slim disk geometry, a number of thick
disk and pseudo-spherical solutions have also been developed, most
notably the advection-dominated accretion flows [ADAF; see
\citet{naray94}], convection-dominated accretion flows [CDAF; see
\citet{naray00}], the advection-dominated inflow-outflow solution
[ADIOS; see \citet{bland99}], and the disk + corona geometry mentioned
above.

As the X-ray observations grew steadily more sensitive throughout the
1980s and 1990s (see Section \ref{history_observ} below), the focus of
theoretical work shifted towards better understanding the data, much
like the trend of most other branches of physics in the 20$^{\rm th}$
century. 
Some approaches have simplified the hydrodynamics in favor of a flat, thin,
steady-state disk and a more detailed treatment of general relativistic
effects \citep{georg91,laor91,karas92,reyno97,dovci04a}. This approach focuses
on calculating the \textit{transfer function} of radiation from a flat
disk to a distant observer, described below in Section
\ref{transfer_function}, and most often applied to observations of
broad iron emission lines and studies of photon polarization. 
More detailed spectra from finite thickness $\alpha$-disk models have
been simulated 
in a series of papers by Hubeny et al., who include non-LTE (local
thermodynamic equilibrium) radiation
transport and the detailed vertical structure of the disk
\citep{huben97,huben98,huben00,huben01,davis04}. Emission and absorption lines
of many other atomic species have been studied in great detail by
\citet{garat01} and applied successfully to high-resolution
\textit{XMM-Newton} observations \citep{garat02}.

To include dynamic effects, essential for modeling QPOs, others
have included magnetohydrodynamics [MHD; \citet{hawle01,armit01}] in a
pseudo-Newtonian potential \citep{paczy80}. These 
MHD calculations are in turn based on detailed local simulations of
the magneto-rotational instability
\citep{balbu91,hawle95,hawle96,stone96}. Another relatively simple
approach is through the use of smoothed particle hydrodynamics
\citep{lanza98,lee02}, but this is a method that has yet to gain full
acceptance in the astrophysical community, largely due to the
difficulty in achieving convergent solutions with different codes. At
the same time, it has the distinct advantage of being able to handle
arbitrary accretion geometries in three dimensions with relative
ease. 

More recently, with the ever-increasing power of parallel computing,
fully relativistic global MHD codes have been developed and tested
\citep{gammi03, devil03a, devil03b}, yet as of this writing, they do
not include the effects of radiation transport or emission, certainly
an important ingredient in accretion disk physics. Furthermore, no
MHD calculation to date has yet predicted the existence of QPOs at any
particular frequency, much less those that have been observed. Again
we face the disconnect between theory and experiment- even in the
most sophisticated simulations, there remains no mechanism for
producing spectra and light curves with real physical units with which
to compare observations. We hope that the work presented in this
Thesis will help bridge that gap. 

\subsection{Observations}\label{history_observ}
As mentioned above in the previous section, theoretical predictions of
the existence of black holes far preceded any observational evidence
of them. As late as the 1960s, when most of the theoretical community
was convinced at least of the possibility of their existence, there
was still no known observation of such an object. Led by the theories
of Zel'dovich and Novikov, a search began for bright X-ray sources
produced by accreting black holes. The first major success in that
search was the
discovery of the accreting neutron star Sco X-1 by Riccardo Giacconi
in 1962 \citep{giacc62} and later Cyg X-1 \citep{bowye65}. The
identification of Cyg X-1 as a black hole candidate was not
proposed until after the launch of \textit{Uhuru}, which also
discovered periodic pulsations in the X-ray light curve
\citep{giacc71, webst72, bolto72}. 

Despite the excitement from the steady early successes of X-ray
astronomy, in reality it was the radio astronomers who really
discovered the first black holes, in the form of quasi-stellar
objects, or \textit{quasars}, a few years earlier. It was the
identification of these compact radio sources with optical point
sources (from which a redshift could be measured) that led to the
realization that they were extremely luminous objects, orders of
magnitude brighter than any known star \citep{schmi63, green63,
matth63}. The intrinsic luminosity of the quasar 3C273 is roughly
equal to the Eddington limit (see Section \ref{vertical_structure}
below) for a mass of $6\times 10^7$ $M_\odot$,
and thus the first evidence for supermassive black holes was
found. The association of such objects with extended radio jets led to
the likely possibility that these supermassive black holes were also
rapidly spinning. 

The birth of X-ray timing astronomy came with a series of satellites
launched during the 1970s that discovered a large number of
X-ray pulsars, attributed to magnetized, spinning, accreting neutron
stars [for a review, see \citet{rappa83}]. \textit{Einstein}, the
first X-ray telescope with real imaging capabilities, was launched in
1978 and discovered the remarkable
precessing jets of SS-433 and measured its spectral properties
\citep{seaqu82,watso83}. The subsequent missions of \textit{EXOSAT},
\textit{Tenma}, 
\textit{Ginga}, \textit{ROSAT}, \textit{BeppoSAX}, and \textit{ASCA}
steadily improved in imaging and spectral resolution, sensitivity, and
energy range. 

\begin{figure}
\begin{center}
\scalebox{1.3}{\includegraphics*[120,250][500,630]{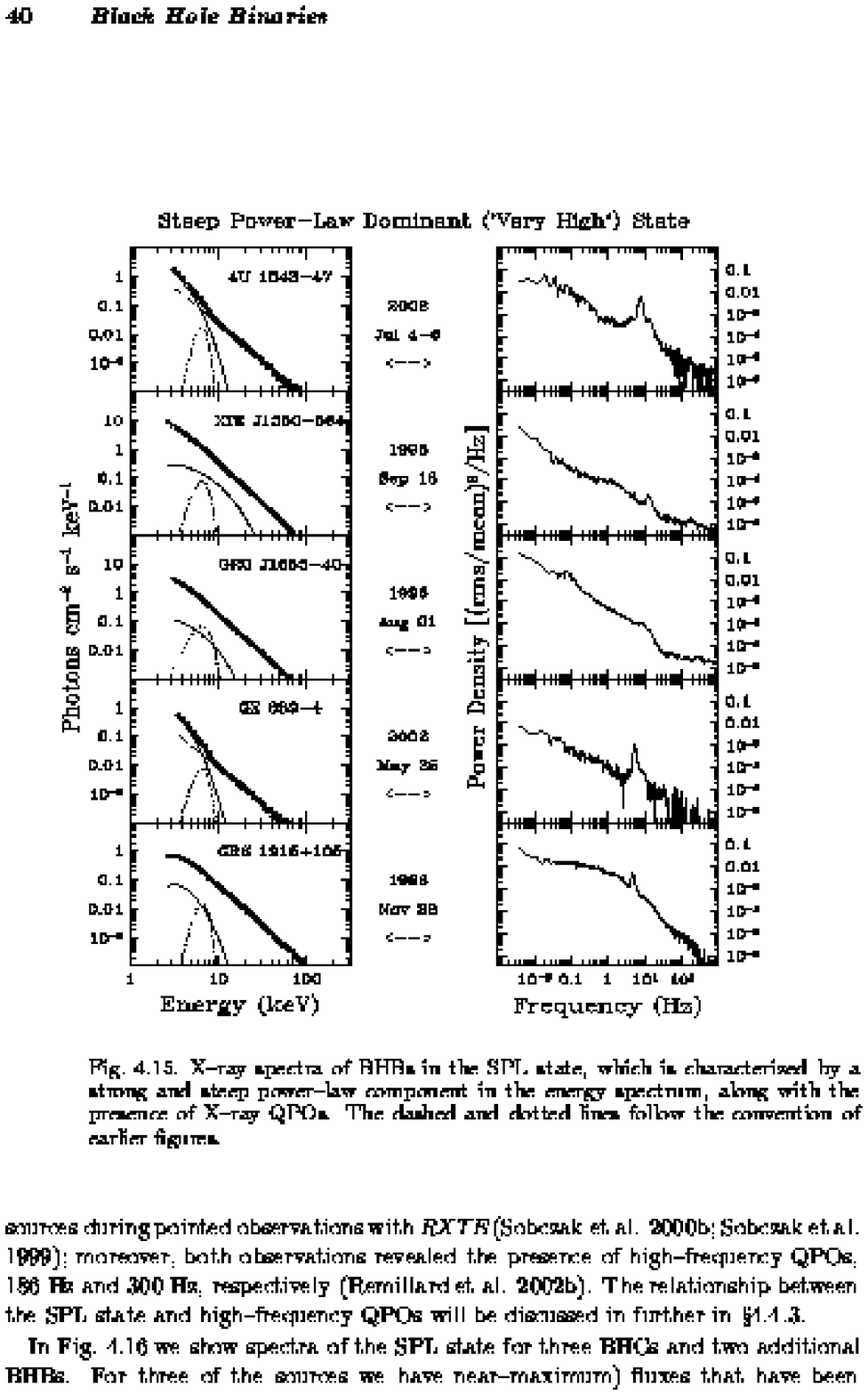}}
\caption[Sample of observations from black holes with
QPOs]{\label{mcclintock_fig} On the left is shown a collection of 
\textit{RXTE} photon energy
spectra from black hole binaries in the ``Very High'' state,
characterized by a dominant steep power-law (dashed curve), a weaker
thermal component (solid), and a faint broad iron emission line
(dotted). On the right, \textit{RXTE} power spectra from X-ray timing
observations of the same 
sources show the presence of QPOs from $\sim 5-500$
Hz. [Reproduced from \citet{mccli04} with permission]}
\end{center}
\end{figure}

The current generation of X-ray observatories is
comprised of the three major satellites mentioned above: \textit{RXTE,
Chandra}, and \textit{XMM-Newton}, as well as the recently launched
\textit{INTEGRAL} for high-energy observations, and the upcoming
launch of \textit{Astro-E2}, which will provide even higher resolution
spectroscopy. An excellent review of the observations of
black hole binaries is given by \citet{mccli04}, and is the motivation
for much of the work in this thesis. A few of the highlights of the
last twenty years are outlined below:
\begin{itemize}
\item Application of the multicolor disk (MCD) model to explain the
thermal spectra observed from accreting low-mass X-ray binaries
(LMXBs) \citep{mitsu84}. Determined disk temperatures in the range
$1-2$ keV. 

\item Broad iron emission lines were detected in a large variety of
sources: Cyg X-1 with
\textit{EXOSTAT} \citep{barr85}, V404 Cyg with \textit{Ginga}
\citep{zycki99a, zycki99b}, the Seyfert 1 galaxy MCG--6-15-30 with \textit{ASCA}
\citep{tanak95}, V4641 Sgr with \textit{BeppoSAX} \citep{mille02b},
and XTE J1650-500 with \textit{XMM-Newton} \citep{mille02}.

\item \citet{esin97} performed a comprehensive classification of
multiple spectral states as a function of the total mass accretion
rate (scaled to Eddington units).

\item The discovery of ultraluminous X-ray sources (ULXs) that exceed
the Eddington luminosity for a typical $10 M_\odot$ black hole
\citep{fabbi89, makis00, fabbi01, humph03}. One popular explanation for these
ULXs is that they are accreting, intermediate-mass black holes with
$100 \lesssim M/M_\odot \lesssim 1000$ \citep{mille04}. Alternatively,
they may be ``normal'' black holes that are exhibiting super-Eddington
luminosities due either to relativistic beaming or very high
mass-transfer rates for short periods when the system is not in
hydrostatic equilibrium \citep{rappa04}.

\item By analyzing the orbits of stars in the galactic center,
\citet{ghez03} and \citet{schod03} have determined the radio source
Sgr A$^\ast$ to be a supermassive black hole with mass in the range
$(3-4)\times 10^6 M_\odot$. 

\item Rapid X-ray flares from Sgr A$^\ast$ with time scales less than
an hour suggest the existence of hot gas near the inner-most stable
circular orbit of the galactic center black hole
\citep{bagan01}. There is even some evidence that the X-ray
light curve of this source has QPO variability in the power spectrum
\citep{asche04a}. 

\item Shortly after the launch of \textit{RXTE} at the end of 1995,
\citet{stroh96} discovered a pair of high frequency QPOs (HFQPOs) in neutron
star binary system. \citet{remil96} found a broader, weaker HFQPO in
the black hole binary GRO J1655--40, soon followed by similar features
in the power spectra of 4U 1543--47, XTE J1858+226, and XTE J1550--564
\citep{remil98, markw99, remil99, homan01}.

\item \citet{mccli04} compiled a representative sample of photon
energy spectra and light curve power spectra for many of these black
hole systems, reproduced here in Figure \ref{mcclintock_fig}. The QPOs
typically are seen in the ``Very High'' or ``Steep Power Law''
spectral state, described in more detail later in this Thesis.

\item \citet{mille01} first identified the 3:2 ratio of HFQPO pairs in
XTE J1550--564, followed by identical ratios in GRO J1655--40, GRS
1915+105, and H1743--322 \citep{remil02,homan04,remil04}. Recently,
similar ratios have been reported in intermediate-mass and even
super-massive black holes
\citep{abram04,fiori04,asche04a,asche04b,torok04}, but
these results are still quite preliminary and widely open to other
interpretations.

\item \citet{mille05} have recently found evidence linking the phase
of a low frequency QPO with the shape of the iron emission line in GRS
1915+105, giving a promising connection between the two leading
methods of probing strong field gravity.
\end{itemize}

\section{Outline of Methods and Results}
\subsection{Ray-tracing in the Kerr Metric}
The central framework on which the results in this Thesis are based is
a relativistic ray-tracing code that calculates the trajectories of
photons in the Kerr spacetime of a spinning black hole. By tracing
the photons from either the emitter to the observer or backwards in
time from the observer to the emitter, we can reconstruct
time-dependent images and spectra of the accretion region. When
starting at the observer, the image plane is divided into pixels of
equal solid angle, each corresponding to a single ray pointing in a
slightly different direction, not unlike a classical
telescope. Following the sample rays backward in time,
we tabulate the spacetime position and momentum at multiple points on
the trajectory, which are then used in conjunction with an emission
model to solve the radiative transfer equation along the photon
path. The gravitational lensing and magnification by the black hole 
is performed automatically by the geodesic integration of these evenly
spaced photon 
trajectories, so that high magnification occurs in regions where
nearby points in the disk are projected to points with large
separation in the image plane.

In Chapter 2 we describe in detail the Hamiltonian methods used to
numerically integrate the photon trajectories. The code primarily uses
Boyer-Lindquist coordinates, but we also include a discussion of the
Doran coordinate system and their relative benefits and
drawbacks. Similarly, while we employ an adaptive-step Runge-Kutta
integrator, Section \ref{analytic_methods} describes some of the
alternative analytic methods for calculating trajectories with
separable equations of motion. 

To calculate physical processes such as emission, absorption, and
scattering along the photon path, it is convenient to define locally
orthonormal reference frames called ``tetrads'' at each point in
coordinate space. Then the transformation from any tetrad basis to
another defined at the same point can be carried out by a special
relativistic Lorentz transformation, as described in Section
\ref{tetrads}. The Zero Angular Momentum Observer (ZAMO) tetrad is
particularly useful in the Kerr metric, since unlike the
Boyer-Lindquist coordinate basis, the ZAMO time coordinate is
in fact time-like even inside the static limit (ergosphere). Based on
the special relativistic discussion in \citet{rybic79}, Section
\ref{radiative_transfer} describes how we solve the classical
radiative transfer equation numerically in general relativity by using
the tetrad
formalism. By combining these results with an independent hydrodynamic
calculation of the accretion disk, we can in principle use the
ray-tracing code as 
a post-processor analysis tool that allows direct comparison of
simulations and observations.

Section \ref{num_methods} presents a description of the numerical techniques
used to integrate the Hamiltonian equations of motion, both for
photons and also massive test particles. Most of the results
presented in this thesis are based on relatively benign computations,
easily carried out in a few minutes on a personal computer. For the
more time-intensive calculations, a parallel version of the code was
developed to run on the Astrophysics Beowulf Cluster at MIT. Since
the light rays can be treated as non-interacting, the problem can be split
up trivially into multiple processes and thus scales extremely well
with the number of processors.

Chapter 2 concludes with the application of the ray-tracing code to a
flat disk made up of test particles on circular orbits in the plane
normal to the black hole spin axis. Assuming each particle is a
monochromatic, isotropic emitter in its rest frame, we calculate the
``transfer function'' from the disk to the observer. This function is
a measure of how the disk emission is relativistically redshifted, beamed,
and lensed, and is a classic means of simulating the shape of
broadened iron emission lines seen by many X-ray observations. We
show how the transfer function is sensitive to the disk inclination,
but not the black hole spin. Only when truncating the disk at the
ISCO \textit{and} scaling the emission by a power law (e.g.\ $g(r) \sim
r^{-\alpha}$), thus giving more weight to the inner regions, are the
line profiles noticeably different for different spin values. However,
since both the ISCO-truncation and $r^{-\alpha}$ scaling are only
conjectures as of
this writing, the broadened emission lines do not seem to be an
unambiguous means for measuring black hole spin. Hence we turn our
attention in the direction of timing observations and QPOs.

\subsection{The Hot Spot Model}
Recent observations of commensurate integer ratios in the
high-frequency QPOs of black hole accretion disks \citep{mille01,remil02}, as
well as the longstanding puzzles of the frequency variability of
low-frequency QPO peaks and their correlations with X-ray flux and
energy, motivate more detailed study of the QPO phenomenon as a means
to determining the black hole parameters [for reviews, see
\citet{lamb02} and \citet{psalt04}]. We
have developed a model that is a combination of many of the above
approaches (see Section \ref{history_theory}), in which additional
physics ingredients can be added
incrementally to a framework grounded in general relativity. The model
does not currently include radiation pressure, magnetic
fields, or hydrodynamic forces, instead treating the emission region
as a collection of cold test particles radiating isotropically
in their respective rest frames. The dynamic model uses the
geodesic trajectory of a massive particle as a guiding center for a
small region of excess emission, a ``hot spot,'' that creates a 
time-varying X-ray signal, in addition to the steady-state background
flux from the disk.

An early prototype of the hot spot model was originally proposed by
\citet{sunya72} as a means for identifying the black hole horizon (as
opposed to a NS surface) as the emitter spirals in
towards the horizon and then fades away to infinity. \citet{bao92}
calculated light curves and power spectra for a collection of random
hot spots in an AGN disk to model the variability seen on time scales
of hours or days. Our version of
the hot spot model described in Chapter 3 is motivated by the
similarity between the QPO frequencies and the black hole (or neutron
star) coordinate frequencies near the ISCO \citep{stell98,stell99a} as
well as the suggestion of a resonance leading to 3:2 integer
commensurabilities between these coordinate frequencies
\citep{abram01,abram03b,kluzn01,rebus04,horak04}. \citet{stell99a}
investigated primarily the
QPO frequency pairs found in low-mass X-ray binaries (LMXBs) with a
neutron star (NS) accretor, but their basic methods can be applied to
black hole systems as well. Both NS and BH
binaries also show strong low frequency QPOs (LFQPOs; $\nu \approx 5-10
\mbox{ Hz}$) at frequencies that vary between observations. One 
critical difference between these systems is the variability of the
HFQPOs $(\nu > 50 \rm{Hz})$ in NS systems as opposed to the generally
constant frequencies of the black hole HFQPOs \citep{mccli04}. If anything, it
seems more appropriate to apply the
geodesic hot spot model to the black hole systems since they lack the
complications of magnetic fields and X-ray emission from the rotating
neutron star surface, which confuse the interpretation of coordinate
frequencies. In fact, the QPOs from the two different types of compact
binaries may be caused by two completely different physical
mechanisms. 

\citet{marko00} have presented a thorough
analysis of this hot spot model for a collection of NS binaries for
which pairs of QPOs have been observed. Based on a number of
observational and theoretical arguments, they conclude that the
geodesic hot spot model is not a physically viable explanation for the
observed neutron star
QPOs. For low to moderate eccentricity orbits, the coordinate
frequencies simply do not agree with the QPO data. For highly
eccentric geodesics, they argue that the relative power in the
different frequency modes are qualitatively at odds with the
observations. Furthermore, they show that hydrodynamic
considerations place strong constraints on the possible size, luminosity,
coherency, and trajectories of the hot spots. 

Many of these points are addressed in our version of the hot spot
model. Also, by including full three-dimensional (3D) relativistic
ray-tracing, we can
quantitatively predict how much QPO power will be produced by a hot
spot of a given size and emissivity moving along a geodesic orbit near
the ISCO. In Section \ref{overbrightness_amp} we compute the
hot spot overbrightness necessary to produce a given amplitude
modulation in the light curve, as a function of disk inclination and
black hole spin. Along with the special
relativistic beaming of the emitted radiation, we find that strong
gravitational lensing can cause high-amplitude modulations in the
light curves, even for relatively small hot spots. To match the
observed 3:2 frequency ratios, we used closed rosette orbits with
$\nu_\phi=3\nu_r$ to give Fourier power at the beat modes $\nu_\phi \pm
\nu_r$, and in Section \ref{noncircular_orbits} show the dependence of
this power on the orbital eccentricity. The issues of
differential rotation and shearing of the emission region as well as
the possible connection to LFQPOs are
addressed in Section \ref{nonplanar_orbits} when we consider the
generalization of the hot spot model to include arcs and non-planar
geometries.

Perhaps the most powerful feature of the hot spot model is the
facility with which it can be developed and extended to more general
accretion disk geometries. 
In addition to providing a possible explanation for the
commensurate HFQPOs in at least three systems (XTE J1550--564, GRO
J1655--40, and H1743--322), the hot spot model with full general relativistic 
ray-tracing is a useful building block toward any other viable model of
a dynamic 3D accretion disk. Within the computational
framework of the Kerr metric, we can
investigate many different emission models and compare their
predicted X-ray spectra and light curves with observations.

Chapter 4 introduces the first simple extension of the hot spot model,
moving from a single periodic hot spot orbiting the black hole
indefinitely at a single radius, to a collection of hot spots each
with a finite lifetime, distributed over a range of radii and random
phases. The corresponding power spectrum changes from the set of
delta-functions described in Chapter 3 to a set of broad peaks
characteristic of a \textit{quasi}-periodic oscillation. We present a
number of analytic models for the amplitudes and widths of the
different peaks and confirm these results with direct ray-tracing
calculations of multiple hot spots. In particular, we show how the
addition of many hot spots with random phases will broaden every peak
in the power spectrum by the same amount, while a distribution of
geodesic orbits with a range of coordinate frequencies will broaden
different peaks by different
amounts. This ``differential peak broadening'' turns out to be a
promising method for probing the spacetime structure near the ISCO,
testing the assumptions of the hot spot model, and ultimately 
measuring the black hole spin.

\begin{figure}[tp]
\begin{center}
\includegraphics[width=0.8\textwidth]{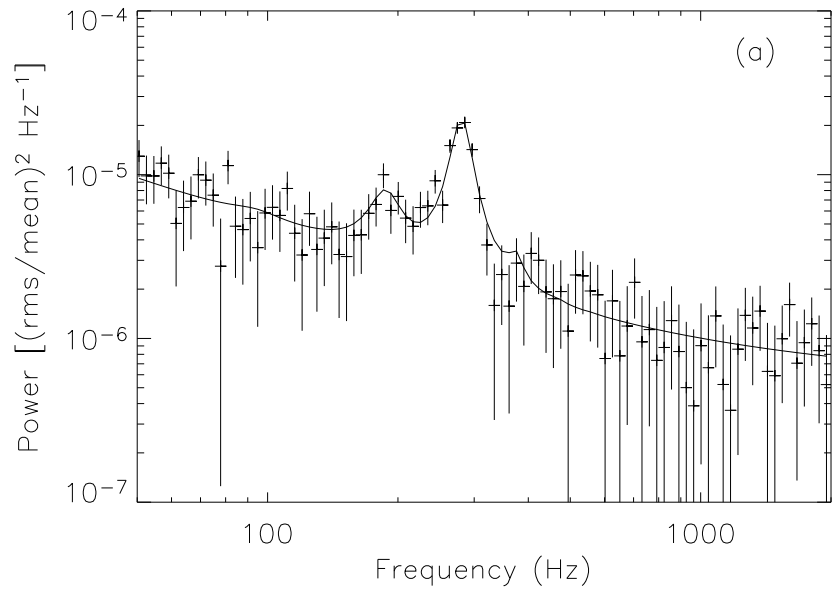}
\caption[Power spectrum from hot spot model and data from
\textit{RXTE}]{\label{intro_datafit} Comparison of hot spot model power 
spectrum (line) with data (crosses) from XTE J1550--564 [taken from
\citet{remil02}]. Details of 
the model are given in Chapters 3 and 4.}
\end{center}
\end{figure}

In Section \ref{scatter} we present a simplified model for electron
scattering in a uniform density corona around the black hole. By
giving a random added path length (and thus time delay) to each
ray, the light curve gets smoothed out in time as each photon is
assigned to a different time bin. This process does not contribute to
broadening the QPO peaks, but can significantly damp out the higher
harmonic modes in the power spectrum, much like the effect of shearing
the hot spot into an arc. Finally, all the
pieces of the model are brought together in Section \ref{data} and used to
interpret the power spectra from a number of observations of
XTE J1550--564. Figure \ref{intro_datafit} shows the remarkable
success this simple hot spot model has in fitting the \textit{RXTE}
data with only a few free parameters.

Based on a recent paper by \citet{macca04}, Section \ref{bispectrum}
introduces the use of
higher-order statistics as an observational tool for distinguishing
between the various peak broadening mechanisms. We apply these
statistical methods to two different hot spot light curves that give
similar power spectra: one broadened by random phases and one by a
finite range of orbital frequencies. Not only can the bispectrum
distinguish between these two models, but like the differential peak
broadening method, it also can be used to map out the spacetime
around the black hole.

\subsection{Steady-state Disks}
To gain more insight into the structure and continuum spectrum of the
steady-state accretion flow, in Chapter 5 we develop a relativistic
$\alpha$-disk model, based largely on the work of 
\citet{shaku73} and \citet{novik73}. Beginning with the Novikov-Thorne
equations for radial structure, we derive a set of boundary
conditions for the vertical structure equations at each radius in the
disk. These vertical structure equations for the density, temperature,
pressure, and energy flux closely resemble the classical stellar
structure equations for hydrostatic equilibrium \citep{hanse94}. The
only difference is that here, the gravitational force is caused by the
relativistic tidal force in the plane of the disk, and the thermal energy is
generated not by nuclear fusion, but by turbulent viscosity. We also
present an analytic ``two-zone'' model that gives the temperature and
density at the mid-plane as well as the disk's surface. With this
analytic approach, we are able to derive a modified expression for the
Eddington luminosity, giving an estimated upper limit to the
accretion rate for the thin disk geometry.

To self-consistently model the torque on the inner edge of the disk,
in Section \ref{geodesic_plunge} we show how the accreting gas expands
along plunging geodesic trajectories inside of the ISCO. By matching
the radial scale length of the plunge $l_{\rm plunge}$ to the turbulent scale
length of the disk $l_{\rm turb} \approx h(R_{\rm ISCO})$, we can
solve for the integrated stress at the ISCO, which in turn gives the
initial conditions for the infalling radial velocity. Following
a column of gas in the frame of the plunging particle,
we can model the time-dependent vertical structure of the innermost
disk with one-dimensional Lagrangian hydrodynamics. We find that the
plunging disk temperature and density fall off rapidly inside the
ISCO, contributing little to the thermal emission of the
disk. However, we also show that even a small torque at the inner edge
can significantly change the flux and surface temperature outside the
ISCO, as well as the total accretion efficiency of the disk.

We outline the numerical methods
used to solve for the disk structure in Section \ref{implicit_scheme},
both for inside and outside of the
ISCO. The Lagrangian hydrodynamics is based on an implicit scheme
described in \citet{bower91}. This implicit scheme is especially
useful since the sound speed of the radiation pressure-dominated gas
would otherwise demand an extremely small Courant step for an explicit
approach. Given the temperature and scale
height of the disk atmosphere, in Section \ref{ntdisk_spectra} we
use the relativistic ray-tracing code described above to calculate
a ``multi-colored'' spectrum of the disk. This modified thermal
spectrum is characterized by a slow rise with $I_\nu \sim \nu$ at low
energies, a broad peak around $0.5-2$ keV, followed by a steep cutoff
around 10 keV. The location of the thermal peak is a function of the
black hole mass and accretion rate (and thus might ultimately be used
to identify intermediate-mass black holes), while the cutoff frequency
appears to be sensitive to the black hole spin and the inclination of
the disk.

\subsection{Electron Scattering}
While the simple scattering model introduced in Section \ref{scatter}
ignored a number of important physical details, in Chapter 6 we
revisit the topic of electron scattering with a Monte Carlo code
including angular dependence, multiple scatterings, and relativistic
effects. We begin with a derivation of electron scattering in
the low-energy photon regime ($h\nu \ll m_ec^2$), following the
treatment of \citet{rybic79}. As with the treatment of the radiative
transfer equation in Section \ref{radiative_transfer}, the classical
results can easily be applied to a general relativistic model by
transforming to a tetrad basis at the point of scattering. The major
difference is that in Chapter 6 we trace the photons \textit{forward}
in time from the emitter to the observer. While conceptually simpler,
this approach is computationally more intensive, but even that minor
drawback can be somewhat mitigated, as will be explained in the main
text.

The coronal electron density and temperature profiles can be
approximated by a self-similar ADAF distribution \citep{naray94},
which gives $\rho_e \sim r^{-3/2}$ and $T_e \sim r^{-1}$. Thus most of
the scattering events occur close to the hot spot emitter in the inner
region of the disk/corona, where the scattering electrons are
hottest. We assume an isotropic Maxwellian velocity
distribution in the ZAMO frame, giving a random 4-velocity to each
electron. Boosting from the ZAMO frame to the rest frame of the
electron, we can use the Thomson differential cross section to
determine the new photon momentum, which is then transformed back to
the coordinate frame and then continues to propagate along its
new geodesic path. 

The electron scattering has two major observable effects: the thermal
photon spectrum is modified by the inverse-Compton process [which can
be approximated by solving the Kompaneets equation
\citep{kompa57}], giving a significant high energy power-law tail, and
the integrated light curve is smoothed out in time, effectively
damping the higher harmonic modes in the power spectrum. We also see
some evidence of phase lags between different \textit{RXTE} energy
bands, since photons that experience more scattering events tend to
have higher energy and a greater time delay to the observer
\citep{galee79}. Finally,
in Section \ref{implications_QPO}, we review the results of these
scattering calculations in view of the observations and discuss the
implications for various QPO models. Our tentative conclusion is that
the HFQPO power is primarily produced by very \textit{hot} hot spots
with $T_{\rm hs} \gtrsim 5$ keV (as opposed to the relatively cool
thermal disk at $T_{\rm disk} \sim 1$ keV), surrounded by a hot
corona with moderate optical depth $\tau_{\rm es} \sim 1$ and temperature $T_e
\sim 100$ keV. As in Chapter 3, the higher harmonics are more likely damped
by ``arc-shearing,'' and not repeated scatterings, which tend to damp
out power in all harmonic modes of the light curve. Another
possibility, which we have not yet thoroughly explored, is that the
seed photons are not from a hot spot, but a more global, isotropic
emitter like an oscillating torus. This would allow the electron
corona to scatter the radiation field isotropically yet still maintain
significant amplitude modulations in time.
 
\section{Alternative QPO Models}\label{alt_models}

In this Thesis, we have focused primarily on the geodesic hot spot
model for QPOs for 
a few reasons: (1) It is conceptually simple and not very
computationally intensive to simulate spectra and light curves; (2) It
is quite successful in fitting the data from
\textit{RXTE}, matching the QPO frequencies, peak widths, and
amplitudes (or lack thereof); (3) It is easily expanded and can be
used as a building block to construct more complex disk
models. However, it also has its shortcomings: First and foremost,
there is no clear physical explanation for how or why the hot spots should
form around one special radius. Also, as we will see in Chapter 6, it is
not clear why the QPOs should be more significant in the higher energy
bands. Related to both of these questions is a marked lack of
understanding of what the geometry of the accretion disk/corona is
like in the Steep Power Law state, where most QPOs are seen. 

Thus one of the major goals of this research is to develop a
generalized analysis tool that can be applied to \textit{any} disk
model and compare simulations directly to observations. Here we
give a brief summary of some of the more popular dynamic disk models
in the literature today, along with a few representative references.
In addition to the global GR-MHD simulations mentioned above
\citep{gammi03, devil03a, devil03b}, there exist at least five basic 
concepts for producing the high frequency quasi-periodic oscillations
seen from accreting black holes.  In historical order (to the best of
our limited knowledge), these are (1) magnetic flares \citep{galee79,
haard94, stern95b, dimat98, belob99, pouta99}, (2)
diskoseismology \citep{okaza87,nowak97,kato98,wagon99,kato01,wagon01},
(3) resonances at geodesic
frequencies \citep{stell99a,stell99b,kluzn01,abram03a,rebus04,horak04},
(4) oscillating axisymmetric tori
\citep{lee02,rezzo03a,rezzo03b,zanot03,lee04}, and (5) Rayleigh-Taylor
instabilities \citep{titar02,titar03,li04}. The hydrodynamic studies
of \citet{psalt00} and \citet{psalt01a} combine the global MHD techniques
with a semi-analytic treatment of the resonance frequencies of
approach (3). 

The magnetic flare model was introduced in the earliest days of X-ray
timing astronomy to explain the high frequency variability of Cyg
X-1 \citep{galee79}. Similar to the magnetic flares in the solar
atmosphere, they can 
produce short-lived, tightly confined regions of overbrightness on or
above the accretion disk surface. Magnetic flares are particularly
promising for explaining the formation and subsequent destruction of
hot spots, as well as the phase lags between soft and hard X-rays
\citep{pouta99}, but at this point cannot explain the frequency
locations of the QPO peaks or their integer commensurability.

The next model proposed was diskoseismology --- the excitation of
various trapped modes in an accretion disk [see \citet{kato98,wagon99}
for reviews]. This model seems not to be directly
applicable to the data, at least for the cases where small integer
ratios of frequencies exist; it would require considerable fine tuning
in the different mass and spin values for the black holes to produce
routinely a 3:2 frequency ratio. Global perturbations may also take
the form of spiral density waves in the disk, much like the arms of
the milky way \citep{gottl02}. However, like the classical
diskoseismic modes, it may require fine tuning to produce the
appropriate frequencies in the observed light curve.

Related to diskoseismology is the
oscillating torus model of \citet{rezzo03a}. This model also
computes the frequencies of $p$-modes (i.e.\ sound waves), but in a
geometrically thick, pressure supported torus (as expected at high
accretion rates like those where the HFQPOs are seen), rather than in
a geometrically thin, Keplerian
accretion disk. In this case, the different overtones are found to be
approximately in a series of integer ratios, starting from 2, so the
model is compatible with existing data on high frequency QPOs in black
holes. We have recently begun to apply the ray-tracing code to the
torus model with mixed results. Assuming a simplified emission model
and an optically thin torus, we can produce light curves with the
output data from the hydrodynamic simulations. These light curves have
significant modulation in the fundamental mode, but it seems very
difficult to produce appreciable amplitude power in the
overtones. Furthermore, the present version of the torus model is
axisymmetric, but accretion flows have been shown to be susceptible to
a number of hydrodynamic instabilities that would compromise this
symmetry [see, e.g.\ \citet{papal84,papal85,hawle00}].

The model most recently applied to high frequency QPOs from black hole
candidates is that of Rayleigh-Taylor instabilities, although the same
basic idea had been applied earlier to QPOs from accreting neutron
stars earlier \citep{titar02,titar03}.  In this picture,
non-axisymmetric structures can grow
unstably at the magnetospheric radius (presumed to exist also for
black holes, as their accretion disks can become magnetically
dominated) with frequencies of integer ratios of the angular
frequency at that radius, though the lowest mode will be stable for
relatively low gas pressures \citep{li04}. Related to this model is that of 
\citet{wang03}, who propose a magnetic  
coupling between the rotating black hole and the accretion disk as a
means of producing high-frequency QPOs, analogous to the
Blandford-Znajek process \citep{bland77} sometimes used to explain the
behavior of relativistic jets.

After the first indications that small integer ratios between HFQPO
frequencies were likely, it was noted by \citet{abram01}
that if the relativistic coordinate frequencies determined the
frequencies of the quasi-periodic oscillations,
then resonances between these different frequencies
(e.g.\ vertical and radial epicyclic frequencies) might occur at
locations in the accretion disk where these frequencies have small
integer ratios. The excitation of these resonances could very well
produce the regions of overdensity and overbrightness that we treat as
geodesic hot spots. 

In addition to these models for the high frequency QPOs, there are
also a large number and variety of different models to describe the low
frequency QPOs in black hole systems. Here too, global diskoseismic
modes are used to explain the observed oscillations. These include the
``normal disk'' mode, where the entire disk is displaced by a small
perturbation normal to the rotation plane \citep{titar00}, and the
``corrugation'' or ``c-mode'' oscillations where the inner regions of
an inclined disk precess at roughly the Lense-Thirring rate
\citep{silbe01}. A recent model by \citet{tagge04} employs ``magnetic
floods,'' which in turn lead to the accretion-ejection instability
\citep{tagge99}, 
to explain the LFQPOs and also transfer energy from a cool
disk to a hot corona. While this Thesis primarily focuses on HFQPOs,
many of the techniques and models presented herein could easily be
applied to observations and models of LFQPOs as well.

\chapter{Ray-Tracing in the Kerr Metric}
\begin{flushright}
{\it
Do not worry about your difficulties in Mathematics. \\
I can assure you mine are still greater.\\
\medskip

God does not care about our mathematical difficulties. \\
He integrates empirically.\\
\medskip
}
-Albert Einstein
\end{flushright}
\vspace{1cm}

The results presented in the next two chapters are based largely on
the paper ``The Harmonic Structure of High Frequency Quasi-periodic
Oscillations in Accreting Black Holes,'' by Schnittman \& Bertschinger
(2004), ApJ {\bf 606}, 1098.
 
\section{Equations of Motion}\label{Hamiltonian_eom} 
To simulate the appearance of a distant black hole and surrounding
accretion disk, we begin by dividing the image plane into regularly
spaced ``pixels''
of equal solid angle in the observer's frame, each corresponding to a
single ray. Following the sample rays backward in time,
we calculate the original position and direction that a photon emitted
from the disk would require in order to arrive at the appropriate
position in the detector. The gravitational lensing and magnification of 
emission from the plane of the accretion disk is performed automatically
by the geodesic integration of these evenly spaced photon
trajectories, so that high magnification occurs in regions where
nearby points in the disk are projected to points with large
separation in the image plane. To model the time-varying emission from
the disk, each photon path is marked with the time delay along
the path from the observer to the emission point in the disk. Then
coupled with a dynamic model for the accretion disk, we can
reconstruct the observed X-ray ``movies.''

To integrate the geodesic trajectories of photons or massive
particles, we use a Hamiltonian
formalism that takes advantage of certain conserved quantities in the
dynamics \citep{berts99,berts01}. The resulting equations of motion do not
contain any sign
ambiguities from turning points in the orbits, as are introduced
by many classical treatments of the geodesic equations in the Kerr
metric (see below, Section \ref{analytic_methods}). We define a
Hamiltonian function of eight phase space
variables $(x^\mu, p_\mu)$ and an integration variable (affine
parameter) $\lambda$ along the path length. For a general spacetime
metric $g_{\mu \nu}(\mathbf{x})$ with inverse $g^{\mu
  \nu}(\mathbf{x})$, we can define a Hamiltonian $H_2$ quadratic in
the momenta as
\begin{equation}\label{H_2}
H_2(x^\mu,p_\mu;\lambda) = \frac{1}{2}g^{\mu \nu}(\mathbf{x})p_\mu
p_\nu = -\frac{1}{2}m^2,
\end{equation}
where the rest mass $m$ is a constant ($m=0$ for photons, $m=1$ for
massive particles).

Applying Hamilton's equations of motion from classical mechanics, we
reproduce the geodesic equations:
\begin{subequations}
\begin{equation}
\frac{dx^\mu}{d\lambda} = \frac{\partial H_2}{\partial p_\mu} = g^{\mu
\nu}p_\nu = p^\mu,
\end{equation}
\begin{equation}
\frac{dp_\mu}{d\lambda} = -\frac{\partial H_2}{\partial x^\mu} =
-\frac{1}{2}\frac{\partial g^{\alpha \beta}}{\partial x^\mu} p_\alpha
 p_\beta = g^{\gamma \beta}\Gamma^\alpha_{\mu \gamma}p_\alpha p_\beta.
\end{equation}
\end{subequations}

For any metric, the Hamiltonian $H_2$ is independent of the affine
parameter $\lambda$, allowing us to use one of the coordinates as the
integration parameter and reduce the dimensionality of the phase space
by two. We use the coordinate $t=x^0$ as the independent time
coordinate for the
six dimensional phase space $(x^i,p_i)$. The corresponding Hamiltonian
(now homogeneous of degree 1 in the momenta) is 
\begin{equation}\label{H_1_general}
H_1(x^i,p_i;t) \equiv -p_0 = \frac{g^{0i}p_i}{g^{00}} +
\left[\frac{g^{ij}p_ip_j + m^2}{-g^{00}}
+\left(\frac{g^{0i}p_i}{g^{00}}\right)^2 \right]^{1/2}
\end{equation}
with equations of motion
\begin{subequations}
\begin{equation}\label{hameq_1}
\frac{dx^i}{dt} = \frac{\partial H_1}{\partial p_i},
\end{equation}
\begin{equation}\label{hameq_2}
\frac{dp_i}{dt} = -\frac{\partial H_1}{\partial x^i}.
\end{equation}
\end{subequations}
We have thus reduced the phase space to the six-dimensional tangent
bundle $(x^i,p_i)$. Moreover, because the Kerr metric is independent
of $t=x^0$ and $\phi=x^3$, $H_1=-p_0$ and $p_\phi$ are also integrals
of motion. These two
integrals of motion correspond to the \textit{Killing vectors}
$\mathbf{\xi}_{(t)} = \partial_t$ and $\mathbf{\xi}_{(\phi)} =
\partial_\phi$. A Killing vector field $\xi_\mu$ satisfies the
equation 
\begin{equation}
\nabla_\mu \xi_\nu + \nabla_\nu \xi_\mu = 0,
\end{equation}
with $\xi_\mu p^\mu = {\rm const}$ along a geodesic trajectory
\citep{mtw73}. 

Just as in classical mechanics, we can alternatively take the
Lagrangian approach to the equations of motion, working with the
coordinate velocities instead of the momenta [see e.g.\
\citet{shapi83}]. In fact, one possible
Lagrangian takes the same form as the original Hamiltonian in equation
(\ref{H_2}):
\begin{equation}\label{L_2}
L_2(x^\mu,p^\nu;\lambda) = \frac{1}{2}g_{\mu \nu}(\mathbf{x})p^\mu
p^\nu = -\frac{1}{2}m^2.
\end{equation}
The Euler-Lagrange equations of motion are
\begin{equation}\label{euler_lagrange}
\frac{d}{d\lambda}\left(\frac{\partial L_2}{\partial p^\alpha}\right)
-\frac{\partial L_2}{\partial x^\alpha} = 0.
\end{equation}
Plugging equation (\ref{L_2}) into (\ref{euler_lagrange}) gives
\begin{equation}\label{e_l2}
\frac{d}{d\lambda}\left(g_{\alpha \nu}p^\nu\right) - \frac{1}{2}p^\mu p^\nu
g_{\mu \nu, \alpha} = 0.
\end{equation}
Writing
\begin{equation}
\frac{d}{d\lambda}g_{\alpha \nu} = \frac{\partial g_{\alpha
\nu}}{\partial x^\mu} \frac{dx^\mu}{d\lambda},
\end{equation}
we get
\begin{equation}
g_{\alpha \nu} \frac{dp^\nu}{d\lambda} + \left(g_{\alpha \nu, \mu}
-\frac{1}{2}g_{\mu \nu, \alpha} \right) p^\mu p^\nu = 0.
\end{equation}
From symmetry in the indices, we can write
\begin{equation}
g_{\alpha \nu, \mu}p^\mu p^\nu = \frac{1}{2}
(g_{\alpha \nu, \mu}+g_{\alpha \mu,\nu})p^\mu p^\nu,
\end{equation}
and then multiplying through by the inverse metric $g^{\alpha \gamma}$
gives the Lagrangian geodesic equation:
\begin{equation}\label{e_l_geodesic}
\frac{d^2x^\gamma}{d\lambda^2} + \Gamma^\gamma_{\mu\nu} p^\mu p^\nu = 0,
\end{equation}
where $\Gamma^\gamma_{\mu \nu}$ is the Christoffel symbol with its
standard definition.

Despite the relative compactness of equation (\ref{e_l_geodesic}), the
Lagrangian approach comes with the increased algebraic and
computational cost of calculating all the Christoffel symbols (of
which there are at least 20 different non-zero terms for the Kerr metric).
Recently \citet{dovci04} has compiled a complete list of the
Christoffel terms in Kerr ingoing coordinates, as well as their first
derivatives, which are necessary for integrating the geodesic
deviation equation \citep{rauch94}.

\subsection{Boyer-Lindquist Coordinates}\label{BL_coord}

By far the most common implementation of the Kerr solution for a
neutral, spinning black hole spacetime
is the Boyer-Lindquist coordinate system \citep{boyer67}.
In Boyer-Lindquist coordinates $(t,r,\theta,\phi)$, the Kerr metric
may be written 
\begin{equation}\label{BL_metric}
g_{\mu \nu} = \left(\begin{array}{cccc}
-\alpha^2+\omega^2\varpi^2 & 0 & 0 & -\omega\varpi^2 \\
0 & \rho^2/\Delta & 0 & 0 \\
0 & 0 & \rho^2 & 0 \\
-\omega\varpi^2 & 0 & 0 & \varpi^2 \end{array}\right),
\end{equation}
giving a line element 
\begin{equation}\label{BL_lineelement}
ds^2 = -\alpha^2 dt^2 +\varpi^2(d\phi -\omega dt)^2 
+\frac{\rho^2}{\Delta}dr^2 +\rho^2d\theta^2.
\end{equation}
This allows a relatively simple form of the inverse metric
\begin{equation}
g^{\mu \nu} = \left(\begin{array}{cccc}
-1/\alpha^2 & 0 & 0 & -\omega/\alpha^2 \\
0 & \Delta/\rho^2 & 0 & 0 \\
0 & 0 & 1/\rho^2 & 0 \\
-\omega/\alpha^2 & 0 & 0 &
1/\varpi^2-\omega^2/\alpha^2 \end{array}\right).
\end{equation}
For a black hole of mass $M$ and specific angular momentum $a=J/M$,
we have defined (in geometrized units with $G=c=1$)
\begin{subequations}
\begin{eqnarray}
\rho^2 &\equiv& r^2 +a^2 \cos^2\theta \\
\Delta &\equiv& r^2 -2Mr + a^2 \\
\alpha^2 &\equiv& \frac{\rho^2 \Delta}{\rho^2 \Delta+2Mr(a^2+r^2)} \\
\omega &\equiv& \frac{2Mra}{\rho^2\Delta + 2Mr(a^2+r^2)}
\label{BL_omega} \\
\varpi^2 &\equiv& \left[\frac{\rho^2\Delta +
2Mr(a^2+r^2)}{\rho^2}\right] \sin^2\theta.
\end{eqnarray}
\end{subequations}
As a check, we see that equation (\ref{BL_lineelement}) reduces to the
well-known
Schwarzschild metric in the limit $a\to 0$. In the limit $M\to 0$
(holding $a/M$ constant),
it reduces to flat spacetime with hyperbolic-elliptical coordinates.

The horizon can be defined as the surface where
\begin{equation}\label{kerr_horizon}
r_\pm=M\pm\sqrt{M^2-a^2},
\end{equation}
where, unlike the Schwarzschild solution, there are two distinct
horizons, corresponding to the two roots of the equation
$\Delta(r)=0$. But as in the Schwarzschild case, no information or
particles can escape once they cross the outer horizon, so we will
define that as the effective surface of our black hole. From equation
(\ref{kerr_horizon}), it is also evident that the spin $a$ must be
less than or equal to $M$, or else the horizon will not exist and the
result would be a ``naked singularity.'' 

Another interesting feature of the Boyer-Lindquist coordinates is that
the metric component $g_{tt}$ can be greater than zero outside of the
horizon. Solving the equation
\begin{equation}\label{kerr_ergo}
g_{tt}=-\left(1-\frac{2Mr}{r^2+a^2\cos^2\theta}\right)=0
\end{equation}
gives us a formula for the surface of the ``ergosphere'':
\begin{equation}\label{r_ergo}
r_{\rm erg} = M+\sqrt{M^2-a^2\cos^2\theta}.
\end{equation}
Inside of the ergosphere (where $g_{tt} > 0$), there can exist no
coordinate stationary
observers, no matter how hard they fire their rockets. Thus the
ergosphere is sometimes also referred to as the static limit. While it is
always good to be cautiously skeptical about results based on the
choice of coordinates in GR (the most common
example is of course the coordinate singularity at the horizon of the
Schwarzschild metric), this particular feature does have some physical
significance. As shown by \citet{penro69}, particles can exist inside
the ergosphere with negative energy trajectories, get captured by the
horizon, and effectively transfer angular momentum away from the black
hole. 

With the form of the metric given in equation (\ref{BL_metric}), the
Hamiltonian $H_1$ can be written in Boyer-Lindquist coordinates
\begin{equation}\label{H_1_BL}
H_1(r,\theta,\phi,p_r,p_\theta,p_\phi;t) = \omega p_\phi
+\alpha\left(\frac{\Delta}{\rho^2}p_r^2 
+\frac{1}{\rho^2}p_\theta^2 +\frac{1}{\varpi^2} p_\phi^2
+m^2\right)^{1/2}.
\end{equation}
Since $H_1$ is independent of $t$, it can be thought of as the
conserved energy at infinity $E_0=H_1=-p_t$.
This new Hamiltonian is also independent of $\phi$ (azimuthally
symmetric spacetime), giving the conjugate momentum $p_\phi$ as
the second integral 
of motion for $H_1$. We are now left with five coupled equations for
$(r, \theta, \phi, p_r,p_\theta)$. The third integral of
motion, ``Carter's constant'' \citep{carte68a}
\begin{equation}\label{carter}
\mathcal{Q} \equiv p_\theta^2
+\cos^2\theta \left[a^2(m^2-p_0^2)+p_\phi^2/\sin^2\theta\right],
\end{equation}
is used as an independent check of the accuracy of the numerical
integration. In Appendix A, we include all the relevant derivatives
and formulas for solving equations (\ref{hameq_1}, \ref{hameq_2}, and
\ref{H_1_BL}). 

\subsection{Doran Coordinates}\label{doran_coord}

The Boyer-Lindquist coordinate system is relatively compact and easy
to visualize in flat space, but shares with the
Schwarzschild metric the problem of a coordinate singularity at the
horizon. One way around this problem is presented by \citet{doran00},
who defines a new coordinate system in terms of observers freely falling
from rest at infinity. This approach can be seen most clearly in the
spherically symmetric Schwarzschild case, where the metric can be
written 
\begin{equation}
ds^2 = -d\tau^2+\left[dr+\left(\frac{2M}{r}\right)^{1/2}d\tau\right]^2
+r^2(d\theta^2+\sin^2\theta d\phi^2).
\end{equation}
In these coordinates, the time $d\tau$ is the same as the proper time
measured by a free-falling observer, whose trajectory is given simply by
\begin{subequations}
\begin{equation}\label{xdot_doran1}
p^\mu = [1,-\sqrt{2M/r},0,0]
\end{equation}
and
\begin{equation}
p_\mu = [-1,0,0,0].
\end{equation}
\end{subequations}
This solution is well-behaved at and inside the horizon, consistent
with the fact that an observer crossing the horizon should feel
nothing particularly special, which makes these coordinates especially
useful for ``flight simulators'' that image the extreme gravitational
lensing of observers falling into black holes \citep{hamil04a}. In
fact, the trajectory defined by
equation (\ref{xdot_doran1}) is consistent with the Newtonian law of
gravity: 
\begin{equation}
\frac{d^2r}{d\tau^2}=-\frac{M}{r^2}.
\end{equation}

Doran's achievement was extending this approach to the Kerr metric,
following the trajectories of free-falling observers as they get swept
into the swirling spacetime around the black hole. This approach has
recently been explained in detail with the \textit{River Model} of
\citet{hamil04b}. Following \citet{berts01}, it is convenient to
define
\begin{equation}
b \equiv (r^2+a^2)^{1/2}, \hspace{2cm} c \equiv (2Mr)^{1/2}.
\end{equation}
In Doran coordinates, the metric of a spinning black hole has the form
\begin{equation}
ds^2 = -d\tau^2+\left[\frac{\rho}{b}dr +
\frac{c}{\rho}(d\tau-a\sin^2\theta d\bar{\phi})\right]^2
+\rho^2 d\theta^2+b^2\sin^2\theta d\bar{\phi}^2,
\end{equation}
where $r$ and $\theta$ have the same meaning as in the Boyer-Lindquist
metric, $\tau$ is the free-falling observer's proper time, and
$\bar{\phi}$ is defined by the trajectory of a free-falling
particle with zero angular momentum at infinity. Thus, unlike the
Boyer-Lindquist case, in Doran coordinates particles can fall in
``radially'' along paths
of constant $\theta$ and $\bar{\phi}$. The Boyer-Lindquist $t$ and
$\phi$ can easily be recovered via the transformations
\begin{subequations}
\begin{equation}
t = \tau +\int_r^\infty \frac{bc}{\Delta}dr
\end{equation}
and
\begin{equation}
\phi = \bar{\phi} +\int_r^\infty
\frac{ac}{b\Delta}dr.
\end{equation}
\end{subequations}

As above, the equations of
motion are perfectly well-behaved at the horizon, making it an
attractive coordinate system for calculating physical processes
there. Furthermore, the inverse metric also has a very convenient
form:
\begin{equation}
g^{\mu \nu} = \left(\begin{array}{cccc}
-1 & bc/\rho^2 & 0 & 0 \\
bc/\rho^2 & \Delta/\rho^2 & 0 & ac/b\rho^2 \\
0 & 0 & 1/\rho^2 & 0 \\
0 & ac/b\rho^2 & 0 & 1/b^2\sin^2\theta \end{array}\right).
\end{equation}
The same Hamiltonian approach to the equations of motion works even
better in Doran coordinates, since there are no imaginary roots in the
Hamiltonian even inside the event horizon. Equation
(\ref{H_1_general}) can be expressed in the Doran metric as
\begin{equation}\label{H_1_doran}
H_1(r,\theta,\bar{\phi},p_r,p_\theta,p_{\bar{\phi}};\tau) =
-\frac{bc}{\rho^2}p_r +D,
\end{equation}
where the determinant $D^2$ is given by
\begin{equation}\label{D2_doran}
D^2 =
\left(\frac{b^2}{\rho^2}+\frac{a^2c^2\sin^2\theta}{\rho^4}\right)p_r^2
+\frac{2ac}{b\rho^2}p_r p_\theta +\frac{p^2_\theta}{\rho^2}
+\frac{p^2_{\bar{\phi}}}{b^2\sin^2\theta}+m^2. 
\end{equation} 
Since $D^2>0$ for all $r\ge 0$ and $\rho >0$, it is particularly easy
to parameterize trajectories (even inside the horizon) with the Doran
time coordinate $\tau$. As we mentioned above, particles falling from
rest at infinity will follow paths of constant $\theta$ and
$\phi$. Thus, like the Schwarzschild case, the geodesics can be
defined by
\begin{subequations}
\begin{equation}\label{xdot_doran2}
p^\mu = [1,-bc/\rho^2,0,0]
\end{equation}
and
\begin{equation}
p_\mu = [-1,0,0,0].
\end{equation}
\end{subequations}

Despite the many attractive features of the Doran coordinates, for the
majority of the calculations in this thesis, we will generally use the
more traditional Boyer-Lindquist coordinate system. First, we will not
be considering processes on or inside the horizon, so need not worry
too much about the coordinate singularities there. Second, since for
the most part we are interested in comparing theory with experiment,
it is particularly convenient to use the Boyer-Lindquist $t$
coordinate to parameterize trajectories, since $t$ corresponds with
the distant observer's proper time. Lastly, to a certain degree, we must
be slaves to convention, especially when comparing with previously
published results, and the Boyer-Lindquist coordinates are by far the
predominant coordinates for modeling the Kerr metric in the
literature.

\subsection{Analytic Methods}\label{analytic_methods}

The Hamiltonian equations of motion described above can be reduced
from eight to five coupled, first-order differential equations by
employing the symmetries in $t$ and $\phi$. For ease of implementation,
we have not employed Carter's constant $\mathcal{Q}$ in the equations
of motion, but rather use it as an independent check for the accuracy
of our numerics. Many traditional schemes to calculate
trajectories in the Kerr metric use this additional integral of motion
to further reduce the dimensionality of the problem
and even reduce the problem to one of quadrature
integration [e.g.\ \citet{rauch94}]. While potentially increasing
the speed of the computation, this approach also introduces
significant complications in the form of arbitrary signs in the
equations of motion corresponding to turning points in $r$ and
$\theta$. 

As \citet{carte68b} first showed, the Lagrangian equations of motion
can be written in separable form as [in Boyer-Lindquist coordinates,
following \citet{merlo99}]:
\begin{subequations}
\begin{eqnarray}
\rho^2\frac{dr}{d\lambda} &=& \pm \sqrt{R(r)} \label{carter_a}\\
\rho^2\frac{d\theta}{d\lambda} &=& \pm \sqrt{\Theta(\theta)}
\label{carter_b} \\ 
\rho^2\frac{d\phi}{d\lambda} &=&
\frac{p_\phi}{\sin^2\theta}+ap_t+\frac{aP}{\Delta}  \label{carter_c} \\
\rho^2\frac{dt}{d\lambda} &=&
a(p_\phi+ap_t\sin^2\theta)+(r^2+a^2)\frac{P}{\Delta} 
\label{carter_d} 
\end{eqnarray}
\end{subequations}
where
\begin{subequations}
\begin{eqnarray}
\Theta(\theta) &=&
\mathcal{Q}-\cos^2\theta[a^2(1-p_t^2)+p_\phi^2/\sin^2\theta] \\
P(r) &=& -p_t(r^2+a^2)-ap_\phi \\
R(r) &=& P^2-\Delta[m^2r^2+\mathcal{Q}+(p_\phi+ap_t)^2],
\label{carter2_c}
\end{eqnarray}
\end{subequations}
with $p_t$, $p_\phi$, and $\mathcal{Q}$ constants of the motion as
described above. 

Now here is where things start to get
complicated. A single trajectory can have both positive and negative
signs in equations (\ref{carter_a}) and (\ref{carter_b}) along
different segments of its path. So the first thing that needs to be
done is solve for the turning points of $R(r)$ and
$\Theta(\theta)$ at $r_0$ and $\theta_0$, respectively. Then $r$ and
$\theta$ can be solved parametrically by equating
\begin{equation}
\rho^2 = \frac{dr}{\pm\sqrt{R(r)}} =
\frac{d\theta}{\pm\sqrt{\Theta(\theta)}},
\end{equation}
or
\begin{equation}
\int_{r_0}^r \frac{dr'}{\sqrt{R(r)}} = (\rm{sgn}_r)(\rm{sgn}_\theta)
\int_{\theta_0}^\theta \frac{d\theta'}{\sqrt{\Theta(\theta)}},
\end{equation}
where the signs (sgn$_r$) and (sgn$_\theta$) are equal to $\pm 1$ and
change whenever a turning point is reached for either
variable. \citet{rauch94} show how these solutions can be written in
terms of elliptic integrals, further accelerating the speed of
computation. 

Once the values of $r$ and $\theta$ are known along the trajectory,
$\rho^2$ is known and then equations (\ref{carter_c}) and
(\ref{carter_d}) can be computed directly, and the entire trajectory
is known. The momentum components $p_r$ and $p_\theta$ can be
reproduced trivially from $p^\mu$ and the metric, as in the Lagrangian
approach. 

For equatorial orbits with $\theta=\pi/2$ and $p^\theta=0$, it is
relatively straightforward to frame the equations of motion in terms
of an effective potential, as is often done for Schwarzschild
orbits. In this case, equations (\ref{carter_a}-\ref{carter_b}) reduce
to \citep{shapi83}
\begin{subequations}
\begin{eqnarray}
\frac{dt}{d\lambda} &=& -\frac{1}{r^2\Delta}[(r^4+a^2r^2+2Ma^2r)p_t
+2aMp_\phi r] \\
\frac{d\phi}{d\lambda} &=& \frac{1}{r^2\Delta}[(r^2-2Mr)p_\phi -2aMp_t]
\label{shapiro_b} \\
\left(\frac{dr}{d\lambda}\right)^2 &=&
\frac{1}{r^4}R(r,p_t,p_\phi), \label{shapiro_c} 
\end{eqnarray}
\end{subequations}
with $R(r,p_t,p_\phi)$ defined as in equation (\ref{carter2_c}).
In equation (\ref{shapiro_c}), the right hand side $R/r^4$ can be
thought of as an effective potential for radial motion in the
equatorial plane. Stable circular orbits for massive particles exist
when
\begin{equation}\label{isco_reff}
R = 0; \hspace{1.5cm} \frac{\partial R}{\partial r} = 0;
\hspace{1.5cm} \frac{\partial^2 R}{\partial r^2} \le 0.
\end{equation}
The inner-most stable circular orbit (ISCO) occurs at the smallest
possible $r$ where a solution to equations (\ref{isco_reff})
exists. From \citet{barde72}, this radius is given by
\begin{equation}
r_{\rm ISCO}/M = 3+Z_2\mp [(3-Z_1)(3+Z_1+2Z_2)]^{1/2},
\end{equation}
where 
\begin{subequations}
\begin{equation}
Z_1 \equiv 1+\left(1-\frac{a^2}{M^2}\right)^{1/3}
\left[\left(1+\frac{a}{M}\right)^{1/3}+\left(1-\frac{a}{M}\right)^{1/3}
\right]
\end{equation}
and
\begin{equation}
Z_2 \equiv \left(3\frac{a^2}{M^2}+Z_1^2\right)^{1/2}.
\end{equation}
\end{subequations}
Here the upper signs refer to prograde orbits (particles orbiting in
the same direction as the black hole angular momentum) and the lower
signs correspond to retrograde orbits. For the Schwarzschild case with
$a/M=0$, the ISCO is located at $r=6M$, while for a maximally spinning
Kerr black hole with $a/M=1$, the prograde ISCO is at $r=M$ and the
retrograde ISCO is at $r=9M$ [$R_{\rm ISCO}(a/M)$ for prograde orbits
is plotted as a dashed line in Fig.\ \ref{plotnine} below].

\section{Geodesic Ray-tracing}
The initial conditions for the photon or particle geodesics are
determined in the local orthonormal frame of a ``Zero Angular Momentum
Observer'' (ZAMO). The ZAMO basis is defined such that the spatial
axes are aligned with the coordinate axes and then the time axis is
determined by orthogonality (see Section \ref{tetrads}). 

At a point far away from the black hole, the spacetime is nearly flat so
Euclidean spherical geometry gives the spatial direction of the photon
$n^{\hat{i}}\mathbf{e}_{\hat{i}}$, from which the initial momentum in
the coordinate basis is calculated:
\begin{subequations}
\begin{eqnarray}
p_t &=& -E_{\rm obs}(\omega\varpi n^{\hat{\phi}}+\alpha) \\
p_r &=& E_{\rm obs}\sqrt{\frac{\rho^2}{\Delta}}\, n^{\hat{r}} \\
p_\theta &=& E_{\rm obs}\sqrt{\rho^2}\, n^{\hat{\theta}} \\
p_\phi &=& E_{\rm obs}\sqrt{\varpi^2}\, n^{\hat{\phi}},
\end{eqnarray}
\end{subequations}
where the photon energy measured by the distant ZAMO is $E_{\rm obs}$.

\begin{figure}[ht]
\begin{center}
\includegraphics[width=0.7\textwidth]{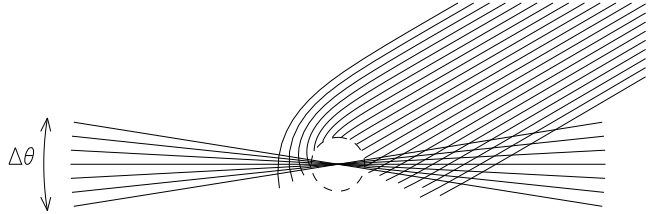}
\caption[Schematic picture of ray-tracing method]{\label{plotone}
Schematic picture of ray-tracing method
  from distant observer through a disk of angular thickness
  $\Delta\theta$. The rays either terminate at the black hole horizon
  (dashed circle) or
  pass through the disk, with each point of intersection labeled with
  the photon position and momentum $(x^\mu,p_\mu)$.} 
\end{center}
\end{figure}

The photon trajectories are integrated backward in time from the
image plane oriented at some inclination angle $i$ with respect to the
axis of rotation for the black hole, where $i=0^\circ$ corresponds to
a face-on view of the disk and $i=90^\circ$ is an edge-on view. The
accretion disk is confined to a finite region of latitude with angular
thickness $\Delta \theta$, oriented normal to the
rotation axis. The photons terminate either at the event horizon or
pass through the surfaces of colatitude ($\theta={\rm const}$), as
shown in Figure \ref{plotone}. As trajectories pass through
the disk, the photon's position and momentum $(x^\mu, p_\mu)$ are
recorded at each plane intersection in order to later reconstruct an
image of the disk.

\begin{figure}[ht]
\begin{center}
\includegraphics[width=0.7\textwidth]{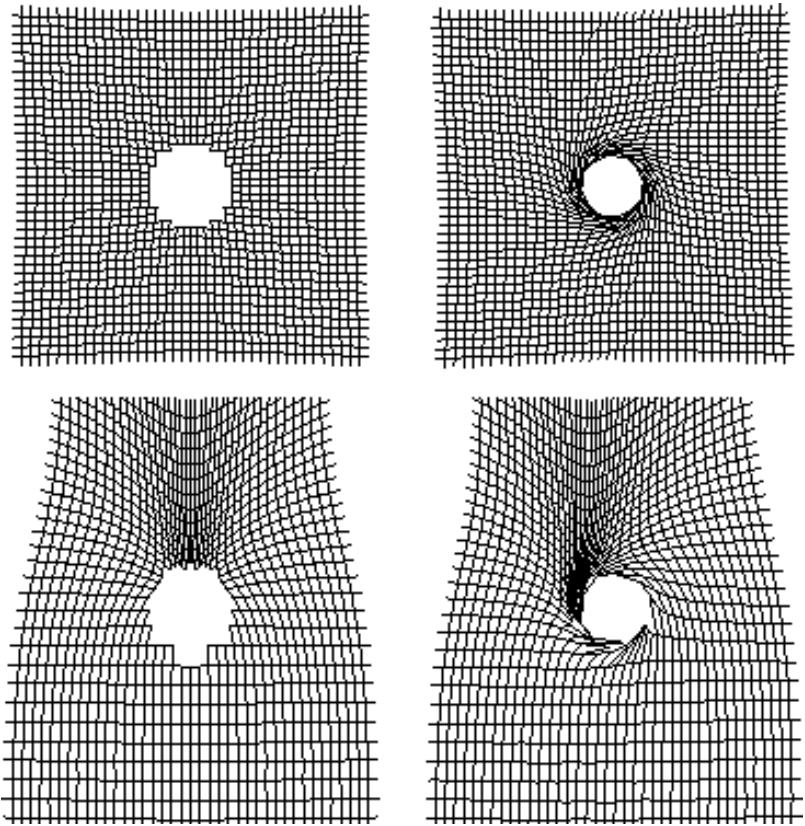}
\caption[Projection of image plane onto disk plane]{\label{plottwo}
Projection of a uniform Cartesian grid in the image
plane onto the source plane of the accretion disk $(\theta=\pi/2)$. 
Inclination angles 
are $i=0^\circ$ $(top)$ and $i=60^\circ$ $(bottom)$ and spin parameters are
$a/M=0$ $(left)$ and $a/M=0.95$ $(right)$. The region inside the horizon
is cut out from each picture.}
\end{center}
\end{figure}

For an infinitely thin disk $(\Delta \theta \to 0)$, it is easy to
show how the image plane 
maps onto the source plane. Taking an evenly spaced grid of initial
photon directions, Figure \ref{plottwo} plots the positions of
intersection with the source plane, in pseudo-Cartesian coordinates
defined by 
\begin{subequations}
\begin{eqnarray}
x &=& \sqrt{r^2+a^2}\cos \phi \label{cart_boyera}\\
y &=& \sqrt{r^2+a^2}\sin \phi. \label{cart_boyerb}
\end{eqnarray}
\end{subequations}
Photons that cross the black hole event horizon before intersecting
the plane are not shown. For $i>0^\circ$, as the rays are deflected 
by the black hole, they tend to be focused on the far side, giving a
strong magnification by mapping a large area in the image plane onto a
small area of the source plane, as seen here by a higher
density of lattice grid points. For the flat disk geometry, rays
are not allowed to pass through the plane defined by $\theta=0$, so we
do not see multiple images of sources ``behind'' the black hole, as is
often observed in the strong gravitational lensing of distant quasars by
intervening galaxies \citep{hewit88}. However, for sufficiently high
inclinations and spin values,
single points in the equatorial plane can be mapped to different regions of
the image plane, creating multiple images of certain regions of the
disk. This effect is seen in the folding of the image map onto itself
near the horizon in the bottom right of Figure \ref{plottwo}.

The disk itself is modeled as a collection of mass 
elements moving along geodesic orbits around the black hole, emitting
isotropic, monochromatic light with energy $E_{\rm em}$ in the emitter's
rest frame. For each photon with 4-momentum $p_\mu(\mathbf{x}_{\rm
  em})$ intersecting a particle trajectory with 4-velocity 
$v^\mu(\mathbf{x}_{\rm em})$, the measured redshift at the observer is
given by 
\begin{equation}
\frac{E_{\rm obs}}{E_{\rm em}} =
\frac{p_\mu(\mathbf{x}_{\rm obs})v^\mu(\mathbf{x}_{\rm obs})}
{p_\mu(\mathbf{x}_{\rm em})v^\mu(\mathbf{x}_{\rm em})},
\end{equation}
where for a distant observer at $r\to \infty$, we take
$v^\mu(\mathbf{x}_{\rm obs}) =
[1,0,0,0]$. 

For disk models with finite thickness, the radiative transfer
equation can be solved as the ray passes through the disk.
While the classical transfer equation is applicable in
the locally flat frame of the emitting gas, the spectral intensity at
a given frequency also evolves as the photons are gravitationally
red-shifted through the spacetime around the black hole,
maintaining the Lorentz invariance of $I_\nu/\nu^3$. This
Lorentz factor also accounts for the special relativistic beaming that
is especially important in the hot spot model. The coupling of the
geodesic ray-tracing and the radiation transfer equation
is described in greater detail below in Section
\ref{radiative_transfer}. 

For most of the calculations presented in this Section, we are primarily
concerned with radiation coming from a limited region of the disk, treated as
a monochromatic source with zero opacity. When calculating the
emission from a flat, steady-state disk, the plane defined by $\cos
\theta=0$ is taken to be totally opaque so that
rays cannot curve around and see the ``underside'' of the accretion disk.
In the ``thick disk'' case, for each pixel $(i,j)$ in the image plane,
an observed photon 
bundle spectrum $I_\nu(t_{obs},i,j)$ is given for each time step $t_{\rm
obs}$ by integrating the contribution of the hot spot and the disk
through the computational grid. This collection of incident photons
can then be summed to give time-dependent light curves, spectra, or spatially
resolved images. 

As mentioned above in the Introduction, one of the most promising
applications of this approach is the ability 
to use the ray-tracing code as a post-processor to analyze other,
more detailed simulations of the accretion disk. For example, we could
take the tabulated output of a three-dimensional hydrodynamic
calculation such as those by \citet{devil03b} or \citet{zanot03} and,
with a given emission mechanism, produce simulated images and
spectra. These simulations generally follow the hydrodynamic
variables (e.g.\ density, temperature, velocity and magnetic fields)
within a collection of volume elements. For the most part, these
variables are defined with respect to an observer's locally
orthonormal reference frame. Such a frame is often referred to as a
``tetrad,'' and is an important tool for analyzing any physical
process in general relativity. 

\subsection{Tetrads}\label{tetrads}
One of the cornerstones of general relativity is the principle that,
on small enough scales, spacetime can be treated as locally flat. In
these locally Minkowski reference frames, physics appears (to first
order) to follow
the laws of special relativity. For much of the above discussion in
this chapter, we have primarily used a global coordinate basis, which
allows for a relatively straightforward metric and equations of motion
for geodesic trajectories. Yet when modeling physical events such as
photon-electron scattering or solving the radiative transfer equation
(\ref{rad_trans_eq}), it is more convenient to define a locally
flat, orthonormal coordinate basis, conventionally called a
\textit{tetrad}. 

One of the simplest examples of such a tetrad is that of the
\textit{coordinate stationary observer} (CSO) in the Schwarzschild
metric. This tetrad is exactly what is sounds like: an orthonormal
basis fixed to an observer instantaneously at rest with coordinate
4-velocity $v^\mu({\rm CSO}) \propto [1,0,0,0]$. The basis axes in such a
tetrad are parallel to those of the coordinate basis, but normalized
so that the metric appears to be locally Minkowski. Denoting tetrad
vectors by ``hat'' indices $\hat{\mu}$, the CSO tetrad
$\mathbf{e}_{\hat{\mu}}$ is 
\begin{subequations}
\begin{eqnarray}\label{CSO_tetrad}
\mathbf{e}_{\hat{t}} &=& \left(1-\frac{2M}{r}\right)^{-1/2}
\mathbf{e}_t \\
\mathbf{e}_{\hat{r}} &=&\left(1-\frac{2M}{r}\right)^{1/2} \mathbf{e}_r \\
\mathbf{e}_{\hat{\theta}} &=&\frac{1}{r} \mathbf{e}_\theta \\
\mathbf{e}_{\hat{\phi}} &=&\frac{1}{r\sin\theta} \mathbf{e}_\phi.
\end{eqnarray}
\end{subequations}

To transform between bases, we follow the approach employed
in any standard vector analysis text: writing the invariant vector
$\mathbf{p}$ as a linear combination of basis vectors:
\begin{equation}
\mathbf{p} = \mathbf{e}_\mu p^\mu = \mathbf{e}_{\hat{\mu}}
p^{\hat{\mu}},
\end{equation}
the transformation can be written as a matrix operation
\begin{equation}
p^\mu = E_{\hat{\mu}}^\mu \, p^{\hat{\mu}} \hspace{1cm}
p^{\hat{\mu}} = E_\mu^{\hat{\mu}} \, p^\mu
\end{equation}
with 
\begin{equation}
E_{\hat{\mu}}^\mu = \left(\begin{array}{cccc}
1/\sqrt{1-2M/r} & 0 & 0 & 0 \\
0 & \sqrt{1-2M/r} & 0 & 0 \\
0 & 0 & 1/r & 0 \\
0 & 0 & 0 & 1/r\sin\theta \end{array}\right)
\end{equation}
and
\begin{equation}
\left[E_\mu^{\hat{\mu}}\right] = \left[E_{\hat{\mu}}^\mu\right]^{-1}.
\end{equation}
While it does not make a difference for diagonal or symmetric
transformations, in general we will write matrix components with the
lower index labeling the matrix row and the upper index labeling the
column.

As mentioned above in Section \ref{BL_coord}, there is a region around
a Kerr black hole called the ergosphere, where no coordinate stationary
observers can exist. In other words, there is no physical acceleration
that can 
give a time-like trajectory with $v^\mu \propto [1,0,0,0]$. Since
spacetime itself appears to be rotating faster than the speed of
light, if we want to create a locally Minkowski coordinate basis, the
only option is to ``go with the flow.'' In this approach, we consider
the observer orbiting the black hole on a non-geodesic orbit at the
frequency $\omega$ as defined in equation (\ref{BL_omega}) (not to
be confused with an actual massive particle orbiting on a circular
orbit at the Kepler frequency $\Omega_\phi$). Since this observer's 4-momentum
component in the $\mathbf{e}_\phi$ direction is $p_\phi =0$, it is
called a ``Zero Angular Momentum Observer,'' or ZAMO. Due to the
non-diagonal components of the Kerr metric, the ZAMO 4-velocity has
$p^\phi \ne 0$, and is thus not a CSO. 

The ZAMO tetrad $\mathbf{e}_{\hat{\mu}}$ is derived in \citet{barde72}
in Boyer-Lindquist coordinates:
\begin{subequations}
\begin{eqnarray}\label{ZAMO_tetrad}
\mathbf{e}_{\hat{t}} &=& \frac{1}{\alpha}\mathbf{e}_t +
\frac{\omega}{\alpha} \mathbf{e}_\phi \label{ZAMO_tetrada}\\
\mathbf{e}_{\hat{r}} &=&
\sqrt{\frac{\Delta}{\rho^2}}\, \mathbf{e}_r \label{ZAMO_tetradb}\\
\mathbf{e}_{\hat{\theta}} &=&
\sqrt{\frac{1}{\rho^2}}\, \mathbf{e}_\theta \label{ZAMO_tetradc}\\
\mathbf{e}_{\hat{\phi}} &=&
\sqrt{\frac{1}{\varpi^2}}\, \mathbf{e}_\phi. \label{ZAMO_tetradd}
\end{eqnarray}
\end{subequations}
The corresponding change of basis is given by
\begin{equation}
E_{\hat{\mu}}^\mu = \left(\begin{array}{cccc}
1/\alpha & 0 & 0 & \omega/\alpha \\
0 & \sqrt{\Delta/\rho^2} & 0 & 0 \\
0 & 0 & 1/\rho & 0 \\
0 & 0 & 0 & 1/\varpi \end{array}\right).
\end{equation}
As mentioned above, one advantage of the ZAMO basis is that
the basis vector $\mathbf{e}_{\hat{t}}$ is time-like
($g_{\hat{t}\hat{t}} < 0$) everywhere outside of the horizon. For a
coordinate stationary observer, on the other hand, the time basis
vector $\mathbf{e}_t$ becomes space-like ($g_{tt} > 0$) inside the
ergosphere. For sufficiently large values of the spin parameter
$a$, the inner-most stable circular orbit (often taken for the inner
edge of the accretion disk) extends within the ergosphere, emphasizing
the advantage of using the ZAMO basis.

Another useful feature of tetrads is that any tetrad basis with
time-like $\mathbf{e}_{\hat{t}}$ can be transformed to another tetrad
through a Lorentz transformation (boost + rotation). By definition,
this new basis will also have a time-like axis
$\mathbf{e}_{\hat{t}'}$. In Chapter 6, we will use this feature to
calculate scattering cross-sections and angles for Compton scattering
of photons in the corona around the black hole. To do so requires
first a transformation from the coordinate basis used for calculating
geodesics to a ZAMO frame, then a Lorentz boost to the rest frame of
the electron, where the scattering can be treated classically. After
the scattering gives the photon a new direction in the electron frame,
we do a boost back to the ZAMO frame, and finally
an inverse transform to the coordinate basis, where the photon can
continue along its new geodesic path until the next scattering event. 

\subsection{The Radiative Transfer Equation}\label{radiative_transfer}

Following the approach of \citet{rybic79}, we write the
\textit{radiative transfer equation} for the intensity $I_\nu$ along
a given ray's path length $ds$:
\begin{equation}\label{rad_trans_eq}
\frac{dI_\nu}{ds} = j_\nu-\alpha_\nu I_\nu,
\end{equation}
where $ds$ is the differential path length and $I_\nu$, $j_\nu$, and
$\alpha_\nu$ are respectively the radiation intensity, the emissivity,
and the absorption coefficient of the plasma at a frequency 
$\nu$. The absorption coefficient is related to the opacity
$\kappa_\nu$ through the density $\rho$: $\alpha_\nu =
\rho\kappa_\nu$.
In this form, basic emission and absorption is included, but not
scattering, which involves more complicated angular terms and takes
the form of an integrodifferential equation, which in general must be
solved using more advanced numerical techniques \citep{rybic79}. In
most of the results presented in this Thesis, we use extremely simple
models for emission and do not include absorption. However, since the
ultimate goal of this work is to produce a relativistic post-processor
for any hydrodynamic simulation, in this Section we include both
emission and absorption terms for generality.

We can rewrite the transfer equation by defining the optical depth
$\tau_\nu$
\begin{equation}\label{tau_nu}
d\tau_\nu \equiv \alpha_\nu ds.
\end{equation}
Now the transfer equation can be written as
\begin{equation}\label{rad_trans_eq2}
\frac{dI_\nu}{d\tau_\nu} = S_\nu - I_\nu,
\end{equation}
where $S_\nu \equiv j_\nu/\alpha_\nu$ is called the \textit{source
function}, which is often more convenient to work with than the
emissivity $j_\nu$. Over regions of constant source function $S_\nu$,
equation (\ref{rad_trans_eq2}) has the simple solution
\begin{equation}\label{rad_solution}
I_\nu(\tau_\nu) = S_\nu + e^{-\tau_\nu}[I_\nu(0)-S_\nu].
\end{equation}

As mentioned above, the term $I_\nu/\nu^3$ is a Lorentz invariant (in
fact proportional to the photon phase space density $f$). A simple
proof of this invariance can be given as follows: Consider a small
volume of particles in phase space $dV=d^3\mathbf{x} d^3\mathbf{p}$
moving in the $x$-direction in the unprimed frame with velocity
$\beta=v_x/c$. In the comoving
frame, the proper volume is $dV'=d^3\mathbf{x}'
d^3\mathbf{p}'$. Due to relativistic length contraction along the
$x$-axis, $dy=dy'$, $dz=dz'$, and $dx=dx'/\gamma$, giving
\begin{equation}\label{dx_contract}
d^3\mathbf{x} = \gamma^{-1}d^3\mathbf{x}'.
\end{equation}
The momentum components transform according to
\begin{subequations}
\begin{eqnarray}
dp_x &=& \gamma(dp_x'+\beta dp_t'), \\
dp_y &=& dp_y', \\
dp_z &=& dp_z',
\end{eqnarray}
\end{subequations}
but to first order in the energy, $dp_t'=0$ in the fluid frame, giving
\begin{equation}\label{dp_contract}
d^3\mathbf{p} = \gamma d^3\mathbf{p}'
\end{equation}
and thus the invariant phase space volume
\begin{equation}
dV = dV' = \mbox{Lorentz invariant}. 
\end{equation}
Similarly, the phase space density $f=dN/dV$ is invariant, since $dN$
is simply a number and therefore also invariant. The angular spectral
energy density $U_\nu(\Omega)=I_\nu/c$ can be expressed in terms of
the phase space density $f$:
\begin{equation}
U_\nu(\Omega)d\Omega d\nu = h\nu fd^3\mathbf{p} = 
h\nu f p^2dp d\Omega.
\end{equation}
Writing $p=h\nu/c$, we have
\begin{equation}
\frac{I_\nu}{\nu^3} = \frac{h^4}{c^2}f= \mbox{Lorentz invariant}. 
\end{equation}
Since the source function $S_\nu$ appears in equations
(\ref{rad_trans_eq2}) and 
(\ref{rad_solution}) as the difference $I_\nu - S_\nu$, it must have
the same transformation properties as $I_\nu$, so we can write
\begin{equation}\label{S_nu}
\frac{S_\nu}{\nu^3} = \mbox{Lorentz invariant}. 
\end{equation}

\begin{figure}[tp]
\begin{center}
\includegraphics[width=0.7\textwidth]{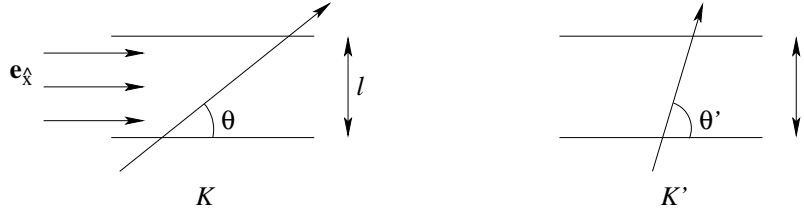}
\caption[Relativistic transformation of an absorbing
medium]{\label{ref_frames} Two reference frames for a finite volume of
matter flowing parallel to the $x$-axis. On the left is the ``lab''
frame $K$, and on the right is the material's local rest frame $K'$. A
ray propagates through the
medium at respective angles $\theta$ and $\theta'$ in the two
frames. Reproduced from \citet{rybic79}.}
\end{center}
\end{figure} 

Another Lorentz invariant is the optical depth, since the fraction of
photons passing through a finite medium is given by $e^{-\tau}$, which
is just a number, and thus the same in any reference frame. From this
feature, we can calculate the absorption coefficient in a relativistic
medium. Consider a small volume of matter flowing in the
$\mathbf{e}_{\hat{x}}$ direction with respect to the lab frame $K$, as
in Figure \ref{ref_frames}. The temperature and density and thus the
emissivity $j_\nu'$ is typically given
in the rest frame of the material $K'$. Since the motion is in the $x$
direction, the slab thickness $l$ is the same in both reference
frames. The optical depth $\tau_\nu$ can be written
\begin{equation}\label{tau_nu1}
\tau_\nu = \frac{l\alpha_\nu}{\sin\theta} =
\frac{l}{\nu\sin\theta}\nu\alpha_\nu = \mbox{Lorentz invariant}.
\end{equation}
Since $\nu\sin\theta$ is proportional to the $p_y$ component of the
photon 4-momentum, it must be the same in both frames because the
boost is in a perpendicular direction. Thus $\nu\sin\theta$ is another
Lorentz invariant, and we find 
\begin{equation}\label{nu_alpha_nu}
\nu\alpha_\nu = \mbox{Lorentz invariant}.
\end{equation}
Recalling the definition of the source function from $j_\nu =
\alpha_\nu S_\nu$, we can combine equations (\ref{S_nu}) and
(\ref{nu_alpha_nu}) to find
\begin{equation}\label{j_nu1}
\frac{j_\nu}{\nu^2} = \mbox{Lorentz invariant}
\end{equation}
or
\begin{equation}\label{j_nu2}
j_\nu = \left(\frac{\nu}{\nu'}\right)^2 j_\nu'.
\end{equation}

Now we can proceed to solve the radiative transfer equation along a
geodesic path through an arbitrary medium with emission and
absorption. Earlier in this Section, we showed a schematic view (see
Fig.\ \ref{plotone}) of the rays being traced through a fixed
coordinate grid. That method is particularly well suited for thin,
optically thick disks, where the photons cannot pass through the disk
and will generally intersect each surface of constant $\theta$ only
once. For more general geometries and optically thin emission regions,
it is more reasonable to tabulate the photon's momentum and position
at many points along its trajectory. Due to the adaptive step size
used in integrating the equations of motion (see below, Section
\ref{num_methods}), the points of tabulation conveniently tend to be
closer together in regions of higher curvature, which will also
generally correspond to regions of higher density and temperature.

\begin{figure}[tp]
\begin{center}
\includegraphics[width=0.5\textwidth]{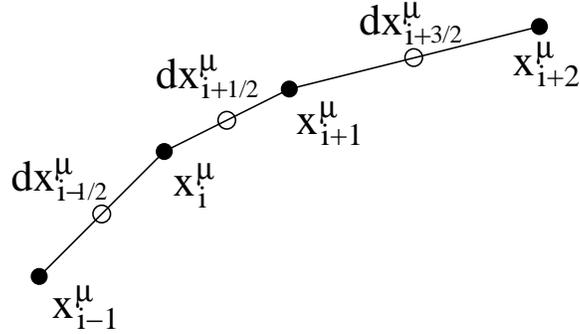}
\caption[Schematic of photon path for radiative transfer
equation]{\label{path_steps} Photon positions and momenta are
tabulated along its geodesic paths at coordinate points $x^\mu_i$. The
calculation of the proper distances $dl^2_i$ is described in the
text.}
\end{center}
\end{figure}

Denoting the photon's spacetime position at step $i$ by $x^\mu_i$,
the differential vectors between tabulated points are given by
\begin{equation}\label{dx_i12}
dx^\mu_{i+1/2} = x^\mu_{i+1}-x^\mu_i.
\end{equation}
These are the distances between solid circles in Figure
\ref{path_steps}. Then we can define a differential path length around
$x^\mu_i$ by the average
\begin{equation}\label{dx_i}
dx^\mu_i = \frac{1}{2}(dx^\mu_{i-1/2}+dx^\mu_{i+1/2}),
\end{equation}
which is the distance between the empty circles in Figure
\ref{path_steps}. 

Next, we transform from the coordinate basis to the ZAMO basis defined
at $x^\mu_i$, giving the differential $dx^\mu_i \to dx^{\hat{\mu}}_i$
and the momentum $p_{\mu,i} \to p_{\hat{\mu},i}$. In the ZAMO frame
(here it can be thought of as the lab frame), the photon spatial path
length is given by
\begin{equation}\label{ds2_i}
ds^2_i = \eta_{\hat{j}\hat{k}}dx^{\hat{j}}_i dx^{\hat{k}}_i.
\end{equation}
In principle, we know the fluid velocity at a collection of fixed points
in spacetime from
another tabulated set of data produced by an independent hydrodynamics
simulation. Using multi-linear extrapolation, the fluid variables
(4-velocity, density, temperature) can be determined at the point
$x^\mu_i$. The 4-velocity of the fluid in the ZAMO basis
$u^{\hat{\mu}}_i$ gives the angles $\theta$ and $\theta'$ from Figure
\ref{ref_frames}, and the tabulated density and temperature (and thus
absorption $\alpha_\nu'$ and emissivity $j_\nu'$) are typically given in the
rest frame of the fluid. 

To calculate the special relativistic redshift between the photons in
the ZAMO frame and fluid frame, we define a null 4-vector parallel to
the photon momentum in the ZAMO frame:
\begin{equation}
n_{\hat{\mu}} = [-1,\vec{\mathbf{n}}],
\end{equation}
where $\vec{\mathbf{n}}=n_{\hat{j}}$ is a normalized 3-vector in
standard Cartesian coordinates. Writing the fluid velocity 
\begin{equation}
u^{\hat{\mu}} = [\gamma,\gamma \vec{\mathbf{v}}],
\end{equation}
with $\vec{\mathbf{v}}=v^{\hat{j}}$ having magnitude $|\vec{\mathbf{v}}|
= \beta=v/c$, the
frequency ratio is then given as 
\begin{equation}
\frac{\nu}{\nu'} = \gamma(1+\beta\cos\theta') =
\frac{1}{\gamma(1-\vec{\mathbf{v}}\cdot\vec{\mathbf{n}})},
\end{equation}
where $\gamma \equiv 1/\sqrt{1-\beta^2}$ as usual. Now we have enough
information to solve the radiative transfer equation in a relativistic
flow \citep{rybic79}: 
\begin{equation}\label{rad_trans_eq3}
\frac{dI_\nu}{ds} = \left(\frac{\nu}{\nu'}\right)^2 j_\nu' -
\left(\frac{\nu'}{\nu}\right) \alpha_\nu' I_\nu.
\end{equation}
In the first order finite difference form along the path $dx^\mu_i$,
equation (\ref{rad_trans_eq3}) can be written
\begin{equation}
I_{\nu,i+1} = I_{\nu,i}+ ds_i \left[\left(\frac{\nu}{\nu'}\right)_i^2
j_{\nu, i}' -
\left(\frac{\nu'}{\nu}\right)_i \alpha_{\nu,i}' I_{\nu,i} \right],
\end{equation}
where $I_{\nu,i}$ is the spectrum of the photon beam entering the
small volume around $x^\mu_i$ and $I_{\nu,i+1}$ is the spectrum of the
beam upon leaving the volume element.

The above analysis, while quite useful for special relativistic flows
in the ZAMO basis, ignores all general relativistic effects of curved
spacetime around the black hole. To include these effects, we need
only consider the invariant $I_\nu/\nu^3$ along the geodesic path of
the photons. This is particularly straightforward from a computational
point of view, where the spectrum is stored as a finite array $I^j$,
evaluated at the frequencies $\nu^j$. These frequencies are redshifted
from one zone to the next due solely to gravitational effects. Since
all the frequencies are shifted the same way, only a single fiducial
redshift must be calculated. Let $V^\mu_i$ be the coordinate
4-velocity of a ZAMO at position $x^\mu_i$. Then we can define the dot
product with the photon 4-momentum as
\begin{equation}\label{alpha_pv}
\chi_i \equiv p_{\mu, i}V^\mu_i.
\end{equation}
Then the array of frequencies is redshifted along the photon path
according to 
\begin{equation}\label{nu_j_ip1}
\nu^j_{i+1} = \nu^j_i \left(\frac{\chi_{i+1}}{\chi_i}\right).
\end{equation}
Similarly, the spectral intensity defined at each frequency point
scales as
\begin{equation}\label{I_j_ip1}
I^j_{i+1} = I^j_i \left(\frac{\chi_{i+1}}{\chi_i}\right)^3.
\end{equation}
While these methods are ideally suited for our implementation of the
ray-tracing code, it should be noted that purely covariant approaches
also exist for solving the radiative transfer equation in curved
spacetime \citep{fuers04}. 

Because of this invariant scaling, a source with a blackbody spectrum
will appear to a distant observer as a blackbody with temperature
scaled as the redshift
\begin{equation}
(1+z)^{-1} = \frac{\nu_{\rm obs}}{\nu_{\rm em}}.
\end{equation}
Since the differential frequency $d\nu$
is also scaled by this factor, the total flux observed $\int
I_\nu d\nu$ will also scale as $(1+z)^{-4}$, just like a blackbody
with temperature $T_{\rm obs} = T_{\rm em}(1+z)^{-1}$. 

To summarize, the radiative transfer equation (\ref{rad_trans_eq}) is
solved in full general relativity with the following steps:
\begin{itemize}
\item The geodesic photon trajectory is integrated backwards in time
from a distant observer to the black hole, through the emission
region, and is either captured by the horizon or escapes to infinity.
\item At each point along the photon's path, the spacetime position
$x^\mu_i$ and momentum $p^\mu_i$ are tabulated. 
\item At the beginning (ray-tracing ``end'') of the photon path, we
set the spectrum $I(\nu^j)=0$ for all $\nu^j$. 
\item The spectrum $I(\nu^j)$ and the frequencies $\nu^j$ are
transformed according to equations (\ref{nu_j_ip1}) and
(\ref{I_j_ip1}) from one tabulated position to the next.
\item At each tabulated position in the emission region, after the
spectrum is adjusted for general relativistic effects, the special
relativistic radiative transfer equation (\ref{rad_trans_eq3}) is used
to update the spectrum $I^j_i$. The emission and absorption
coefficients at that point are interpolated from another tabulated set
of data (e.g.\ from a hydrodynamic simulation). The results presented
in this Thesis generally treat the gas as an optically thin emitter
with finite thickness and zero opacity or as flat disk of infinite
opacity (Section \ref{transfer_function}).
\item The spectrum is transformed to the next tabulated position, the
transfer equation is applied again, and so on until the ray reaches
the end-point at infinity (ray-tracing ``start''), where the spectrum
is observed.
\item This entire procedure is done for each ray in the image plane
and repeated for each time step of the simulation, creating a
time-dependent spectrum and images of the accretion region.
\end{itemize}

\section{Numerical Methods}\label{num_methods}
The calculations as described so far can be divided into two major
pieces (and in practice, they are carried out by two separate
programs). The first, what we call ``ray-tracing,'' integrates the
geodesic equations of motion in a vacuum, tabulating the position and
momentum of each photon along its path. The second step, which is more
accurately described as the ``radiation transport'' part of the
calculation, requires an independent model for the disk emissivity at
each point in the computational grid. 

The ray-tracing calculation is carried out by numerically integrating
equations (\ref{hameq_1}) and (\ref{hameq_2}) with a fifth-order
Runge-Kutta algorithm with adaptive time stepping. This provides high
accuracy over a large range of scales as the photon follows a long
path through the relatively flat spacetime between the observer and the
black hole, and then experiences strong curvature over a small region
close to the horizon. The integrator was written from scratch, roughly
following the methods described in \citet{press97}. For completeness
(and pedagogy), the basic algorithm is described here.

We begin with the classical fourth-order Runge-Kutta
algorithm. Combining all position and momentum variables
$[x^\mu,p_\mu]$ into a single vector $\mathbf{y}_n$, the next
iterative value for $\mathbf{y}_{n+1}$ is given to first order by
\begin{equation}\label{first_rk}
\mathbf{y}_{n+1}=\mathbf{y}_n + h \mathbf{f}(\mathbf{y}_n).
\end{equation}
Here $\mathbf{f}(\mathbf{y})$ is the first derivative of the vector
$\mathbf{y}$ with respect to the independent variable $\lambda$
(generally taken 
to be the coordinate time in our Hamiltonian approach), evaluated
exactly according to equations (\ref{hameq_1})
and (\ref{hameq_2}). Equation (\ref{first_rk}) is only accurate to
first order in $h$ [i.e.\ the error term is $\mathcal{O}(h^2)$], but
by evaluating the function $\mathbf{f}$ at a few points between
$\mathbf{y}_n$ and $\mathbf{y}_{n+1}$, we can achieve fourth order
accuracy:
\begin{eqnarray}\label{fourth_rk}
\mathbf{k}_1 &=& h \mathbf{f}(\mathbf{y}_n) \nonumber\\
\mathbf{k}_2 &=& h \mathbf{f}(\mathbf{y}_n+\frac{1}{2}\mathbf{k}_1) \nonumber\\
\mathbf{k}_3 &=& h \mathbf{f}(\mathbf{y}_n+\frac{1}{2}\mathbf{k}_2) \nonumber\\
\mathbf{k}_4 &=& h \mathbf{f}(\mathbf{y}_n+\mathbf{k}_3) \nonumber\\
\mathbf{y}_{n+1} &=& \mathbf{y}_n + \frac{\mathbf{k}_1}{6} +
\frac{\mathbf{k}_2}{3}+ \frac{\mathbf{k}_3}{3} + \frac{\mathbf{k}_4}{6}
\hspace{0.5cm}\left[+ \mathcal{O}(h^5)\right].
\end{eqnarray}

While we have written the error term as $\mathcal{O}(h^5)$, we can
actually be a bit more precise and say that it is approximately
$h^5\phi$, where $\phi$ is proportional to the Taylor series expansion
term $\mathbf{y}^{(5)}/5!$. Since $\phi$ remains constant over the step
size (at least to order $h^5$), the error term in equation
(\ref{fourth_rk}) scales directly as $h^5$. Thus if we compare two
iterations, one with a single step of size $2h$, and one with two
small steps $h+h$, the difference should give a quantitative value for
$\phi$. By means of Richardson extrapolation, we can then arrive at a
more accurate estimate for $\mathbf{y}$. The two
solutions can be written 
\begin{eqnarray}\label{two_steps}
\mathbf{y}(\lambda+2h) &=& \mathbf{y}_1 + (2h)^5\phi +
\mathcal{O}(h^6) \nonumber\\
\mathbf{y}(\lambda+2h) &=& \mathbf{y}_2 + 2(h)^5\phi +
\mathcal{O}(h^6),
\end{eqnarray}
where $\mathbf{y}(\lambda)=\mathbf{y}_0$, $\mathbf{y}_1$ is the
solution to equation (\ref{fourth_rk}) when taking a single step of
size $2h$, and $\mathbf{y}_2$ is the solution to taking two small
steps, each of size $h$. Writing the difference as
\begin{equation}\label{delta_y}
\mathbf{\Delta} = \mathbf{y}_2 - \mathbf{y}_1 = 30 h^5 \phi,
\end{equation}
we can now derive a fifth-order estimate for $\mathbf{y}(\lambda+2h)$:
\begin{equation}\label{fifth_rk}
\mathbf{y}(\lambda+2h) = \mathbf{y}_2 + \frac{\mathbf{\Delta}}{15} +
\mathcal{O}(h^6). 
\end{equation}
Not only do we increase the order of the solution (and thus usually,
but not always, increase the accuracy), but this approach also gives a
good estimate for the absolute error in the solution, i.e.\ the
difference between the exact solution for $\mathbf{y}(\lambda)$ and
the numerical approximation. Particularly for the problem of
ray-tracing, where the photons encounter strongly curved space around
the black hole and large regions of nearly flat space on their way to
the observer, we would like to be able to take the largest steps
possible while maintaining reasonable accuracy. In practice, the size
of these steps will vary by many orders of magnitude along the photon
path. 

Let $\epsilon$ be the desired fractional error for the numerical
solution. Then the magnitude of the desired error should be, to
leading order in $h$,
$|\mathbf{\Delta}_0|=\epsilon|\mathbf{y}|$. Since $\mathbf{\Delta}$
scales as $h^5$, if a step size of $h_1$ produces an error vector
$\mathbf{\Delta}_1$, then the step size $h_0$ that would have given
the desired error $\mathbf{\Delta}_0$ can be approximated as
\begin{equation}\label{h_Delta}
h_0 \approx h_1 \left(\frac{|\mathbf{\Delta}_0|}{|\mathbf{\Delta}_1|}
\right)^{1/5}.
\end{equation}
If the error $\mathbf{\Delta}_1$ was too large, the step size will be
adjusted accordingly in order to achieve the required
accuracy. Similarly, if $\mathbf{\Delta}_1$ is smaller than the
acceptable error, the step size will be increased in order to maximize
efficiency. 

Following \citet{press97}, we actually use a slightly different
prescription when the attempted step is too large, scaling the new
step by an exponent of $1/4$ instead of $1/5$. We also include a
``safety factor'' $S$ to ensure that the estimate in equation
(\ref{h_Delta}) errs on the side of caution. It is much more
computationally efficient to take steps that are $\sim 10 \%$ smaller
than necessary, rather than trying to match exactly the target error,
and thus going over half of the time, thus taking many steps that turn
out to be wasted. So equation (\ref{h_Delta}) is replaced by
\begin{equation}\label{h_Delta2}
h_0 = \left\{ \begin{array}{lc}
Sh_1 \left(\frac{|\mathbf{\Delta}_0|}{|\mathbf{\Delta}_1|}
\right)^{1/5} & |\mathbf{\Delta}_0| \ge |\mathbf{\Delta}_1| \\
Sh_1 \left(\frac{|\mathbf{\Delta}_0|}{|\mathbf{\Delta}_1|}
\right)^{1/4} & |\mathbf{\Delta}_0| < |\mathbf{\Delta}_1| 
\end{array} \right. ,
\end{equation}
where the safety factor is typically $S=0.9-0.95$.

We typically maintain an accuracy of one part in
$10^8-10^{10}$, which can be independently confirmed by monitoring
$\mathcal{Q}$, Carter's constant from equation (\ref{carter}). The
images and spectra are formed by
ray-tracing a set of photon paths, usually of the order $500\times
500$ grid points in $(i,j)$ with $\sim 20$ latitudinal zones in
$\theta$ and spectral resolution of $\nu/\Delta \nu \sim 200$. When
tracing photons originating at the emitter (see Chapter 6), only
a fraction of the trajectories will actually end at the
observer. Therefore, to get a comparable angular and energy resolution,
a much larger number of rays must be traced, typically around $10^8$.

Fortunately, both methods of ray-tracing, whether from the observer to
the source or vice versa, are quite suitable for
parallelization. Since the photons are non-interacting, virtually no
communication is necessary between different processors. Thus the
problem is also extremely scalable and easily load-balanced. While the
majority of the calculations in this thesis are small enough to carry
out on a single-processor computer, some of the higher resolutions
runs were performed on the MIT Astrophysics Beowulf cluster. 

\section{Broadened Emission Lines from Thin Disks}
\subsection{Transfer Function}\label{transfer_function} 
There have been a number of calculations of the relativistic
broadening of spectral lines from a steady-state accretion disk
[\citet{laor91, georg91, reyno97, broml97, parie01}; for a detailed
review, see \citet{reyno03}]. As a check of the
ray-tracing code and trajectories of massive particles, we have
reproduced the results published in these papers for a variety of
black hole spins and disk inclinations.

A steady-state disk can be made of a collection of massive particles 
moving in concentric planar circular orbits (in reality, these orbits
will have a small inward radial velocity in order to satisfy mass
conservation with a
steady-state accretion flow; see Chapter 5 below). For orbits at a radius
$r$ in a plane orthogonal to the spin axis, a particle's specific
energy and angular momentum are given analytically by \citet{barde72}:
\begin{subequations}
\begin{equation}
-p_0 = \frac{r^2-2Mr\pm a\sqrt{Mr}}{r(r^2-3Mr\pm 2a\sqrt{Mr})^{1/2}}
\end{equation}
and
\begin{equation}
p_\phi = \pm\frac{\sqrt{Mr}(r^2\mp 2a\sqrt{Mr}+a^2)} {r(r^2-3Mr\pm
 2a\sqrt{Mr})^{1/2}}.
\end{equation}
\end{subequations} 
Here the top sign is taken for prograde orbits (particle angular
momentum parallel to black hole angular momentum) and the bottom sign
for retrograde orbits. Combining these equations gives the circular
orbital frequency
\begin{equation}\label{omega_phi}
\Omega_\phi \equiv \frac{p^{\phi}}{p^0} = 
\frac{g^{\mu \phi}p_\phi}{g^{\mu 0}p_0} = 
\frac{\pm \sqrt{M}}{r^{3/2}\pm a\sqrt{M}}.
\end{equation}
For the large part of the disk, the orbits have nearly Keplerian 
frequencies, as measured in coordinate time $t$.

Inside the ISCO, the particles follow plunging trajectories with
constant energy and angular momentum determined at the
ISCO. Traditionally, when calculating emission from a steady-state
disk, many approaches take
the inner edge of the disk to be the ISCO radius $R_{\rm ISCO}$. For
larger values of $a$, the ISCO extends in closer to the event horizon,
increasing the radiative area of the disk. However, due to the strong
gravitational redshift in this inner region, the observed intensity is
reduced by a significant factor of $\nu^3_{\rm obs}/\nu^3_{\rm em}$,
resulting in a weak dependence on spin for disks with uniform
emission.

\begin{figure}[tb]
\begin{center}
\scalebox{0.5}{\includegraphics*[54,360][414,720]{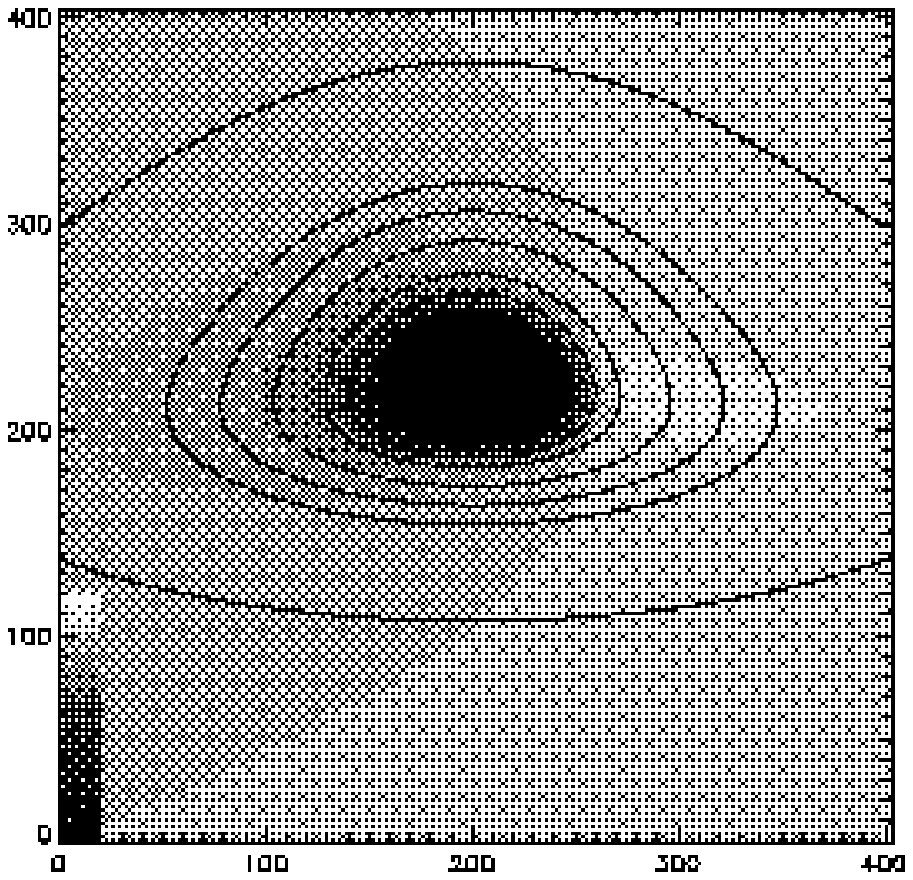}}
\scalebox{0.5}{\includegraphics*[54,360][414,720]{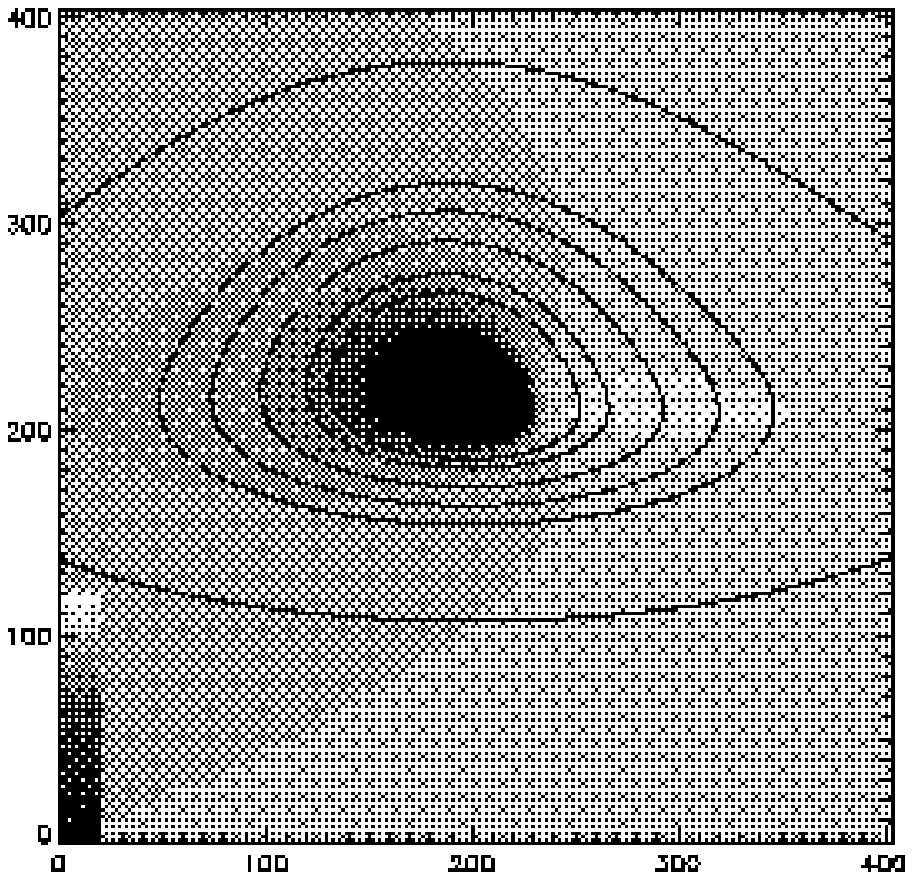}}
\caption[Redshift transfer function for flat disks]{\label{chap2_f3_1}
Projection of the accretion disk plane onto the plane of the
observer. The color corresponds to a logarithmic scale of the redshift
$\nu_{\rm obs}/\nu_{\rm em}= 0.1 \to 2$. On the left is a
Schwarzschild black hole with $a/M=0$ and on the right is a Kerr black
hole with $a/M=0.95$. The contours correspond to constant values of
$r/M=[3,4,6,8,10,20]$. Both disks are seen from an observer inclination
angle of $i=70^\circ$.}
\end{center}
\end{figure}

Figure \ref{chap2_f3_1} shows a projection of the disk plane onto the
image plane, color-coded by the observed redshift of an isotropic,
monochromatic emitter. This projection is
sometimes referred to as the transfer function, describing a map from
the emission regions of the disk to the detector plane of the
observer. The color scale in the lower left represents a logarithmic
scale from $\nu_{\rm obs}/\nu_{\rm em} = 0.1 \to 2$. Contours of
radius are also plotted as solid black lines, spaced at
$r/M=[3,4,6,8,10,20]$. Both Figures \ref{chap2_f3_1}a and
\ref{chap2_f3_1}b correspond to an observer at inclination
$i=70^\circ$ with respect to the disk axis, and the disk is rotating
in a clockwise direction. Thus the blueshifted emission on the right
side of the image is caused by gas moving towards the observer at
roughly half the speed of light. The high redshift of the inner
regions is caused by the deep gravitational potential well near the
black hole.

On the left, Figure \ref{chap2_f3_1}a shows
the transfer function for a Schwarzschild black hole with
$a/M=0$. While the ISCO for such a black hole is at $6M$, it is clear
that the emission extends continually all the way into the horizon at
$2M$. On the right, Figure \ref{chap2_f3_1}b is a black hole with
near-maximal spin of $a/M=0.95$. Qualitatively, the transfer function
is nearly identical, right down to the inner-most regions of the
disk, where the ISCO approaches the horizon at $r/M \approx 1.3$. For
this reason, the observed spectra from these two disks are nearly
identical. 

For a disk made up of massive particles on circular orbits emitting
isotropically from a 
region between $R_{\rm in}$ and $R_{\rm out}$, the Doppler broadening of an
emission line (typically iron K$\alpha$ with $E_{\rm em} = 6.4$ keV)
may be used to determine the inclination of the disk with respect to the
observer. Disks at higher inclination will have an intense blue-shifted
segment of the spectrum corresponding to the Doppler-boosted photons
emitted from gas moving toward the observer. The higher intensity for
the blue-shifted photons is caused by relativistic beaming, determined
by the Lorentz invariance of $I_\nu/\nu^3$ along a photon bundle:
\begin{equation}
I_\nu(\rm obs) = I_\nu(\rm em) \frac{\nu^3_{\rm obs}}{\nu^3_{\rm em}}.
\end{equation}

\begin{figure}[tp]
\begin{center}
\includegraphics[width=0.7\textwidth]{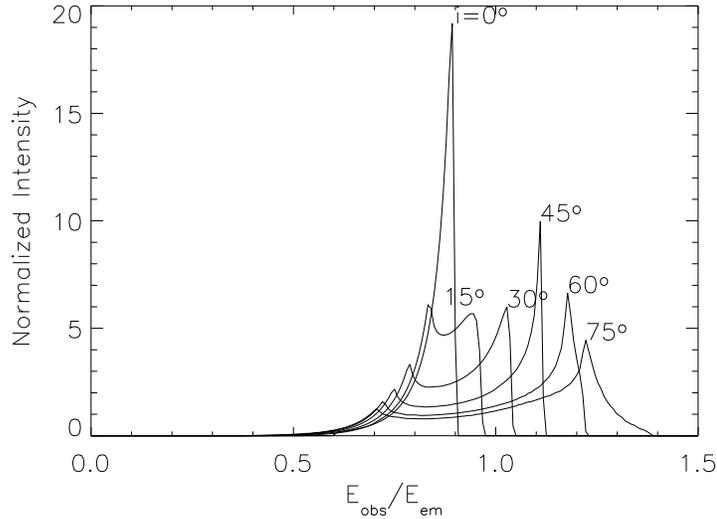}
\caption[Line broadening as a function of
inclination]{\label{chap2_f3_2} Normalized spectra of a monochromatic 
  emission line from steady-state
  accretion disks of varying inclination. An inclination of $i=0^\circ$
  corresponds to a face-on view of the disk while $i=90^\circ$ would be
  edge-on. The emissivity is taken to be uniform between $R_{\rm in} =
  R_{\rm ISCO}$ and $R_{\rm out} = 15M$. The spin is taken to be
  $a/M=0.5$ but the dependence on $a$ is negligible for
  uniformly emitting disks.} 
\end{center}
\end{figure}

\begin{figure}[tb]
\begin{center}
\scalebox{0.45}{\includegraphics*[74,360][540,720]{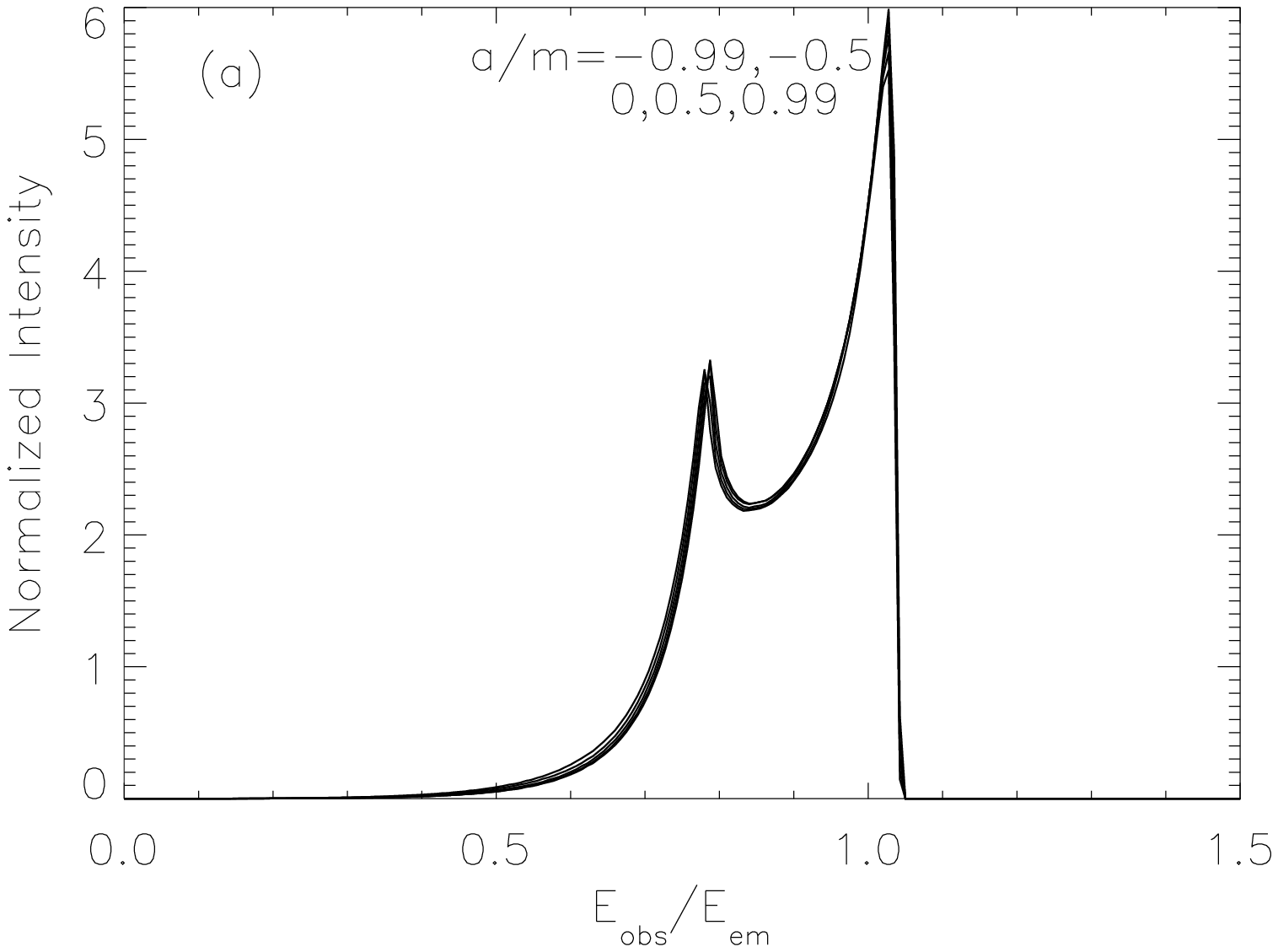}}
\scalebox{0.45}{\includegraphics*[74,360][540,720]{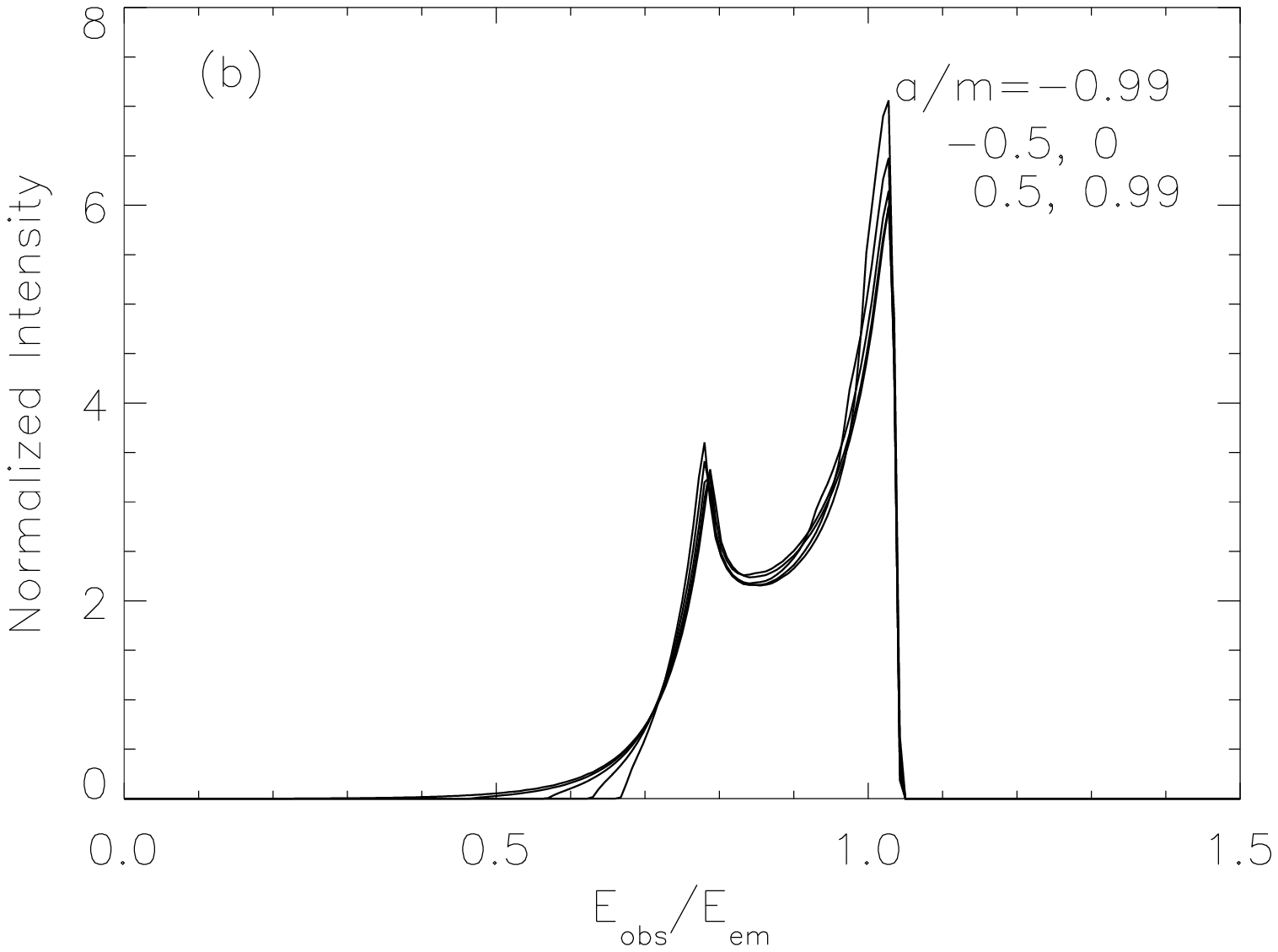}}
\caption[Line broadening as a function of spin; constant
emission]{\label{chap2_f3_3} Normalized spectra of a monochromatic
  emission line from steady-state accretion disks with inclination
  $i=30^\circ$ and varying spin parameter $a/M$ (negative spin values
  correspond to retrograde disk rotation). The emissivity is taken as
  constant between $R_{\rm in}$ and $R_{\rm out}=15M$. The
  disk extends all the way into the horizon
  in (a), with plunging trajectories inside of the ISCO, as described
  in the text. In (b), all emission is truncated inside of $R_{\rm
  ISCO}$. For this uniform emission model, the various theoretical
  spectra are nearly indistinguishable, even when truncating the disk
  at the ISCO.}
\end{center}
\end{figure}

Figure \ref{chap2_f3_2} shows the integrated spectra from a set of
accretion disks with outer radius $R_{\rm out}=15M$ and inner radius at
$R_{\rm in}=R_{\rm ISCO}$ for a spin parameter $a/M = 0.5$, normalized
such that 
\begin{equation}
\int I(E_{\rm obs}/E_{\rm em}) d(E_{\rm obs}/E_{\rm em}) =1. 
\end{equation}
All spectra are 
assumed to come from a flat, opaque disk with uniform
emission (this is not quite physically accurate; below we include the
likely possibility of increased emissivity in the inner disk). Repeating this
calculation for a range of spin parameters (and thus a range of
$R_{\rm ISCO}$), we find that the dependence on disk \textit{inclination} is
quite strong, while the dependence on black hole \textit{spin} is almost
insignificant, as shown in Figure \ref{chap2_f3_3} for an inclination
of $i=30^\circ$ and various spins of
$a/m=[-0.99,-0.5,0,0.5,0.99]$. Figure \ref{chap2_f3_3}a assumes
constant emission all the way down to the horizon, while the emission
in Figure \ref{chap2_f3_3}b is cut off at the ISCO. Clearly, the
different spectra are virtually indistinguishable, certainly for the
present levels of observational sensitivity. This 
is reasonable because, except for very close to the horizon, the spin has
little effect on the orbital velocity for circular orbits, as seen
from equation (\ref{omega_phi}). Furthermore, the emission from the innermost
regions is weighted much less due to its significant redshift and
smaller area. However, as we will see below, many models predict
a much higher emissivity in the inner disk, countering this
redshift effect and in turn reviving the possibility of measuring
black hole spin with relativistic line profiles.

\subsection{Observations of Iron Emission Lines}
As we mentioned in the Introduction, one of the most important
astronomical measurements of strong field GR would be the successful
determination of a black hole's spin. Since the leading order
curvature terms scale as $\sim M/r^3$ for the mass and $\sim a/r^4$
for the spin contributions, any observable that is sensitive to the spin
parameter will presumably originate from the regions closest to the
black hole.

\begin{figure}[ht]
%\begin{center}
\scalebox{1.0}{\includegraphics*[210,433][460,660]{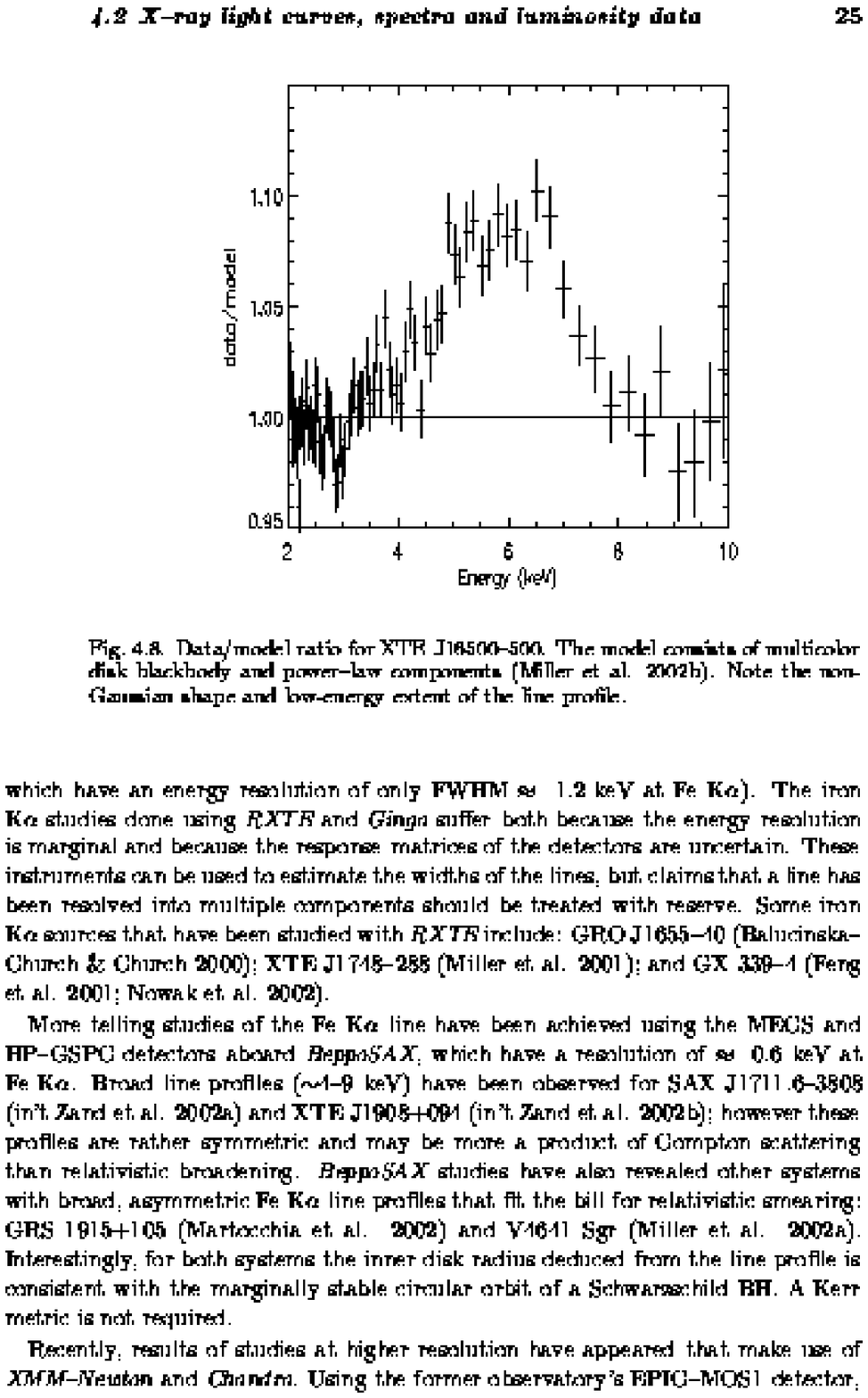}}
\scalebox{0.89}{\includegraphics*[340,190][580,450]{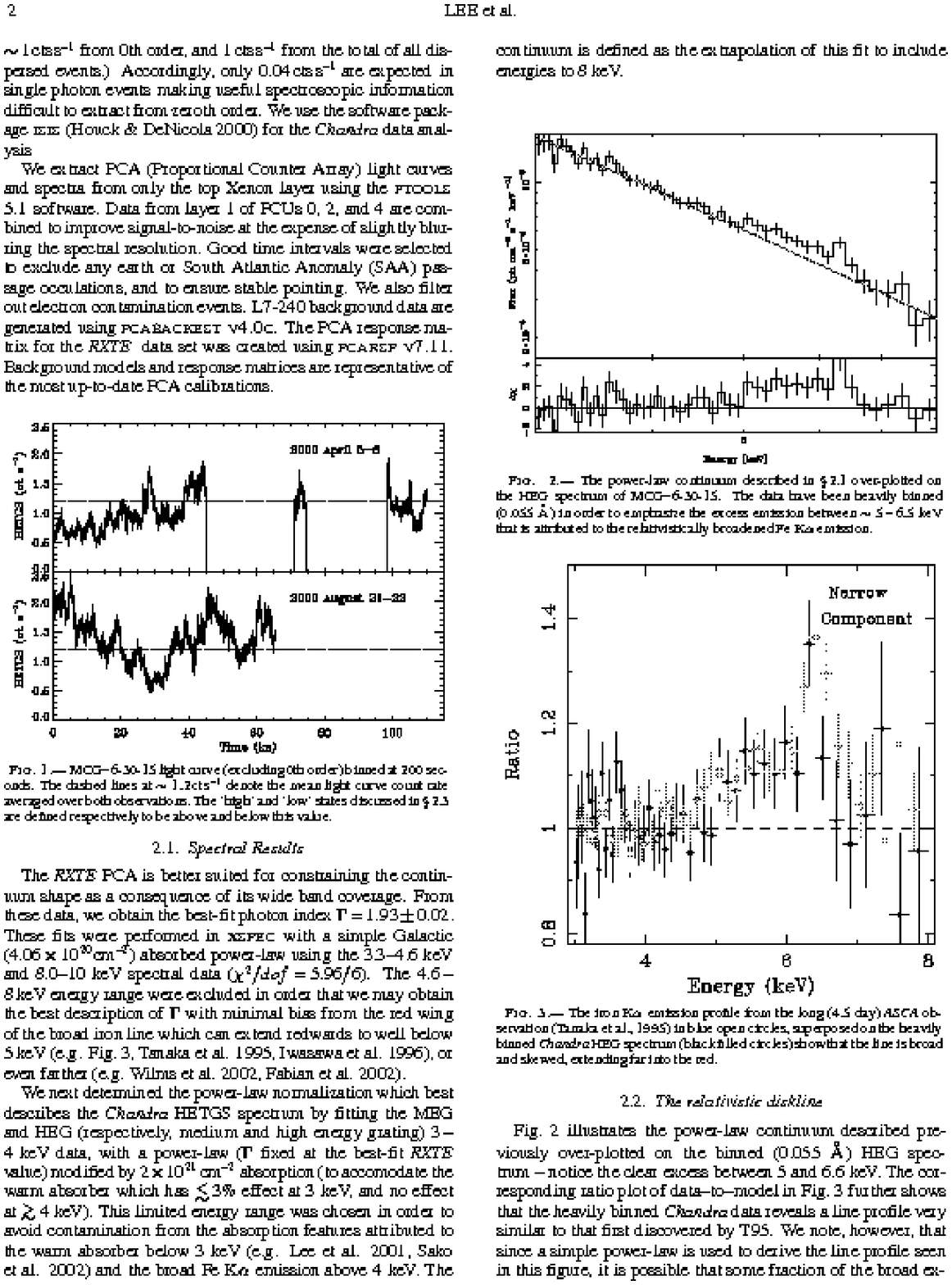}}
\caption[Observations of broad iron emission lines]{\label{broadiron}
(\textit{left}) A broadened Fe K$\alpha$
line from the black hole binary XTE J1650--500, observed with
\textit{XMM-Newton}. (\textit{right}) A similar line
from the Seyfert 1 galaxy MCG--6-30-15, observed with \textit{ASCA}
(blue) and \textit{Chandra} (black). Both plots show the excess
emission with respect
to a background model with blackbody and power-law components for a
multicolor disk. The lines extend well below the rest energy of
6.4 keV, suggesting emission for highly relativistic regions of the
inner accretion disk. [Reproduced from \citet{mille02} and
\citet{lee02b} with permission]}
%\end{center}
\end{figure}

It is actually quite possible that such an observation has already
been made, yet we currently lack the confidence in our
theoretical models necessary to interpret the results as an
unambiguous measurement of black hole spin. One of the most promising
observations is that of the relativistically broadened Fe K$\alpha$
emission line seen in both stellar-mass black holes and active
galactic nuclei (AGN), easily seen with the remarkable spectral
resolution and large collecting areas of \textit{Chandra} and
\textit{XMM-Newton}. An example of such a line is shown in Figure
\ref{broadiron}a from the black hole binary XTE J1650--500, reproduced
from \citet{mille02}. Similar lines have been seen in
the Seyfert 1 galaxy MCG--6-30-15 [\citet{tanak95,wilms01,lee02b}; see
Fig.\ \ref{broadiron}b], and both have
been interpreted as consistent with a near-maximal black hole spin
($a/M = 0.998$). 

This interpretation is heavily dependent on the
assumption that the accretion disk has a relatively sharp edge
at the inner-most stable circular orbit (ISCO). But many relativistic
magnetohydrodynamic (MHD) simulations find no such cut-off
\citep{gammi03,devil03b}, with significant pressure and density (and
thus emission) all the way in to the horizon. This point has been made
by \citet{reyno97}, but has unfortunately not been fully appreciated
by much of the high-energy astrophysics community. Using simplified
yet reasonable physical estimates (and without putting undue emphasis
on the ISCO), \citet{reyno04b} are able to confirm some sort of
spinning black hole in MCG--6-30-15 as well as the galactic black hole
binary GX 339--4, but they still cannot provide a clear measurement of
that spin.

\begin{figure}[ht]
\begin{center}
\scalebox{1.3}{\includegraphics*[110,540][460,700]{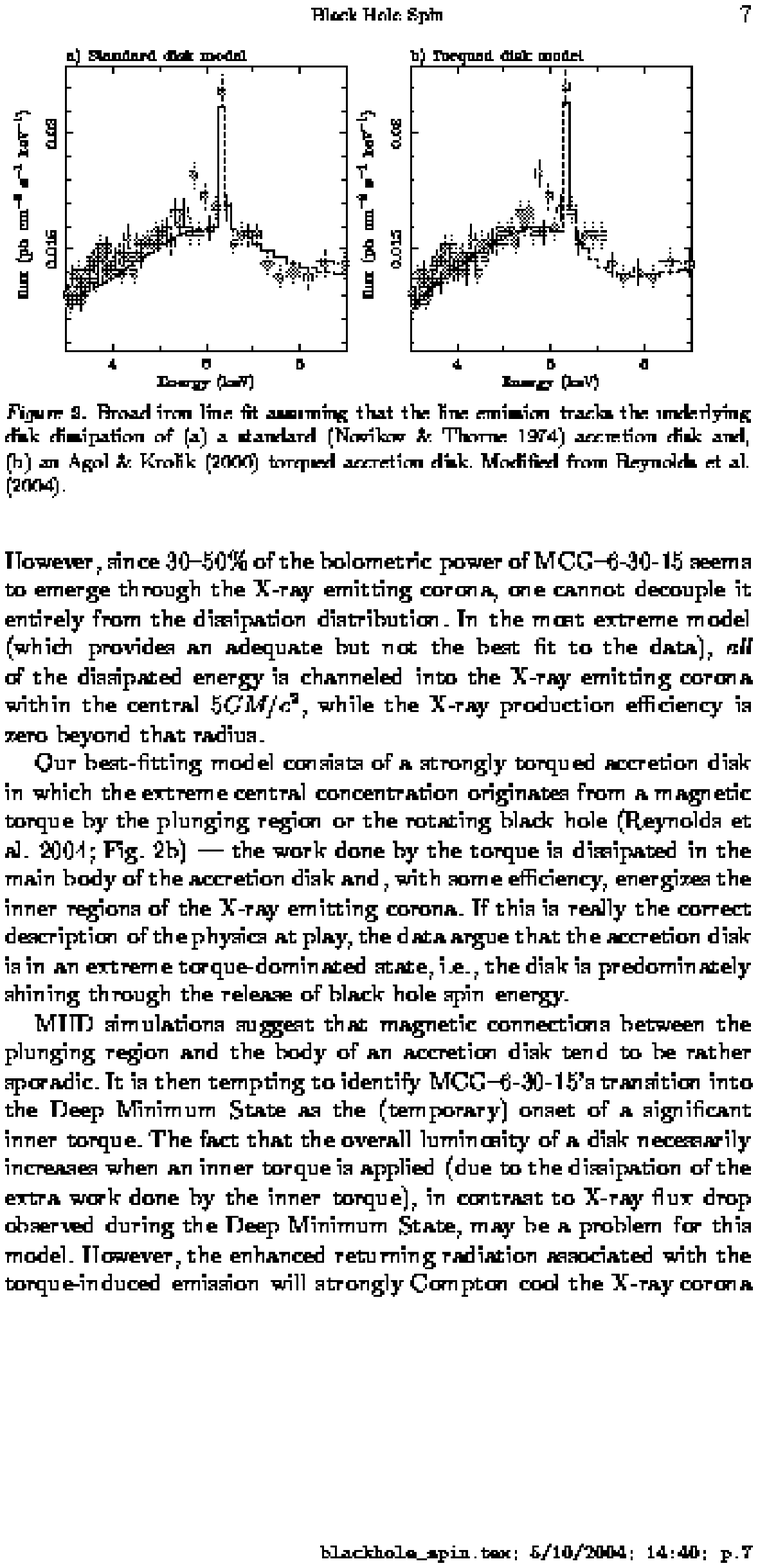}}
\caption[Observational evidence for the Blandford-Znajek
process]{\label{reynolds_fig} \textit{XMM-Newton} observation of the
broadened Fe K$\alpha$ line from the Seyfert 1 galaxy
MCG--6-30-15. The theoretical fits assume that the line emission
is proportional to the total local flux, as determined by a
steady-state relativistic $\alpha$-disk model. On the left is the
standard Novikov-Thorne model with near-maximal spin $a/M=0.998$ and
zero torque (and thus zero emission) at the ISCO. On the right is a
model with non-zero torque at the ISCO, as in \citet{agol99}, which
transfers significant energy into the inner disk, highlighting the
iron emission there. [Reproduced from \citet{reyno04a} with permission]}
\end{center}
\end{figure}

In addition to the uncertainty around the treatment of the disk
boundary conditions at the ISCO, there is also not an unambiguous
illumination mechanism that would
cause the disk to produce a high-energy emission line such as the Fe
K$\alpha$ at 6.4 keV. One likely possibility is that the iron emission
is produced by hard X-rays from a hot electron corona reflecting off
the relatively cool disk [see, e.g.\ \citet{mccli04}]. Another
option is that it simply follows the intensity of the thermal emission
from the disk itself \citep{agol99}. If the line emission indeed
tracks the total flux at each point in the disk, it may be possible to
measure more exotic processes in the disk, including magnetic torques
at the ISCO. In observations of MCG--6-30-15, \citet{reyno04a} claim
to find evidence of a torque on the inner edge of the disk, presumably
caused by some version of the Blandford-Znajek process, which provides
a mechanism for extracting energy from the spin of a black hole
through magnetic fields that thread the accretion disk as well as the
black hole horizon \citep{bland77}. This effect can be seen in Figure
\ref{reynolds_fig}, reproduced from \citet{reyno04a}. The added stress
on the inner disk puts a greater weight on the portions of the iron
line spectrum produced there, generally highlighting the broader
features caused by the strong relativistic effects near the ISCO.

For lack of a clear picture of the disk+corona geometry, many
accretion disk models include an emissivity that simply scales as a 
power of the radius. Following \citet{broml97}, we apply an emissivity
factor proportional to $r^{-2}$, giving an added weight to the
inner, presumably hotter, regions. However, unlike the model of
\citet{reyno04a}, where the iron line emission traces that of the
thermal disk, here we should note that the
emission is coming essentially from the corona, but fluorescing off
the much cooler disk. Thus, even though we will see in Chapter 5 that
almost no \textit{thermal} emission comes from inside of the ISCO,
there is still enough matter in that region to reflect the high energy
photons from the hot corona and contribute significantly to the iron
emission line profile.

\begin{figure}[tb]
\begin{center}
\scalebox{0.65}{\includegraphics*[74,360][540,720]{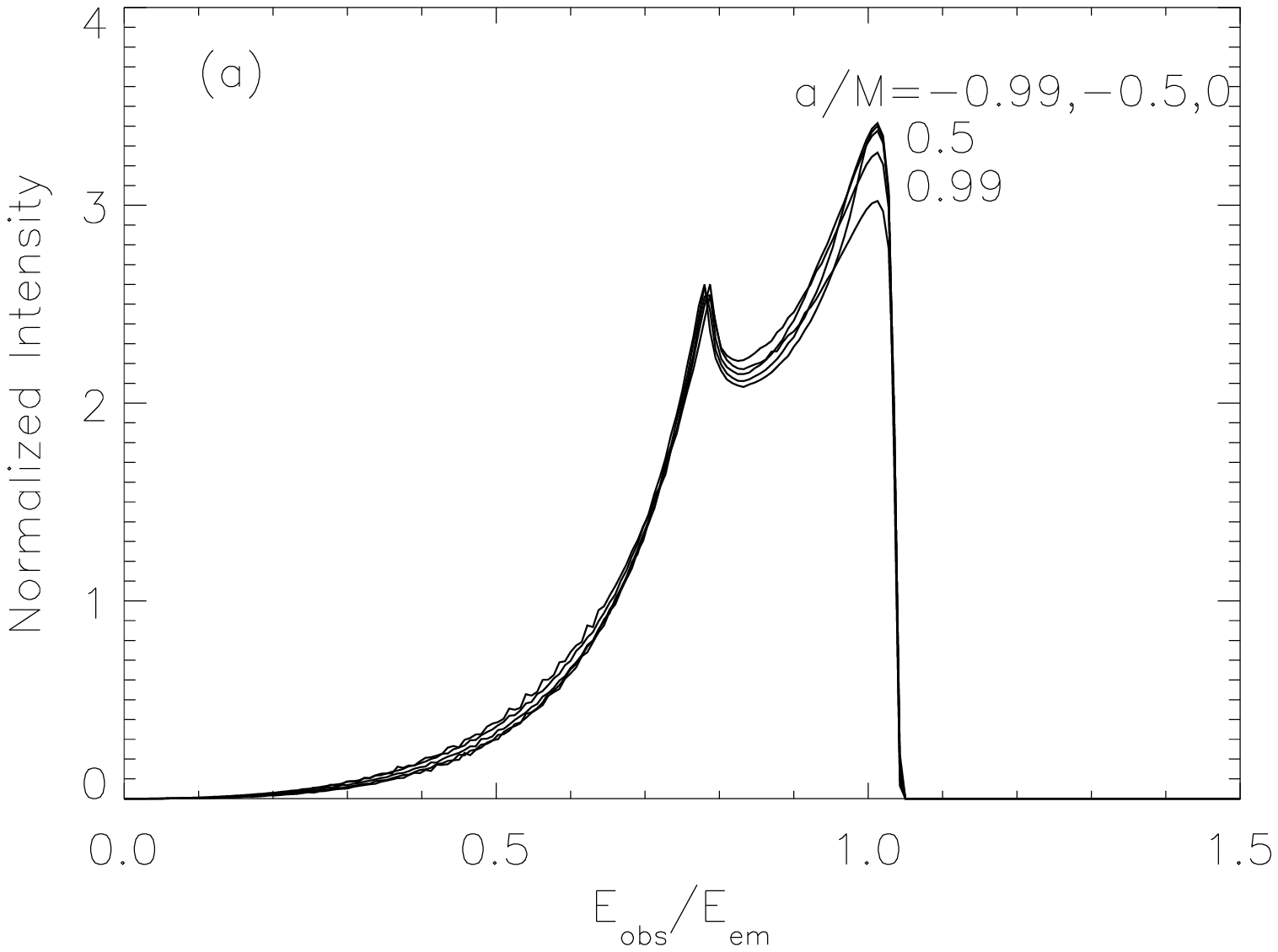}}
\scalebox{0.65}{\includegraphics*[74,360][540,720]{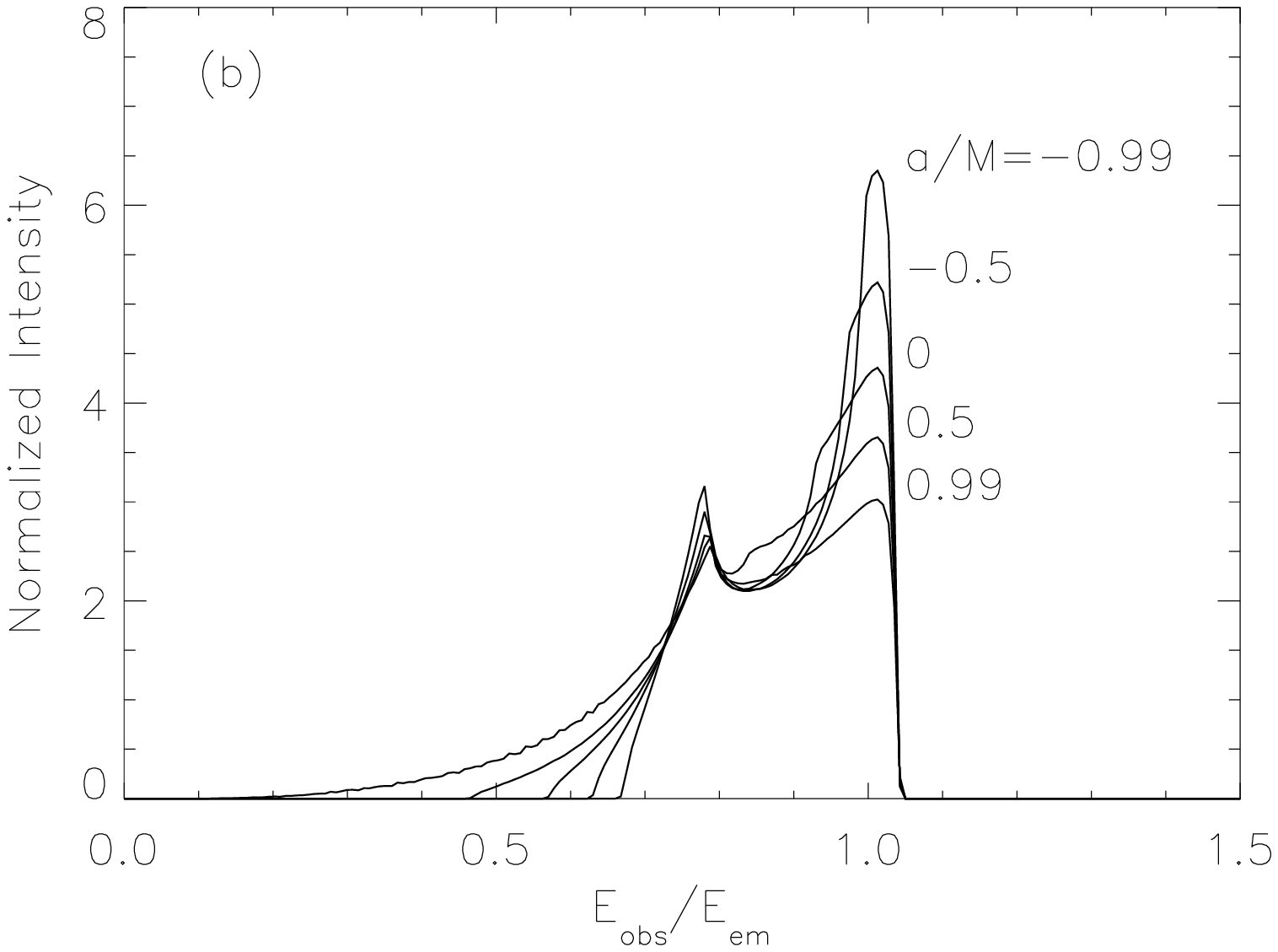}}
\caption[Line broadening as a function of spin; power-law
emission]{\label{plotthree}
  Normalized spectra of steady-state
  accretion disks with inclination $i=30^\circ$ and varying spin
  parameter $a/M$ (negative spin values correspond to retrograde disk
  rotation). The emissivity is taken to be proportional to $r^{-2}$ 
  between $R_{\rm in}$ and $R_{\rm out}=15M$. The
  disk extends all the way into the horizon
  in (a), with plunging trajectories inside of the ISCO, as described
  in the text. In (b), all emission is truncated inside of $R_{\rm
  ISCO}$. Black holes with
  higher values of $a/M$ allow the inner disk to extend in closer to the
  horizon, giving a greater weight to the high-redshift radiation
  emitted there. Yet even with the added emission, for disks that are
  not truncated at the ISCO, the spectrum is quite insensitive to the
  black hole spin.} 
\end{center}
\end{figure}

As can be seen in Figure \ref{plotthree}, this extra emission from
close to the black hole can serve to break the otherwise weak dependence
on spin, but only if we assume all emission is truncated at the
ISCO. For an inclination of $i=30^\circ$, five different spin
values are shown: $(a/M=-0.99, -0.5, 0, 0.5, 0.99)$, corresponding to
inner disk boundaries at $(R_{\rm ISCO}/M = 8.97, 7.55, 6.0, 4.23,
1.45)$. Since the sign of $a$ is defined with respect to the angular
momentum of the accretion disk, negative values of $a$ imply
retrograde orbits that do not survive as close to the black hole,
plunging at larger values of $R_{\rm ISCO}$. The disks that extend in
closer produce more low-energy red-shifted photons, giving longer
tails to the spectra at $E_{\rm obs}/E_{\rm em} < 0.7$ and smaller
relative peaks at $E_{\rm obs}/E_{\rm em} \approx 1$. These photon
energy spectra closely reproduce previously published results; in
particular compare with Figure 3 in \citet{broml97}.

Assuming the ``best case scenario'' where the disk \textit{does} in
fact get cut off at the ISCO, \textit{and} if we can determine
the inclination of the disk independently (e.g.\ through spectroscopic
observations of the binary companion), the spin \textit{might} be
inferred from the broadening of an iron emission line. However, since
the plane of the disk tends to align normal to the black hole spin
axis close to the ISCO, the binary inclination may not coincide
with the inclination of the inner disk. The problem of inclination and
unknown illumination mechanisms, along with other complications, such as
additional emission lines and other causes of scattering and line
broadening, motivates us to look more closely at QPO power spectra as a
method for determining black hole spin.

\chapter{The Geodesic Hot Spot Model}
\begin{flushright}
{\it
Everything should be made as simple as possible, but not simpler.\\
\medskip
}
-Albert Einstein
\end{flushright}
\vspace{1cm}

\section{Hot Spot Emission}
Given the ray-tracing map from the accretion disk to the image plane,
with each photon bundle labeled with a distinct 4-momentum and time
delay, we can reconstruct time-dependent images of the disk based on
time-varying emission models. The simplest model we consider is a
single region of isotropic,
monochromatic emission following a geodesic trajectory:
the ``hot spot'' or ``blob'' model \citep{sunya72,bao92,stell98,stell99a}. 

The hot spot is a small region with finite radius and
emissivity $j(x)$ chosen to have a Gaussian distribution in local
Cartesian space:
\begin{equation}\label{I_exp}
j(\mathbf{x}) \propto \exp\left[-\frac{|\mathbf{\vec{x}}-
\mathbf{\vec{x}}_{\rm spot}(t)|^2}{2 R_{\rm spot}^2}\right]. 
\end{equation}
The spatial position 3-vector $\mathbf{\vec{x}}$ is given in pseudo-Cartesian
coordinates by the transformation defined by equations
(\ref{cart_boyera},\ref{cart_boyerb})
and $z=r\cos\theta$. Outside a distance of $4R_{\rm spot}$ from the
guiding geodesic trajectory, there is no emission. We typically take
$R_{\rm spot}=0.25-0.5M$, but find the normalized light curves and QPO power
spectra to be rather independent of spot size. We have
also explored a few different hot spot shapes, ranging from spherical
to an ellipsoid flattened in the $\mathbf{e}_\theta$ direction and
similarly find no significant dependence of the spectra on spot
shape.

Because we assume all points
in the hot spot have the same 4-velocity as the geodesic guiding
trajectory, one must be careful not to use too large a spot or the
point of emission $\mathbf{x}_{\rm em}$ can be spatially far enough away from
the center $\mathbf{x}_{\rm spot}$ to render the inner product
$p_\mu(\mathbf{x}_{\rm em})v^\mu(\mathbf{x}_{\rm spot})$
unphysical. One way to quantify the size of this physical region is
through the use of Riemann normal coordinates \citep{mtw73}, where the
metric is locally flat and the Cartesian dot product is well
behaved. The quadratic deviations from flat space scale according to
the local curvature scale, which is of order $\mathcal{R}\sim 10M$ at the
ISCO of a Schwarzschild black hole. Thus as long as the hot spot is
within 
\begin{equation}
\frac{|\mathbf{\vec{x}}-\mathbf{\vec{x}}_{\rm
spot}(t)|^2}{\mathcal{R}^2} \ll 1,
\end{equation}
equation (\ref{I_exp}) should be reasonably well behaved.

After calculating and tabulating the hot spot trajectory as a function
of coordinate time $t$, the ray-tracing map between the disk and the
observer is used to construct a time-dependent light curve from the
emission region. For each photon bundle intersection point there is a
time delay $\Delta t_{i,j,k}$ (where $i,j$ are the coordinate indices
in the image plane and $k$ is the latitude index in the disk) so for
the observer time $t_{\rm obs}$, we first determine where the hot spot
was at coordinate time $(t_{\rm em})_{i,j,k} =
t_{\rm obs}-\Delta t_{i,j,k}$. If the spot centroid is close enough (within
$4R_{\rm spot}$) to the disk intersection point $(r,\theta,\phi)_{i,j,k}$,
then the redshifted emission is added to the pixel spectrum
$I_\nu(t_{\rm obs},i,j)$, weighted by equation (\ref{I_exp}). 

In this way, a movie can be produced that shows
the blob orbiting the black hole, including all relativistic
effects. Such a movie shows a few
immediately apparent special relativistic effects such as the Doppler
shift and beaming as the spot moves toward and then away from the
observer. For a hot spot orbiting in the clock-wise direction as seen
from above ($v^\phi <0$ with $\phi=270^\circ$ toward the observer),
the point of maximum blue
shift actually occurs at a point where $\phi > 0$ because of the
gravitational lensing of the light, beamed in the forward direction of
the emitter and then bent toward the observer by the black
hole. Gravitational lensing also causes significant magnification
of the emission region when it is on the far side of the black hole,
spreading the image into an arc or even an Einstein ring for $i
\approx 90^\circ$, much like distant quasars are distorted
by intervening matter in galaxy clusters \citep{hewit88}.

\begin{figure}[tb]
\begin{center}
\scalebox{0.7}{\includegraphics*[54,300][540,740]{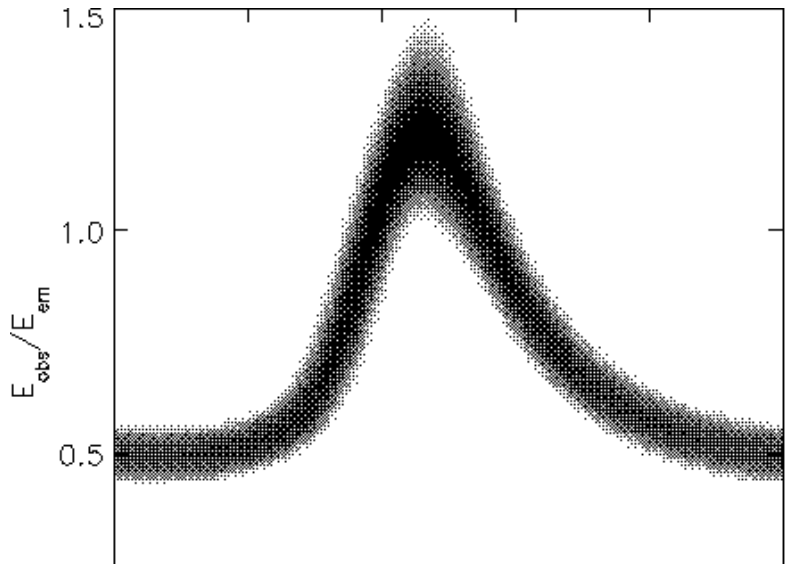}}
\caption[Spectrogram of hot spot on circular orbit]{\label{plotfour}
Spectrogram of a circular hot spot with
  radius $R_{\rm spot} = 0.5M$ orbiting a
  Schwarzschild black hole at the ISCO $(R_{\rm ISCO}=6M$), viewed at an
  inclination of $60^\circ$. The spot is moving in the
  $-\mathbf{e}_{\hat{\phi}}$ direction with $\phi(t=0)=180^\circ$ and
  the observer at $\phi=270^\circ$. The maximum
  redshift occurs when $\phi \approx 160^\circ$ and the maximum blueshift
  occurs when $\phi \approx 20^\circ$.}
\end{center}
\end{figure}

A simulated time-dependent spectrum or \textit{spectrogram} for this
hot spot model is shown 
in Figure \ref{plotfour}, for an inclination $i=60^\circ$ and black
hole spin $a/M=0$. The  horizontal axis measures time in the
observer's frame, with $t=0$ corresponding to the time at which the
spot center is moving most directly away from the observer $(\phi =
180^\circ)$. As mentioned above, this is \textit{not} quite the same as the
point of maximum redshift, which occurs closer to $\phi = 160^\circ$ due
to gravitational deflection of the emitted light.

\subsection{Overbrightness and QPO Amplitudes}\label{overbrightness_amp}
The spectrogram shown in Figure \ref{plotfour} can be integrated in time
to give a spectrum similar to those shown in Figures \ref{chap2_f3_2}
and \ref{chap2_f3_3},
corresponding to something like a very narrow circular emitting region with
$R_{\rm out} \approx R_{\rm in} \approx R_{\rm ISCO}$. By integrating
over photon energy, we get the total X-ray flux as a 
function of time, i.e.\ the light curve $I(t)$. This time-varying signal
can be added to a background intensity coming from the inner regions
of a steady state disk described below. By definition the hot
spot will have a higher temperature or density and thus greater
emissivity than the background disk, adding a small modulation to the
total flux. \textit{RXTE} observations find the HFQPO X-ray modulations to have
typical amplitudes of 1-5\% of the mean flux during the outburst
\citep{remil02,remil04a}. \citet{marko00} present a
first-order argument that a 1\% amplitude modulation requires a hot
spot with 100\% overbrightness extending over an area of 1\% of the
steady-state region of the disk. For $R_{\rm out}=15M$, this requires a
hot spot with radius $R_{\rm spot} \approx 1.5M$, which they argue is too
large to survive the viscous shearing of the disk.

Hydrodynamic stability aside, this reasoning ignores the
important effects of disk inclination, relativistic beaming and
gravitational lensing of the hot spot emission. Assuming a
Shakura-Sunyaev type disk with steady-state emissivity $g(r) \propto
r^{-2}$ and a similar scaling for the hot spot emission, we find that
hot spots with significantly smaller size or overbrightness are
capable of creating X-ray modulations on the order of 1\% rms, defined
by
\begin{equation}\label{rms}
{\rm rms} \equiv \sqrt{\frac{\int\left[I(t)-\bar{I}\right]^2dt}
  {\int I^2(t)dt}}.
\end{equation} 

\begin{figure}[tb]
\begin{center}
\includegraphics[width=0.7\textwidth]{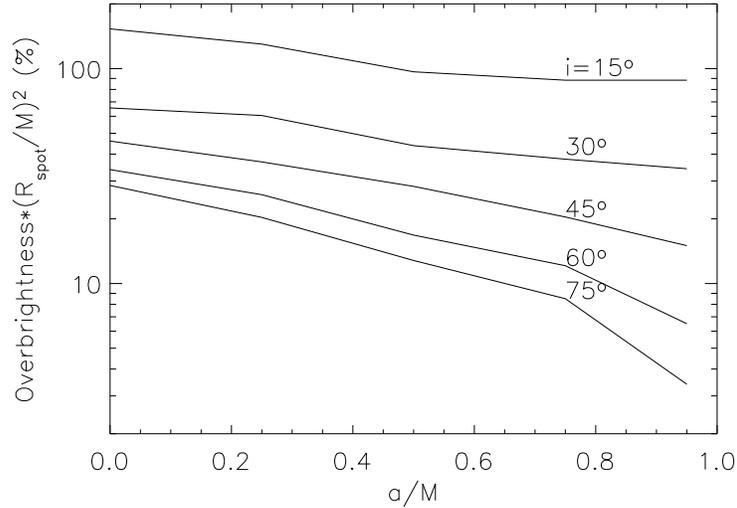}
\caption[Required hot spot overbrightness for 1\% rms
  modulation]{\label{plotfive} Overbrightness required of a hot spot
  on a circular orbit to produce a 1\% rms modulation in X-ray flux when
  added to a steady-state
  disk with $R_{\rm in}=R_{\rm ISCO}$, $R_{\rm out}=15M$ and emissivity
  $g(r)\propto r^{-2}$. An overbrightness of 100\% means the peak hot
  spot emissivity is twice that of the steady-state disk with no hot
  spot. The spot size $R_{\rm spot}$ is measured in gravitational
  radii $M$, so for a black hole with $a/M=0.5$ and $i=60^\circ$, the
  required overbrightness for a hot spot with $R_{\rm spot}=0.5M$
  would be 67\%.}
\end{center}
\end{figure}

Figure \ref{plotfive} shows the required overbrightness of a flattened
Gaussian hot spot orbiting near the ISCO to produce a modulation with
rms amplitude of 1\% for a range of
inclinations and black hole spin parameters. In the limit of
a face-on accretion disk $(i=0^\circ)$, even an infinitely bright spot on
a circular orbit
will not produce a time-varying light curve. As the inclination
increases, the required overbrightness decreases, since the special
relativistic beaming focuses radiation toward the observer from a
smaller region of the disk, \textit{increasing} the relative
contribution from the hot spot. As the spin of the black hole
increases, the ISCO moves in toward the horizon and the velocity of a
trajectory near that radius increases, as does the gravitational
lensing, further magnifying the contribution from the hot spot. This result
seems to predict an observational preference for high-inclination,
high-spin systems when detecting HFQPOs. As the number of black hole
LMXB observations increases, the growing data set seems to confirm
this prediction with regard to binary inclination and possibly spin as
well [see \citet{mccli04} and references therein].

Understandably, the required overbrightness is inversely proportional
to the area of the hot spot so we should expect
\begin{equation}
[{\rm overbrightness}]*R_{\rm spot}^2 = \mbox{ const}.
\end{equation}
For example, from Figure \ref{plotfive} we see that a black
hole binary with inclination $i = 60^\circ$ and spin $a/M=0.5$ would
require a spot size of $R_{\rm spot}= 0.5M$ with 67\%
overbrightness (e.g.\ 14\% temperature excess for blackbody emission)
to produce a 1\% rms modulation in the light curve. This is well
within the range of the typical size and magnitude of fluctuations
predicted by MHD calculations of 3-dimensional
accretion disks \citep{hawle01,devil03a}. The hot spot model is
well-suited for simplified calculations of the X-ray emission from
these random fluctuations in the accretion disk. By adding the
emission from small, coherent hot spots to the flux from a steady-state
disk, we can interpret the amplitudes and positions of
features in the QPO spectrum in terms of a model for the black hole
mass, spin, and inclination.

\begin{figure}[tb]
\begin{center}
\includegraphics[width=0.7\textwidth]{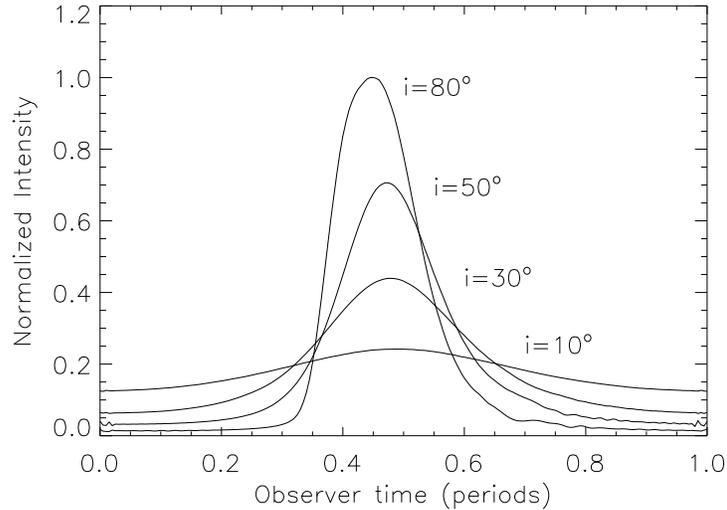}
\caption[Light curves for a variety of inclinations]{\label{plotsix}
  Frequency-integrated light curves of an 
  orbiting hot spot at the ISCO of a Schwarzschild black hole for
  different disk inclination angles. The spot is moving in the
  $-\mathbf{e}_{\hat{\phi}}$ direction as in Figure \ref{plotfour}
  with $\phi(t=0)=180^\circ$. For high
  inclination angles, the special relativistic beaming causes the light
  curve to become sharply peaked as the hot spot moves toward the
  observer.}
\end{center}
\end{figure}

\subsection{Harmonic Dependence on Inclination and Spin}\label{harm_inc_spin}
Considering the X-ray flux from the hot spot alone, the
frequency-integrated light curves for a variety of inclinations are 
shown in Figure \ref{plotsix}. All light curves are shown for one
period of a hot spot orbiting a Schwarzschild black hole at the
ISCO. As the inclination increases, the light curve goes from nearly
sinusoidal to being sharply peaked by special relativistic
beaming. Thus the shape of a hot spot light curve may be used to
determine the disk inclination. With current observational
capabilities, it is not possible even for the brightest sources to get
a strong enough X-ray signal over individual periods as short as 3-5
msec to be able to differentiate between the light curves in Figure
\ref{plotsix}. Instead, the Fourier power spectrum can be used to
identify the harmonic features of a periodic or quasi-periodic light
curve over many orbits. Disks with higher inclinations will produce
more power in the higher harmonic frequencies,
due to the ``lighthouse'' effect, as the hot spot emits a
high-power beam of photons toward the observer once per orbit,
approximating a periodic delta-function in time. 
 
\begin{figure}[ht]
\begin{center}
\includegraphics[width=0.45\textwidth]{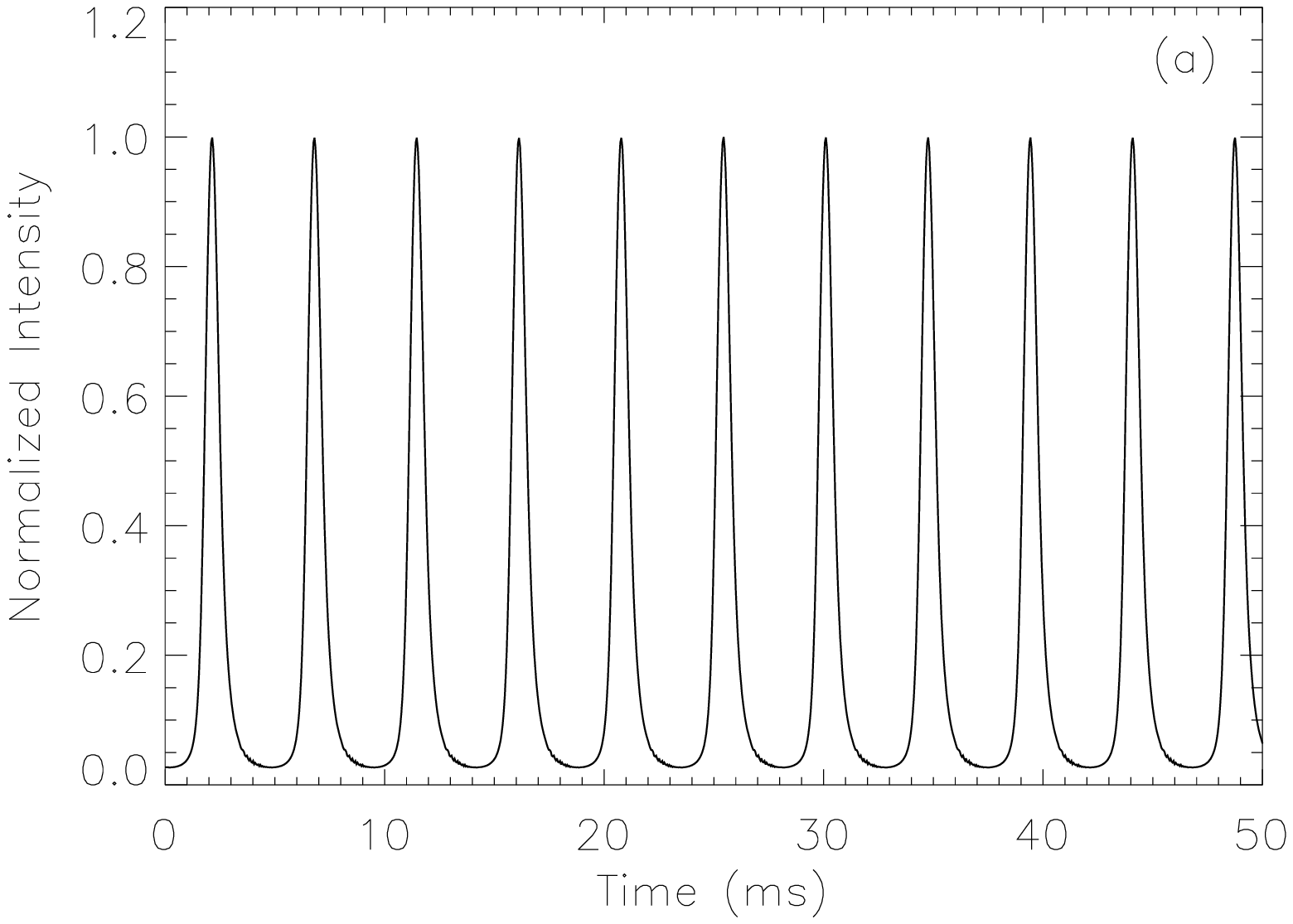}
\includegraphics[width=0.45\textwidth]{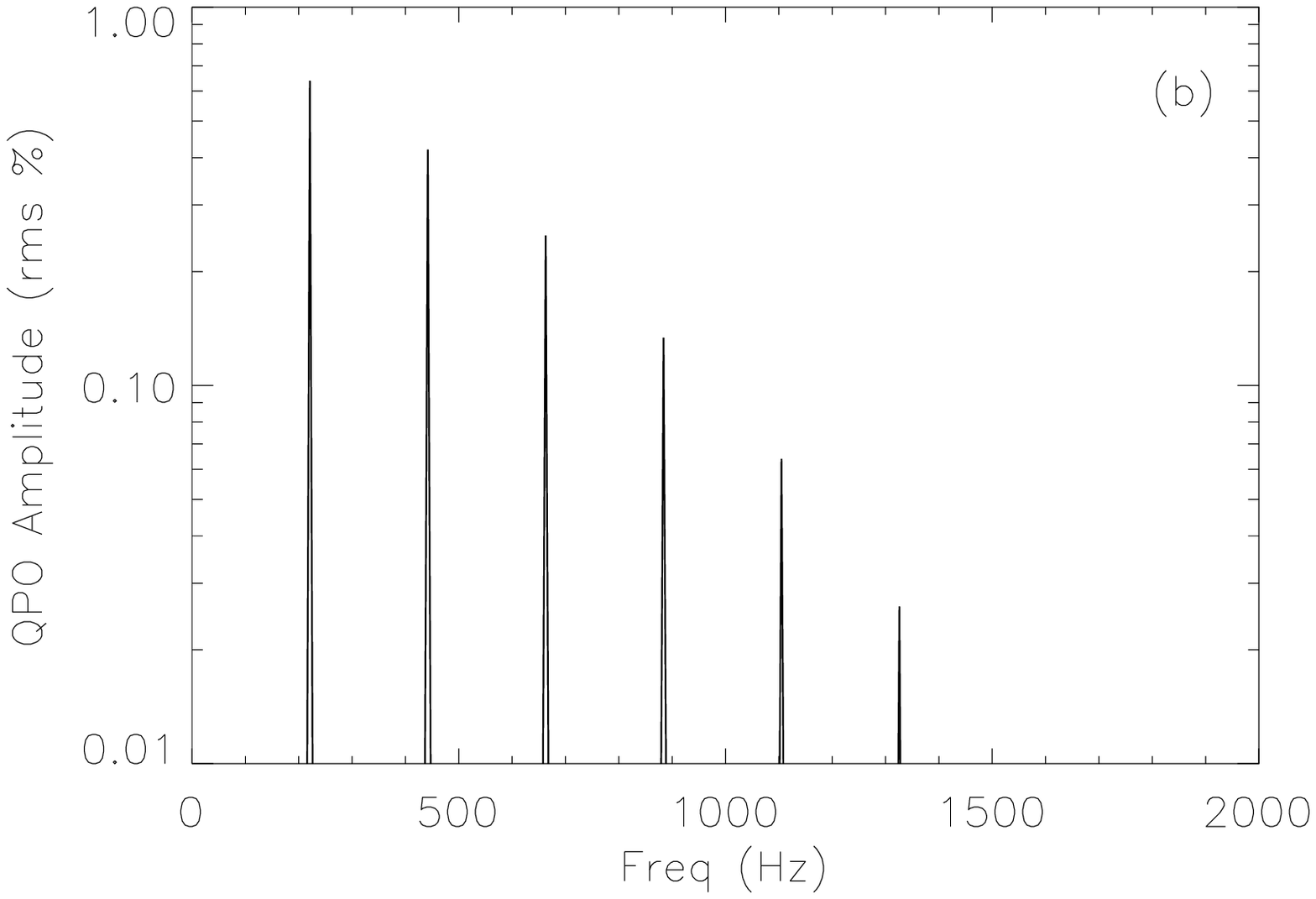}
\caption[Light curve and power spectrum for circular hot spot
orbit]{\label{plotseven} (a) X-ray light curve of an
  orbiting hot spot with same parameters as in Figure
  \ref{plotfour}. (b) Fourier amplitude $a_n({\rm rms})$ of the
  above light curve with overbrightness of unity, normalized to
  the flux from a steady-state disk as in equation (\ref{arms}), showing the
  fundamental Kepler frequency at 220 Hz for $M=10M_\odot$. The
  non-sinusoidal shape of the light curve, due largely to beaming
  effects, is characterized by the declining power in the higher harmonic
  frequencies at $n220$ Hz, where $n > 1$ is an integer.}
\end{center}
\end{figure}

Figure \ref{plotseven}a shows a sample section of such a light curve,
including only the X-ray flux from the hot spot, subtracting out the
steady-state flux from the disk. The sharp peaks in the light curve,
while unresolvable in the time domain, will give a characteristic
amount of power in the higher harmonics, shown in Figure
\ref{plotseven}b. Here we have normalized the rms amplitudes to
the background flux from the disk with a hot spot size $R_{\rm
spot}=0.5M$, overbrightness of 100\%, and inclination of
$60^\circ$. For a signal $I(t)$ with Fourier components $a_n$:
\begin{equation}
I(t) = \sum_{n=0}^\infty a_n \cos(2\pi n t),
\end{equation}
we define the rms amplitude $a_n({\rm rms})$ in each mode $n>0$ as
\begin{equation}\label{arms}
a_n({\rm rms}) \equiv \frac{a_n}{\sqrt{2a_0}}.
\end{equation}
With this normalization, the rms defined in equation (\ref{rms}) can be
conveniently written 
\begin{equation}
{\rm rms} = \sqrt{\sum_{n>0} a_n^2({\rm rms})}.
\end{equation}

\begin{figure}[ht]
\begin{center}
\includegraphics[width=0.45\textwidth]{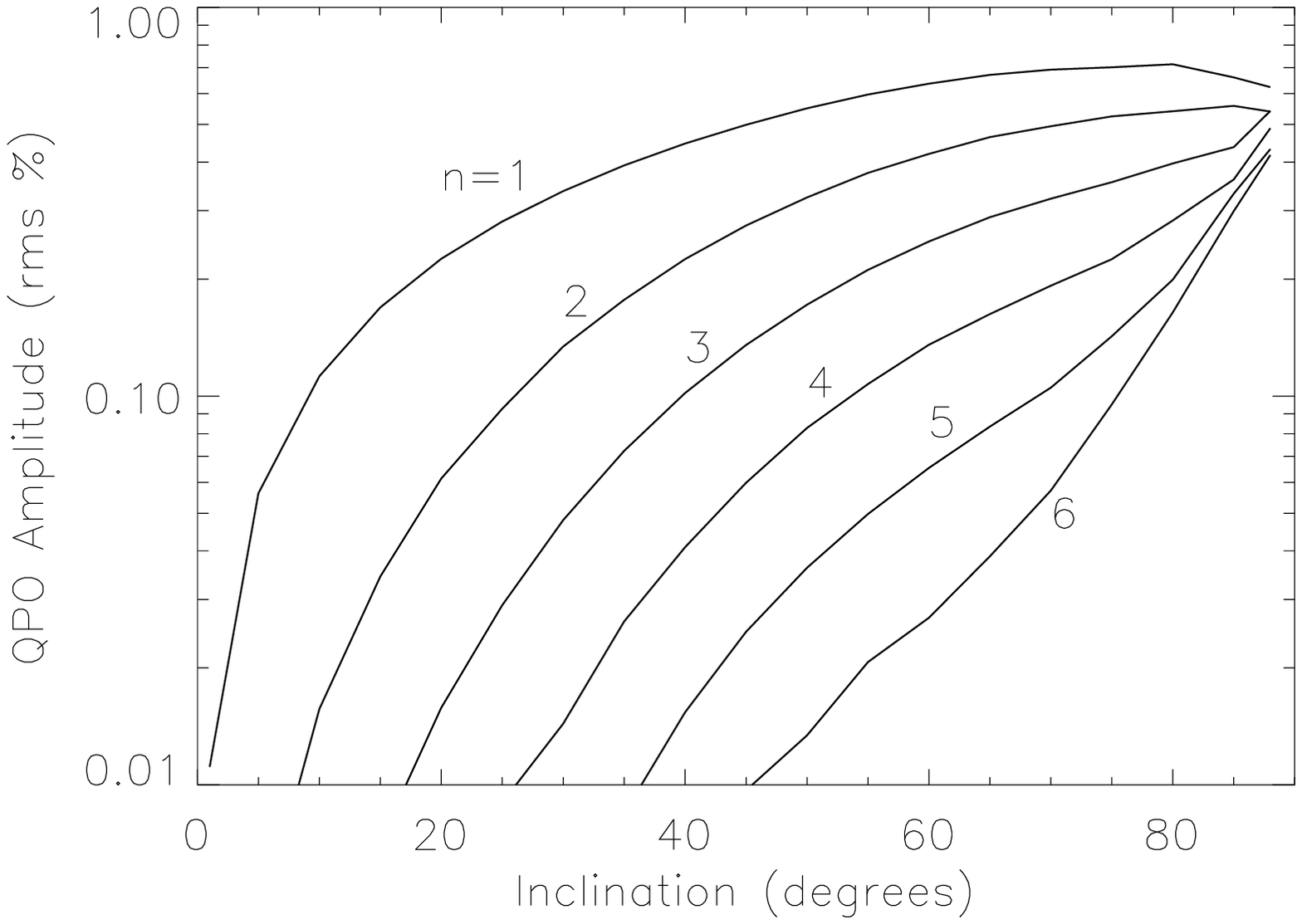}
\includegraphics[width=0.45\textwidth]{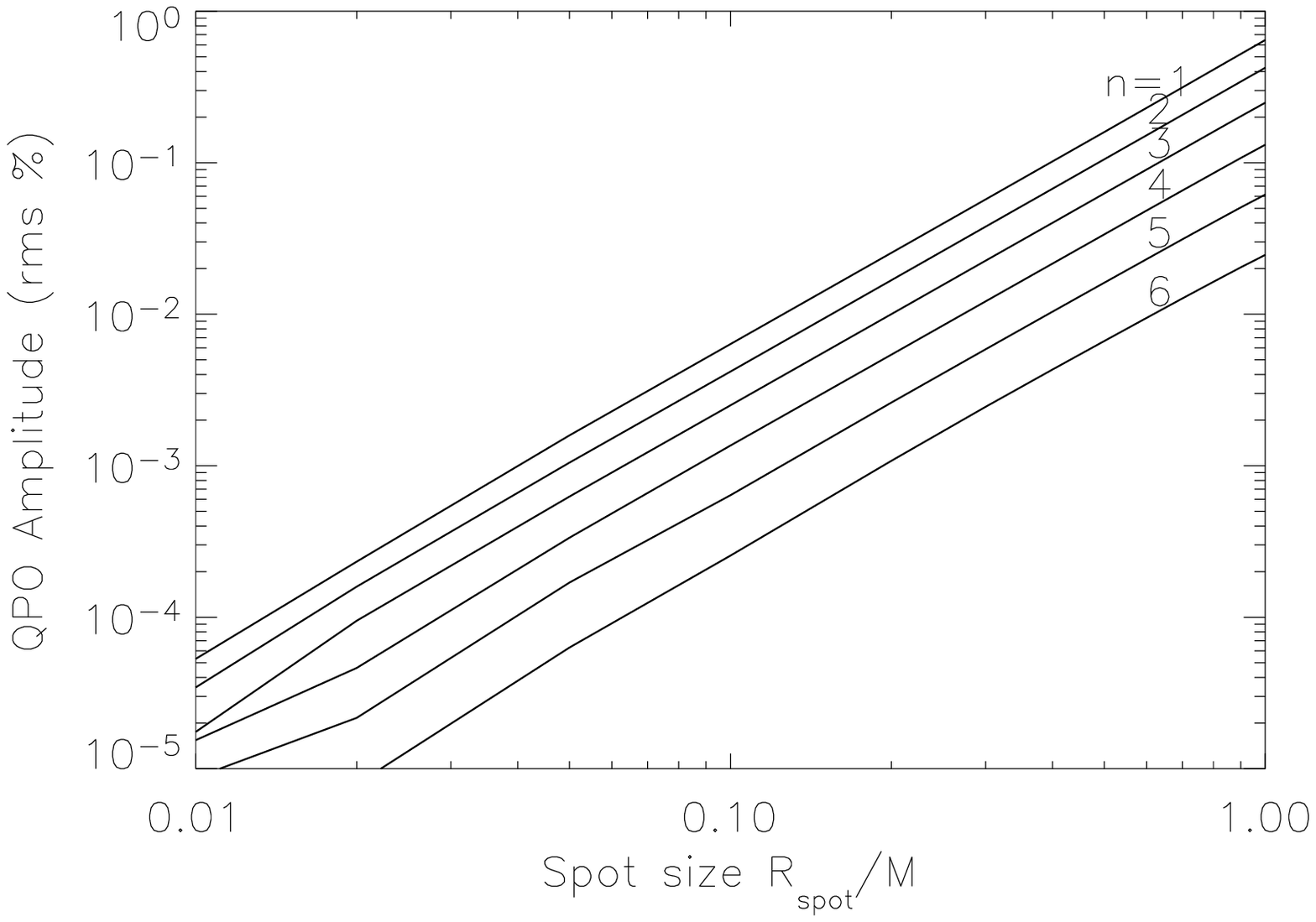}
\caption[Fourier amplitude $a_n({\rm rms})$ as a function of
inclination and hot spot size]{\label{ploteight} (a) Fourier amplitude
$a_n({\rm rms})$ in higher harmonic
  frequencies $\nu_n=n\nu_\phi$ as a function of orbital inclination
  to the observer, normalized as in equation (\ref{arms}). The hot spot
  has size $R_{\rm spot}=0.5M$, an overbrightness factor of 100\%, and
  is in a circular orbit at $R_{\rm ISCO}$ around a Schwarzschild 
  black hole. (b) The same harmonic amplitude $a_n({\rm rms})$ as a
  function of spot size $R_{\rm spot}$ with $i=60^\circ$ and a constant
  overbrightness factor of 100\%. The harmonic amplitude scales directly
  with hot spot area $a_n({\rm rms}) \propto R_{\rm spot}^2$.}
\end{center}
\end{figure}

In Figure \ref{plotseven}b, the main peak at $f=220$ Hz corresponds to
the azimuthal frequency for
an orbit at the ISCO of a $10M_\odot$ Schwarzschild black
hole. In the limit where the light curve is a periodic
delta-function in time, there should be an equal amount of power
in all harmonic modes, because the Fourier transform of a periodic
delta-function is a periodic delta-function. However, even in the
case of edge-on inclination $(i=90^\circ)$, unless the hot spot is
infinitesimally small and ultra-relativistic, the light curve will
always be a continuous function with some finite width
and non-zero minimum, thus contributing less and less power to the
higher harmonics. The harmonic dependence on inclination for a hot
spot orbiting a Schwarzschild black hole is shown in 
Figure \ref{ploteight}a. Predictably, as the inclination increases,
we see that both the absolute and relative amplitudes of the higher
harmonics increase, almost to the limit of a periodic delta-function
when $i\to 90^\circ$.

Interestingly, we find very little dependence of the relative
harmonic structure on hot spot size or shape. This emphasizes
the robustness of the simple hot spot model in interpreting an X-ray
power spectrum, without needing to include the detailed physics of the
disk perturbations. To show clearly the independence of harmonic
structure on spot size, Figure \ref{ploteight}b plots the rms amplitude
in each mode $a_n({\rm rms})$, defined as above. The overbrightness of
$100\%$ is held constant as the spot size varies. As expected, for
constant overbrightness, the amplitude in each harmonic scales
linearly with $R_{\rm spot}^2$. Thus if the combination
$[\mbox{overbrightness}]*R_{\rm spot}^2$ is held constant, the rms
amplitudes would also be constant. The small amount of numerical noise
as $R_{\rm spot}\to 0$ is 
caused by the finite resolution of the ray-tracing grid; as the hot
spot size approaches the grid size, it becomes more difficult to
accurately calculate the light curve and associated power spectrum.

\begin{figure}[ht]
\begin{center}
\scalebox{0.35}{\includegraphics*[54,300][485,740]{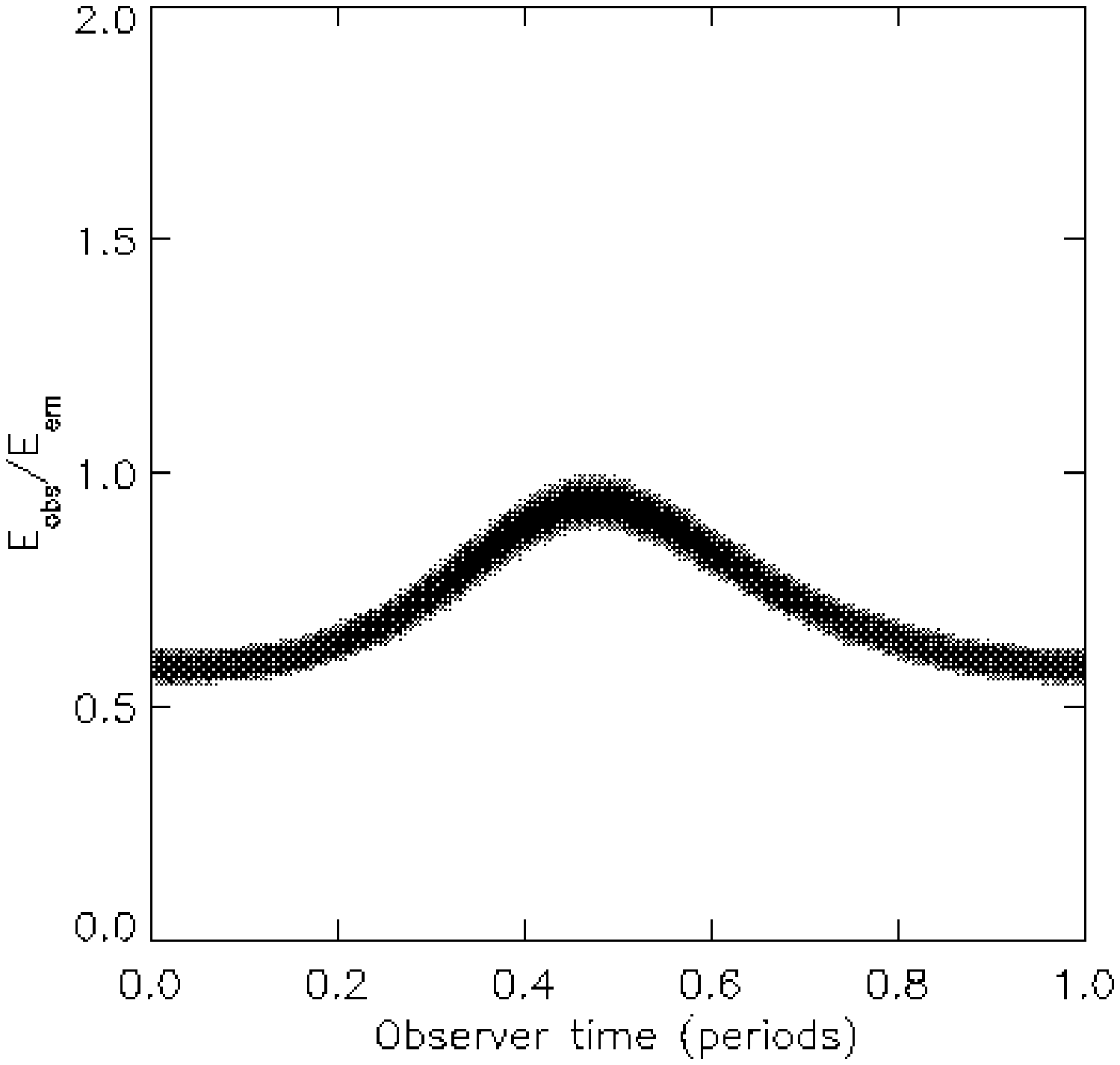}}
\scalebox{0.35}{\includegraphics*[108,300][485,740]{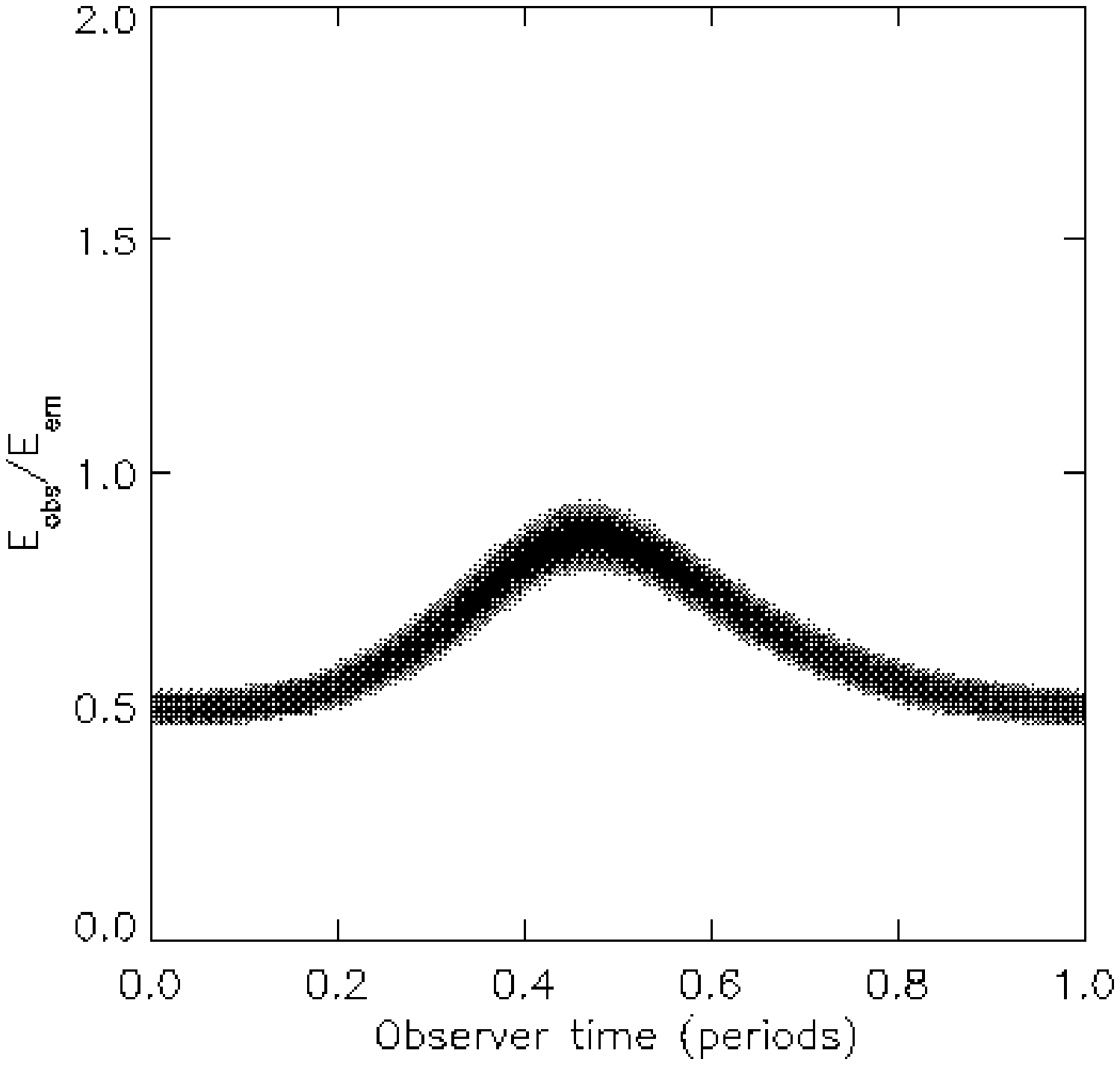}}
\scalebox{0.35}{\includegraphics*[108,300][485,740]{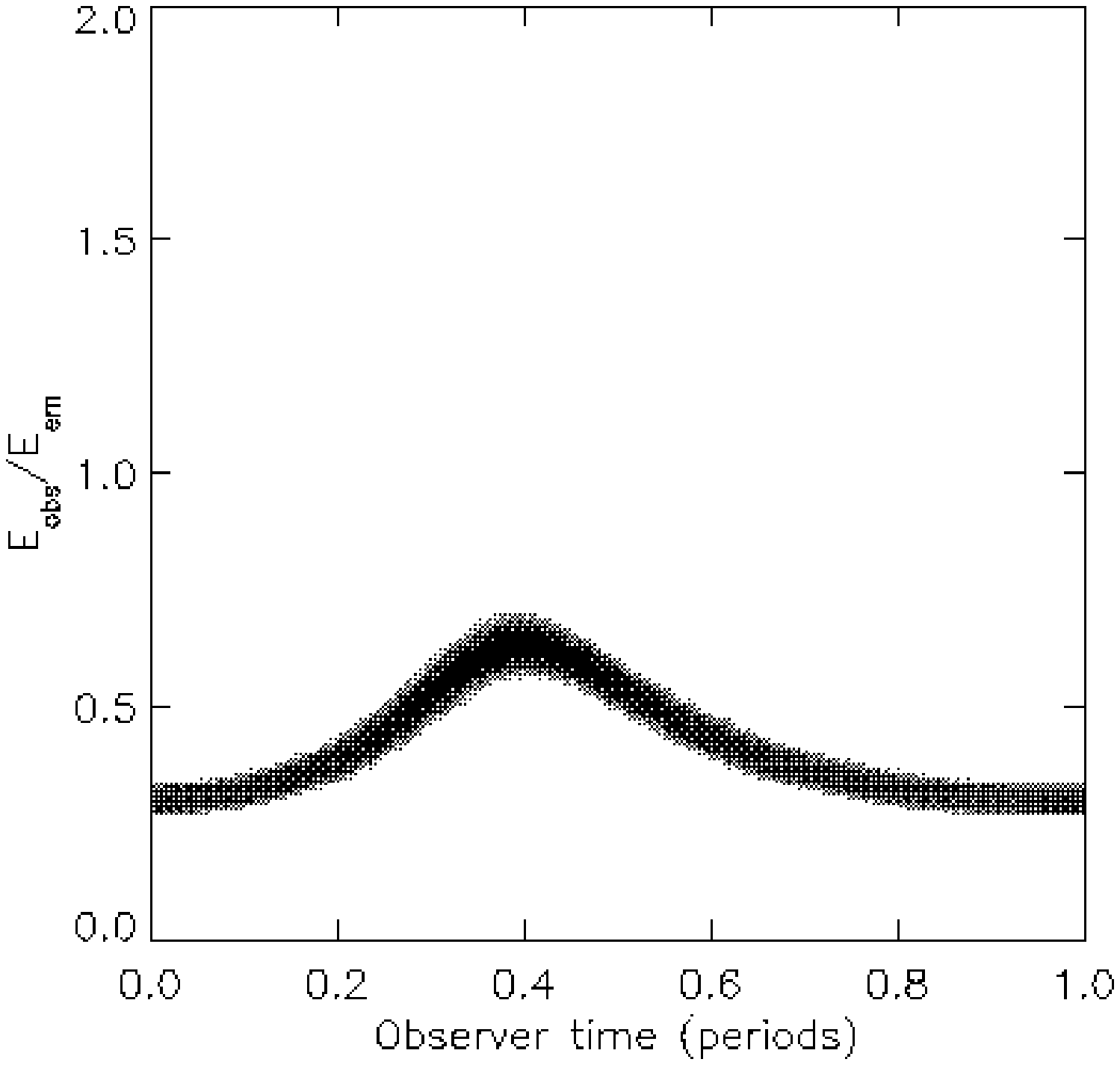}}
\caption[Spectrograms for a variety of spins,
$i=30^\circ$]{\label{spectrogram30} Spectrograms for hot spots on
circular orbits at $R_{\rm ISCO}$ for $i=30^\circ$ and a variety of
spins: $a/M=0,0.5,0.9$. As the spin increases, the ISCO moves in
towards the horizon, so the gravitational redshift is greater,
lowering the observed frequency from a monochromatic emitter.}
\end{center}
\end{figure}

\begin{figure}[ht]
\begin{center}
\scalebox{0.35}{\includegraphics*[54,300][485,740]{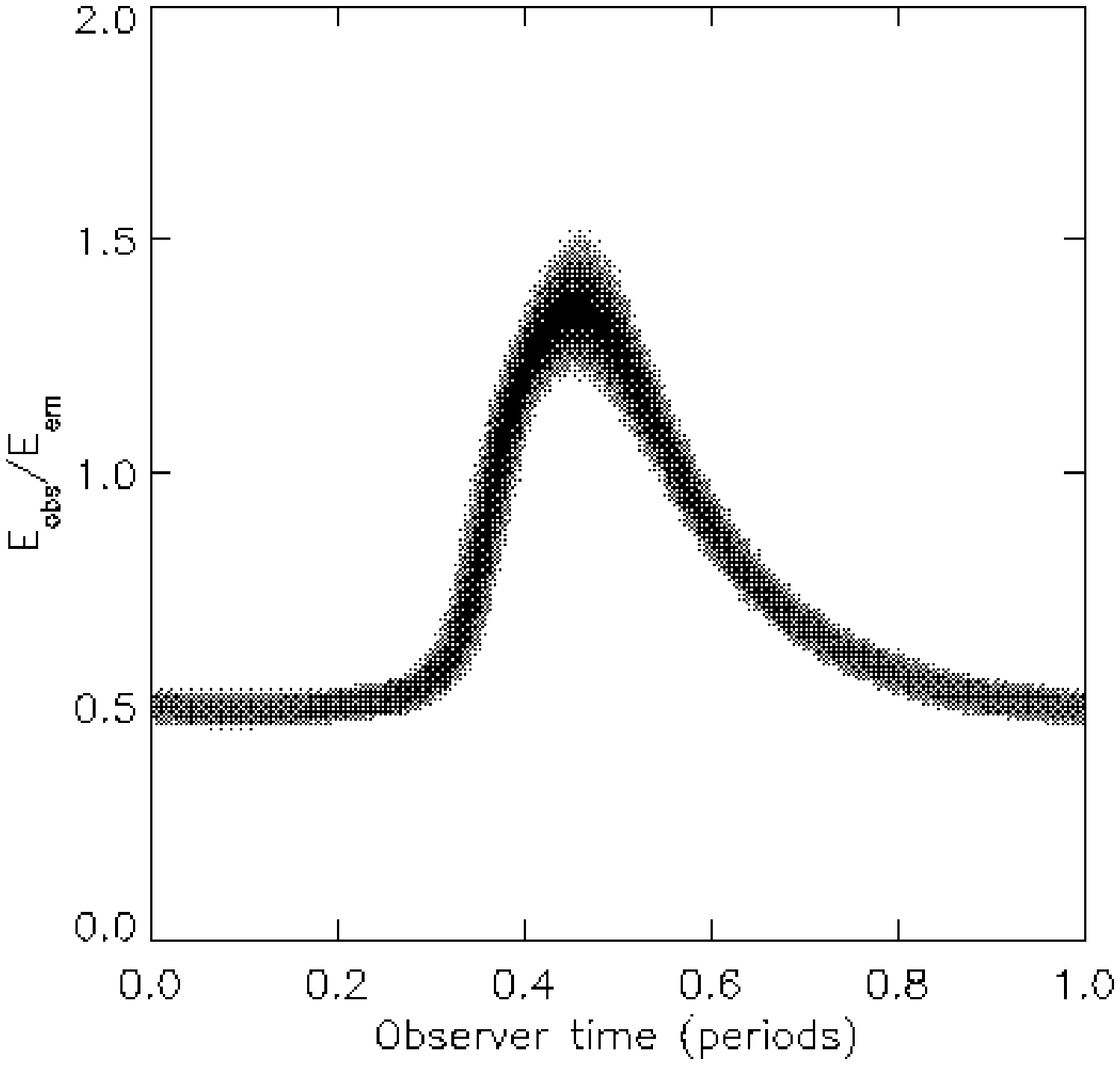}}
\scalebox{0.35}{\includegraphics*[108,300][485,740]{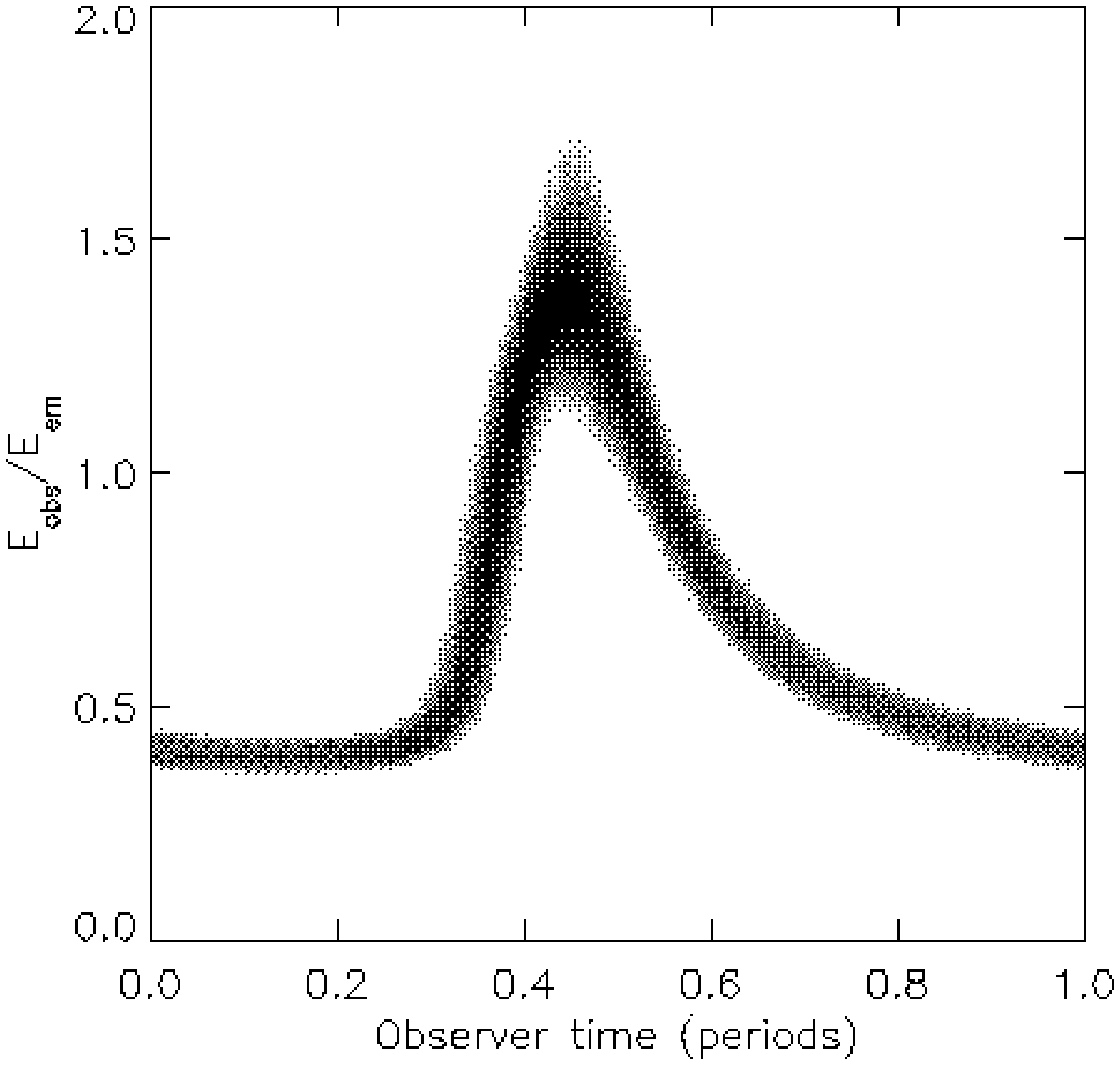}}
\scalebox{0.35}{\includegraphics*[108,300][485,740]{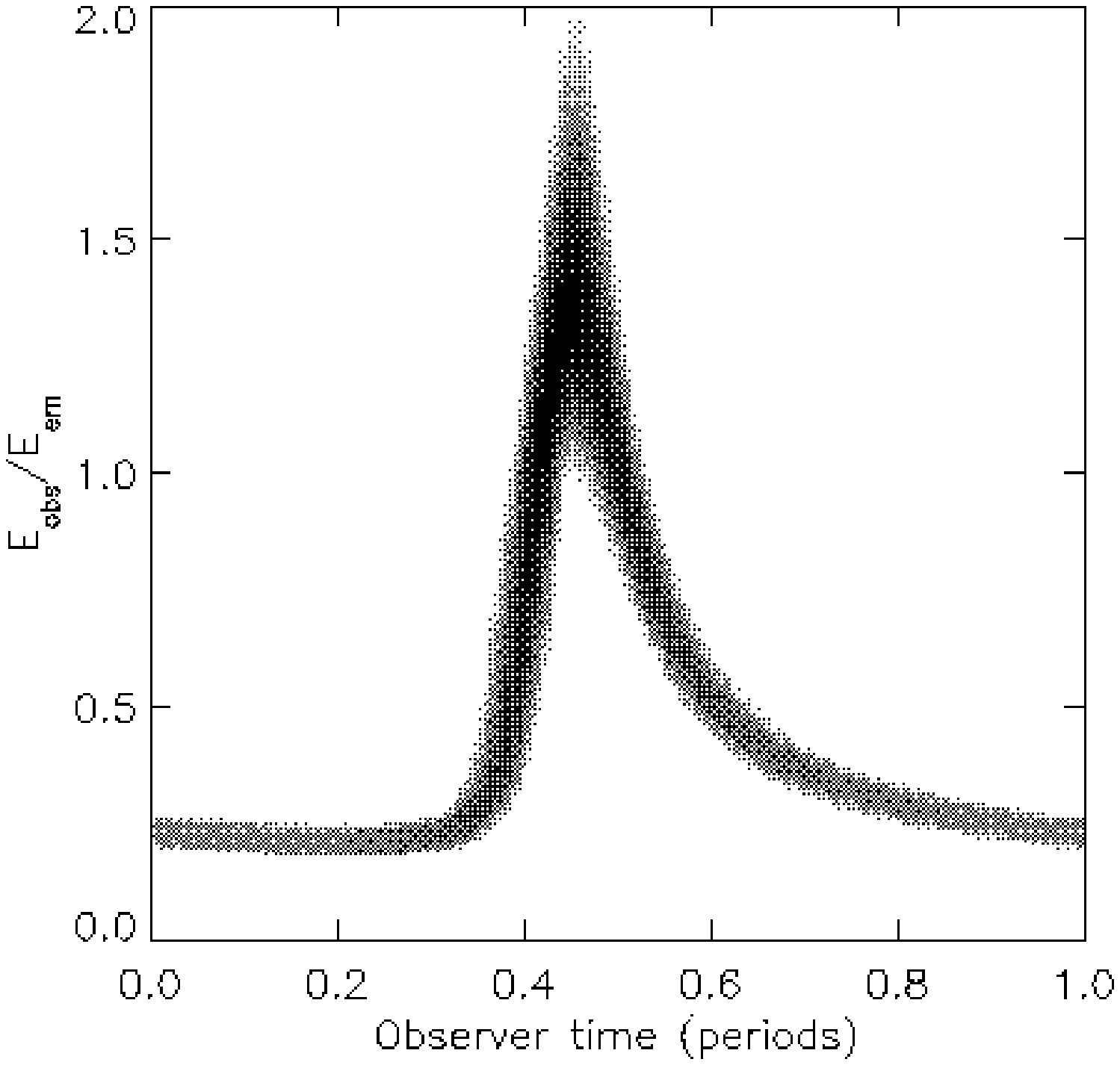}}
\caption[Spectrograms for a variety of spins,
$i=75^\circ$]{\label{spectrogram75} Spectrograms for hot spots on
circular orbits at $R_{\rm ISCO}$ for $i=75^\circ$ and a variety of
spins: $a/M=0,0.5,0.9$. In addition to the increased gravitational
redshift for higher spin, the orbital velocity increases, in turn
increasing the relativistic beaming and blueshift as the hot spot
moves towards the observer.}
\end{center}
\end{figure}

\begin{figure}[ht]
\begin{center}
\includegraphics[width=0.7\textwidth]{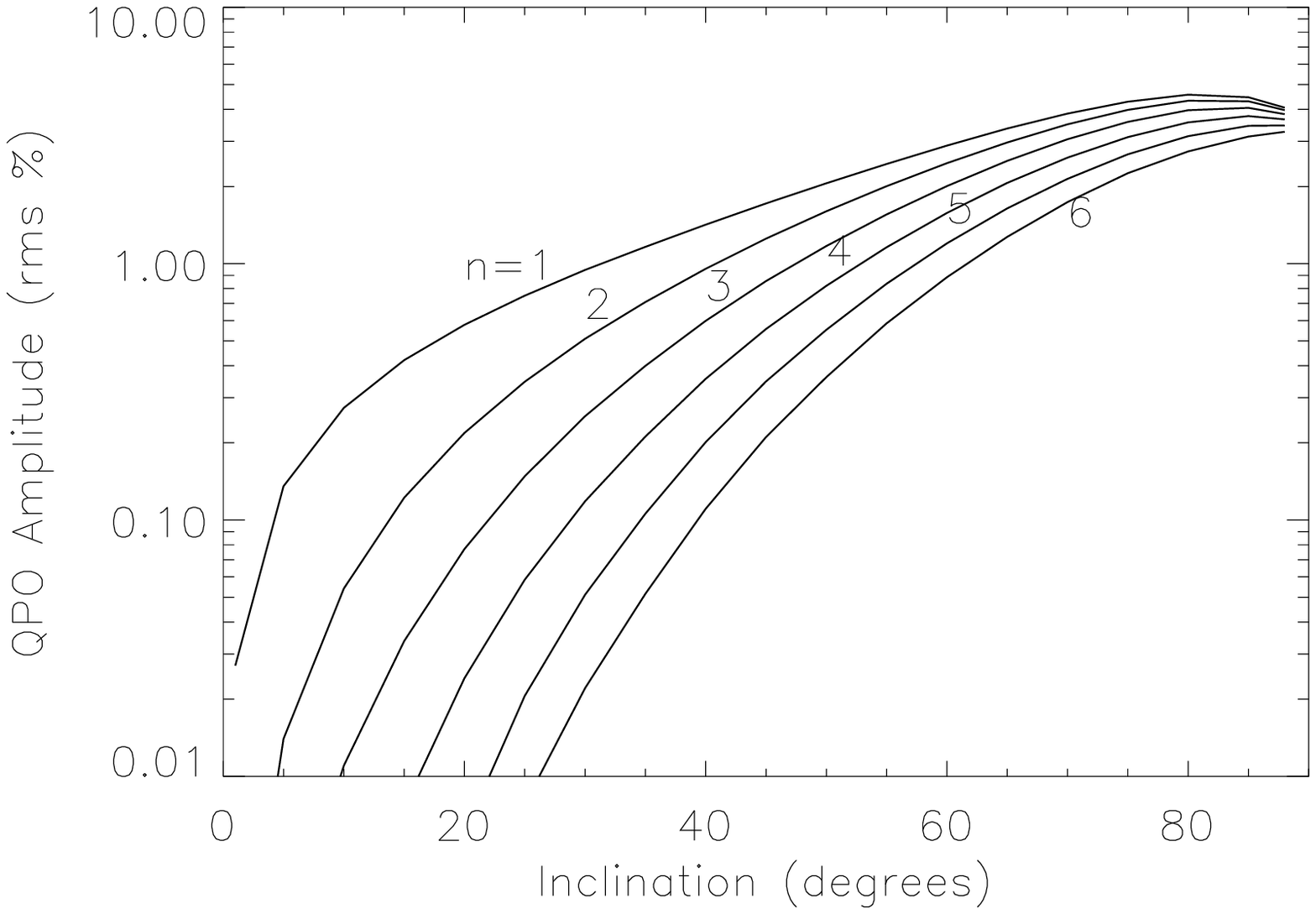}
\caption[Fourier amplitude $a_n({\rm rms})$ as a function of
inclination for $a/M=0.9$]{\label{harmonics9} (a) Fourier amplitude
$a_n({\rm rms})$ in higher harmonic
  frequencies $\nu_n=n\nu_\phi$ as a function of orbital inclination
  to the observer, normalized as in equation (\ref{arms}). The hot spot
  has size $R_{\rm spot}=0.15M$, an overbrightness factor of 100\%, and
  is in a circular orbit at $R_{\rm ISCO}$ around a black hole with
$a/M=0.9$.}
\end{center}
\end{figure}

As the spin parameter increases for Kerr black holes, the ISCO moves
closer to the horizon, increasing the circular velocities of particles
on the ISCO and thus the Doppler shifts, giving broader
photon energy spectra, as seen in Figure \ref{plotthree}b. The phase
lag in time of the peak 
blueshift with respect to angular phase of the hot spot is
also amplified for these smaller values of $R_{\rm ISCO}$, giving light
curves that are asymmetric in time. Figure \ref{spectrogram30} shows
spectrograms as in Figure \ref{plotfour}, now for inclination of
$i=30^\circ$ and spin values of $a/M=0,0.5,0.9$. In each case the hot
spot is on a circular orbit at the ISCO. Figure
\ref{spectrogram75} repeats these results for a higher inclination of
$i=75^\circ$. Clearly the beaming and redshift effects increase for
higher spin values. In Figure \ref{harmonics9}, we plot the harmonic
power as a function of inclination for $a/M=0.9$, as in Figure
\ref{ploteight}a. For the higher spin, the lighthouse effects are
amplified, showing the high harmonic power affiliated with a periodic
delta function.

\section{Non-circular Orbits}\label{noncircular_orbits}
One of the major unsolved puzzles motivating theoretical models of black
hole QPOs is the observation of multiple peaks in the high frequency
power spectrum \citep{mccli04}. As discussed above, any non-sinusoidal
light curve 
will contribute to Fourier power in harmonics at integer multiples of
the fundamental orbital frequency. However, for at least three X-ray
binary systems (XTE J1550--564, GRO J1655--40, and H1743--322;
possibly also GRS 
1915+105), peaks are found with rational (but non-integer) frequency
ratios \citep{mille01,stroh01b,remil02,remil04,homan04}. In these
particular examples, significant power is measured around the
frequencies (184, 276 Hz) for XTE J1550--564, (160, 240 Hz) for
H1743--322, and (300, 450 Hz) for GRO
J1655--40, almost exactly a 2:3 commensurability in
frequencies, while GRS 1915+105 has peaks at 40 and 67 Hz. Following
the work of \citet{merlo99}, we investigate the 
possibility of these commensurabilities coming from integral
combinations of the radial and azimuthal coordinate frequencies of
nearly circular geodesics around a Kerr black hole. 

In a Newtonian point mass potential, the radial, azimuthal, and vertical
(latitudinal) frequencies $\nu_r,
\nu_\phi,$ and $\nu_\theta$ are identical, giving closed planar
elliptical orbits. For the Schwarzschild metric the vertical and
azimuthal frequencies are identical, giving planar rosette orbits that
are closed only for a discrete set of initial conditions. The Kerr
metric allows three unique coordinate frequencies, so geodesic orbits
in general can fill out a 
torus-shaped region around the black hole spin axis. When these
coordinate frequencies are rational multiples of each other, the
trajectories will close after a finite number of orbits. 

While there
is currently no clear physical explanation for why hot spots may
tend toward such trajectories, some recent theoretical work suggests
the possible
existence of nonlinear resonances near geodesic orbits with integer
commensurabilities \citep{abram03a,horak04,rebus04}. Another important clue
may come from the fact that these special orbits are closed,
perhaps enhancing the
stability of non-circular trajectories. The quasi-periodic nature of
the oscillations suggest the continual formation and subsequent
destruction of hot spots \textit{near}, but not exactly at, the
resonant orbits (see Chapter 4 below). For the purposes of this
Thesis, we will take the 
apparent preference for such orbits as given and concern ourselves
primarily with calculating the resulting light curves and power spectra. 

In geometrized units with $G=c=M=1$, coordinate time is measured in
units of $4.9\times 10^{-6} (M/M_\odot)$ sec. For example, an orbit
with angular frequency $\Omega_\phi=2\pi\nu_\phi=0.1$ around a $10M_\odot$
black hole would have an observed period of 3.1 ms, whereas the analogous
orbit around a supermassive black hole with mass $10^9M_\odot$ would
have a period of 86 hours. In these units, the
three fundamental coordinate frequencies for nearly
circular orbits are given by \citet{merlo99} [following earlier work by
\citet{barde72,perez97}]:
\begin{subequations}
\begin{equation}\label{omega_phi2}
\Omega_\phi = 2\pi\nu_\phi = 
\frac{1}{r^{3/2}\pm a},
\end{equation}
\begin{equation}\label{omega_theta}
\Omega_\theta = 2\pi\nu_\theta = \Omega_\phi \left[1\mp
\frac{4a}{r^{3/2}}+\frac{3a^2}{r^2}\right]^{1/2},
\end{equation}
and
\begin{equation}\label{omega_r}
\Omega_r = 2\pi\nu_r = \Omega_\phi
\left[1-\frac{6}{r} \pm
\frac{8a}{r^{3/2}}-\frac{3a^2}{r^2}\right]^{1/2},
\end{equation}
\end{subequations} 
where the upper sign is taken for prograde orbits and the lower sign
is taken for retrograde orbits. The radial frequency approaches zero
at $r\to R_{\rm ISCO}$, where geodesics can orbit the black hole many
times with steadily decreasing $r$. In the limit of zero spin and large
$r$, the coordinate frequencies reduce to the degenerate
Keplerian case with $\Omega_\phi = \Omega_\theta = \Omega_r = r^{-3/2}$.

\begin{figure}[ht]
\begin{center}
\includegraphics[width=0.7\textwidth]{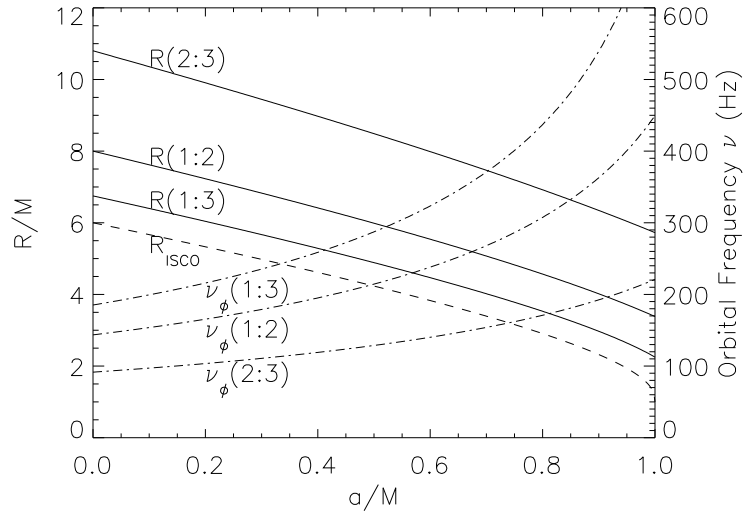}
\caption[Radius of orbits with commensurate
frequencies as a function of spin] {\label{plotnine} Radius of
  prograde orbits with
  commensurate frequencies $\nu_r:\nu_\phi = $(1:3, 1:2, 2:3) (solid
  lines) as a function of dimensionless spin parameter $a/M$. The ISCO
  (dashed line) corresponds to $\nu_r:\nu_\phi = 1:\infty$. Also
  shown are the respective orbital frequencies $\nu_\phi$
  at these radii for a black hole with mass $10M_\odot$ (dot-dashed
  lines).
}
\end{center}
\end{figure}

To model the 2:3 frequency commensurability, we begin by looking for
perturbed circular planar orbits where the radial frequency $\nu_r$ is
one-third the azimuthal frequency $\nu_\phi$. Since the orbits are
nearly circular, the fundamental mode of the light curve should peak
at the azimuthal frequency with additional power in beat modes at
$\nu_\phi \pm \nu_r$. For $\nu_r$:$\nu_\phi$ = 1:3, the power spectrum
should have a triplet of peaks with frequency ratios 2:3:4. These
commensurate orbits can be found easily from equations (\ref{omega_phi2})
and (\ref{omega_r}) and solving for $r$:
\begin{equation}
\left[1-\left(\frac{\nu_r}{\nu_\phi}\right)^2\right]r^2
-6r\pm8ar^{1/2}-3a^2=0.
\end{equation}
Figure \ref{plotnine} shows the radius (solid lines) of these special
orbits as a function of spin parameter. The orbital frequencies
$\nu_\phi$ are plotted against the right-hand axis (dot-dashed lines)
for a black hole with mass $10M_\odot$. Also shown (dashed line) is
the inner-most stable circular orbit for prograde trajectories. The
position of the 1:3 commensurate orbits follows 
closely outside the ISCO, suggesting a connection
between the high frequency QPOs and the black hole ISCO. However,
other integer commensurabilities such as 1:2, 2:5, or 1:4 also closely
follow the ISCO curves for varying $a$, so the proximity to the ISCO
alone is probably not enough to explain the hot spot preference for
these specific coordinate frequencies. It is important to note that
any given black hole source will have a constant value of $a/M$,
certainly over the lifetime of our observations. Thus, we may need to
observe many more sources like XTE J1550--564 and GRO J1655--40 in order
to better 
sample the parameter space of Figure \ref{plotnine} and thus the
connection between certain preferred orbits and the black hole
ISCO. 

\begin{figure}[ht]
\begin{center}
\includegraphics[width=0.7\textwidth]{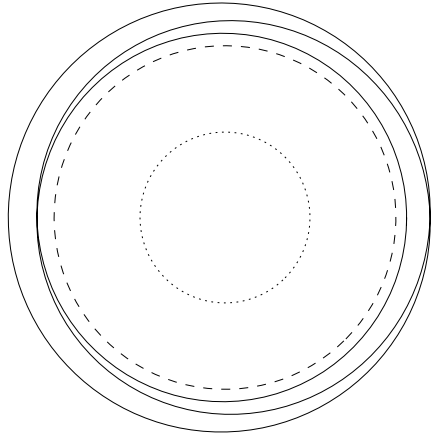}
\caption[Trajectory of hot spot with closed rosette orbit]
{\label{part_traj31} Overhead view ($i=0^\circ$) of hot spot
trajectory with eccentricity $e=0.1$ and commensurate coordinate
frequencies $\nu_\phi=3\nu_r$, giving a closed rosette orbit where
the hot spot circles the black hole three times between subsequent
pericenter passages. The dotted and dashed lines are the horizon and
ISCO, respectively, for a spin of $a/M=0.5$.}
\end{center}
\end{figure}

A 1:3 commensurate trajectory moves through three revolutions in
azimuth for each radial period,
forming a closed rosette of three ``layers.'' For such rosettes, the
eccentricity can be defined as 
\begin{equation}
e \equiv \frac{r_{max}-r_{min}}{r_{max}+r_{min}} = \frac{\Delta r}{r_0}.
\end{equation}
A ``birds-eye view'' of such an orbit is shown in Figure
\ref{part_traj31} for $a/M=0.5$, where the horizon is shown as a
dotted line, the ISCO is a dashed line, and the commensurate
rosette trajectory is plotted as a solid line. 

\begin{figure}[ht]
\begin{center}
\includegraphics[width=0.45\textwidth]{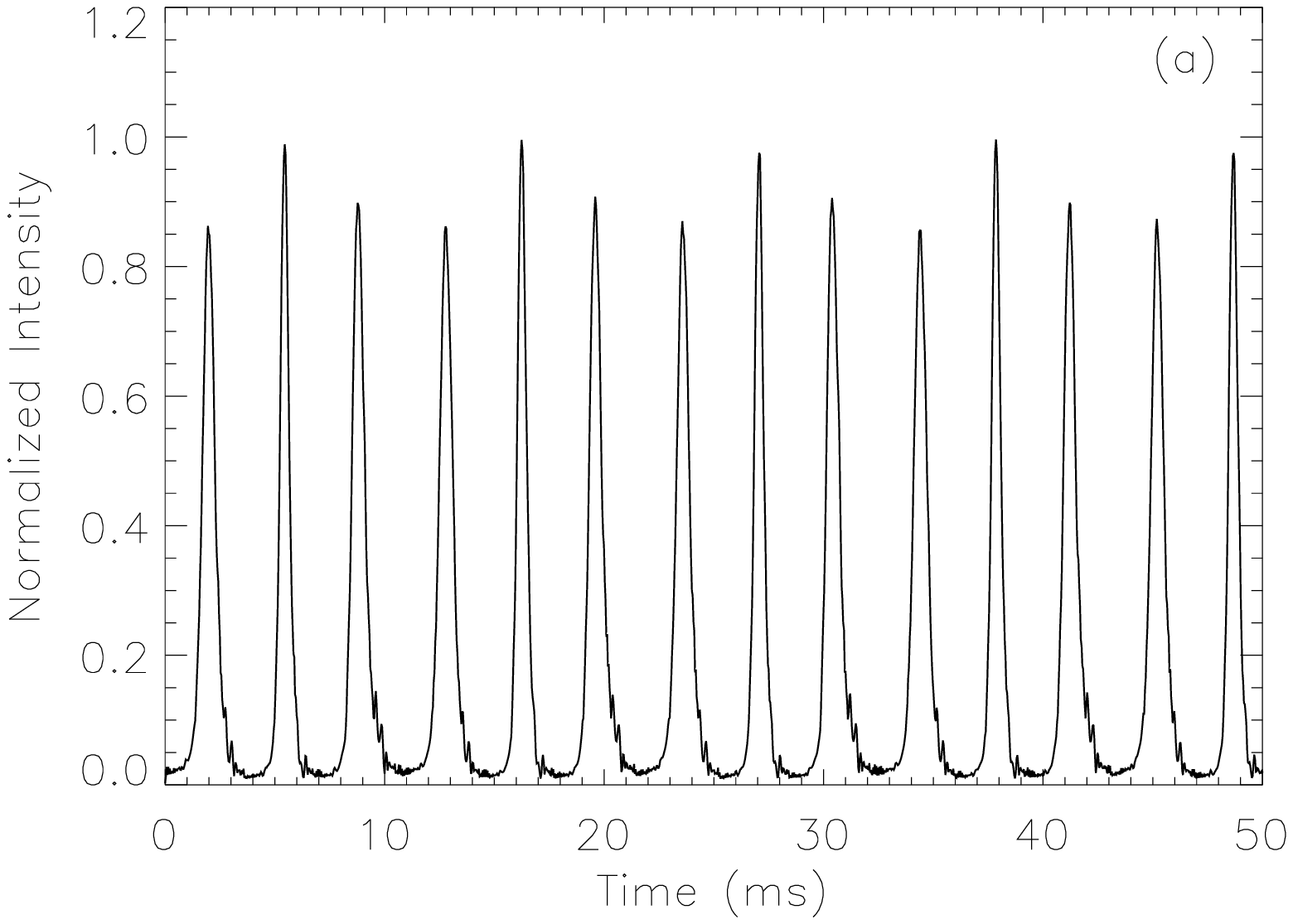}
\includegraphics[width=0.45\textwidth]{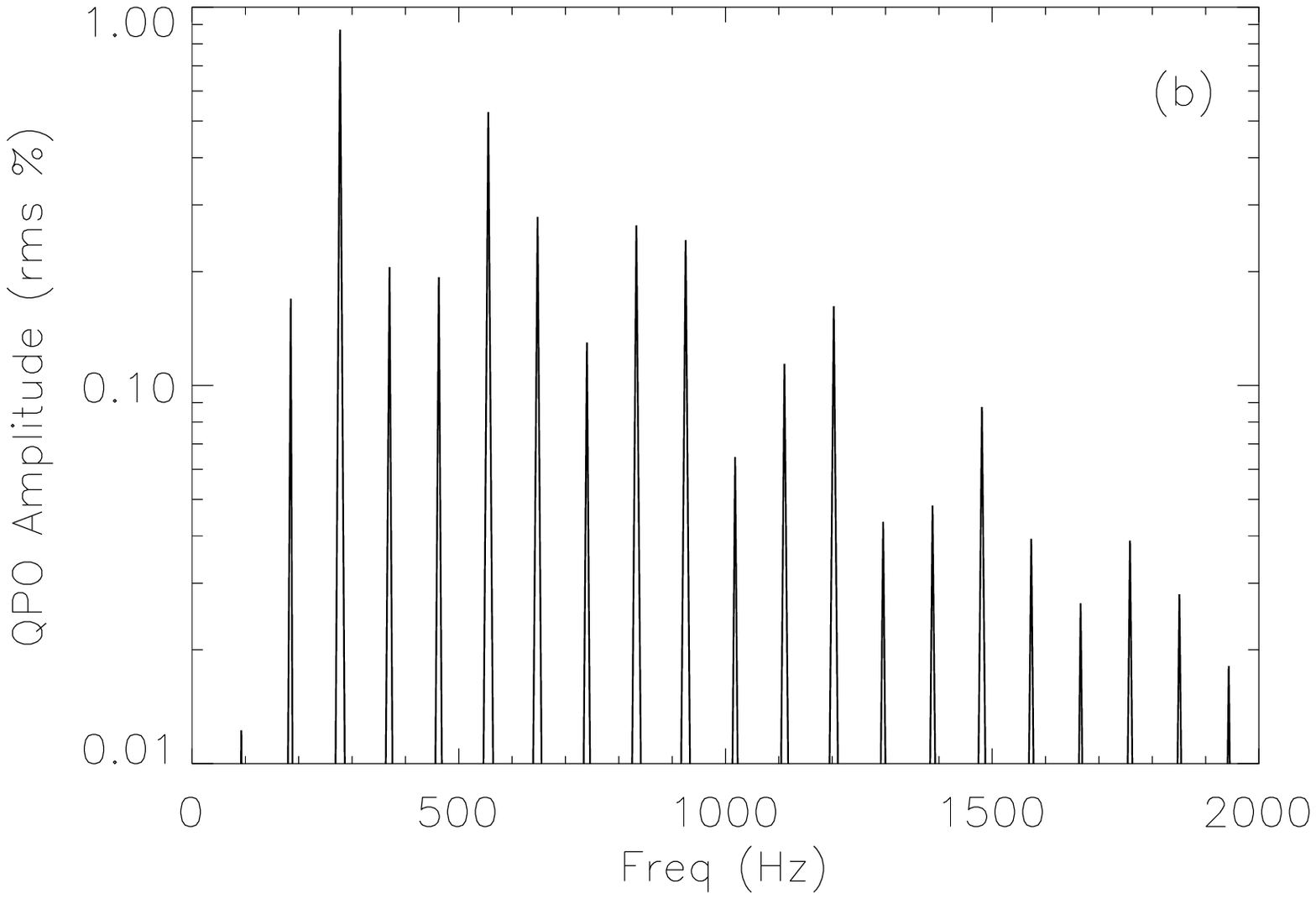}
\caption[Light curve and power spectrum for eccentric hot spot
orbit]{\label{plotten} (a) X-ray light curve of 
  a hot spot orbit with $\nu_\phi=3\nu_r$, $e=0.089$, $M=10M_\odot$,
  $a/M=0.5$, $i=60^\circ$, and $R_{\rm spot}=0.5M$. (b) The Fourier
  amplitude $a_n({\rm rms})$ of the above light curve, normalized as
  in Figure \ref{plotseven}b, showing the fundamental Kepler frequency at
  $\nu_\phi=285$ Hz and beat modes at $\nu = n\nu_\phi\pm \nu_r$.}
\end{center}
\end{figure}

The time-dependent light curve for a prograde orbit with eccentricity
$e=0.089$, spin $a/M=0.5$, and inclination $i=60^\circ$ is shown in Figure 
\ref{plotten}a. The time axis begins at the point when the hot
spot is at apocenter, moving away from the observer. Thus the first
and third peaks come from the hot spot moving toward the observer at
a relatively larger radius, while 
the second, higher peak is caused by the emitter moving toward the
observer through pericenter at a higher velocity, giving a larger
blueshift and thus
beaming factor. The combined Doppler beaming and gravitational lensing
causes the peak following the pericenter peak to be slightly larger,
as the emitter is moving away from the black hole yet the light is
focused more toward the observer.

The power spectrum for this light curve is shown in Figure
\ref{plotten}b, with the strongest peaks at the azimuthal
frequency of $\nu_\phi = 285$ Hz and its first harmonic at $2\nu_\phi =
570$ Hz for $M=10M_\odot$.  Even for this modest deviation from
circularity, there is significant power in the frequencies
$\nu_\phi \pm \nu_r$. The beating of the fundamental $\nu_\phi$ with
the radial frequency $\nu_r = (1/3)\nu_\phi= 95$ Hz gives the
set of secondary peaks at $(2/3)\nu_\phi$ and
$(4/3)\nu_\phi$. Additional peaks 
occur at beats of the harmonic frequencies $n\nu_\phi \pm \nu_r$. 

It is interesting to note that there is not significant power in the
radial mode at $\nu = 95$ Hz, but only in the beats with the
fundamental azimuthal frequency and its harmonics. However, in the
limit of a face-on orientation $(i\to 0^\circ)$, the radial frequency should
dominate the light curve variation as the gravitational and transverse
Doppler redshift modulate the intensity as a function of the hot
spot's radial coordinate. The radial mode should also be present in
the limit of an edge-on orientation $(i\to 90^\circ)$, as gravitational
lensing becomes more important, and the hot spot will experience more
magnification when closer to the black hole. These two effects
actually depend on the inclination in opposite ways. The gravitational
redshift and transverse Doppler shift will make the light curve have a
minimum at pericenter, where the gravitational lensing is stronger,
giving a relative maximum. 

\begin{figure}[ht]
\begin{center}
\includegraphics[width=0.7\textwidth]{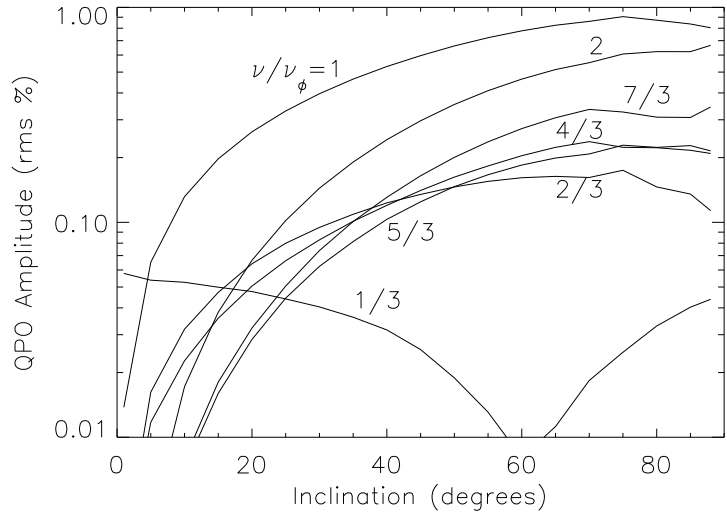}
\caption[Harmonic power in beat modes as a function of
inclination]{\label{ploteleven} Power in low-order harmonics and beat
  modes with frequencies $\nu = n\nu_\phi \pm \nu_r$, as a function of
  disk inclination angle. The hot spot trajectory is the same as in
  Figure \ref{plotten}, with $a_n({\rm rms})$ normalized as in
  Figure \ref{ploteight}.
  The curves are labeled by the ratio $\nu/\nu_\phi$.}
\end{center}
\end{figure}

In Figure \ref{ploteleven}, which shows the dependence on inclination
of the lower order harmonics and beat modes, these two
competing effects are clearly evident in the power at $\nu_r$,
canceling each other out and producing a net minimum for $i \approx
60^\circ$. At low
inclinations, the radial frequency contributes significant
power, while at higher inclinations, the first harmonic of the
azimuthal mode begin to dominate with similar behavior
to the circular orbits shown in Figure \ref{ploteight}a. Along with
increasing power at $2\nu_\phi$, there is also increasing power in the
radial beats of the first harmonic at $2\nu_\phi \pm \nu_r =
(5/3)\nu_\phi, (7/3)\nu_\phi$. For a $10M_\odot$ black
hole, all these frequencies should be observable within the timing
sensitivity of \textit{RXTE}.

\begin{figure}[ht]
\begin{center}
\includegraphics[width=0.7\textwidth]{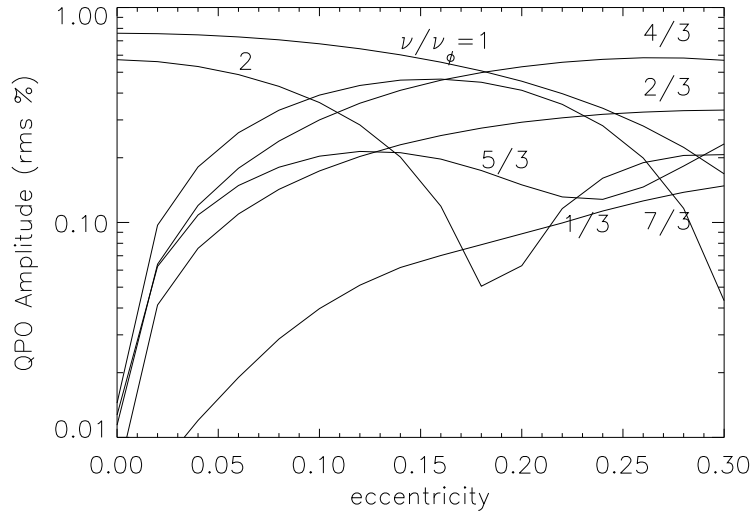}
\caption[Harmonic power in beat modes as a function of
eccentricity]{\label{harmonics5_e} Power in low-order harmonics and beat
  modes with frequencies $\nu = n\nu_\phi \pm \nu_r$, as a function of
  orbital eccentricity. The power in each mode $a_n({\rm rms})$ is
  normalized as in Figure \ref{ploteight}.
  The curves are labeled by the ratio $\nu/\nu_\phi$.}
\end{center}
\end{figure}

To further explore the constraints of our model, we
investigated the effect of orbital eccentricity on the QPO
power. Maintaining a 3:1 commensurability between azimuthal and
radial frequencies, we calculated the light curves for a range of
eccentricities $0\le e \le 0.3$, shown in Figure
\ref{harmonics5_e}. As expected, the beat modes at $\nu=\nu_\phi
\pm \nu_r$ have more power for more eccentric orbits, as the radial
variation of the emitter becomes larger. At the same time, the first
harmonic at $\nu=2\nu_\phi$ provides relatively less power with
increasing eccentricity. This is best understood as the ``picket
fence'' character of the light curve becomes modulated in amplitude
and frequency from peak to peak, i.e. for each 3-peak cycle, the time
between peaks 1-2, 2-3, and 3-1 are not all the same, damping the
harmonic overtone. 

The Fourier power in the beat modes
$\nu_\phi\pm \nu_r$ appears to saturate at a moderate eccentricity of
$e\approx 0.15$, while the fundamental power at $\nu_\phi$ continues to
decrease. One clear conclusion from this calculation is that without
the full ray-tracing calculation, these results would be difficult if
not impossible to derive by simple physical intuition. Since we still
do not have a strong physical explanation for why hot spots might form
at this special commensurate radius $r_0$, we also do not have a clear
means for determining the eccentricity of such orbits. One possibility
is to limit the inner-most extent of the pericenter to the ISCO
radius. With this approach, the maximum eccentricity for 3:1 orbits
when $a/M=0.5$ would be
\begin{equation}\label{max_eccentricity}
e_{\rm max} = 1-\frac{r_{\rm ISCO}}{r_0} \approx 0.13,
\end{equation}
similar to the typical values used throughout this Thesis. Another
limit on the eccentricity is given by the coordinate frequencies
themselves. Equations (\ref{omega_phi2}, \ref{omega_theta}, and
\ref{omega_r}) only apply for small deviations from circularity. As
the eccentricity increases, the radius for which $\nu_\phi=3\nu_r$
changes slightly, as do the frequencies. Thus for a given black hole
mass and spin, if the hot spot eccentricity grows too large, the 3:1
commensurable frequencies will no longer agree with the observed
location of the QPO peaks.

While there is some evidence for higher frequency harmonic
and beat modes in the QPO power spectrum of XTE J1550--564, the Fourier
power is clearly dominated by the two frequencies $184$ and $276$ Hz
\citep{remil02},
corresponding to $\nu_\phi-\nu_r$ and $\nu_\phi$ in our model. What
are the physical mechanisms that could damp out the higher
frequency modes? One possible explanation is in the geometry of the
hot spot. As explained in Section \ref{overbrightness_amp}, in order
to produce the
power observed in QPOs, the total X-ray flux coming from the hot
spot must be some finite fraction of that of the disk (typically
$10^{-3}-10^{-2}$ for a QPO amplitude of $1-5\%$), so the hot
spot must have some minimum size or it would not produce enough
emission to be detected above the background disk. Yet if the
hot spot is too large, it would be sheared by differential rotation in
the accretion disk and not survive long enough to create the coherent
X-ray oscillations that are observed. As mentioned above,
we find that the relative QPO power in different modes is not
sensitive to the size of the hot spot $R_{\rm spot}$, as long as the
hot spot remains roughly circular. Three-dimensional MHD simulations
\citep{hawle01,devil03a} show
a range of density and temperature fluctuations consistent with
the hot spot size and overbrightness factor predicted by our model
in conjunction with the observations.

\begin{figure}[ht]
\begin{center}
\includegraphics[width=0.45\textwidth]{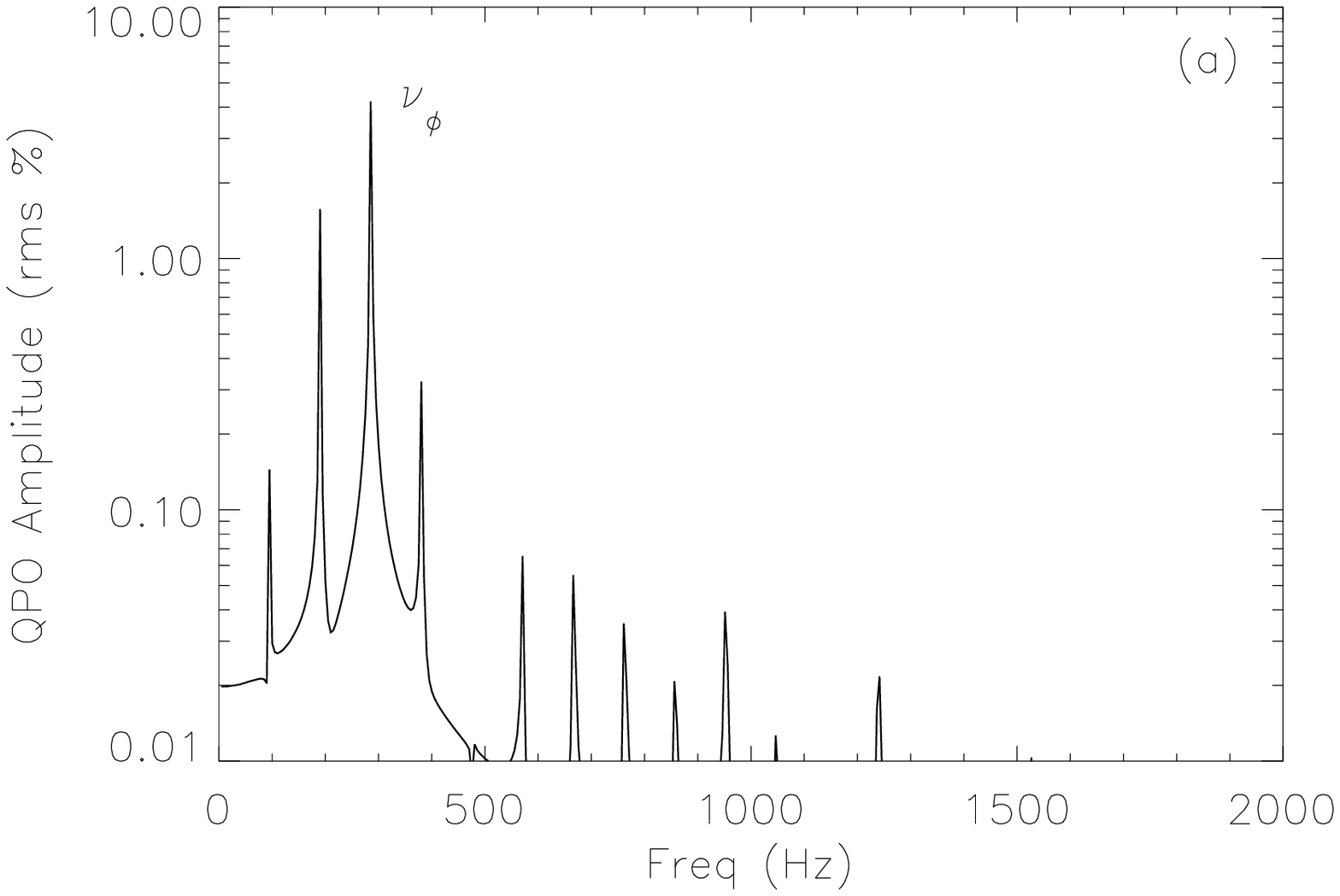}
\includegraphics[width=0.45\textwidth]{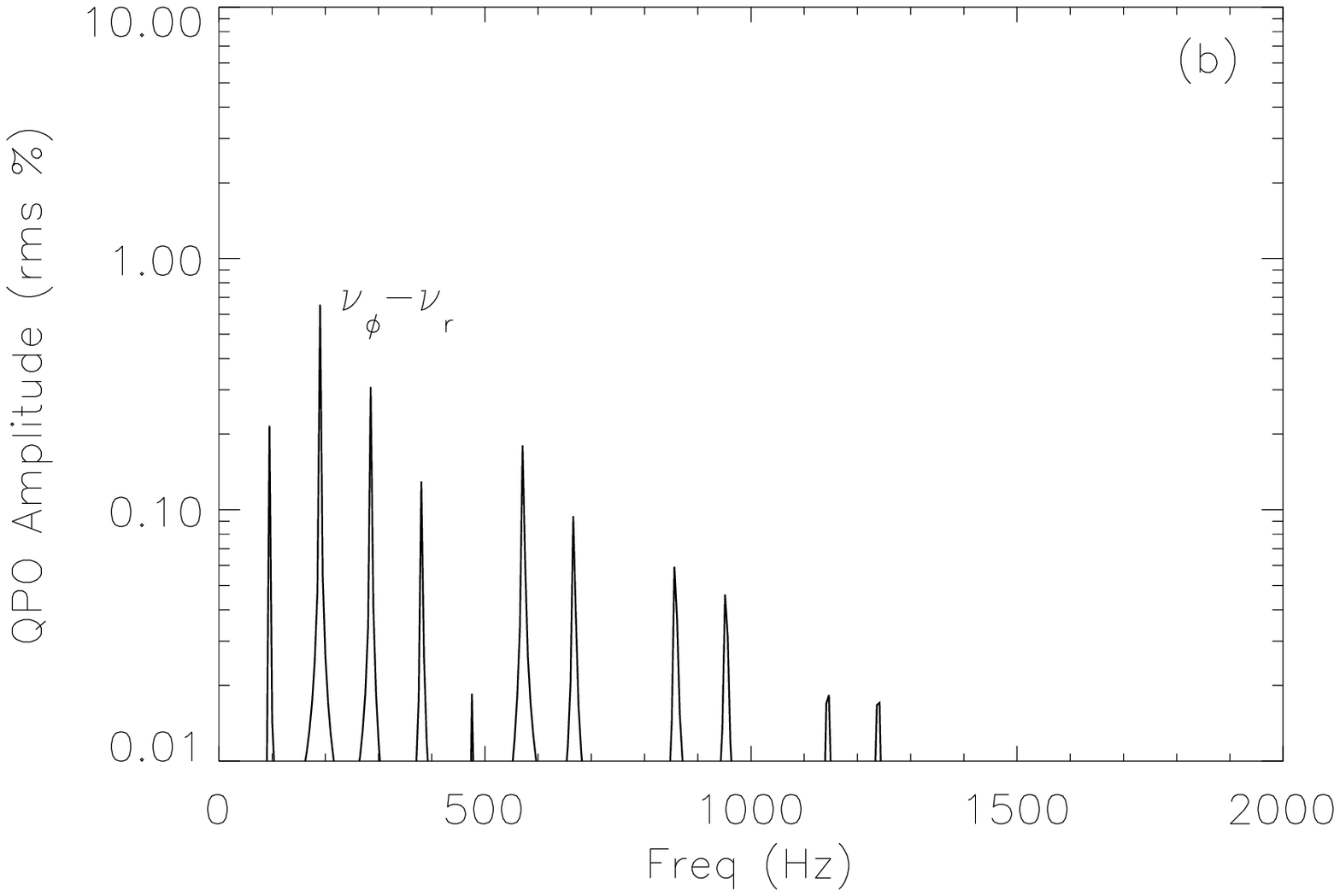}
\caption[Power spectra for sheared hot spot arcs]{\label{plottwelve}
Power spectrum for a hot spot with
  same trajectory as in Figure \ref{plotten}, with the emission
  region sheared along the geodesic into an arc of length (a) $180^\circ$
  and (b) $360^\circ$. For the shorter arc (a), the power is still peaked
  at the fundamental frequency $\nu_\phi=285$ Hz, while the extended
  arc (b) produces more power in the beat frequency $\nu_\phi-\nu_r =
  190$ Hz.}
\end{center}
\end{figure}

It also appears from simulations that as the hot spot is formed in the disk,
differential rotation will tend to shear a finite region of gas as
it follows a geodesic orbit around the black hole, modifying the shape
of the hot spot into an arc in azimuth. In the limit that
the emission region could be sheared into a ring of arc length $\Delta
\phi = 360^\circ$, the fundamental mode and its harmonics would be
essentially removed, leaving power only in the radial
modulation. Indeed, as shown in Figure
\ref{plottwelve}a, for an arc length of $\Delta\phi=180^\circ$, the higher
frequency modes at $\nu = 2\nu_\phi$ and $\nu=2\nu_\phi \pm \nu_r$ are
strongly suppressed, while still maintaining a significant amount of
power in the fundamental beat modes $\nu_\phi\pm\nu_r$. The total QPO
power also increases as the area of the emission region increases
relative to the circular hot spot geometry. 

However, when we
allow the arc to be sheared into a ring with $\Delta\phi=360^\circ$, the
total QPO power is actually decreased as the differential beaming is
essentially eliminated by the extended emission region: there
is always some portion of the arc moving toward the
observer. The resulting modulation is then more weighted to
the first radial beat mode at $\nu_\phi-\nu_r$, as seen in Figure
\ref{plottwelve}b. It is not intuitively obvious why the
$\nu_\phi-\nu_r$ mode is dominant while the $\nu_\phi+\nu_r$ mode
($\nu$ = 380 Hz) is much weaker in the arc geometry. If anything, this
is a strong argument for the necessity of a full ray-tracing calculation
of the hot spot light curves when predicting QPO power spectra, as it
clearly gives information unavailable to simple analysis of the
geodesic coordinate frequencies. In Chapter 4 below, we will give a
more physical explanation for why these lower frequency beat modes are
amplified while the higher frequency modes are suppressed. 

This behavior offers a plausible explanation for the two major types
of QPOs described in \citet{remil02}, initially distinguished by the
properties of their simultaneous LFQPOs. For XTE J1550--564, the type A
power density spectra have
more total power in the HFQPOs, with a major peak at 276 Hz and a
minor peak at 184 Hz. Type B spectra have most of the QPO power
around 184 Hz and a smaller peak around 276 Hz and less overall
power in the high frequency region of the spectrum. Thus we propose
that type A QPOs are coming from more localized
hot spot/arc regions, while type B QPOs come from a more extended ring
geometry.  

\section{Non-planar Orbits}\label{nonplanar_orbits}

\begin{figure}[ht]
\begin{center}
\includegraphics[width=0.45\textwidth]{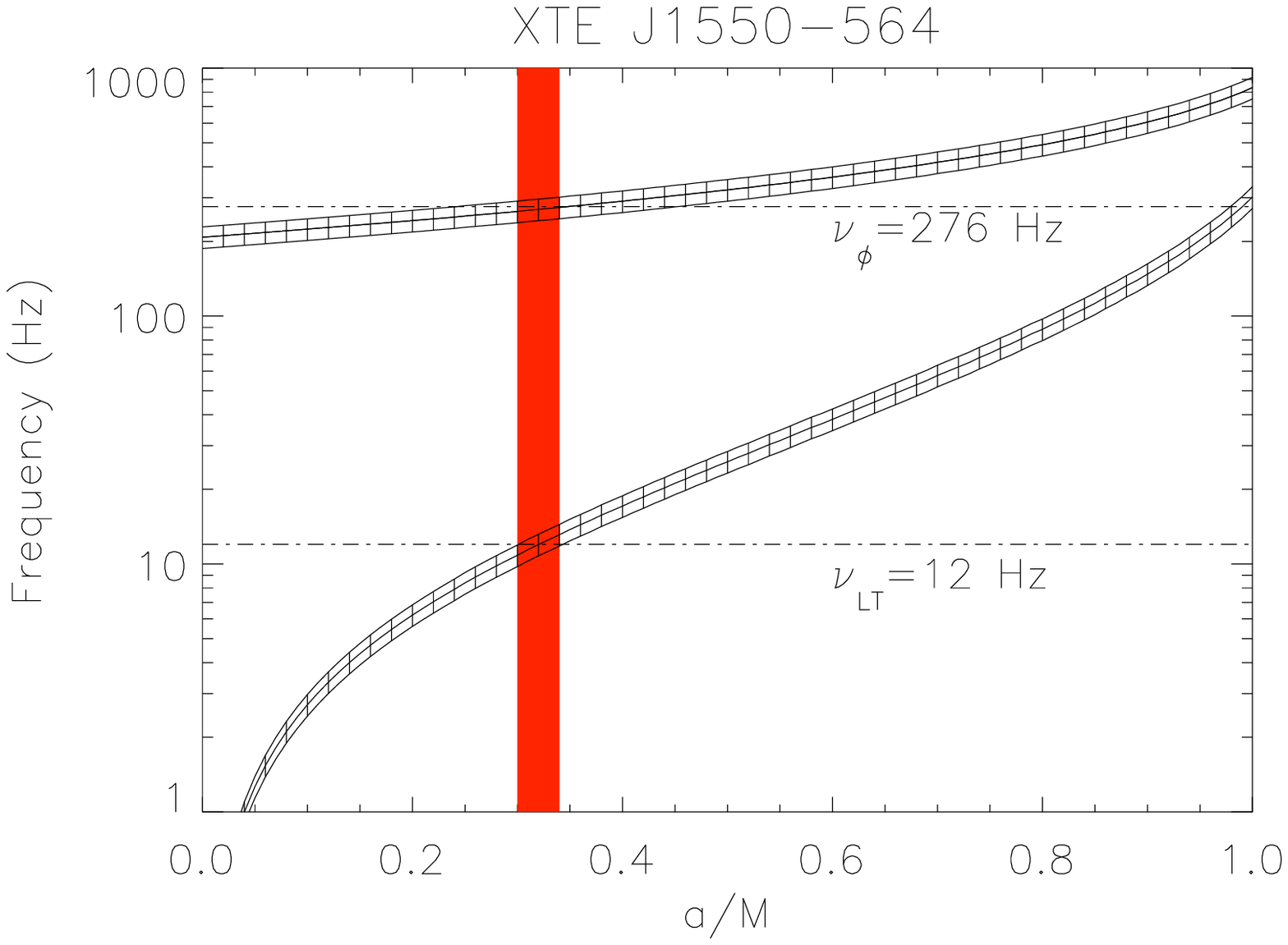}
\includegraphics[width=0.45\textwidth]{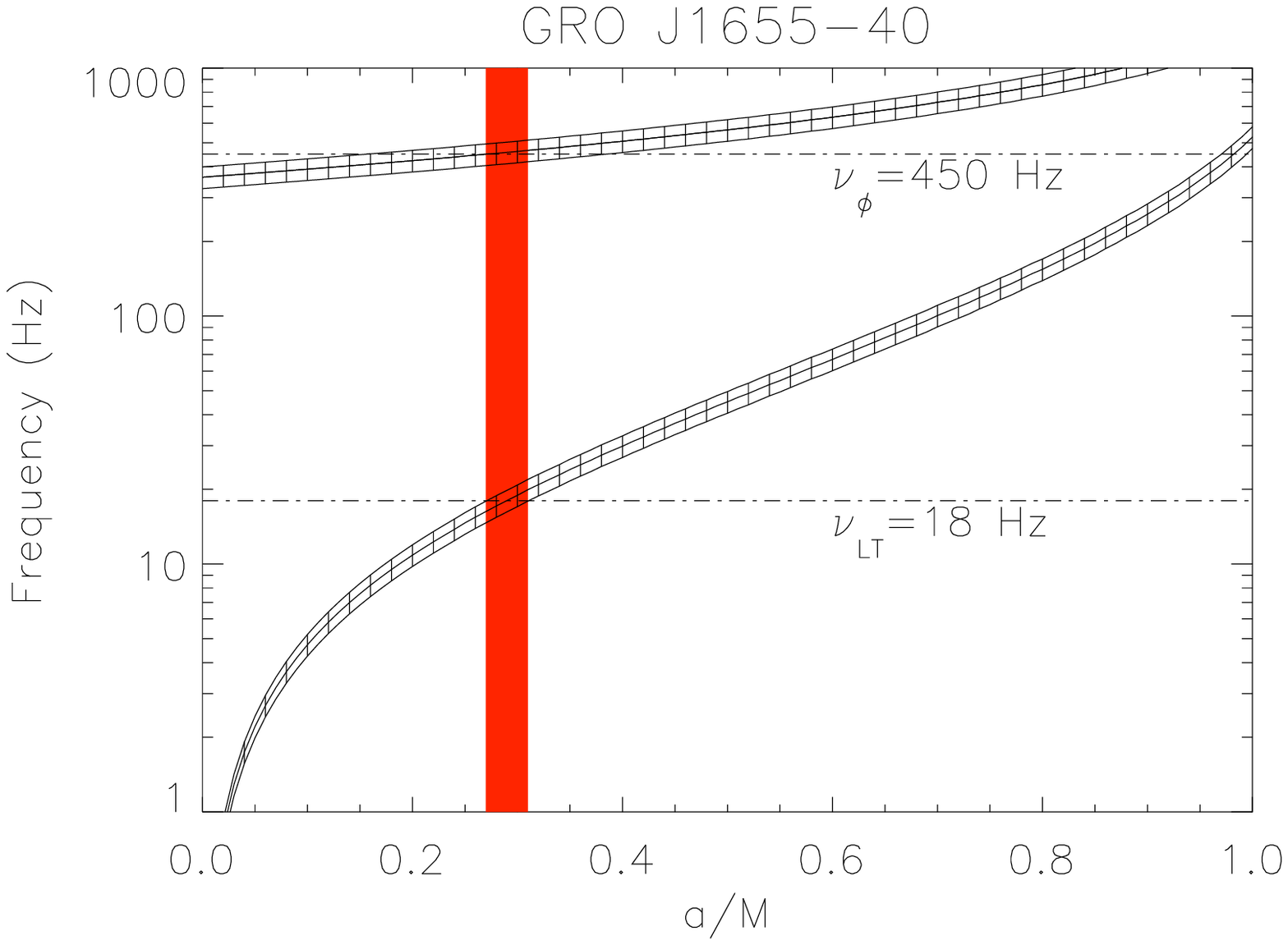}
\caption[Matching LFQPOs and HFQPOs as a function of spin]{\label{QPO_match}
Coordinate frequencies $\nu_{LT}$ and $\nu_\phi$ as a function of spin
for a small range of black hole masses (hashed bands), plotted over
the observed low- and high-frequency QPO locations for XTE J1550--564
and GRO J1655--40. The orbital radius is picked such that
$\nu_\phi=3\nu_r$, and the spin is determined by the solution region
marked by a vertical red band, corresponding to $M=8.9\pm 0.9 M_\odot$
for XTE J1550 and $M=5.1\pm 0.5 M_\odot$ for GRO J1655. [Compare with
Figure 8 in \citet{remil02}]}
\end{center}
\end{figure}

In addition to the commensurate high-frequency QPOs observed in
sources like XTE J1550--564, there are also strong low-frequency QPOs
observed at the same time with frequencies in the range
$5-15$ Hz. There have been suggestions that these low-frequency QPOs
may be caused by the Lense-Thirring precession of the disk near the
ISCO, also known as ``frame dragging'' \citep{marko98, merlo99, abram01,
remil02}. For geodesic orbits out of the plane perpendicular to the
black hole
spin, the latitudinal frequency $\Omega_\theta$ of massive particles
is not equal to the azimuthal frequency $\Omega_\phi$ [see eqs.\
(\ref{omega_theta}) and (\ref{omega_phi})], leading to a
precession of the orbital plane with frequency
\begin{equation}
\Omega_{LT} \equiv |\Omega_\theta - \Omega_\phi|.
\end{equation}
Figure \ref{QPO_match} shows the bands of coordinate frequencies
$\nu_{LT}(r_0)$ and $\nu_\phi(r_0)$ as a function of spin, for a small range of
possible black hole masses for each of XTE J1550--564 and GRO
J1655--40. Where the coordinate frequencies match the observed QPO
frequencies, a possible solution for the spin exists.
For black holes with the mass and spin used
above $(M = 10M_\odot, a/M = 0.5)$, the frame-dragging
frequency, as calculated at the radius corresponding to the
commensurability $\nu_r$:$\nu_\phi$=1:3, is somewhat higher than that
observed in the low-frequency QPOs from XTE J1550--564. The type A QPO
peaks at 12 and 276 Hz appear to be consistent with a black hole
mass of $8.9 M_\odot$ and spin parameter of $a/M = 0.35$
[\citet{orosz02} give a (1$\sigma$) estimate of $9.7-11.6 M_\odot$], quite
similar to the values used throughout much of this paper. For the BH
binary GRO J1655--40, we can fit the QPOs at 18 and 450 Hz with a mass
of $5.1 M_\odot$ and spin $a/M=0.28$, also slightly less than the published
mass range of $5.5-7.9 M_\odot$ \citep{shahb99}. These results are
shown in Table \ref{tableone}. 

\begin{table}[ht]
\caption[Black hole parameters, matching LFQPOs and HFQPOs to geodesic
frequencies]{\label{tableone} Black hole parameters for the XTE J1550--564
and GRO J1655--40, matching low- and high-frequency QPOs to geodesic
  coordinate frequencies}
\begin{center}
\begin{tabular}{llcc}
  Black Hole Parameters &  & XTE J1550--564 & GRO J1655--40 \\
  \hline
  & BH Mass & $8.9M_\odot$ & $5.1M_\odot$ \\
  & BH Spin & $0.35M$ & $0.28M$ \\
  & $R_{\rm ISCO}$ & $4.8M$ & $5.05M$ \\
  & Inclination & $70^\circ$ & $65^\circ$ \\
  \hline\hline
  Geodesic Frequencies & & \\
  \hline
  & $r_0$ & $5.54M$ & $5.77M$ \\
  & $\nu_{\rm LT}$ & 12 Hz & 18 Hz \\
  & $\nu_r$ & 92 Hz & 150 Hz \\
  & $\nu_\phi$ & 276 Hz & 450 Hz 
\end{tabular}
\end{center}
\end{table}

If we relax 
the requirement of matching the LFQPOs and only fit the HFQPOs with a 1:3
coordinate frequency commensurability, there remains
a 1-dimensional degeneracy in the mass-spin parameter space. Based
solely on the HFQPOs, for XTE
J1550--564 with $9.5<M/M_\odot<11.5$, the range of spin parameters
would be $0.42<a/M<0.6$, and for GRO J1655--40, the spin would be in the
range $0.35<a/M<0.66$. 

To get a more quantitative feel for the effect of Lense-Thirring
precession on the power spectrum, we investigated hot spot
orbits with initial trajectories inclined to the plane of the disk:
$v^\theta \ne 0$. This is much like changing the observer's
inclination with a period of $2\pi/\Omega_{LT}$. Thus we see
additional modulation in the hot spot light curve at the
``double-beat'' modes $\nu_\phi \pm \nu_r \pm \nu_{LT}$. We find that,
for modestly inclined hot spot orbits $(i_0 = \pm 5^\circ)$,
the contribution to the power spectrum at Lense-Thirring frequencies is
quite small ($<1\%$ of total power) for the basic circular hot spot
geometry. This relative contribution increases with arc 
length as the spot becomes a ring precessing about the spin axis,
consistent with the relative power in LFQPOs and HFQPOs
in the type A (more high frequency power than low frequency) and type
B (more low frequency power) sources described
above. 

Under the premise that the HFQPO commensurate frequencies are
caused by the geodesic motion of a sheared, overbright region in the
disk, in Table \ref{tabletwo} we show the best fit parameters for the
type A and type B power density
spectra from XTE J1550--564 [cf. Table 1 in \citet{remil02}]. Guided
also by the (somewhat speculative) assumption that the LFQPOs come
from the Lense-Thirring precession of the hot spot orbital plane, we
predict a black hole mass and spin. Using a fixed inclination of
$70^\circ$, we can match the frequencies and amplitudes of the observed
HFQPO peaks (and, equally important, the lack of power at certain
frequencies) for both type A and type B QPOs. Setting constant the
eccentricity $e=0.1$, geodesic inclination to the disk $i_0=5^\circ$, and
the overbrightness to be a factor of unity, we fit the hot spot radius
and arc length to match the observations. Being able to match the QPO
rms amplitudes of the 
peaks (or lack thereof) at 92, 184, 276, and 368 Hz, for at least two
different types of X-ray outburst, shows the robustness of our simple
model in explaining these phenomena.

\begin{table}[ht]
\caption{\label{tabletwo} QPO amplitudes of the hot spot/arc model for XTE
  J1550--564}
\begin{center}
\begin{tabular}{lccc}
  Parameter & & Type A & Type B \\
  \hline
  $R_{\rm spot}$ & & $0.3M$ & $0.5M$ \\
  arc length & & $200^\circ$ & $320^\circ$ \\
  eccentricity & & $0.1$ & $0.1$ \\
  inclination to disk & & $5^\circ$ & $5^\circ$ \\
  overbrightness & & 100\% & 100\% \\
  \hline
  Amplitude (mode) & Frequency (Hz) & rms(\%) & rms(\%) \\
  \hline
  $a(\nu_{\rm LT})$ & 12 & 0.63 & 2.1 \\
  $a(\nu_r)$ & 92 & 0.48 & 0.89 \\
  $a(\nu_\phi-\nu_r)$ & 184 & 1.3 & 2.2 \\
  $a(\nu_\phi)$ & 276 & 3.2 & 0.42 \\
  $a(\nu_\phi+\nu_r)$ & 368 & 0.20 & 0.23 \\
\end{tabular}
\end{center}
\end{table}

However, even with the sheared arc emission, many observations
still show significantly more power in the LFQPOs (typically $\sim
20\%$ rms) than can be
explained solely from the Lense-Thirring precession of a geodesic near
the ISCO \citep{remil04a}. Coupled with the difficulty in
simultaneously fitting the
mass and spin to three coordinate frequencies in a manner consistent
with spectroscopic mass predictions, it seems likely that the LFQPOs
may be caused by some other mechanism in the disk that is related only
indirectly to the high-frequency hot spot emission. Another likely
possibility is that the thin, warped disk model breaks down near the ISCO,
allowing more complicated emission geometries and thus amplifying the
effects of Lense-Thirring precession \citep{marko98}.

A recent paper by \citet{mille05} shows evidence for a correlation
between the phase of the low-frequency QPO light curve (easily
resolvable by \textit{RXTE} for $\nu \approx 1-2$ Hz) and the shape of
the time-varying iron emission line in the black hole binary GRS
1915+105. One possible explanation for this
behavior is a simple precessing ring made of geodesic particles
orbiting out of the plane of the disk. For high observer
inclinations, the total flux and the iron emission line will both be
significantly modulated as the ring rotates due to Lense-Thirring
precession. The edge-on disk gives a broader emission line, with a
higher overall flux due to the greater relativistic beaming,
while the face-on edge gives a minimum in the light curve,
with a narrower emission line, just as reported by
\citet{mille05}. Preliminary ray-tracing calculations of such a
geometry suggest that the ring is centered around $r\approx 10M$ with
an inclination to the plane of $\pm 20^\circ$, with a black hole spin
of $a/M=0.1-0.2$ and observer inclination $i=70^\circ$. 

\section{Summary}

The hot spot model makes a number of general predictions of the Fourier power
of the X-ray light curve as a function of inclination and black hole
spin, and is also able to explain QPO observations from the black hole
binaries XTE J1550--564 and GRO J1655--40. Simply by matching the
locations of the low-frequency and high-frequency QPOs with the
coordinate frequencies (under the condition $\nu_\phi=3\nu_r$), we can
determine the black hole mass and spin. Relaxing the LFQPO constraint,
the spin can still be determined uniquely for a given mass, which in turn
could be measured independently with the inclination and radial
velocity of the companion star. 

By matching the amplitudes of various QPO peaks observed in XTE
J1550--564, we have explored the model parameters such as the hot spot
size, shape, and the overbrightness relative
to a steady-state background disk. The predicted magnitude of these
fluctuations are within the range predicted by 3D MHD calculations of the
accretion disk. Future work will investigate the effect of multiple
hot spots of various size, emissivity, and lifetime, as guided by MHD
calculations. Observations of additional sources
with commensurate frequency QPOs may help us further constrain the
hot spot model and better understand the connection between the
LFQPOs and HFQPOs. 

Some of the physical problems with the original hot spot model have
been raised by \citet{marko00}, as discussed above.
Many of these points are addressed in our model. First, unlike
\citet{stell98,stell99a}, we only
attempt to explain a set of QPO data from \textit{black hole} binaries,
which differ qualitatively from neutron star binaries in many ways,
e.g.\ lacking strong global magnetic fields and thermonuclear
activity. And perhaps most significant, black
holes have no rotating surface to interact with the accreting matter
and provide additional confusion to the QPO power spectrum.
Our model produces light curves with power spectra consistent with
black hole observations even with low eccentricity hot spot
orbits. 

Because they do not include ray tracing in their calculations,
\citet{marko00} are unable to model many relativistic effects,
including the gravitational lensing of
the hot spot source, which can be quite significant for systems with
moderate to high inclination angles $(i \ge 60^\circ)$. Since we calculate
the actual X-ray modulation from the orbiting hot spots, we predict
both the location and amplitude of every peak in the light curve power
spectrum, which cannot be done by analyzing the BH coordinate
frequencies alone. By introducing a perturbation on circular orbits
near the ISCO, additional peaks begin to appear in the power
spectrum, caused by beats of the azimuthal and radial frequencies
$\nu_\phi$ and $\nu_r$. The dependence of the relative power in the
different peaks on inclination and spin helps to constrain the
details of the hot spot model in explaining the HFQPOs, particularly the
2:3 commensurability observed in the power spectra from XTE
J1550--564 and GRO J1655--40. 

As an additional parameter, we introduce
a finite arc length for the emission region, motivated by the
shearing of the hot spot by differential rotation in the disk. The
spreading of the hot spot in azimuth leads to suppression of the
higher QPO modes, in agreement with observations.
We have also examined the possibility of Lense-Thirring precession
for non-planar orbits as an explanation for the low-frequency QPOs
that have been observed coincident with the HFQPOs yet often with even
stronger Fourier power. The predicted power spectra from non-planar
precessing arcs are consistent with observations of XTE J1550--564 if we
associate type A QPOs with hot spot arcs of $\Delta \phi \approx
180^\circ$ and type B QPOs with hot spot rings of $\Delta \phi
\approx 360^\circ$. However, the difficulty in matching the LFQPOs
amplitude and frequency with a single hot spot geodesic suggests the
low-frequency modulations may be caused by a different mechanism or
perhaps our disk geometry is too simplistic.

One major remaining issue with the hot spot model is the preferred
location of the geodesic that gives rise to 1:3 coordinate
frequencies. Why should the orbital frequencies favor integer ratios,
and why should the preferred ratio be 1:3 and not 1:2 or 1:4? It is
possible that detailed global radiation-MHD 
calculations with full general relativity will be required to answer this
question. Perhaps the non-circular orbits can only survive along
closed orbits such as these to somehow avoid destructive
intersections \citep{abram03b}. Or there may be magnetic interactions
with the black
hole itself, analogous to the Blandford-Znajek process, that lock the
accreting gas into certain preferred
trajectories \citep{wang03}. For now, we are forced to leave this as
an open question unanswered by the geodesic hot spot model.

A less difficult problem is the explanation of
the widths of the QPO peaks. As it stands, our hot spot model predicts
purely periodic light curves and thus power spectra made up of
delta functions. If there \textit{is} some physical mechanism 
that preferentially focuses accreting material onto eccentric orbits
at specific radii, then it is likely that these hot spots are forming
and then being destroyed as a continual process. The superposition
of many hot spots around the same orbit, all with slightly different initial
trajectories, could explain the quasi-periodic nature of the power
spectrum: the phase decoherence of the hot spots would cause a natural
broadening of the strictly periodic signal from a single spot
\citep{schni04b,schni05}.
With the computational framework in place, this question can be
answered by modeling a whole collection of hot spots and arcs
continually forming and evolving in shape and emissivity.

\chapter{Features of the QPO Spectrum}
\begin{flushright}
{\it
As far as the laws of mathematics refer to reality, they are not \\
certain, as far as they are certain, they do not refer to reality. \\
\medskip

If the facts don't fit the theory, change the facts.\\
\medskip
}
-Albert Einstein
\end{flushright}
\vspace{1cm}

The results presented in this Chapter are based on the extension of
the hot spot model, as described in the papers ``Interpreting the High
Frequency QPO Power Spectra of Accreting Black holes,''
\citep{schni05} and ``The Bicoherence as a Diagnostic for Models of
High Frequency QPOs,'' \citep{macca04}.

\section{Introduction}\label{intro}
One of the most exciting results from the \textit{Rossi X-Ray Timing
Explorer} (\textit{RXTE}) has been the discovery of high frequency
quasi-periodic oscillations (HFQPOs) from
neutron star and black hole binaries [\citet{stroh96,vande96,stroh01a};
see \citet{lamb02} for a review]. For
black hole systems, these HFQPOs are observed repeatedly at more or less
constant frequencies, and in a few cases with integer ratios
\citep{mille01,remil02,homan04,remil04}. These discoveries give the
exciting prospect of
determining a black hole's mass and spin, as well as testing general
relativity in the strong-field regime [see e.g.\,
\citet{kluzn90,dedeo04,psalt04}].

To understand these observations more quantitatively, we have developed a
ray-tracing code to model the X-ray light curve from a collection of ``hot
spots,'' small regions of excess emission moving on geodesic orbits
\citep{schni04a,schni04b}. 
Similar ray-tracing models of multiple hot spots with a range of
geodesic orbits in a Schwarzschild metric have been proposed by
\citet{karas92} and \citet{karas99}, and used to produce theoretical
light curves and power spectra.
The hot spot model is largely motivated by the similarity between
the QPO frequencies and the black hole coordinate frequencies near the
inner-most stable circular orbit (ISCO)
\citep{stell98,stell99a}, as well as the suggestion of a resonance leading to
integer commensurabilities between these coordinate frequencies
\citep{abram01,abram03b,rebus04}. \citet{stell99a} investigated
primarily the QPO
frequency pairs found in accreting neutron star binary systems, but their
approach can be applied to black hole systems as well. The hot spots
themselves could be formed and destroyed in any number of ways,
including magnetic flare avalanches \citep{pouta99,zycki02}, vortices
and flux tubes \citep{abram92}, or magnetic instabilities \citep{balbu91}. 

The basic geodesic hot spot model (see above, Chapter 3) is
characterized by the black hole mass and
spin, the disk inclination angle, and the hot spot size, shape, and
overbrightness. Motivated by the 3:2 frequency commensurabilities observed in
QPOs from XTE J1550--564, GRO J1655--40, and H1743--322
\citep{remil02,homan04,remil04}, we pick a radius for the geodesic
orbits such that the coordinate frequencies $\nu_\phi$ and $\nu_r$
will have a 3:1 ratio. Thus the 3:2 commensurability is interpreted as the
fundamental orbital frequency $\nu_\phi$ and its beat mode with the
radial frequency at $\nu_\phi-\nu_r$. Conversely,
when the orbital and radial frequencies have a 3:2 ratio, the
corresponding power spectrum shows the strongest peaks at
$\nu_\phi$:$(\nu_\phi-\nu_r) = $ 3:1, in disagreement with the data. For
this reason, in the discussion
below, we will focus primarily on the 3:1 coordinate frequency
resonance proposed by \citet{abram01,abram03b}. Furthermore, we relax
the low frequency QPO criterion, described above in Section
\ref{nonplanar_orbits}, leaving a one-dimensional degeneracy in
the mass-spin parameter space which can be broken by an independent
determination of the binary system's inclination and radial velocity
measurements of the low-mass companion star [see e.g.,
\citet{orosz02,orosz04}].

Given the black hole mass, spin, inclination, and the radius of the
geodesic orbit, the parameters of the hot spot model (i.e.\ the hot
spot size, shape, and overbrightness, and the orbital eccentricity)
are determined by fitting to the amplitudes of the peaks in the
observed power spectrum \citep{schni04a}. However, the model as
described so far produces a perfectly 
periodic X-ray light curve as a single hot spot orbits the 
black hole indefinitely. Such a periodic light curve will give a power
spectrum composed solely of delta-function peaks,
unlike the broad features seen in the observations.

In this Chapter we introduce two simple physical models to account for
this broadening of the QPO peaks. The models are based on analytic results,
then tested and confirmed by comparison with three-dimensional
ray-tracing calculations of multiple hot spots \citep{schni04b}. 
We find the power spectrum can be accurately modeled by
a superposition of Lorentzian peaks, consistent with the standard
analysis of QPO data from neutron stars and black holes
\citep{olive98,nowak00,bello02}. Many of the methods and results presented
here are equally valid for other QPO models such as diskoseismology
\citep{wagon99}, vertically-integrated disk oscillations
\citep{zanot03,rezzo03a,rezzo03b}, toroidal perturbations
\citep{lee02,lee04}, and magnetic resonances \citep{wang03}.

The previous Chapters outlined most of the general features of the
basic hot spot model and the ray-tracing code used to produce periodic
light curves. The specific parameters used in this Chapter are briefly
summarized in Section \ref{hotspot}. In Section \ref{phase} we
explain the effect of summing the light curves from 
multiple hot spots with random phases and different lifetimes to give
Lorentz-broadened peaks in the power spectrum. Section \ref{freq} shows how a
finite width in the radii of the geodesic orbits produces a
corresponding broadening of the QPO peaks. In Section \ref{scatter} we
develop a simple model for photon scattering in the corona,
which affects other features of the power spectrum such as the
continuum noise and the damping of high frequency harmonics, but does
not contribute to the broadening of the QPO peaks. Finally, all the
pieces of the model are brought together in Section \ref{data} and used to
interpret the power spectra from a number of observations of
XTE J1550--564. Section \ref{bispectrum} introduces the use of
higher-order statistics as an observational tool for distinguishing
between the various peak broadening mechanisms.

\section{Parameters for the Basic Hot Spot Model}\label{hotspot}

In \citet{schni04a} we developed a geodesic hot spot model (see
Chapters 2 and 3 above) to explain
the 3:2 frequency commensurabilities seen in the QPO power spectra of
XTE J1550--564, GRO J1655--40, and H1743--322. The results of this
extended model are based on the fully relativistic ray-tracing framework
described in that paper. Starting from a distant observer, a
collection of photon trajectories are integrated backwards in time to
a fixed coordinate grid surrounding the black hole. With the
spacetime position and momentum recorded at each point in the
computational grid, time-dependent images of the dynamic disk can be
created with ease. While this technique is quite general and can be
used to analyze various QPO models, for simplicity we restrict our
discussion in this Chapter to the geodesic hot spot model.

The hot spots are treated as monochromatic, isotropic emitters in
their rest frames, moving along the geodesic orbits of massive test
particles. For the Kerr geometry, these orbits generally have three
non-degenerate frequencies (azimuthal, radial, and vertical). As
explained above in Section \ref{intro}, we focus our analysis on closed orbits
with $\nu_\phi = 3\nu_r$, giving the strongest peaks in the power
spectrum at the fundamental orbital frequency $\nu_\phi$ and the beat
mode $\nu_\phi-\nu_r$. The relative damping of the upper beat mode at
$\nu_\phi+\nu_r$ is explained below in Section \ref{scatter}. 

For a given black hole mass, the spin is determined uniquely by
matching the coordinate frequencies to the observed QPO peaks. As
shown in \citet{schni04a}, the rms amplitudes of the various peaks are
determined by the hot spot's orbital inclination, eccentricity, and
overbrightness relative to the steady-state emission from the
disk. For the fiducial example used in much of this Chapter, the black
hole has mass $M=10M_\odot$ and spin $a/M=0.5$ with a disk inclination
of $i=70^\circ$. Each hot spot is on a planar orbit around a radius
of $r_0 = 4.89M$ with $\nu_\phi=285$ Hz,
$\nu_r = 95$ Hz, and a moderate eccentricity of $e = 0.1$. These
figures are similar, but not identical to the best-fit parameters for
observations of XTE J1550--564 presented in Section \ref{data}.

For a single hot spot on a geodesic trajectory, the resulting light
curve will be purely periodic, corresponding to a power spectrum made
up of multiple delta-functions. These delta-function peaks will be
located at linear combinations of the
coordinate frequencies, with their relative amplitudes determined by
the orbital parameters via the ray-tracing calculation. But unlike
these periodic features, the actual
data shows broad peaks in the observed power spectra, hence the term
\textit{quasi}-periodic oscillations. In \citet{schni04b} we showed
how the superposition of many hot spots with finite lifetimes and
random phases could give 
a natural explanation for this broadening, as we will explain in
greater detail below.

\section{Peak Broadening from Hot Spots with Finite Lifetimes}\label{phase}

While the basic model presented above takes as given the existence of
hot spots on certain special orbits, we can also gain some physical
intuition about these hot spots from more detailed calculations.
Three-dimensional magnetohydrodynamic simulations of
accretion disks suggest that in general such hot spots are continually
being formed and destroyed with random phases, with a range of
lifetimes, amplitudes, and orbital frequencies \citep{hawle01,devil03b}.

For now, we consider the contribution from identical hot
spots, assuming that each one forms around the same radius with
similar size and overbrightness and survives for some finite time
before being destroyed. Over this lifetime, the hot spot produces a
coherent periodic light curve as in the single spot model.
Analogous to radioactive decay processes, we assume that
during each time step $dt$, the probability of the hot spot dissolving
is $dt/T_l$, where $T_l$ is the characteristic lifetime of the hot
spots. As derived in Appendix B, if each coherent segment is a
purely sinusoidal function $f(t) = A\sin(2\pi\nu_0 t+\phi),$ the
corresponding power spectrum is a Lorentzian peak centered around $\nu
= \nu_0$ with a characteristic width given by
\begin{equation}\label{dec_width}
\Delta \nu = \frac{1}{2\pi T_l}.
\end{equation}

If this model is a qualitatively accurate description of how hot spots
form and dissolve in the disk, one immediate conclusion is that the
oscillator quality factor $Q\equiv \nu_0/\mbox{FWHM}$ can be fairly
high even for relatively short coherence times:
\begin{equation}
Q=\pi T_l\nu_0 = \pi \times \langle \mbox{\# of orbits} \rangle.
\end{equation}
If, on the other hand, every hot spot has a lifetime of
\textit{exactly} four orbits $(T_l=4/\nu_0)$, the central peak of the
power spectrum $G^2(\nu,T_l)$ has coherence $Q\approx 4.5$, about what
one would expect from a first-order estimate. However, after
integrating over the exponential lifetime distribution to get the
Lorentzian profile of equation (\ref{lorentz1}), the resulting quality
factor is $Q\approx 12.6$, roughly a factor of
three higher. \citet{remil02} observe quality values of $Q\sim 5-10$
for the HFQPOs seen in XTE J1550--564, corresponding to typical hot
spot lifetimes of only 2-3 orbits.

While this result is based on a boxcar sampling function for the hot
spots (i.e.\ instantaneous creation and destruction of each hot
spot, with constant brightness over its lifetime), these results are
quite general for other window functions as well. In Appendix B we
show that, for any set of self-similar sampling functions $w(t;T)$
[and its Fourier pair $W(\nu;T)$], the
exponential lifetime distribution has the effect of narrowing the peak
of the net power spectrum compared with that of a single segment of
the light curve with length $T_l$. This smaller width can be
understood by considering the distribution of hot spot lifetimes
and their relative contribution to the total power spectrum [see eqns.\
(\ref{sin1}) and (\ref{distribution})]. While there are
actually more segments with individual lifetimes shorter than $T_l$,
the few long-lived
segments of the light curves add significantly more weight to the
QPO peaks since $W(\nu=0;T) \propto T$ while $\Delta \nu \propto
1/T$. 

In addition to the boxcar function, another physically
reasonable model for the hot spot evolution is that of a sharp rise
followed by an exponential decay, perhaps caused by magnetic
reconnection in the disk \citep{pouta99,zycki02}. In this case, the
light curve would behave
like a damped harmonic oscillator, for which the power spectrum is
also given by a Lorentzian [see \citet{vande89}, where this
result is presented in the context of an exponential shot
model]. Interestingly, the shape and width of the resulting QPO peak is exactly
the same, whether we use a collection of boxcar functions with an
exponential lifetime distribution, or if we use a set of exponential
sampling functions, each with the same decay time. In the discussion
below and when doing the actual light curve simulations, we will
assume a boxcar sampling function and an exponential
lifetime distribution, with its corresponding Lorentzian power
spectrum. This approach also facilitates a direct comparison with
observations and other theoretical models,
where the QPO data is often fit by a collection of Lorentzian peaks
\citep{nowak00,bello02}.

Due to the linear properties of the Fourier transform, equation
(\ref{dec_width}) and the analysis of Appendix B,
while derived assuming a purely sinusoidal signal with a
single frequency $\nu_0$, can be applied equally well to any periodic light
curve with an arbitrary shape. If each coherent section of the light
curve is written as
\begin{equation}\label{segsum}
f(t) = \sum_j A_j \sin(2\pi \nu_j t + \phi_j),
\end{equation}
then the total power spectrum (integrating over a distribution of coherent
segments with random phase) is simply the sum of the Lorentz-broadened
peaks from equation (\ref{lorentz1}):
\begin{equation}\label{multi_lor}
\tilde{I}^2(\nu) = 2N_{\rm spot}A_j^2 \frac{T_l}{T_f}\sum_j 
\frac{1} {1+4\pi^2 T_l^2(\nu-\nu_j)^2},
\end{equation}
where on average there are $N_{\rm spot}$ hot spots in existence at
any given time and the light curve is integrated over a total time of
$T_f$. Note that every peak in the power spectrum $\tilde{I}^2(\nu)$
has the same characteristic width $\Delta \nu=1/(2\pi T_l)$. 

The sum in equation (\ref{segsum}) can be generalized to a Fourier
integral so that equation (\ref{multi_lor}) becomes the convolution
(denoted by the symbol $\star$) of the segment power spectrum
$F^2(\nu)$ with a normalized Lorentzian $\mathcal{L}(0,\Delta \nu)$
centered on $\nu=0$ with width $\Delta \nu$: 
\begin{equation}\label{conv_lor}
\tilde{I}^2(\nu) = \frac{1}{\pi\Delta\nu} \int_{-\infty}^\infty
\frac{F^2(\nu')d\nu'} {1+\left(\frac{\nu-\nu'}{\Delta \nu}\right)^2} =
[F^2\star \mathcal{L}(0,\Delta \nu)](\nu).
\end{equation}

Now we can apply our results to the light curves as calculated by
the original ray-tracing code for a single geodesic hot spot. First, the  
X-ray light curve over one complete period is calculated to give the
Fourier components $A_j$ in equation (\ref{segsum}). For geodesic
orbits in the Kerr metric, the
power spectrum $F^2(\nu)$ is concentrated at integer combinations of
the black hole coordinate frequencies $\nu_\phi$, $\nu_r$, and
$\nu_\theta$. Given these frequencies $\nu_j$, amplitudes $A_j$, and a
characteristic hot spot lifetime $T_l$, the integrated power spectrum follows
directly from equation (\ref{multi_lor}).

\begin{figure}[tp]
\begin{center}
\includegraphics[width=0.8\textwidth]{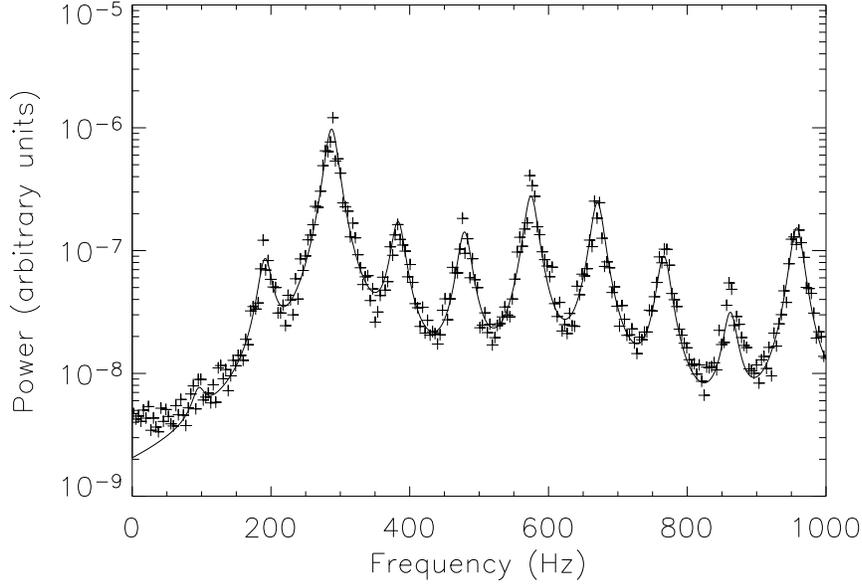}
\caption[Power spectrum of multiple hot spots with random
phases and finite lifetimes]{\label{chap4_f1} Simulated power spectrum
(crosses) from a
ray-tracing calculation of many hot spots on geodesic orbits with random
phases and different
lifetimes, along with an analytic model (solid line) of that power
spectrum. The black hole has mass $M=10M_\odot$ and spin $a/M=0.5$, giving
$\nu_r= 95$ Hz and $\nu_\phi = 285$ Hz. The hot spot orbit has an
eccentricity of 0.1 around a radius of $r_0 = 4.89M$ and an inclination of
$70^\circ$. The peaks have Lorentzian profiles with $\Delta \nu \approx
11$ Hz, corresponding to a characteristic hot spot lifetime of four
orbits.}
\end{center}
\end{figure}

Using the same ray-tracing code, we can also directly simulate the
extended light curve and corresponding power spectrum produced by many
hot spots orbiting with random phases, continually formed and
destroyed over each time step with probability $dt/T_l$
\citep{schni04b}. The power spectrum of such a simulation is shown in
Figure \ref{chap4_f1} (crosses),
along with the analytic model (solid curve). The orbital parameters
for each individual hot spot are the same as those outlined in Section
\ref{hotspot}. The characteristic lifetime $T_l$ is four orbits
(about 14 msec), corresponding to a Lorentzian width of $\Delta \nu
\approx 11$ Hz. We should stress
that this curve is \textit{not a fit to the simulated data}, but an
independent result calculated using the model described above.

The defining characteristic of QPO peaks broadened by the summation of
hot spots with finite lifetimes and
random phases is the uniform width of the individual peaks. For a
power spectrum with multiple harmonics and beat modes, each peak is
broadened by exactly the same amount, determined by the average lifetime of
the individual hot spots. Thus if we can measure the widths of multiple QPO
peaks in the data, the hot spot lifetime can be determined redundantly with a
high level of confidence.

\section{Distribution of Coordinate Frequencies}\label{freq}
In the previous Section, we assumed a single radius for all the hot spot
orbits. This ensures identical geodesic coordinate frequencies for
different hot spots with different phases and lifetimes. However, this
assumption betrays one of the major weaknesses of the geodesic hot
spot model: there still does not exist a strong physical argument for
why these hot spots should form at one special radius or why that
radius should have coordinate frequencies with integer
commensurabilities. For now, we will be forced to leave that question
unanswered, but we can make progress by drawing on intuition gained
from other fields of
physics. If there does exist some physical resonance in the system that
favors these orbits, causing excess matter to ``pile up'' at certain
radii \citep{abram01,abram03b}, then just like any other resonance, there
should be some finite 
width in phase space over which the resonant behavior is
important. The integer commensurability of the QPO peaks
suggests that closed orbits may be playing an important role in the
hot spot formation. If this is so, then some hot spots should also
form along orbits that \textit{almost} close, i.e.\ geodesics with
nearly commensurate coordinate frequencies. These orbits will have
guiding center radii similar to the critical radius $r_0$ for which
the geodesics form closed curves. 

Motivated by other processes in nature such as damped harmonic oscillators
and atomic transitions, we model the resonance strength as a function of
radius with a Lorentzian of characteristic width $\Delta r$. Then the
probability of a hot spot forming at a given radius is proportional to
the strength of the resonance there, giving a distribution of orbits
according to
\begin{equation}\label{lorentz_r}
P(r)dr = \frac{dr/(\pi\Delta r)}{1+\left(\frac{r-r_0}{\Delta r}\right)^2}.
\end{equation}
For a relatively small resonance width $\Delta r$, we can linearize the
coordinate frequencies $\nu_j(r)$ around $r=r_0$ with a simple Taylor
expansion: 
\begin{equation}\label{nu_taylor}
\nu_j(r) \approx \nu_{j0} + (r-r_0) \left. \frac{d\nu_j}{dr} \right|_{r_0},
\end{equation}
in which case the probability distribution in frequency space is also
a Lorentzian:
\begin{equation}\label{lorentz_nu}
P(\nu_j)d\nu_j = \frac{d\nu_j/(\pi\Delta\nu_j)}
{1+\left(\frac{\nu_j-\nu_{j0}}{\Delta \nu_j}\right)^2}.
\end{equation}
Here $\nu_j = \nu_\phi, \nu_\theta, \nu_r$ are the azimuthal, vertical,
and radial coordinate frequencies and $\nu_{j0}=\nu_j(r_0)$ are those
frequencies at the resonance center. 

\begin{figure}
\begin{center}
\includegraphics[width=0.8\textwidth]{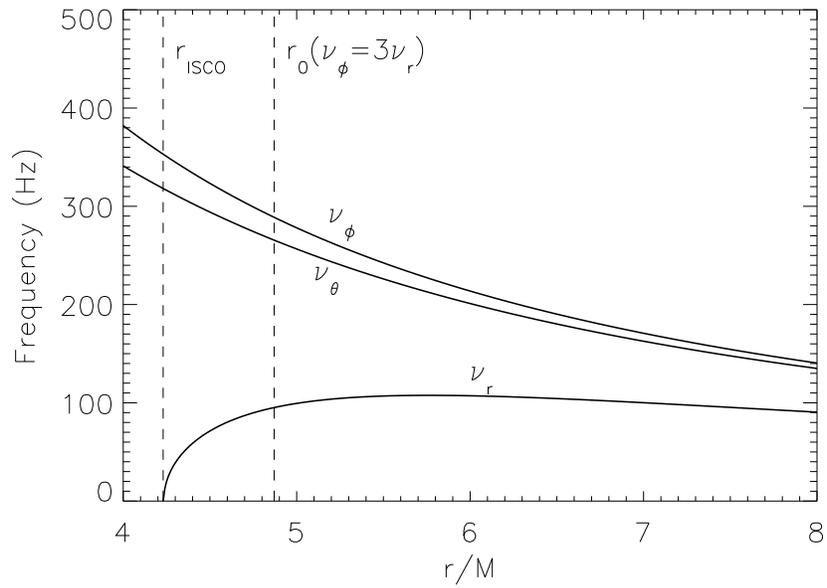}
\caption[Geodesic coordinate frequencies as a function of
radius]{\label{chap4_f2} Geodesic coordinate frequencies as a
function of radius for a black hole with mass $M=10M_\odot$ and spin
$a/M=0.5$. The radius of the inner-most stable circular orbit $r_{\rm
ISCO}$ is where $\nu_r \to 0$. The commensurate radius $r_0$ is where the
ratio of azimuthal to radial coordinate frequencies is 3:1.}
\end{center}
\end{figure}

For nearly circular orbits, the
coordinate frequencies (using geometrized units with $G=c=M=1$) are given by
\citet{merlo99}, as quoted above in Section \ref{noncircular_orbits} 
\begin{subequations}
\begin{equation}\label{nu_phi}
\nu_\phi = \frac{1}{2\pi(r^{3/2}\pm a)},
\end{equation}
\begin{equation}\label{nu_theta}
\nu_\theta = \nu_\phi \left[1\mp
\frac{4a}{r^{3/2}}+\frac{3a^2}{r^2}\right]^{1/2},
\end{equation}
and
\begin{equation}\label{nu_r}
\nu_r = \nu_\phi \left[1-\frac{6}{r}\pm \frac{8a}{r^{3/2}}
-\frac{3a^2}{r^2} \right]^{1/2},
\end{equation}
\end{subequations}
where the upper sign is taken for prograde orbits and the lower sign
is taken for retrograde orbits (the results below assume
prograde orbits, but the analysis for retrograde orbits is essentially
the same). These frequencies are plotted in Figure
\ref{chap4_f2} as a function of $r$ for a representative black hole
with mass $10M_\odot$ and spin $a/M=0.5$.
The radial frequency approaches zero at the ISCO, where geodesics can
orbit the black hole many times with steadily decreasing $r$. In the
limit of zero spin and large $r$, the coordinate frequencies reduce to
the degenerate Keplerian case with $\nu_\phi = \nu_\theta = \nu_r =
1/(2\pi r^{3/2})$. 

Generally, the power spectrum of the periodic light curve from a
single hot
spot orbiting at $r_0$ will be made up of delta-functions located at
the harmonics of the fundamental $\nu_\phi$ and the beat modes with
$\nu_r$ and $\nu_\theta$. Considering for the moment only
planar orbits, the power will be concentrated at the frequencies
$\nu=n\nu_\phi\pm\nu_r$, where $n$ is some positive integer. In fact,
there will be additional peaks at $\nu=n\nu_\phi\pm 2\nu_r$ and even
higher beat-harmonic combinations, but for coordinate frequencies with
$\nu_\phi = 3\nu_r$, these higher modes are degenerate, e.g.\
$\nu_\phi+2\nu_r = 2\nu_\phi-\nu_r$. A careful treatment can
distinguish between these degenerate modes, but in practice we find
the power in the radial double- and triple-beats to be insignificant compared
to the single-beat modes at $n\nu_\phi\pm\nu_r$, so we limit our
analysis to these frequencies. 

From equations (\ref{nu_taylor}) and (\ref{lorentz_nu}), we see that
a QPO peak centered around $\nu=n\nu_\phi\pm\nu_r$ will be a
Lorentzian of width
\begin{equation}\label{broaden1}
\Delta \nu = \Delta r \left(n\frac{d\nu_\phi}{dr}\pm
\frac{d\nu_r}{dr}\right)_{r_0}.
\end{equation}
Unlike in the previous section where the finite lifetimes gave a
single width for every QPO peak, now each peak in the power
spectrum will be broadened by a different but predictable amount. Note in
particular how the peaks at the higher harmonics with $n>1$
will be significantly broader (and thus lower in amplitude) than the
fundamental. Another important feature evident from Figure
\ref{chap4_f2} and equation (\ref{broaden1}) is that, due to the
opposite-signed slopes of $\nu_r(r)$ and $\nu_\phi(r)$ around $r_0$, the beat
mode at $\nu_\phi+\nu_r$ remains very narrow, while the peak at
$\nu_\phi-\nu_r$ is quite broad. These features should play a key role
in using the power spectrum as an observable in understanding the
behavior of geodesic hot spots.

As in Section \ref{phase}, the first step in producing the theoretical
power spectrum is to calculate the Fourier amplitude in each mode with
the full three-dimensional ray-tracing calculation of emission from a
single periodic hot spot at $r_0$. Again, the linear properties of the
problem allow us simply to sum a series of Lorentzians, each
with a different amplitude, width, and location (peak frequency), to
get the total power spectrum. The peak amplitudes $A_j$ are given by the
ray-tracing calculations, the locations $\nu_j$ from the geodesic
coordinate frequencies and their harmonics, and the widths
$\Delta\nu_j$ from equation (\ref{broaden1}). 

Since the QPO peak broadening is most likely caused by a combination
of factors including the hot spots' finite lifetimes as well as
their finite radial distribution, the simulated power spectrum should
incorporate both features in a single model. Now the computational
convenience of Lorentzian peak profiles is clearly evident, since the
net broadening is given by the convolution of both effects and
the convolution of two Lorentzians is a Lorentzian: 
\begin{equation}\label{conv_2lor}
[\mathcal{L}(\nu_1,\Delta \nu_1)\star \mathcal{L}(\nu_2,\Delta
\nu_2)](\nu) = \mathcal{L}(\nu_{\rm tot},\Delta \nu_{\rm tot})(\nu),
\end{equation} 
where the peak centers and widths simply add: $\nu_{\rm tot} = \nu_1 +
\nu_2$ and $\Delta \nu_{\rm tot} = \Delta \nu_1+\Delta \nu_2$. In the
case where one or both of the Lorentzians is \textit{not} normalized,
the amplitude of the convolved function is given as a function of the
individual peak amplitudes and widths:
\begin{equation}
A_{\rm tot} = \pi\frac{A_1 A_2 \Delta\nu_1 \Delta\nu_2}
{\Delta\nu_1 + \Delta\nu_2},
\end{equation}
where $A_1$ and $A_2$ are the peak amplitudes of the respective
Lorentzians [$A_j=1/(\pi\Delta\nu_j)$ corresponds to a
normalized function.]

\begin{figure}[tp]
\begin{center}
\includegraphics[width=0.8\textwidth]{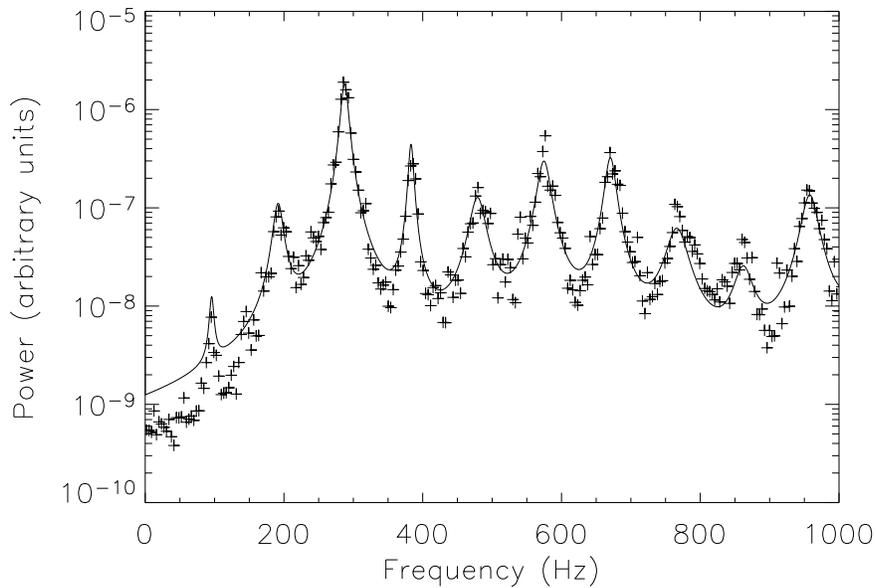}
\caption[Power spectrum of multiple hot spots at different
radii]{\label{chap4_f3} Simulated power spectrum (crosses) from a
ray-tracing calculation of many hot spots on geodesic orbits with 
different radii $r$ and thus different coordinate frequencies, along
with an analytic model (line) of that power
spectrum. The black hole has mass $M=10M_\odot$ and spin $a/M=0.5$,
while the average hot spot orbit has an eccentricity of 0.1 around a radius
of $r_0 = 4.89M$, as in Figure \ref{chap4_f1}. The peaks have Lorentzian
profiles with $\Delta \nu$ given by equations (\ref{dec_width}) and
(\ref{broaden1}) with $T_l=100$ ms and $\Delta r = 0.05M$.}
\end{center}
\end{figure}

Figure \ref{chap4_f3} shows the power spectrum for a collection
of hot spots orbiting near the commensurate
radius $r_0=4.89M$ with a distribution width of $\Delta r =
0.05M$. All other black hole and orbital parameters are identical to
those in Figure \ref{chap4_f1}. 
Both the random phase broadening described in Section \ref{phase} and the
effects of a finite resonance width are included in the model. Again, we
should stress that the solid line is not a fit to the simulated data,
but rather an independent analytic model constructed from the sum of
Lorentzian profiles as
described above. In this example, the hot spots have a typical
lifetime of 30 orbits, so the random phase
broadening contributes only $\Delta \nu \approx 1.5$ Hz for each
peak. While this is rather longer than the expected hot spot lifetime,
it allows us to focus on the effect that a finite resonance
width has on the behavior of the QPO peaks at the coordinate
frequencies and their various beat harmonics. For a resonance width of
$\Delta r = 0.05M$, the peak widths due only to coordinate frequency
broadening are shown in Table \ref{chap4_t1}.

\begin{table}
\caption[Widths of QPO peaks due to finite distribution of
radii]{\label{chap4_t1} Widths of QPO peaks around coordinate
frequency
modes $n\nu_\phi \pm \nu_r$, due to a radial distribution of hot spots with
$\Delta r = 0.05M$, as determined by equation (\ref{broaden1}). For
relatively narrow resonance regions, the QPO
peak widths are linearly proportional to $\Delta r$. The basic black
hole and hot spot model parameters are the same as in Figures
\ref{chap4_f1} and \ref{chap4_f3}.}
\begin{center}
\begin{tabular}{lcc}
Mode & Frequency (Hz) & FWHM (Hz) \\
\hline
$\nu_r$ & 95 & 3.6 \\
$\nu_\phi-\nu_r$ & 190 & 12.2 \\
$\nu_\phi$ & 285 & 8.4 \\
$\nu_\phi+\nu_r$ & 380 & 4.8 \\
$2\nu_\phi-\nu_r$ & 475 & 20.6 \\
$2\nu_\phi$ & 570 & 16.8 \\
$2\nu_\phi+\nu_r$ & 665 & 13.2 \\
\end{tabular}
\end{center}
\end{table}

The narrow peak at $\nu_\phi+\nu_r=380$ Hz and the neighboring broad
peak at $2\nu_\phi-\nu_r=475$ Hz are clearly visible in the simulated
data of Figure \ref{chap4_f3}. Precise measurements of each peak's
amplitude and width may not come until a next generation X-ray timing
mission, but the qualitative behavior shown here should be detectable
with the current observational capabilities of
\textit{RXTE}. Combining equations (\ref{dec_width}) and
(\ref{broaden1}) gives a system of linear equations that can be solved
for the hot spot lifetime $T_l$ and the resonance width $\Delta r$ as
a function
of the QPO peak widths $\Delta \nu_j$. If we could accurately measure the
widths of only two peaks, both $T_l$ and $\Delta r$ could be
determined with reasonable significance. More peaks would give tighter
constraints and thus serve to either support or challenge the assumptions of
the hot spot model.

\begin{figure}[tp]
\begin{center}
\scalebox{0.65}{\includegraphics*[84,360][540,700]{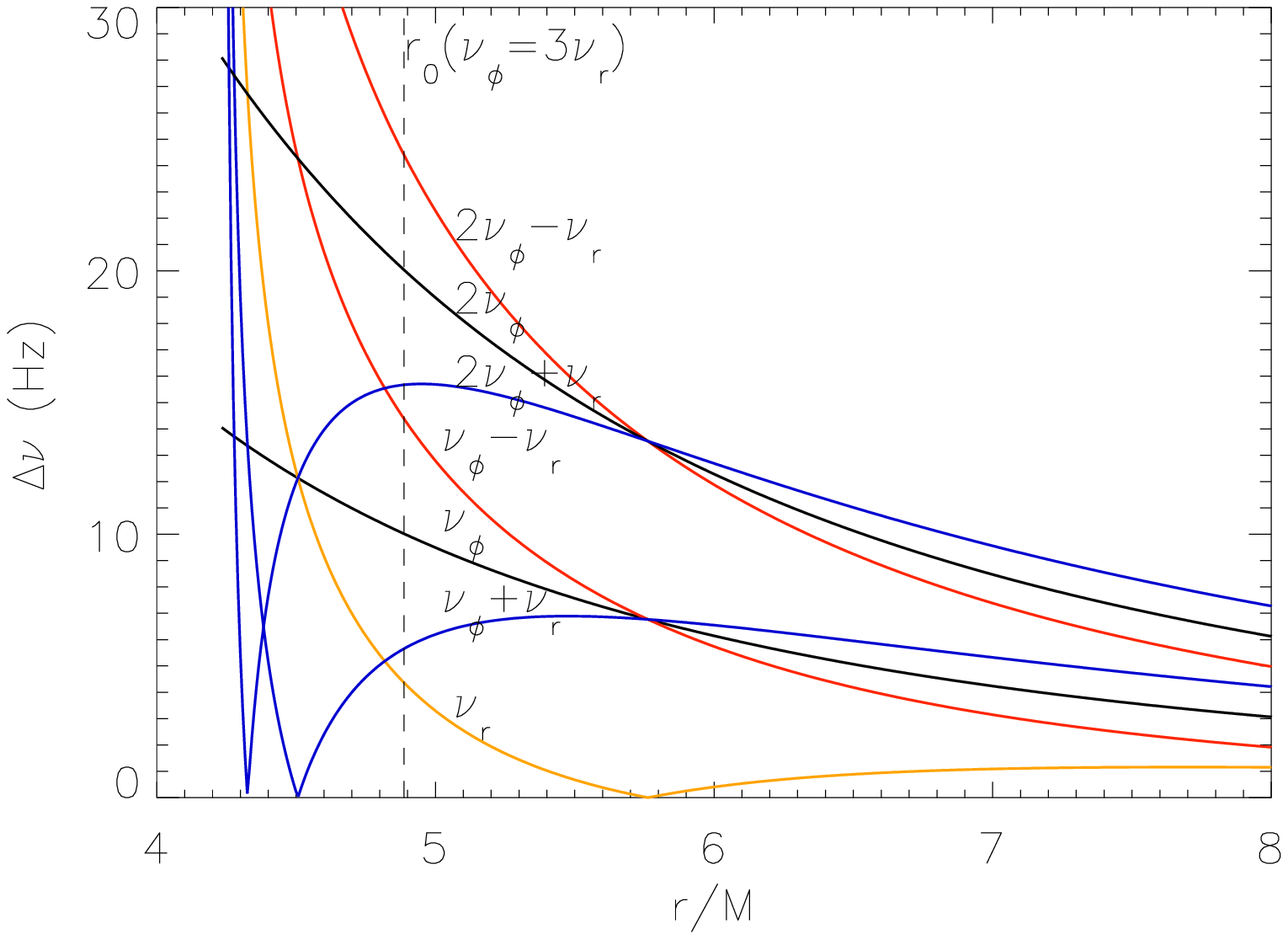}}
\scalebox{0.65}{\includegraphics*[84,360][540,700]{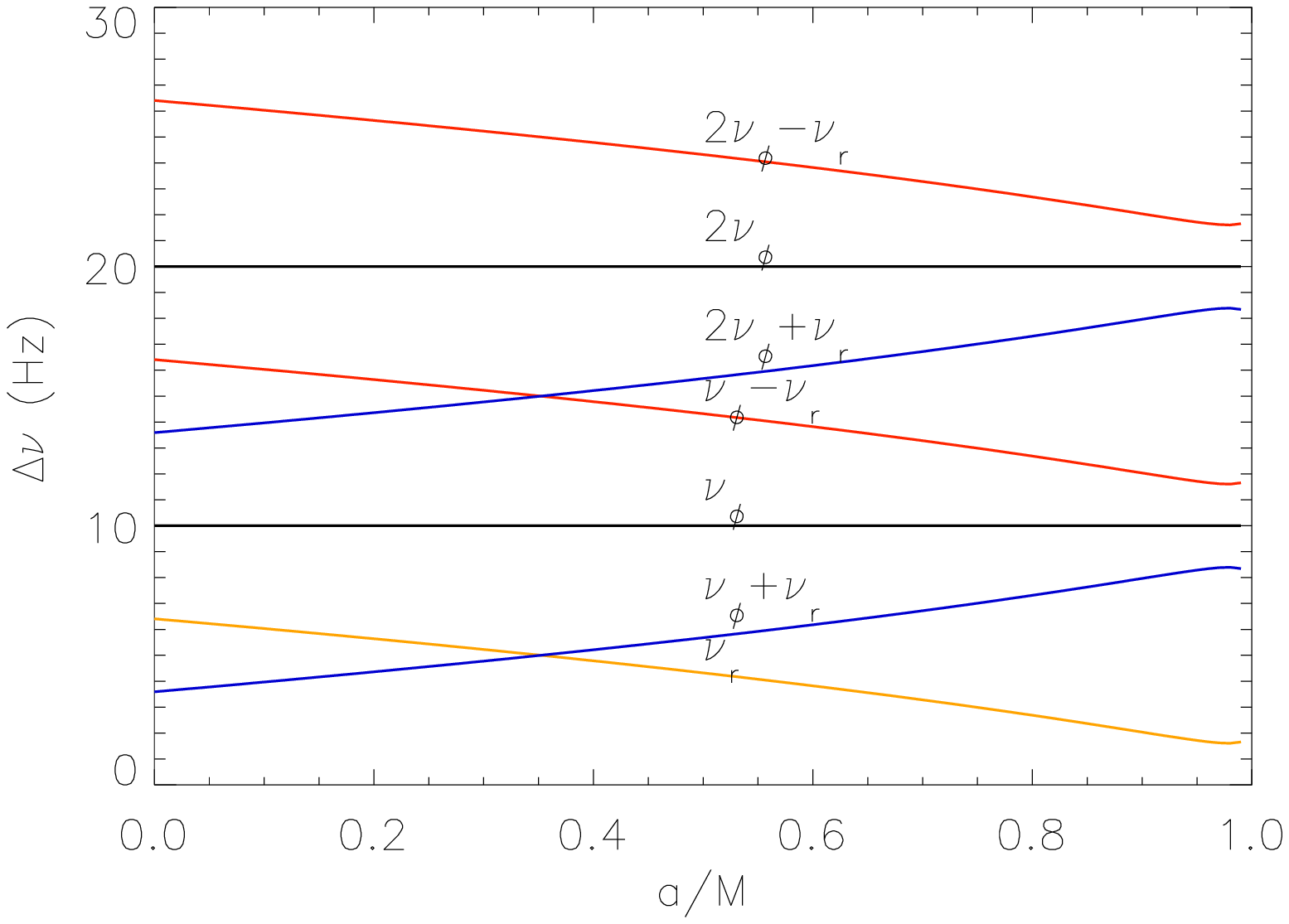}}
\caption[QPO peak widths due to finite resonance
width]{\label{peakwidths_map} Widths of QPO peaks $\Delta \nu_j$
centered at coordinate frequencies $n\nu_\phi\pm\nu_r$ for a black
hole of mass $10M_\odot$, as determined by equation
(\ref{broaden1}). (top) Peak width as a function of hot
spot orbital radius for fixed black hole spin $a/M=0.5$. The vertical
dashed line marks the special radius $r_0$ for which
$\nu_\phi=3\nu_r$. (bottom) Peak width as a function of black hole spin,
assuming resonant orbits around $r=r_0$. To map out the space time
around the hot spot orbit, only the relative widths are important, so
we have normalized $\Delta r$ so that $\Delta\nu_\phi(r_0)=10$ Hz.}
\end{center}
\end{figure}

If the hot spot paradigm is accurate, then by measuring multiple peak
widths we should also be able to map out the spacetime in the
region of the hot spot orbit, thus gaining insight into the specific
resonance mechanisms that may be causing the QPO frequency
commensurability. This technique could conceivably be carried out in
one of (at least) two different ways. If we can assume a given value
for the black hole spin, perhaps by iron line broadening, then by
measuring the QPO peak widths, the specific radius of the preferred hot spot
orbit could be identified. An example of this approach is shown in
Figure \ref{peakwidths_map}a, where the various widths at the
coordinate frequencies $n\nu_\phi\pm \nu_r$ are plotted as a function
of the orbital radius $r$, assuming a black hole spin of
$a/M=0.5$. The absolute vertical scale is set by the width of the
resonance region, but we are generally only interested in
\textit{relative} widths, so here they are normalized to
$\Delta\nu_\phi(r_0)=10$ Hz without any loss of generality. The dashed
vertical line shows the location for the special
commensurate orbit at $\nu_\phi=3\nu_r$. Note how the widths become
degenerate when $\partial \nu_r/\partial r = 0$, around $r\approx
5.75M$ in this case. 

Perhaps the more likely scenario is one in which we do \textit{not}
know the spin value \textit{a priori}, but are reasonably sure that the 3:2
commensurability is forced by a resonance at $r_0$ where
$\nu_\phi=3\nu_r$. For different values of $a/M$, the shape of the
gravitational potential around $r_0$ changes, thus changing the
relative value of the radial epicyclic frequency. In that case,
measuring the widths of multiple
peaks can directly give an estimate for the black hole spin, as shown
in Figure \ref{peakwidths_map}b. As in Figure \ref{peakwidths_map}a,
the vertical scale is normalized so that $\Delta\nu_\phi=10$ Hz, but
only the relative widths between multiple peaks are important. With
high enough precision, this method might even be used to test the
strong-field regime of GR and whether black holes are ``bumpy'' or
indeed ``hairless'' \citep{colli04}.

\section{Electron Scattering in the Corona}\label{scatter}

Another simplified model we have included is that of scattering
photons from the hot spot through a low-density
corona of hot electrons around the black hole and accretion disk. This
is known to be an important process for just about every observed
state of the black hole system
\citep{mccli04}. Unfortunately, it is also an extremely difficult
process to model accurately. Fortunately, for the problem of
calculating light curves and power spectra, a detailed description of
the scattering processes is probably not necessary. The most important
qualitative feature of the coronal scattering is a smearing of the hot
spot image: a relativistic emitter surrounded by a cloud of scattering
electrons will appear blurred, just like a lighthouse shining its beam
through dense fog. The effect is even more 
pronounced in the black hole case, where the hot spot orbital period
is of the same order as the light-crossing time of a small corona,
thus spreading out the X-ray signal in time as well as space. 

Due to the inverse-Compton effect with hot coronal electrons, the
scattered photons are often boosted to higher energies (see Chapter
6). Since each scattering event also adds a time delay to the photon,
a coherent phase lag
in the light curves from different energy channels could be used to estimate
the overall scale length of the corona. \citet{vaugh97} have observed this
effect in neutron star QPOs and infer a scattering length of
$\lambda\sim 5-15M$ for an optical depth of $\tau \sim 5$ in the source 4U
1608-52. \citet{ford99} perform a similar analysis for black holes,
including the possibility for an inhomogeneous corona, and derive a much
larger upper limit for the size of the corona $(\lambda \sim
10^3M)$, although \citet{merlo01} argue for a smaller corona with high
energy density. In either case (large or small scattering length), the
qualitative effect will be the same: the damping of higher harmonic
features in the power spectrum of the X-ray light curve. 

\begin{figure}
\begin{center}
\scalebox{0.45}{\includegraphics*[-20,0][700,520]{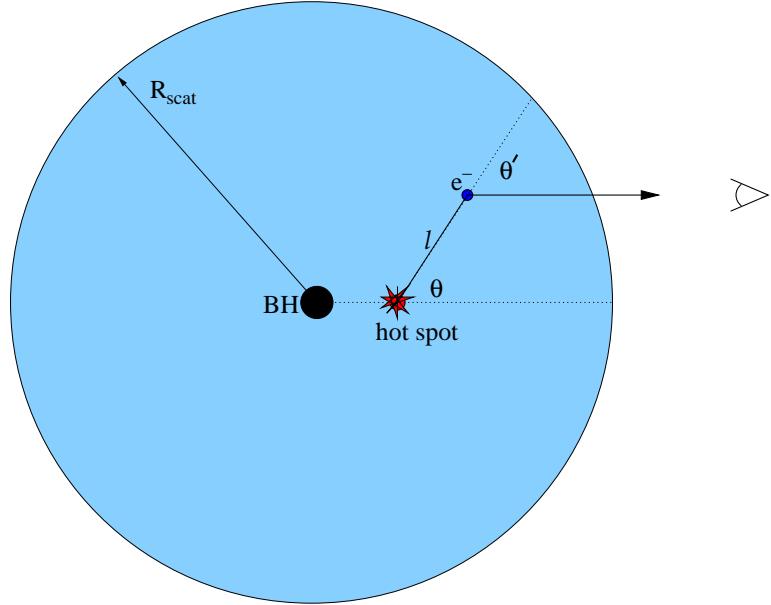}}
\caption[Schematic diagram for coronal scattering]{\label{schem_scat}
Schematic diagram of photon scattering geometry for a hot spot
emitter orbiting a black hole, surrounded by a corona of hot
electrons with length scale $R_{\rm scat}$. The geometry requires
$\theta=\theta'$. The added
photon path length compared to the direct line-of-sight is
$l(1-\cos\theta)$.}
\end{center}
\end{figure}

The simple model we introduce is based on adding a random
time delay to each photon detected from the hot spot. The
distribution of this time delay is computed as follows: we
fix the optical depth to be unity for scattering through a
medium of constant electron density, and for simplicity, each photon
is assumed to scatter exactly once between the emitter and the
observer, thus
determining the length scale of the corona as a function of
density. In Chapter 6 below, we will see that this choice of optical
depth is consistent with the observations of the Steep Power Law
photon energy spectrum. For this constant density model, the
probability of scattering after a distance $l$ is
\begin{equation}
P(l)dl = \frac{dl}{\lambda}e^{-l/\lambda},
\end{equation}
where $\lambda$ is the photon mean free path in the corona. 

Next, due to the
likely existence of an optically thick disk around the black hole equator, we
assume that the photon scattering angle is less than $\pi/2$ (the
scattering angle $\theta'$ is defined as the angle between the
incoming and outgoing wave vectors, so a straight path would correspond to
$\theta'=0$). In
other words, only photons emitted in a hemisphere facing the observer
can ultimately be scattered in the observer's direction. A schematic
view of this geometry is shown in Figure \ref{schem_scat}. For a photon
emitted at an angle $\theta$ to the observer, scattering at a distance
$l$ from the source produces an additional photon path length of
$d=l(1-\cos\theta)$, assuming for simplicity a flat spacetime
geometry. While the photons are emitted with an isotropic
distribution, the scattering distribution is \textit{not}
isotropic. Since the
scattering geometry requires that $\theta=\theta'$, we only detect a
subset of the photons emitted with an angular distribution
in $\theta$ that satisfies this relationship.
In the limit of low-energy photons $(h\nu \ll m_ec^2)$ and elastic
scattering, the classical Thomson cross section $\sigma_T$ for
unpolarized radiation is used \citep{rybic79}:
\begin{equation}
\frac{d\sigma}{d\theta'} =
\frac{3}{8}\sigma_T \sin\theta' (1+\cos^2\theta').
\end{equation}
Integrating this distribution over all forward-scattered photons
$(\theta' < \pi/2)$, we 
find the average additional path length to be $\langle d \rangle = 7l/16$. Since the time
delay is the path length divided by the speed of light $c$, scattering
once in the corona adds a time delay $\Delta t$ to each photon with
probability 
\begin{equation}
P(\Delta t)d(\Delta t) = \frac{d(\Delta t)}{T_{\rm scat}}
e^{-\Delta t/T_{\rm scat}},
\end{equation}
where the average scattering time is given by 
\begin{equation}
T_{\rm scat} = \frac{7}{16}\frac{\lambda}{16c}.
\end{equation} 

Applied to the ray-tracing model, this has the effect of smoothing out
the light curve 
with a simple convolution in the time domain of the original signal
$f(t)$ and the time delay probability distribution function
$P(\Delta t)$. The Fourier transform of the resulting light curve is the
product of the two transforms $F(\nu)$ and $\tilde{P}(\nu)$, where
for notational simplicity, $\tilde{P}(\nu)$ is taken as a
dimensionless Fourier transform of $P(\Delta t)$:
\begin{equation}
\tilde{P}(\nu) = \frac{1}{1+2\pi i T_{\rm scat}\nu}.
\end{equation}
When we square the product to get the
power spectrum $G^2(\nu) = F^2(\nu)\tilde{P}^2(\nu)$, the scaling
factor is another a Lorentzian: 
\begin{equation}\label{scatter_PDS}
G^2(\nu_j) = \frac{A_j^2}{1+(\nu_j/\Delta \nu_{\rm scat})^2},
\end{equation}
where the scale of frequency damping is given by 
\begin{equation}
\Delta\nu_{\rm scat} \equiv \frac{1}{2\pi T_{\rm scat}}
\end{equation}
and $A_j$ are the delta function amplitudes of $F(\nu_j)$ 
as defined above in equation (\ref{segsum}). This analytic result is
perhaps a case where the ends justify the means. Our model for
electron scatting in the corona is extraordinarily simplified,
ignoring the important factors of photon energy, polarization,
non-isotropic emission, multiple scattering events in a
non-homogeneous medium, and all relativistic effects. However,
assuming that almost any analytic model would be equally (in)accurate,
at least the treatment we have applied proves to be computationally
very convenient.

\begin{figure}[tp]
\begin{center}
\includegraphics[width=0.45\textwidth]{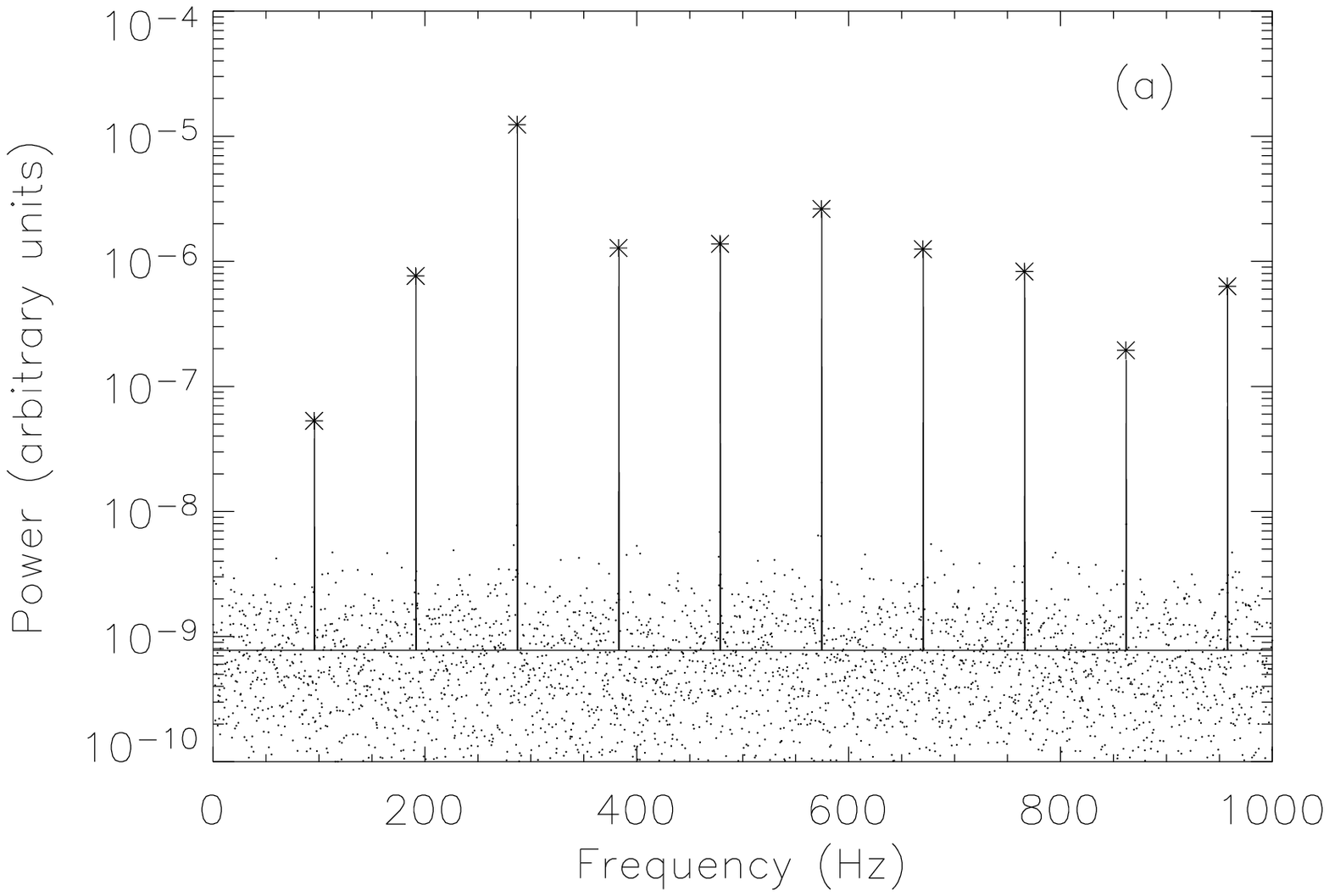}
\includegraphics[width=0.45\textwidth]{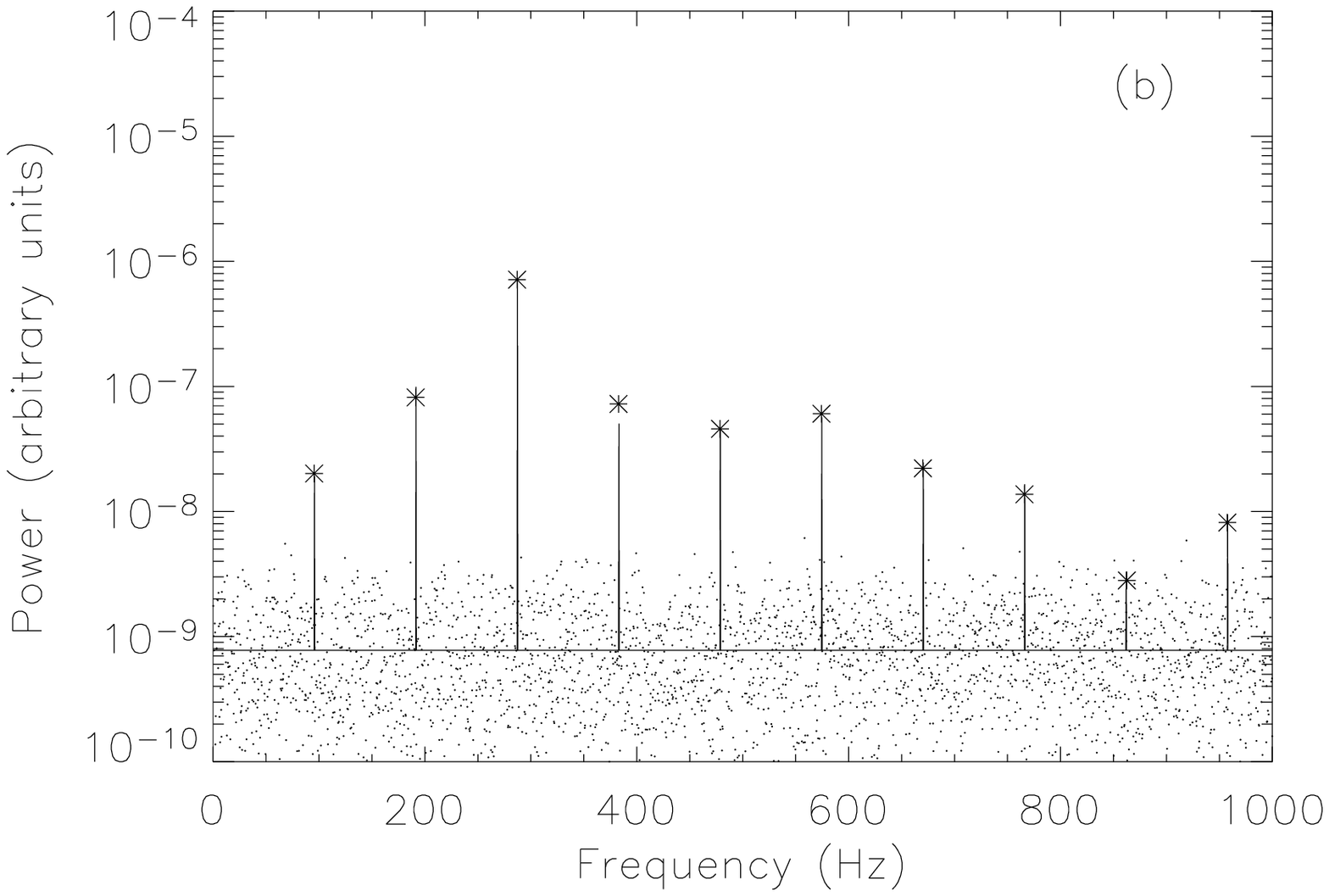}
\caption[Power spectrum from a single hot spot, including coronal
scattering]{\label{chap4_f4} Simulated power spectrum from a single
hot spot light curve where the emitted photons are scattered exactly once
each by a uniform corona of electrons. The simulated spectra
are plotted as dots and asterices, while the analytic model is a solid
line. In (a), the mean free path to scattering is $\lambda = 10M$,
while (b) represents a much larger, low density corona with $\lambda =
100M$.} 
\end{center}
\end{figure}

Equation (\ref{scatter_PDS}) states that the
resulting power spectrum of the scattered light curve is a set of delta
functions, with the higher harmonics damped out by the effective
blurring of the hot spot beam propagating through the coronal
electrons. A simulated power spectrum is shown in Figure \ref{chap4_f4}a for a
scattering length of $\lambda = 10M$, comparable to the size of the
hot spot orbit. Figure \ref{chap4_f4}b shows the effect of a larger,
low-density corona with scale length $\lambda = 100M$, corresponding
to a longer convolution time and thus stronger harmonic damping. The
white background noise (Poisson noise with $\mu$ = 1) in both cases is
due to the statistics of the random scattering of each photon from one
time bin to another. The simulated spectra are plotted as dots
(asterices at $\nu_j$ to highlight the peaks) and the analytic model is a
solid line. 

One significant conclusion from this analysis is that the coronal
scattering alone should not contribute to the broadening of the QPO
peaks. However, it will have a very significant effect on the overall
harmonic structure of the power spectrum, particularly at higher
frequencies. In \citet{schni04a}, we showed a similar result caused by the
stretching of the geodesic blob into an arc along its path, also
damping out the power at higher harmonics. In this context, it is now
clear that the arc damping can be modeled analytically by
interpreting the stretching of the blob in space as a convolution of
the light curve in time. If the stretched hot spot
has a Gaussian distribution in azimuth with length $\Delta \phi$, the
original X-ray light curve will be convolved with a Gaussian window
of characteristic time $T = \Delta \phi/(\pi\nu_\phi)$. A
Gaussian window in time gives a Gaussian profile in frequency space:
\begin{equation}\label{gauss1}
w(t)=\exp\left(\frac{-t^2}{2T^2}\right) \Leftrightarrow 
W(\nu) = \sqrt{2\pi}\frac{T}{T_f}
\exp\left(\frac{-\nu^2}{2\Delta \nu^2}\right),
\end{equation}
where again the characteristic width is given by $\Delta \nu = 1/(2\pi
T)$. The exponential damping of the Gaussian $W(\nu)$ is stronger than
the Lorentzian factor [eqn.\ (\ref{scatter_PDS})] at higher
frequencies, but both effects (coronal scattering and hot spot stretching)
are probably important in explaining the lack of significant power
in the harmonics above $\sim 500$ Hz in the \textit{RXTE}
observations. From the central limit theorem, in the limit of
many scattering events, the time delay distribution should also
approach that of a Gaussian, further damping out the higher frequency
power. Regardless of the precise shape of the convolution window in
time, this simple analytic model shows how the scattering time scale
can be understood as another expression of the causality limits on the
size of the emission region. For an optically thick corona with length
scale $R_{\rm scat}$, all frequency modes above $\nu \sim c/R_{\rm
scat}$ should be damped out significantly.

\section{Fitting QPO Data from XTE J1550--564}\label{data}
In this Section we combine all the pieces of the model
developed above and apply the results to the \textit{RXTE} data from type
A and type B QPOs observed in the low-mass X-ray binary XTE
J1550--564. To compare directly with the data from \citet{remil02}, we
need to change slightly our normalization of the power
spectrum. Following \citet{leahy83} and \citet{vande97}, we define the
power spectrum $Q(\nu)$ (not to be confused with the oscillator
quality $Q$ from Section \ref{phase}) so that the total power
integrated over frequency gives the mean square of the discrete light
curve $I_j=I(t_j)$: 
\begin{equation}\label{Q_rms}
\int_{\nu>0}^{\nu_N} Q(\nu) d\nu =
\frac{1}{N_s}\sum_{j=0}^{N_s-1}\left(\frac{I_j-\langle I
\rangle}{\langle I \rangle}\right)^2, 
\end{equation}
where $I_j$ is sampled over $j=0,...,N_s-1$ with average value
$\langle I \rangle$. In terms of the power spectra used in Sections
\ref{phase} and \ref{freq}, $Q(\nu)$ is given by
\begin{eqnarray}
%Q(0) &=& 2 \nonumber\\
Q(\nu) &=& 2 T_f\frac{\tilde{I}^2(\nu)}{\tilde{I}^2(0)},
\end{eqnarray}
which has units of $[(\rm{rms/mean})^2 \rm{Hz}^{-1}]$.

As we described in Section \ref{intro}, the hot spot model is constructed
in a number of steps. These steps result in a first approximation for
the black hole and hot spot model parameters, after which a $\chi^2$
minimization is performed to give the best values for each data set.
\begin{itemize}
\item The black hole mass and the inclination of the disk are given by
optical radial velocity measurements. We take $M=10.5M_\odot$ and
$i=72^\circ$ as fixed in this analysis \citep{orosz02}. Note this is
somewhat higher than the mass used in Chapter 3 to match
simultaneously LFQPOs and HFQPOs to coordinate frequencies.
\item The black hole spin is determined by matching the
frequencies of the HFQPOs to the geodesic coordinate frequencies such
that $\nu_\phi=3\nu_r$ at the hot spot orbit. This identifies the
frequencies of the two
major peaks with a 3:2 ratio as the orbital frequency $\nu_\phi$ and
its lower beat at $\nu_\phi-\nu_r$. Coupled
with the black hole mass of $10.5 M_\odot$, this assumption gives $a/M
\approx 0.5$ for $\nu_\phi \approx 276$ Hz and $\nu_\phi-\nu_r \approx
184$ Hz. The small uncertainties in the measured
value of $\nu_\phi$ can thus be interpreted indirectly as constraints on
the mass-spin relationship.
\item The orbital eccentricity and hot spot size and overbrightness
are chosen to match the total amplitude of the observed
fluctuations. We use a moderate eccentricity of $e=0.1$, consistent
with the simple approximation of Section \ref{noncircular_orbits} and
equation (\ref{max_eccentricity}).
The question of overbrightness is still an area of
much research, since the nature of the background disk is not well known
during the ``Steep Power Law'' state that produces the HFQPOs
\citep{mccli04}. In practice, we set the hot spot emissivity constant
and then fit an additional steady-state background flux $I_B$ to the
variable light curve.
\item The hot spot arc length and the coronal
scattering time scale are chosen to fit the relative amplitudes of the
different QPO peaks. 
\item The hot spot lifetime and the width of the resonance $\Delta r$
around $r_0$ are chosen to fit the widths of the QPO peaks.
\item As a final step, we include an additional power law component
$\propto \nu^{-1}$ to account for the contribution
due to turbulence and other random processes in the disk [e.g.\
\citet{press78,mande99,pouta99}] not accounted for by
the hot spot model. Instrumental effects such as the detector deadtime
and Poisson counting statistics are
combined with the turbulent noise to give a simple two-component
background spectrum: 
\begin{equation}
Q_{\rm noise}(\nu) = Q_{\rm PL}\nu^{-1}+Q_{\rm flat}.
\end{equation}
\end{itemize}

After using the ray-tracing calculation to determine the Fourier
amplitudes $A_j$ [as in eqns.\ (\ref{segsum}, \ref{multi_lor}, and
\ref{scatter_PDS})] for a single periodic light curve segment,
we minimize $\chi^2$ over the following parameters:
orbital frequency $\nu_\phi$, hot spot lifetime $T_l$, resonance width
$\Delta r$, scattering length $\lambda$, hot spot arc length $\Delta 
\phi$, steady state flux $I_B$, and the background noise components
$Q_{\rm PL}$ and $Q_{\rm flat}$. All these parameters can be combined into a
single analytic expression for the power spectrum, making the $\chi^2$
minimization a computationally simple procedure. The best fit
parameters are shown in Table \ref{chap4_t2}, along with $1\sigma$
($68\%$) confidence limits. These confidence
limits are determined by setting $\Delta\chi^2 < 7.04$, corresponding to
six ``interesting'' parameters of the hot spot model, holding the noise
components constant \citep{avni76,press97}. We find that $Q_{\rm PL}$ and
$Q_{\rm flat}$ are almost identical for both data sets, supporting the 
presumption that they are indeed a background feature independent of the hot
spot model parameters. 

\begin{table}
\caption[Best-fit model parameters for QPOs from XTE
J1550--564]{\label{chap4_t2} Best-fit parameters of the hot spot model
for type 
A and type B QPOs from XTE J1550--564. $(1\sigma)$ confidences are shown in
parentheses, following the approach of \citet{avni76}, corresponding
to a subset of six ``interesting'' parameters. In this case, the background
noise components $Q_{\rm PL}$ and $Q_{\rm flat}$ are held constant
while varying the model parameters.}
\begin{center}
\begin{tabular}{lccc}
  Parameter & & Type A & Type B \\
  \hline
  orbital frequency $\nu_\phi$ (Hz) & & 280.1(2.4) & 270.5(12)\\
  mean lifetime $T_l$ (ms) & & 10(2.0) & 5(1.5) \\
  \hspace{1.6cm} (orbits) & & 2.8(0.55) & 1.4(0.4) \\
  resonance width $\Delta r$ ($M$) & & 0.02(0.05) & 0.025(0.12) \\
  scattering length $\lambda$ ($M$)& & 5(10) & 10(20) \\
  arc length $\Delta\phi$ ($^\circ$) & & 155(30) & 285(20) \\
  flux ratio $\frac{I_{\rm{hot spot}}}{I_B+I_{\rm{hot spot}}}$ & &
  0.085(0.025) & 0.38(0.05) \\
  power law noise $Q_{\rm PL}$ & & $3.5\times 10^{-4}$ & $3.5\times
  10^{-4}$ \\
  flat noise $Q_{\rm flat}$ & & $6.5 \times 10^{-7}$ & $5.8 \times
  10^{-7}$
\end{tabular}
\end{center}
\end{table}

\begin{figure}
\begin{center}
\includegraphics[width=0.65\textwidth]{chap4_f5a.ps}
\includegraphics[width=0.65\textwidth]{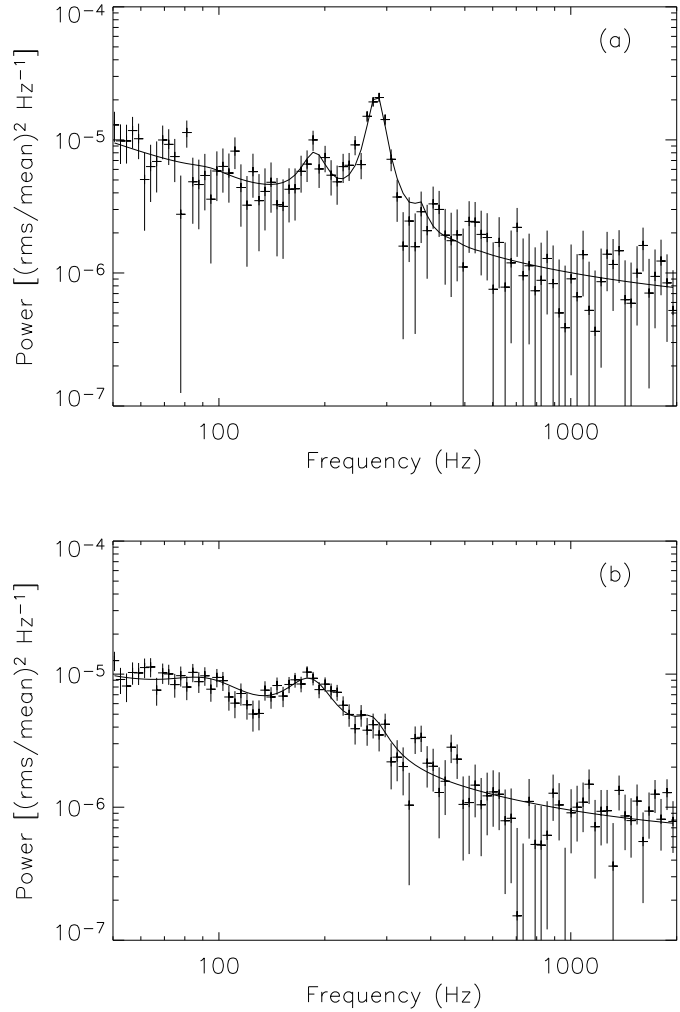}
\caption[Comparison of model spectrum with data from XTE
J1550--564]{\label{chap4_f5} Comparison of hot spot model power
spectrum (line) with data (crosses) from XTE J1550--564 [reproduced
with permission from \citet{remil02}]. (a) Type A QPO, dominated by a
narrow peak at
$\nu_\phi \approx 280$ Hz. (b) Type B QPO, dominated by a broad peak at
$\nu_\phi-\nu_r \approx 180$ Hz. The best fit model parameters for each
data set are shown in Table \ref{chap4_t2} and the resulting QPO
amplitudes and widths are shown in Table \ref{chap4_t3}.} 
\end{center}
\end{figure}

In Figure \ref{chap4_f5} we show the
observed power spectra for type A and type B QPOs, as reported in
\citet{remil02}, along with our best fit models. The type A QPOs are
characterized by a strong, relatively narrow peak at $\nu \approx
280$ Hz, corresponding to $\nu_\phi$ in our model, with a minor peak
of comparable width at $\nu_\phi-\nu_r \approx 187$ Hz. Type B
QPOs on the other hand, have a strong, broad peak around 180 Hz with a
minor peak at 270 Hz. This implies a longer arc for type B,
damping out the higher frequency modes, and a shorter average lifetime,
broadening the peaks. Both types of QPO suggest a narrow
resonance width $\Delta r$, yet the current data does not constrain this
parameter very well. Thus we assume the majority of the peak broadening
is caused by the addition of multiple hot spots with characteristic
lifetimes of $T_l \sim 3$ orbits for the type A QPOs and
about half that for type B.

We performed a covariance analysis of the parameter space near the $\chi^2$
minimum to identify the best-constrained parameters and their relative
(in)dependence. This analysis confirms what the confidence limits suggest:
the best-constrained parameters are the orbital frequency $\nu_\phi$, the hot
spot lifetime $T_l$, the arc length $\Delta \phi$, and the background flux
$I_B$. For the type A QPOs, we find $\nu_\phi$ and
$T_l$ to be independent, while the arc length and background flux are
strongly correlated, so that $\Delta \phi/I_B$ is positive and roughly
constant within our quoted confidence region. This is because, for
shorter arcs with fixed emissivity, increasing the arc length will
increase the amplitude of the light curve modulation, requiring a
larger background flux to give the same QPO amplitude. For the type B
QPOs on the other hand, a longer arc length does not significantly
amplify the modulation, since in the limit $\Delta \phi \to
360^\circ$, the light curve would remain constant, and thus the
parameters $\Delta \phi$ and $I_B$ are relatively independent. For
both type A and type B
QPOs, we find that the resonance width and the coronal scattering
length are independent, yet not very well constrained. This is because
these parameters are most sensitive to the higher frequency peaks,
which appear to be damped out by the Gaussian arcs.

\begin{table}
\caption[Amplitudes and widths of QPO peaks from XTE
  J1550-564]{\label{chap4_t3} Amplitudes and widths of type A and type
  B QPO
peaks from XTE J1550--564, as determined by the best fit parameters
listed in Table \ref{chap4_t2} and equation (\ref{rms_j}). $(1\sigma)$
confidences are shown in parentheses.}
\begin{center}
\begin{tabular}{lclclclcl}
  & & &A& & & &B& \\
  Mode & & rms & & FWHM & & rms & & FWHM \\
  & & (\%) & & (Hz) & & (\%) & & (Hz) \\
  \hline
  $\nu_r$ & & 0.57(0.15) & & 33.1(6.2)& &2.03(0.21)& & 63.6(16.0) \\
  $\nu_\phi-\nu_r$ & & 1.62(0.26) & & 35.7(5.9)& &2.57(0.14)& &67.6(15.5)\\
  $\nu_\phi$ & & 3.35(0.17) & & 34.6(5.5)& &1.48(0.24)& &65.9(15.3)\\
  $\nu_\phi+\nu_r$ & & 0.75(0.19) & & 33.4(5.8)& &0.06(0.02)& &64.1(15.8)\\
\end{tabular}
\end{center}
\end{table}

The resulting amplitudes and widths of the major QPO peaks are shown
in Table \ref{chap4_t3}, along with $1\sigma$ confidence limits. These
amplitudes are given by the analytic model so that
the total rms in the peak centered at $\nu_j$ is 
\begin{equation}\label{rms_j}
{\rm rms}_j = \sqrt{2}\frac{A_j'}{A_0'},
\end{equation}
where $A_0'$ is the mean amplitude of the light curve (including the
background $I_B$) and $A_j'$ are the original Fourier amplitudes $A_j$
given by the ray-tracing code, appropriately scaled according to
equation (\ref{scatter_PDS}). This is more
instructive than measuring the rms directly from
$Q(\nu)$, which includes background power and instrumental effects
uncorrelated to the actual QPO peaks. 

In \citet{schni04a}, the hot spot light curve was added to a steady-state
disk with emissivity that scales as $r^{-2}$, which provides an
estimate of the size and overbrightness of the hot spots required to
produce a given (rms/mean) amplitude in the light curve. Considering that
most high frequency QPOs are observed with the greatest significance
in the 6-30 keV energy band during the steep power-law
spectral state \citep{mccli04}, it seems rather unlikely that the
background flux \textit{is} coming directly from a thermal, optically
thick disk. Even
if the flux is originally produced by such a disk, it clearly 
undergoes significant scattering in a hot corona to give the high
temperature power law observed in the photon energy spectrum.

In the context of the
model presented here, we can only calculate the fraction of the total flux
that is coming from the hot spots, determined by fitting to the QPO
data, without presuming an actual model for the background
emission. For XTE J1550--564, we find that the type A hot spot/arcs
contribute $8.5\%$ of the total observed flux, while the type B
arcs must contribute significantly more $(38\%)$ to give a comparable
amplitude. This is due to the longer arc length described above: in
the limit of an azimuthally symmetric ring, even infinite brightness
would produce no variability.

\section{Higher Order Statistics}\label{bispectrum}
\subsection{The Bispectrum and Bicoherence}
As discussed in Section \ref{alt_models}, a variety of
theoretical models have been proposed to explain the existence and
locations of the high frequency QPOs seen by \textit{RXTE},
particularly the ones with multiple peaks at commensurate
frequencies. Many of these models require a spinning black hole, but
often rather
different values of the spin [compare, e.g.\ \citet{abram04b},
\citet{rezzo03a}, and \citet{asche04b} with $a/M\approx 0.4$, 0.94,
and 0.996 respectively]. Therefore, there is
still much ``astrophysics'' that must be understood before the
fundamental physics can be probed in these systems, but there is
strong cause for optimism that these systems really will ultimately
tell us something profound about spinning black holes.

A key first step to disentangling the astrophysics is, of course,
to develop models which not only match the important frequencies,
but also include radiation mechanisms such that the observed X-rays would
actually be modulated at that frequency. We believe the hot spot model
developed above, motivated physically by the parametric resonance
model of \citet{abram01}, is a very useful building
block for analyzing the observations and understanding the emission
properties of the accretion disk. Ideally, any QPO model should not
only be able to explain current observations, but should also make
predictions for future observations. By doing so, the model can either
be further verified or possibly rejected. One of the results of this
Chapter is that different sets of model parameters can produce roughly
the same Fourier power density spectrum with dramatically different
qualitative appearances to the light curves. In this Section, we will
show that higher order variability statistics can break this
degeneracy.

For now, we will focus on computing the \textit{bispectrum} and the
closely related \textit{bicoherence}. The bispectrum computed from a time
series broken into $K$ segments is defined as:
\begin{equation}
B(k,l)=\frac{1}{K} \sum_{i=0}^{K-1} F_i(k)F_i(l)F^*_i(k+l),
\end{equation}
where $F_i(k)$ is the $k^{\rm th}$ frequency component of the discrete Fourier
transform of the $i^{\rm th}$ time series [see e.g.\ \citet{mende91,fackr96}
and references within].  It is a complex quantity that measures the
magnitude and the phase of the correlation between the phases of a
signal at different Fourier frequencies.  Its expected value is unaffected by
additive Gaussian noise, although its variance will increase for a
noisy signal.

A related quantity, the bicoherence, is the vector magnitude of the
bispectrum, normalized to lie between 0 and 1.  Defined analogously to
the cross-coherence function \citep{nowak96}, it is proportional to the
vector sum of a series of bispectrum measurements, appropriately
normalized as follows: If the biphase (the
phase of the bispectrum) remains constant over time, then the
bicoherence will have a value of unity, while if the phase is random,
then the bicoherence will approach zero in the limit of an infinite
number of measurements.  Linear variability is that in which the
variability on different timescales is uncorrelated.  Thus if the
Fourier phases at different frequencies are not random relative to one
another, the variability is correlated on these frequencies, and hence
it is non-linear.  Mathematically, the bicoherence $b$ is defined as:
\begin{equation}
b^2(k,l) = \frac{\left|\sum{F_i(k)F_i(l)X^*_i(k+l)}\right|^2}{\sum{\left|F_i(k)F_i(l)\right|^2}\sum{\left|F_i(k+l)\right|^2}}.
\end{equation}
This quantity's value \textit{is} affected by Gaussian noise, but it can be
considerably more useful than the bispectrum itself for determining
whether two signals are coupled non-linearly.  In an
astronomical time series analysis context, it has been previously
applied to the broad components in the power spectra of Cygnus~X-1 and
GX~339-4, in both cases finding non-linear variability through the
presence of non-zero bicoherences over a wide range of frequencies
\citep{macca02}.

\subsection{The Bicoherence of the Simulated Data}
We now apply the bicoherence to the simulated data. We consider two
different
model calculations from \citet{schni05} which give similar
power spectra (see Figs.\ \ref{chap4_f1} and \ref{chap4_f3}). In each
case, the quasi-periodic oscillations are produced by a 3:1 resonance
between the orbital frequency and the radial epicyclic frequency.
The parameters have been chosen such that the
orbital frequency is 285 Hz, and the radial epicyclic frequency is 95
Hz (see Section \ref{hotspot}). This corresponds to a black hole mass
of 10 $M_\odot$ and a spin 
$a/M=0.5$, with the resonance occurring at a radius of 4.89$M$; all
these parameters compare reasonably well to those observed in XTE J1550--564
\citep{mille01,remil02}. The disk inclination is also
fixed to be 70 degrees; this parameter does not affect the frequencies
observed, but can affect the amplitudes of the QPOs in the context of
the model we are considering here \citep{schni04a}.
In each case we compute 1000 seconds of simulated data with a binning
timescale of the light curve of 0.1 msec. We then compute Fourier
transforms by breaking the data into 2441 segments of 4096 data points,
making use of 999.84 seconds of the simulated data.

\begin{figure}
\begin{center}
\includegraphics[width=0.6\textwidth]{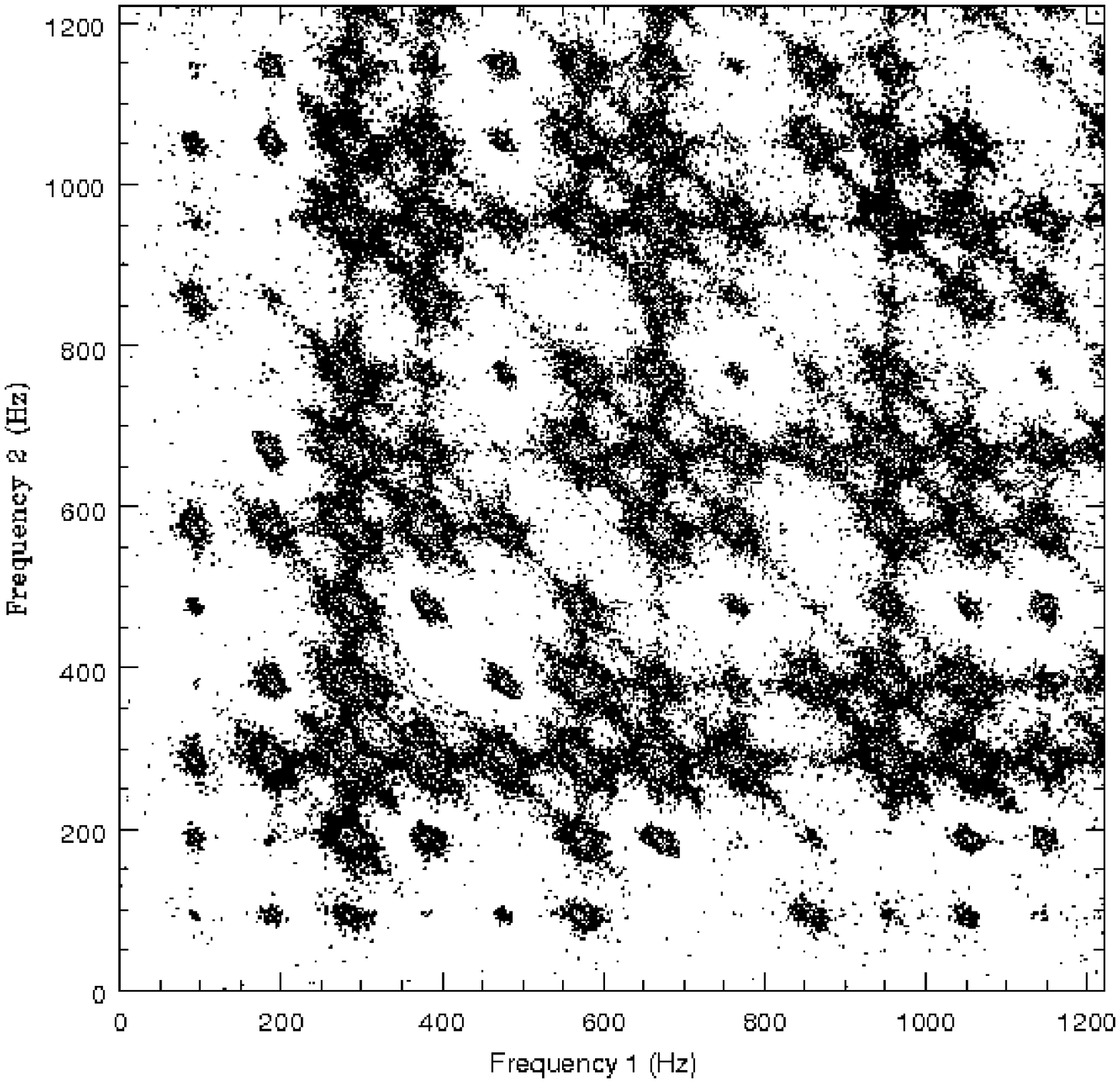}
\includegraphics[width=0.6\textwidth]{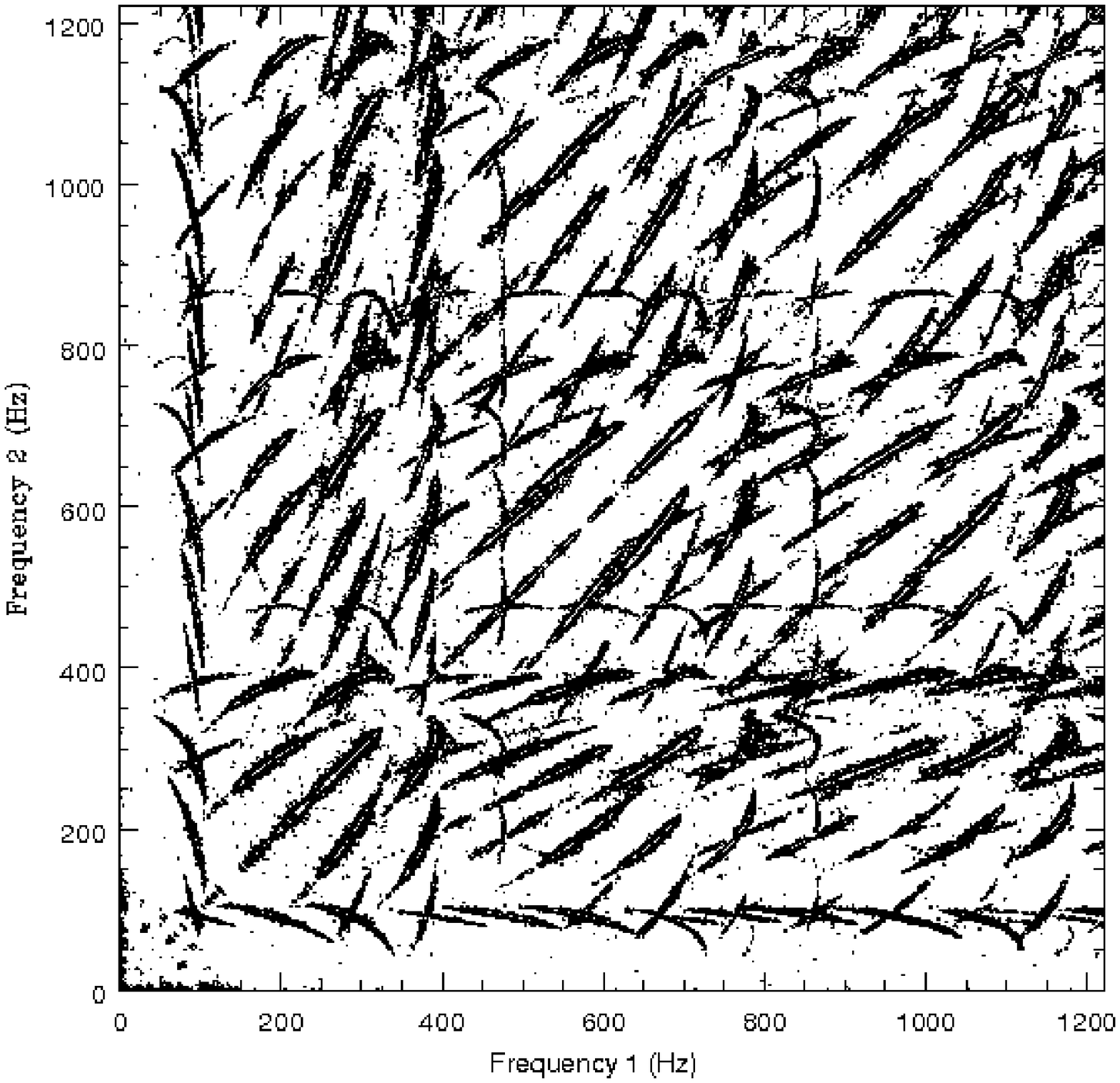}
\caption[Simulated bicoherence for ideal
detector]{\label{sim_bicoherence} The bicoherences $b^2(\nu_1,\nu_2)$
for case 1 (top) and case 2 (bottom) described in the text, with no
Poisson statistics corrections made. The
contour levels for the squared bicoherence $b^2$ are
$10^{-0.5},10^{-0.75},10^{-1.0},10^{-1.25},10^{-1.5},$ and
$10^{-2.0},$ in the colors black, red, green, dark blue, light blue,
and yellow, respectively.  The frequencies correspond to a $10M_\odot$ black
hole with spin parameter $a/M=0.5$.  Note that the symmetry through the
line $x$=$y$ is trivial.}
\end{center}
\end{figure}

In the first case, the emission comes from short lived hot spots with
their orbits all
at a single radius, being continually created and destroyed with a
characteristic lifetime of four orbits. In the second case,
long-lived (lifetimes of 100 msec, or about 30 orbits) hot spots are
distributed over a range of radii ($\Delta r = 0.05M$). In both cases,
the hot spots are on orbits with eccentricities of $e=0.1$. In this
Section, we have not included the extended arcs and coronal scattering
that damp out higher frequency modes. For each
model, the variability appears quasi-periodic, rather than truly periodic, but
for different reasons. In the first case, the creation and
destruction of hot spots on short timescales leads to a phase jitter
in the light curves. These discontinuous, finite lifetimes broaden the observed
periodicity, as described above in Section \ref{phase}. In the second
case, the power spectrum is truly showing 
that there are many periodicities in the system, with coherent phases
as in Section \ref{freq}.
The bicoherence easily detects this difference, as can be seen from
Figure \ref{sim_bicoherence}. In case 1, the bicoherence shows
nearly identical elliptical peaks
at various combinations of frequencies where there is power at $\nu_1$,
$\nu_2$ and $\nu_1+\nu_2$ in the contour plot, essentially delta function
peaks convolved with two-dimensional Lorentzians due to the random
phase broadening.  In case 2, the bicoherence shows thin elongated
peaks, oriented in a variety of directions depending on the
derivatives of $\nu_1$ and $\nu_2$ with respect to $r$.

The reason for this difference is straightforward.  In the first case,
all hot spots have the same geodesic frequencies, so during a hot
spot's lifetime, it is phase locked to all the other hot spots, giving
a collection of delta function peaks at the coordinate
frequencies. The finite lifetimes of the hot spots will broaden the
delta functions into QPOs, with a
similar Lorentzian width as described in Section \ref{phase}.
The hot spots being created
and destroyed in the middle of a Fourier transform window will thus
create leakage in the power of the QPO to frequencies near the central
frequency, but there will be a phase relation between the power in
these frequencies and the phase in the central frequency. This effect
should thus provide a broadening in the bicoherence similar to
that in the power spectrum. 

The shape and orientation of the elliptical peaks can be understood by
inspecting the shapes of the peaks in the power spectrum. Treating the
frequency distribution $\delta \nu = \nu-\nu_{\rm peak}$ around each
peak as a independent random variable with probability
\begin{equation}
P(\delta \nu) \sim \left[1+ \left(\frac{\delta \nu}{\Delta
\nu}\right)^2 \right]^{-1},
\end{equation}
the distribution of the bicoherence can be written as
\begin{equation}\label{Bkl_shape}
b^2(\nu_1,\nu_2) \sim P(\delta \nu_1)P(\delta \nu_2)
P(\delta\nu_1+\delta\nu_2).
\end{equation}
Expanding equation (\ref{Bkl_shape}) around the center of each peak
and defining $x \equiv \delta \nu_1/\Delta \nu$ and $y \equiv \delta
\nu_2/\Delta \nu$, we see that contours of constant bicoherence have
the form
\begin{equation}
(1+x^2+y^2+x^2y^2)(1+x^2+2xy+y^2) = {\rm const}.
\end{equation}
For small deviations $(x,y \ll 1)$, these contours can be written
\begin{equation}
x^2 + y^2 +xy = {\rm const},
\end{equation}
which is the formula for an ellipse with $a/b = \sqrt{3}$, oriented
with the semimajor axis parallel to the line $y=-x$, as can be seen
clearly in Figure \ref{sim_bicoherence}a.

\begin{figure}
\begin{center}
\hspace{2cm} \includegraphics[width=0.75\textwidth]{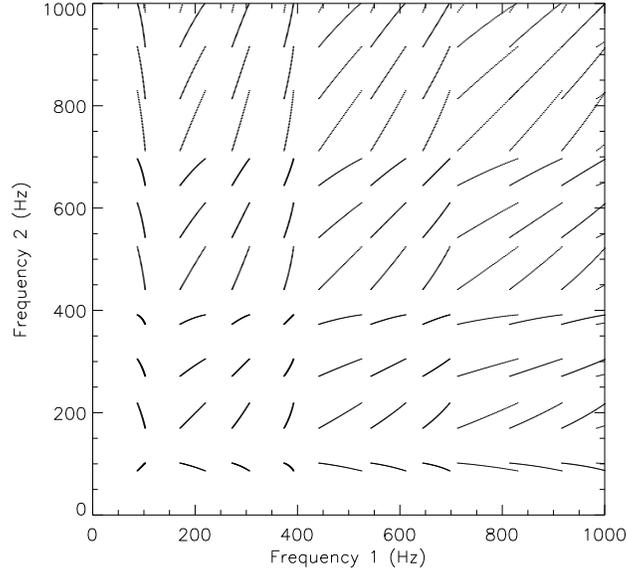}
\caption[Analytic model for bicoherence of hot spot
model]{\label{analytical}The tracks showing how different harmonics of
the QPO vary
with respect to one another when the radius of the hot spot orbit
varies around a central value $r\approx r_0\pm0.2M$.  The
frequencies correspond to a $10M_\odot$ black hole with spin parameter
$a/M=0.5$.}
\end{center}
\end{figure}

In the second case, where there are many frequencies in the power
spectrum due to hot spots found over a range of radii, there will be
phase coherence between the different harmonics of each individual hot
spot, but not with the hot spots at slightly different frequencies.
There will thus be bicoherence between the various harmonic
frequencies found at any individual radius, but not between
frequencies found at different radii. This second case could be
especially interesting. We have calculated
analytically the relationships between different coordinate
frequencies if the radius at which the hot spot occurs is allowed to
vary, and have plotted them in Figure \ref{analytical}.  If in real
data, similar tracks are seen, then, in the context of this model,
they would give the relationships between the different relativistic
geodesic frequencies. In principle
one might expect that, since these tracks trace the coordinate
frequencies as a function of radial distance from the black hole, they
could be used to make precise measurements of the black hole's mass
and spin, plus the central radius of the perturbations, similar to the
approach described in Section \ref{freq}. In practice, the range of
radii allowed is likely to be quite small, as we saw in Section
\ref{data}, Table \ref{chap4_t2}, so this method would probably be of
use only with {\it extremely} high signal-to-noise data.

\subsection{Simulations with Poisson Noise}

To consider whether this observational test is really feasible, we
have performed simulations with the rms amplitude of the oscillations
reduced to realistic levels and with Poisson noise added.  We consider
two count rate regimes---one similar to that detected by {\it RXTE}
for the typical X-ray transients at about 10 kpc, which is about
10,000 counts per second, and another which would be expected from the
same source, but with a 30 m$^2$ detector.  In each case, we allow 6\%
of the counts to come from the variable component and to have,
intrinsically, count rates given by the simulated light curves of
\citet{schni05}, and the remaining 94\% of the counts to come from a
constant component.  We then simulate observed numbers of counts in
100 microsecond segments as Poisson deviates \citep{press97} of
the model count rates.

For the {\it RXTE} count rates, we find that
the bicoherence plots show only noise and only the strongest peak in
the power spectrum is clearly significant in a 1000 second simulated
observation, while marginal detections exist for the QPOs at
two-thirds of and twice this frequency ($\nu_\phi-\nu_r$ and
$2\nu_\phi$).  This is as expected based on
real data, which generally requires exposure times much longer than
1000 seconds to detect these QPOs
\citep{stroh01a,mille01,remil02}. However, since
the signal-to-noise in the bicoherence is generally worse
than the signal-to-noise in the power spectrum, bicoherence
measurements may be possible only when a peak in the power spectrum
is considerably stronger than the Poisson level.

\begin{figure}
\begin{center}
\includegraphics[width=0.6\textwidth]{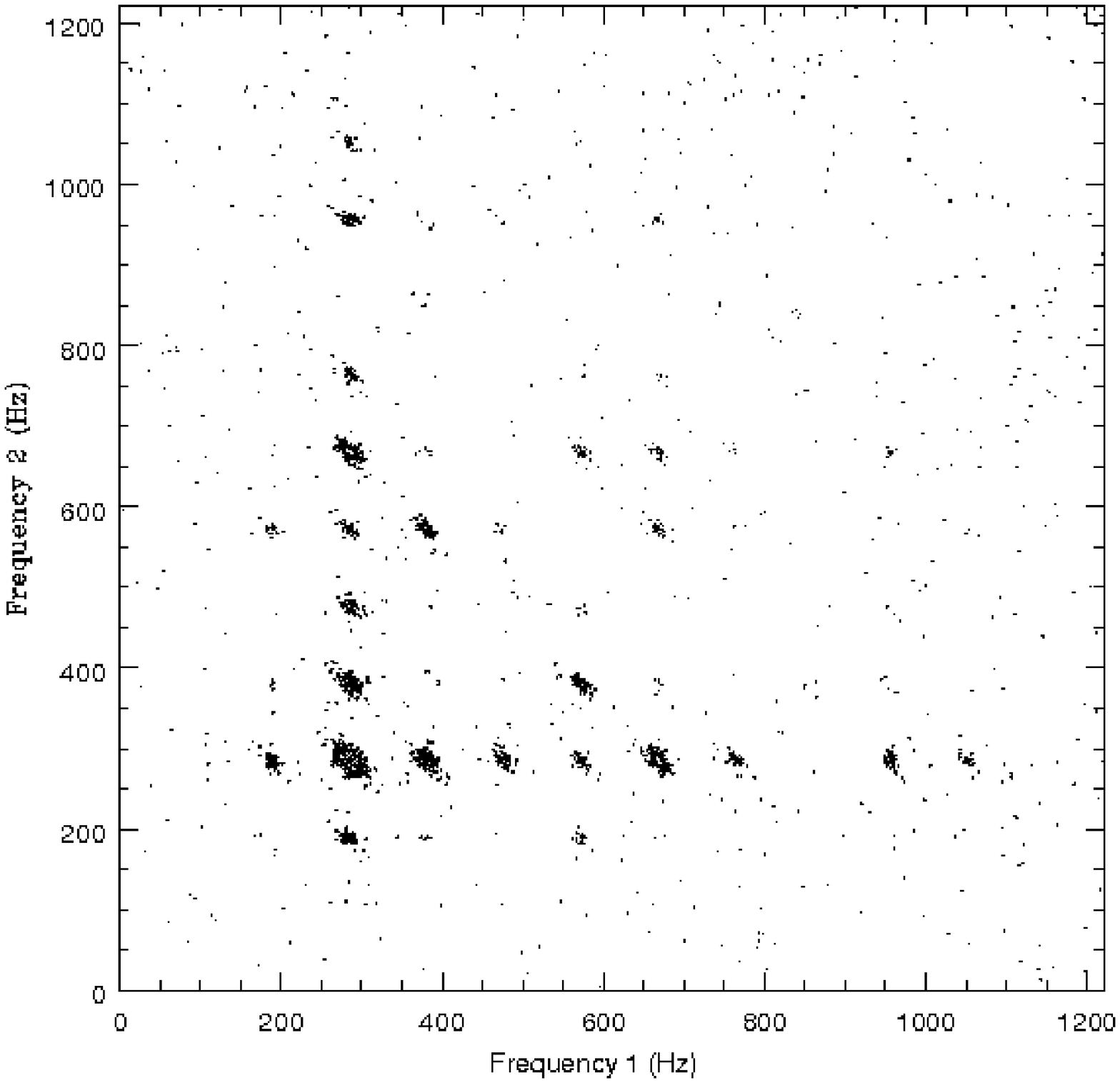}
\includegraphics[width=0.6\textwidth]{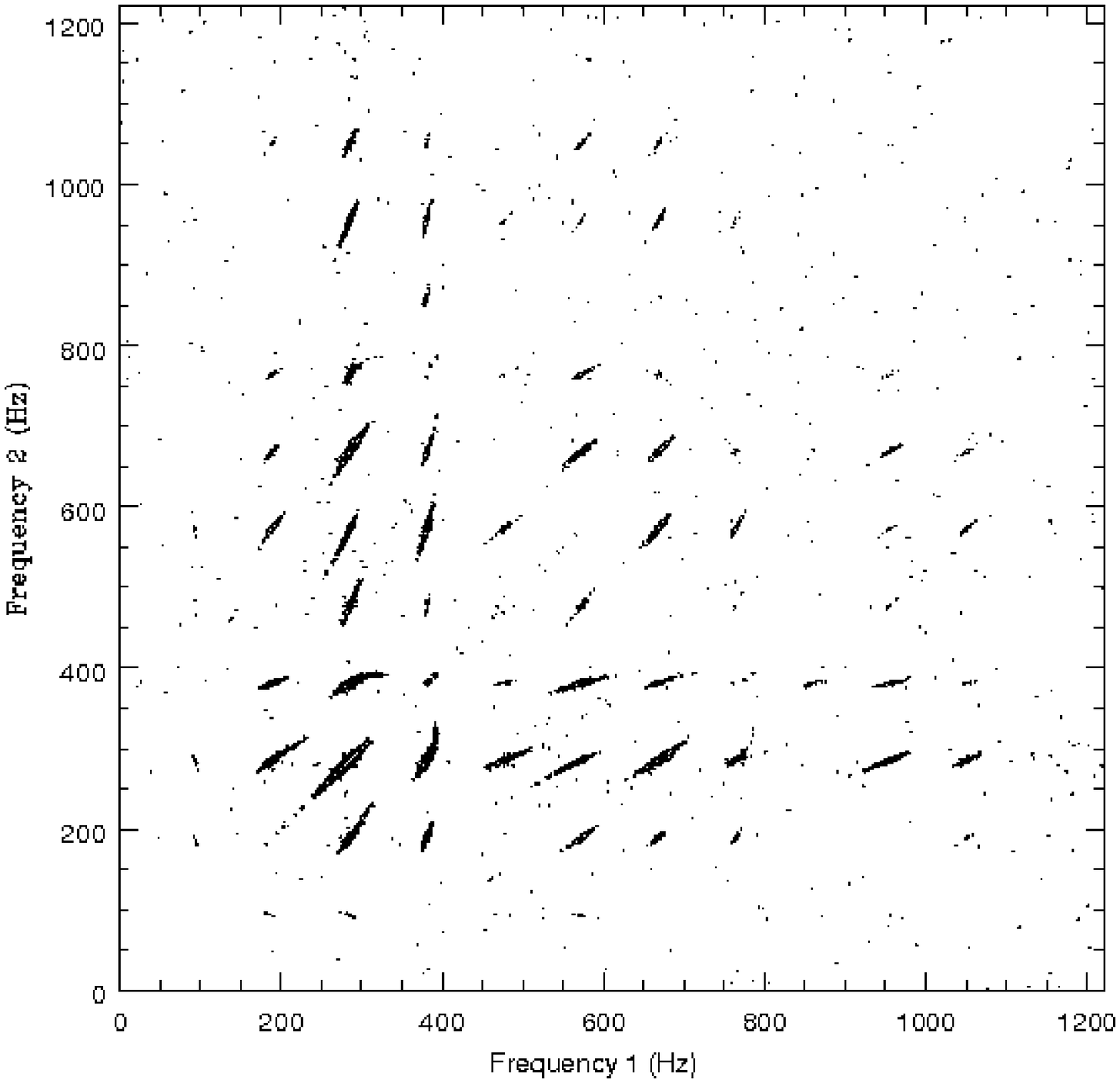}
\caption[Simulated bicoherence for next-generation X-ray timing
mission]{\label{xeus} The bicoherences $b^2(\nu_1,\nu_2)$ for case 1
(top) and case 2 (bottom) described in the text,
assuming 30 m$^2$ area for the detector and a 1000 second integration.
The color scale is the same as in Figure \ref{sim_bicoherence}. The
frequencies correspond to a $10M_\odot$ black hole with spin parameter
$a/M=0.5$.}
\end{center}
\end{figure}

For the count rates expected from a 30 m$^2$ detector, we find that
even within 1000 seconds, several of the higher (i.e. $n>4$) harmonics
are observable in the power spectrum and show the clear elongation in
the bicoherence plot for case 2, indicating that proposed
next-generation timing missions
should be capable of making use of the bicoherence for studying
HFQPOs.  A few very weak peaks are seen in the bicoherence in case 1
even in 1000 seconds.  The simulated bicoherences for a 30 $m^2$
detector are plotted in Figure \ref{xeus}.  We note that these simulations are
a bit over-simplified, in that we have not included the lower
frequency QPOs and low frequency band-limited noise that are typically
observed in conjunction with the HFQPOs, but that these variability
components should not significantly affect the 
phase coupling of the high frequency QPOs.  We also note that it might
be possible to make use of the bicoherence even with {\it RXTE} if a
nearby X-ray transient goes into outburst, but that in such a
case, the deadtime effects we have neglected here might become
important.

\section{Summary}\label{chap4_summary}

In the context of a geodesic hot spot model, we have developed a few
simple analytic methods to interpret the amplitudes and widths of QPO
peaks in accreting black holes. The model combines
three-dimensional ray tracing calculations in full general relativity
with analytic results of basic convolution theory, which are in turn
confirmed by simulating the observed light curves of multiple hot
spots. Given the Fourier amplitudes of a single hot spot light curve, we have
derived a simple formula for the complete QPO power spectrum made
up of Lorentzian peaks of varying amplitudes and widths. This
power spectrum can then be fit to observed QPO data and used to
constrain parameters of the hot spot model, and possibly measure the
black hole mass and spin.

For XTE J1550--564, the locations of the HFQPO peaks are well
constrained, in turn constraining the spin parameter $a/M$ when
combined with radial velocity measurements of the black 
hole mass. Based on the presumption that the 3:2 frequency
ratio is indeed caused by closed orbits with coordinate frequencies in a
3:1 ratio, an observed mass of $M=10.5\pm 1.0 M_\odot$ and orbital
frequency $\nu_\phi=276\pm5$ Hz would predict a spin of $a/M=0.5\pm
0.1$ \citep{orosz02,remil02}. If reliable, this coordinate frequency
method would give one of the best estimates yet for a black hole spin,
although it is admittedly very model dependent.

The amplitudes of the QPO peaks can be used to infer the arc
length of the sheared hot spot and the relative flux contributions
from the hot spot and the background disk/corona. The longer arcs seen
in type B QPOs are also consistent with the broader peaks: if the hot
spots are continually formed and destroyed along special closed
orbits, as the emission region gets stretched into a ring, it is more
likely to be dissolved or disrupted, giving a shorter characteristic
lifetime $T_l$ and thus broader peaks.

Some of the power spectrum features discussed in this paper are unique
to the geodesic hot spot model, while others could be applied to more
general QPO models. Clearly the harmonic amplitudes $A_j$ given by the
ray-tracing calculation are dependent on the hot spot model, as is the
broadening from a finite resonance width, yet both could be
generalized and applied to virtually any perturbed disk
model. Similarly, the peak broadening and the damping of
higher harmonics due to photon scattering will be qualitatively the
same for any emission mechanism that produces periodic light curves
from black holes.

Unfortunately, the quality of the QPO data is not sufficiently
high to confirm or rule out the present hot spot model, leaving a
number of questions unanswered. By fitting only two or three peaks, we are
not able to tightly constrain all the model parameters, particularly the
scattering length scale and the resonance width, both of which are
most sensitive to the higher frequency harmonics. Since the high
temperature electrons in the corona tend to transfer energy into the
scattered photons, measuring the energy spectra of the different QPO
peaks would also prove extremely valuable in understanding the
emission and scattering mechanisms. For this analysis to be most
effective, a more accurate model for the electron
scattering will certainly be necessary (see Chapter 6 below). This has
been done to some degree
with the lower frequency region of the power spectra from black holes
and neutron stars \citep{ford99}, and may even be observable above $\sim
100$ Hz with current \textit{RXTE} capabilities, but very well may have to
wait for a next generation X-ray timing mission.

In the immediate future, however, there is much more to be done with
the \textit{RXTE} data that already exists. Important additional insight might
be gained from new analyses of the X-ray light curves in the time
domain or by using higher-order statistics like the bicoherence and
bispectrum, recovering some of the phase information lost when the power
spectrum is computed in frequency space \citep{macca04}. There is also
an important
message in the relationship between the photon energy spectra and the
QPO power spectra as well as the connection between the low frequency
and high frequency QPOs. Why should the HFQPOs appear in certain spectral
states and not others? The answer to these questions may lie in new
models of the accretion disk and specifically the radiation physics
relating the thermal and power-law emission, as well as broad
fluorescent lines like Fe K$\alpha$. The fact that the HFQPOs are seen
most clearly in the 6-30 keV energy range suggests that standard
models of thin, thermal accretion disks are not adequate for this
problem. This emphasizes the essential role of radiation transport,
particularly through the corona, in any physical model for black hole
QPOs.

\chapter{Steady-state $\alpha$-disks}
\begin{flushright}
{\it
The secret to creativity is knowing how to hide your sources.\\
\medskip
}
-Albert Einstein
\end{flushright}
\vspace{1cm}

As mentioned in Section \ref{history_theory} in the Introduction, the
standard accretion disk model was developed by \citet{shaku73}, followed
shortly by \citet{novik73} (hereafter ``NT'') with a relativistic extension
for the Kerr metric. The defining characteristic for both models is
the ``alpha'' prescription for transporting angular momentum via a
turbulent viscosity that is proportional to the local pressure in the
disk. This shear stress generates heat, which is then radiated from
the top and bottom surfaces of the steady-state disk, as the gas loses
gravitational energy and spirals in towards the central black hole.

While the
original motivation for this Chapter was to develop a disk model which
could be used as a test-bed for the 3-D post-processor, the results
derived below also give important insights into the structure of
$\alpha$-disks, particularly at the ISCO boundary. We also learn a
good deal about the shape of the continuum energy spectrum for the
Thermal-Dominant black hole state.

\section{Steady-state Disks Outside the ISCO}\label{NT_disks}
We begin by presenting an outline of the Novikov-Thorne description of a
steady-state relativistic accretion disk. In addition to
\citet{novik73}, this model is described in more detail in
\citet{page74} and \citet{thorn74}, where the famous value of
$a/M=0.998$ is derived as an upper limit for the spin of an accreting
black hole. 

First, a few definitions to simplify the subsequent algebra:
\begin{subequations}
\begin{equation}
\mathcal{A} \equiv 1+\frac{a_\star^2}{r_\star^2}+2\frac{a_\star^2}{r_\star^3},
\end{equation}
\begin{equation}
\mathcal{B} \equiv 1+\frac{a_\star}{r_\star^{3/2}},
\end{equation}
\begin{equation}
\mathcal{C} \equiv 1-\frac{3}{r_\star}+2\frac{a_\star}{r_\star^{3/2}},
\end{equation}
\begin{equation}
\mathcal{D} \equiv 1-\frac{2}{r_\star}+\frac{a_\star^2}{r_\star^2},
\end{equation}
\begin{equation}
\mathcal{E} \equiv 1+4\frac{a_\star^2}{r_\star^2}-4\frac{a_\star^2}{r_\star^3}
+3\frac{a_\star^4}{r_\star^4},
\end{equation}
\begin{equation}
\mathcal{F}\equiv 1-2\frac{a_\star}{r_\star^{3/2}}+\frac{a_\star^2}{r_\star^2},
\end{equation}
\end{subequations}
where $r_\star \equiv r/M$ and $a_\star \equiv a/M$ are the
dimensionless radius and spin, respectively. In the thin disk
approximation, the angular coordinate $\theta$ can be
replaced by a vertical coordinate $z=r\cos\theta \approx
r(\pi/2-\theta)$. 

\subsection{Radial Structure}\label{radial_structure}
The radial structure of the disk can be described in terms of the
vertically-integrated hydrodynamic variables, as measured in the local
rest frame of the gas (denoted by ``hat'' indices $\hat{\mu}$). This
local frame is simply the tetrad for a massive test particle on a
stable circular orbit at that radius. The integrated shear stress is
given by
\begin{equation}
W(r) \equiv \int T_{\hat{\phi}\hat{r}}(r,z)dz
\end{equation}
and the total radiation flux off either face of the disk is
\begin{equation}
F(r) \equiv T^{\hat{t}\hat{z}}(r,z\to\infty) =
-T^{\hat{t}\hat{z}}(r,z\to-\infty),
\end{equation}
where $\mathbf{T}$ is the stress-energy tensor.
Local conservation of mass gives the accretion rate
\begin{equation}\label{m_dot}
\dot{M} = -2\pi r\Sigma v^{\hat{r}} \mathcal{D}^{1/2}
\end{equation}
as a constant everywhere in the disk. Here $\Sigma(r)$ is the
surface density of the disk in the rest frame of
the orbiting gas:
\begin{equation}\label{sigma_r}
\Sigma(r) = \int_{-\infty}^{\infty} \rho(r,z)dz =
\int_{-\infty}^{\infty} T^{\hat{t}\hat{t}}(r,z)dz, 
\end{equation}
and $v^{\hat{r}} \ll c$ is the average radial velocity of
the slowly inspiraling gas (negative for inward-flowing gas, giving a
positive value for $\dot{M}$).

Conservation of angular momentum gives a first-order differential
equation in $r$ for the stress $W(r)$:
\begin{equation}\label{W_diff}
\frac{d}{dr}\left(-\frac{\dot{M}L}{2\pi}
 +r^2\frac{\mathcal{B}\mathcal{D}}{\mathcal{C}^{1/2}}W\right) +2rLF = 0,
\end{equation}
where
\begin{equation}
L = \sqrt{\frac{GMr}{\mathcal{C}}}\mathcal{F}
\end{equation}
is the specific angular momentum of massive particles on circular
orbits in the equatorial plane. In equation (\ref{W_diff}), the first
term is the rate of angular momentum increase in the gas
[when combined with the mass continuity equation (\ref{m_dot})], the
second is the rate at which the stress $W$ transports angular momentum
outward through the disk,
and the third is the rate at which radiation removes angular momentum
from the two surfaces of the disk. 

The flux $F(r)$ off the face of the disk is given by conservation of
energy: in the steady-state disk, all the energy generated by
turbulent/magnetic stress in the interior must be radiated off the
surface. The energy generated is given by
\begin{equation}\label{F_W}
2F = -s_{\alpha\beta}\int T^{\alpha\beta} dz = -2s_{\hat{\phi}\hat{r}}W,
\end{equation}
where $s_{\alpha\beta}$ is the average shear of the gas. The shear
tensor is defined by
\begin{equation}\label{shear_tensor}
s_{\alpha \beta} \equiv \frac{1}{2}(v_{\alpha; \mu} P^\mu_\beta + v_{\beta;
\mu} P^\mu_\alpha) -\frac{1}{3}\theta P_{\alpha \beta},
\end{equation}
where $v_\alpha$ is the local 4-momentum of the gas, $\theta$ is the
geodesic expansion of the gas defined below in
equation (\ref{expansion}) and $P_{\alpha \beta}$ is the projection
tensor 
\begin{equation}
P_{\alpha \beta} \equiv g_{\alpha \beta} + v_\alpha v_\beta.
\end{equation}
For circular geodesic orbits in the
plane, the only non-zero shear terms are
\begin{equation}\label{s_EG}
s_{\hat{\phi}\hat{r}} = s_{\hat{r}\hat{\phi}} = 
-\frac{3}{4}\sqrt{\frac{GM}{r^3}}
\frac{\mathcal{D}}{\mathcal{C}}.
\end{equation}

The $\alpha$-disk model assumes that all the hydrodynamic turbulence,
molecular viscosity (typically very small), 
magnetic stress, and magnetic heating can be combined into a single
term for the stress tensor
\begin{equation}
T^{\hat{\phi}\hat{r}} = \alpha p,
\end{equation}
where $p$ is the sum of the gas and radiation pressure. The
dimensionless parameter $\alpha$ is generally taken to be between 0.01
and 1. A number of papers have attempted to determine
$\alpha$ directly from MHD simulations, and generally find values
within this range, but also find that $\alpha$ can vary significantly
between different regions in the disk [see, e.g.\
\citet{balbu98,hawle00,hawle01}].

Combining equations (\ref{W_diff}, \ref{F_W}, and \ref{s_EG}), and
defining the function 
\begin{equation}
Z(r) \equiv \frac{2\pi}{\dot{M}} 
\frac{r^2\mathcal{BD}}{M^{1/2}\mathcal{C}^{1/2}}W(r),
\end{equation}
we get the ordinary differential equation
\begin{equation}\label{dZ_dr}
\frac{dZ}{dr} = \frac{dL}{dr}-
\frac{3}{2}\frac{M^{1/2}L}{r^{5/2}\mathcal{BC}^{1/2}}Z.
\end{equation}
For a boundary condition, NT set the integrated stress at the ISCO to
be zero, so $Z(R_{\rm ISCO})=0$. Numerical simulations suggest this is
not quite accurate, so we assume some small, non-zero value for the
stress across the ISCO, typically $Z(R_{\rm ISCO})\sim 2-3\times
10^{-2}$. The exact value is determined self-consistently by matching
the turbulent scale length of the disk with the characteristic size of
the pressure gradient inside the ISCO, as will be described below in
Section \ref{geodesic_plunge}.

Given $Z(r)$, the flux radiated from the each point of the disk can be
calculated from equation (\ref{F_W}):
\begin{equation}
F(r) = \frac{3G\dot{M}M}{8\pi}\frac{Z}{r^{7/2}\mathcal{BC}^{1/2}}.
\end{equation}
This flux is then used as one of the outer boundary conditions for
integrating the one-dimensional equations of vertical structure at
each radial position in the disk.

\subsection{Vertical Structure}\label{vertical_structure}
The accretion disk equations of vertical structure are almost
identical in form to those of steady-state stellar structure
\citep{hanse94}. In the local inertial frame 
of circular geodesic orbits, the gas can be treated entirely
classically. The only relativistic addition necessary is given by the
tidal gravitational acceleration, which comes from the Riemann tensor,
as calculated in the frame of the gas:
\begin{equation}
g = R^{\hat{z}}_{\hat{t}\hat{z}\hat{t}}z.
\end{equation}
Using the Riemann tensor of the 
(ZAMO) $R_{(\alpha)(\beta)(\gamma)(\delta)}$, as calculated by
\citet{barde72}, the components in the gas frame come from a simple
Lorentz transformation of the velocity $v^{(\mu)}$ in the ZAMO frame to
the local inertial frame $v^{\hat{\mu}}=\mathbf{e}_{\hat{t}}$: 
\begin{eqnarray}\label{Riemann}
R_{\hat{z}\hat{t}\hat{z}\hat{t}} &=& 
\Lambda^{(\alpha)}_{\hat{z}}\Lambda^{(\beta)}_{\hat{t}}
\Lambda^{(\gamma)}_{\hat{z}}\Lambda^{(\delta)}_{\hat{t}}
R_{(\alpha)(\beta)(\gamma)(\delta)} \nonumber\\
&=& \Lambda^{(\beta)}_{\hat{t}}\Lambda^{(\delta)}_{\hat{t}}
R_{(z) (\beta) (z) (\delta)} \nonumber\\
&=& v^{(\beta)}v^{(\delta)}R_{(z) (\beta) (z) (\delta)} \nonumber\\
&=& (v^{(t)})^2R_{(z)(t)(z)(t)}+2v^{(t)} v^{(\phi)}
R_{(z)(t)(z)(\phi)}+(v^{(\phi)})^2R_{(z)(\phi) (z)(\phi)}. 
\end{eqnarray}
This term actually appears to be calculated incorrectly in the
NT paper. \citet{riffe95} correctly give it as
\begin{equation}
R^{\hat{z}}_{\hat{t}\hat{z}\hat{t}} = \frac{GM}{r^3} 
\frac{1-4a_\star r_\star^{-3/2}+3a_\star^2 r_\star^{-2}}
{1-3r_\star^{-1}+2a_\star r_\star^{-3/2}},
\end{equation}
which for convenience we will simply call $\mathcal{R}$.

The vertical hydrostatic pressure balance is given by
\begin{equation}\label{press_bal}
\frac{dp}{dz} = -\rho g = -\rho \mathcal{R} z,
\end{equation}
where $\rho$ is the rest mass density of the gas and the acceleration
due to tidal gravity is $g=\mathcal{R}z$. The transport of
energy in the disk will be dominated by radiation diffusion, so the
vertical energy flux $q^z$ is
\begin{equation}\label{q_z}
q^z = -\frac{ac}{3\kappa\rho}\frac{d}{dz}T^4
\end{equation}
or
\begin{equation}\label{rad_trans}
\frac{dT}{dz} = -\frac{3\kappa\rho}{4acT^3}q^z,
\end{equation}
where we assume local thermodynamic equilibrium with radiation energy
density $aT^4$.
Here the opacity $\kappa$ is a combination of free-free opacity and
electron scattering, but for most of the region of interest it is
dominated by electron scattering, so we set
\begin{equation}
\kappa = \kappa_{\rm es} = 0.40 \mbox{ cm}^2 \rm{g}^{-1}.
\end{equation}

As described above, the energy generation in the $\alpha$-disk is
given by the product of the shear and stress tensors:
\begin{equation}\label{enrg_gen}
\frac{dq^z}{dz} = -2s_{\hat{\phi}\hat{r}}t^{\hat{\phi}\hat{r}} =
\frac{3}{2}\sqrt{\frac{GM}{r^3}}\frac{\mathcal{D}}{\mathcal{C}}
\alpha p \equiv \bar{\alpha}p,
\end{equation}
where we have compactified a number of terms into the more convenient
scaling factor $\bar{\alpha}(r)$, which has units of inverse time.
Coupled with the equation of state for an ideal gas of ionized
hydrogen and radiation
\begin{equation}\label{eos}
p = \frac{2k_B T}{m_p}\rho + \frac{a}{3}T^4,
\end{equation}
we have a complete set of coupled first-order differential equations
for the vertical disk structure at each radial position in the
disk. In equation (\ref{eos}) we have assumed a fully ionized hydrogen
gas, where the particle number density is $n=2\rho/m_p$, but any
composition could just as easily be used by substituting the
relationship \citep{hanse94}
\begin{equation}
n =\frac{\rho}{\mu m_p},
\end{equation}
where $\mu$ is called the ``total mean molecular weight.'' For a
hydrogen mass fraction of $X$, $\mu$ can be approximated by 
\begin{equation}
\mu \approx \frac{4}{3+5X}.
\end{equation}

The three equations for $p$, $T$, and $q^z$ require three boundary
conditions for a complete solution. As is often done in solving the
stellar structure equations \citep{hanse94}, we assume an optically
thin, isothermal atmosphere beginning at the photosphere $z=h$ with
surface temperature $T(h)=T_s$ and density $\rho(h)=\rho_s$. All
the flux is generated inside of this point, so $q^z(h)=F$ (given by
the radial structure), and plane symmetry $(z \to -z)$ demands that
$q^z(0) = 0$. To get the third boundary condition, we have to
solve for $T_s$ self-consistently.

The tidal
gravitational force on a mass $m_p$ can be approximated by the
effective potential 
\begin{equation}
\Phi_{\rm eff}(z \approx h) = \frac{\mathcal{R} h}{2}(z-h),
\end{equation}
which produces an isothermal atmosphere with scale height $H$ and
density profile
\begin{equation}\label{expon_atm}
\rho(z>h) = \rho_s \exp\left[-\frac{m_p \mathcal{R} h(z-h)}{2k_B
T_s}\right] \sim e^{-z/H}.
\end{equation}
The density $\rho_s$ at the ``base'' of the atmosphere is defined
such that the integrated optical depth to electron scattering through
the atmosphere is unity (some texts define the photosphere at $\tau =
2/3$, but we find the net results to be nearly identical in either case):
\begin{equation}
\int_h^\infty \kappa_{\rm es}\rho dz = 1,
\end{equation}
which can be solved to give
\begin{equation}\label{rho_s}
\rho_s = \frac{m_p \mathcal{R} h}{2\kappa_{\rm es}k_B T_s}.
\end{equation}

Because the opacity is dominated by electron scattering, and not
free-free absorption, the resulting radiation will have a modified
black-body spectrum, as described in \citet{shaku73}. They give two
basic models for the scattering atmosphere: a constant density with a
sharp cutoff [$\rho(z<h)=\rho_s$ and $\rho(z>h)=0$], or the
exponential distribution we use here. For the half-plane geometry, the
modified spectrum is of the form \citep{shaku72,felte72}
\begin{equation}\label{modified_halfplane}
F(x) \sim \rho^{1/2} T^{5/4} \frac{x^{3/2}e^{-x}}{(1-e^{-x})^{1/2}},
\end{equation}
where $x$ is defined as a dimensionless scaled frequency $x \equiv
h\nu/k_BT$. For an
exponential density distribution with scale length $H$, the modified
spectrum has the form \citep{zeldo69}
\begin{equation}\label{modified_expo}
F(x) \sim H^{-1/3} T^{11/6} \frac{x^{2}e^{-x}}{(1-e^{-x})^{2/3}},
\end{equation}
which is somewhat more similar to the unmodified blackbody spectrum
where 
\begin{equation}\label{unmodified}
F(x) \sim T^3 \frac{x^3e^{-x}}{1-e^{-x}}.
\end{equation}

NT use the first model, which gives the total flux
integrated over frequency as
\begin{equation}\label{flux_mod}
F = 8.05\times 10^{7}\left(\frac{\rho}{{\rm g/cm^3}}\right)^{1/2}
\left(\frac{T_s}{^\circ {\rm K}}\right)^{9/4} \rm{erg}/\rm{cm^2}/\rm{s}.
\end{equation}
When integrating the stellar structure equations in the diffusion
limit, we find the exponential model to more accurately approximate
the atmospheric density profile. In that case, the integrated flux is
given by
\begin{equation}\label{flux_mod2}
F = 1.3\times 10^{4}\left(\frac{H}{\rm{cm}}\right)^{-1/3}
\left(\frac{T_s}{^\circ {\rm K}}\right)^{17/6} \rm{erg}/\rm{cm^2}/\rm{s}.
\end{equation}
Combining equations (\ref{expon_atm}) and (\ref{flux_mod2}) gives the
boundary condition for the disk's surface temperature:
\begin{equation}\label{T_s}
T_s = 0.28F^{2/5}\mathcal{R}^{-2/15}h^{-2/15}.
\end{equation}

To simultaneously satisfy all three boundary conditions, the system of
differential equations is then solved using a ``shooting'' method,
starting at $z=h$ and integrating $p$, $T$, and 
$q^z$ inwards to $z=0$. We iterate this approach for a series of
initial values for $h$: the solution is given by the value of $h$ that
matches the inner boundary condition of $q^z(0)=0$. Repeating this
entire procedure for each value of $r$ gives the complete structure of the
accretion disk outside of the ISCO.

To get a good starting guess for the value of $h$, we derive here
an approximate solution of the vertical structure equations. The
result is somewhat different from that given in NT, as they ignore the
(often significant) gas pressure in the inner disk, and also we use
different methods of averaging the vertical structure over $z$. The
difference in disk thickness turns 
out to be a factor of at least $2-3$ for typical stellar-mass black
holes. Starting with the pressure balance equation (\ref{press_bal}),
with the pressure going to zero at the surface of the disk $p(h)=0$,
the disk thickness can be approximated in terms of the central
pressure $p_c$ and density $\rho_c$:
\begin{equation}
\frac{dp}{dz} \approx \frac{p_c}{h} = \langle \rho z\rangle \mathcal{R}.
\end{equation}
Taking the density profile as roughly linear with $\rho(h) = 0$, 
\begin{equation}
\langle \rho z\rangle = \frac{\rho_c}{h}\int_0^h\left(1-\frac{z}{h}\right)zdz =
\frac{\rho_c h}{6},
\end{equation}
so our first estimate for $h$ is
\begin{equation}\label{h_1}
h = \left(\frac{6p_c}{\rho_c \mathcal{R}}\right)^{1/2}.
\end{equation}

Taking the average pressure as $p_c/2$, the energy generation equation
(\ref{enrg_gen}) gives us an independent expression for $h$:
\begin{equation}\label{h_2}
h = \frac{F}{\alpha s_{\hat{\phi}\hat{r}} p_c}.
\end{equation}

The third independent estimate for $h$ comes from the radiation
transport equation (\ref{rad_trans}):
\begin{equation}
h = \frac{4acT_c^4}{3\kappa_{\rm es}\rho_c F}.
\end{equation}
Coupled with the equation of state (\ref{eos}), we have four
(non-linear) equations
for the four unknowns $\rho_c$, $T_c$, $p_c$, and $h$ at each radius
in the disk. The resulting
value of $h$ gives a remarkably accurate estimate of the disk
thickness as determined by directly integrating the structure equations,
agreeing within about $10-20\%$ for a range of black hole masses, spins,
and accretion rates. 

\begin{figure}
\begin{center}
\scalebox{0.4}{\includegraphics*[40,390][540,700]{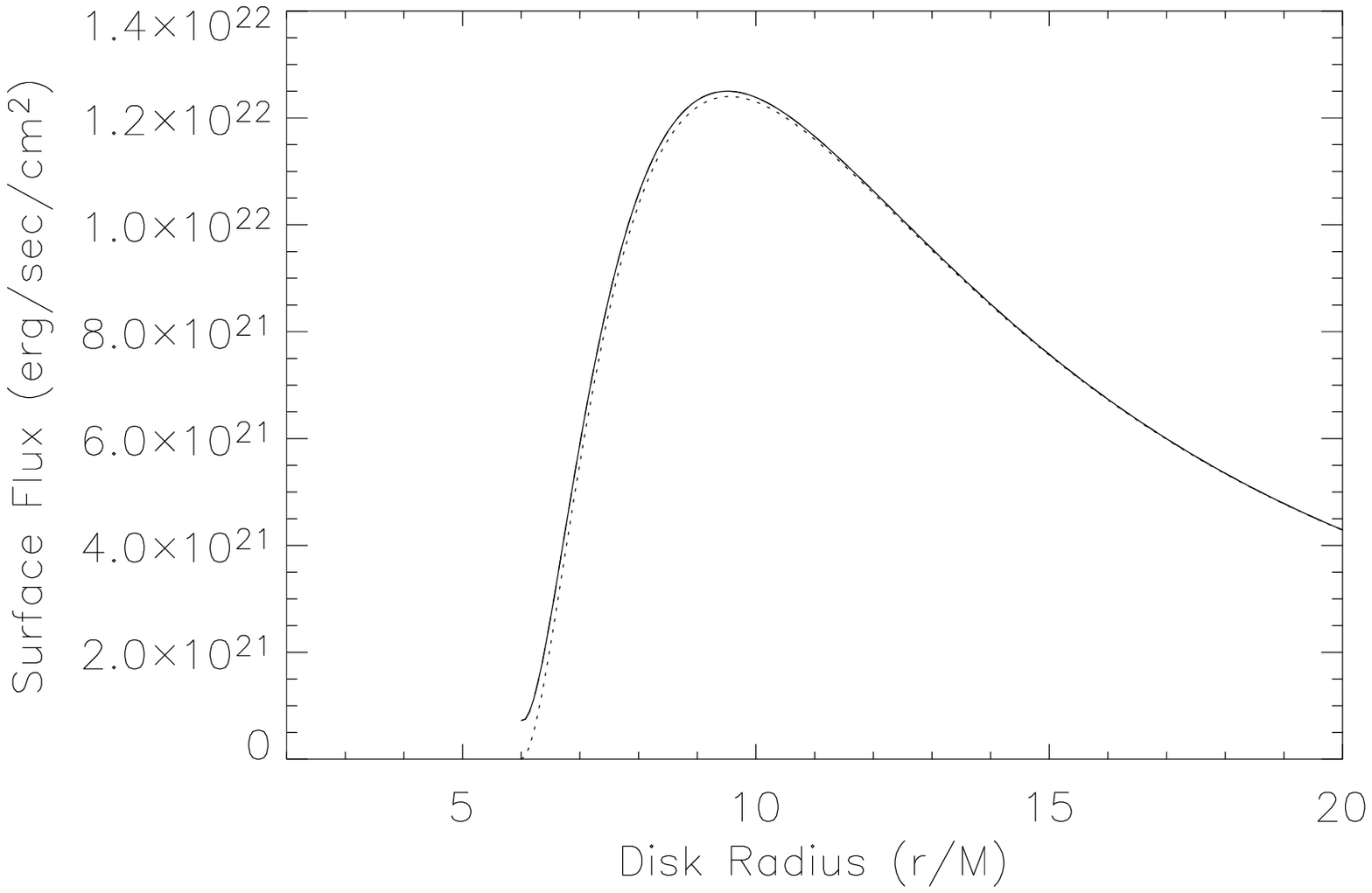}}
\scalebox{0.4}{\includegraphics*[40,390][540,700]{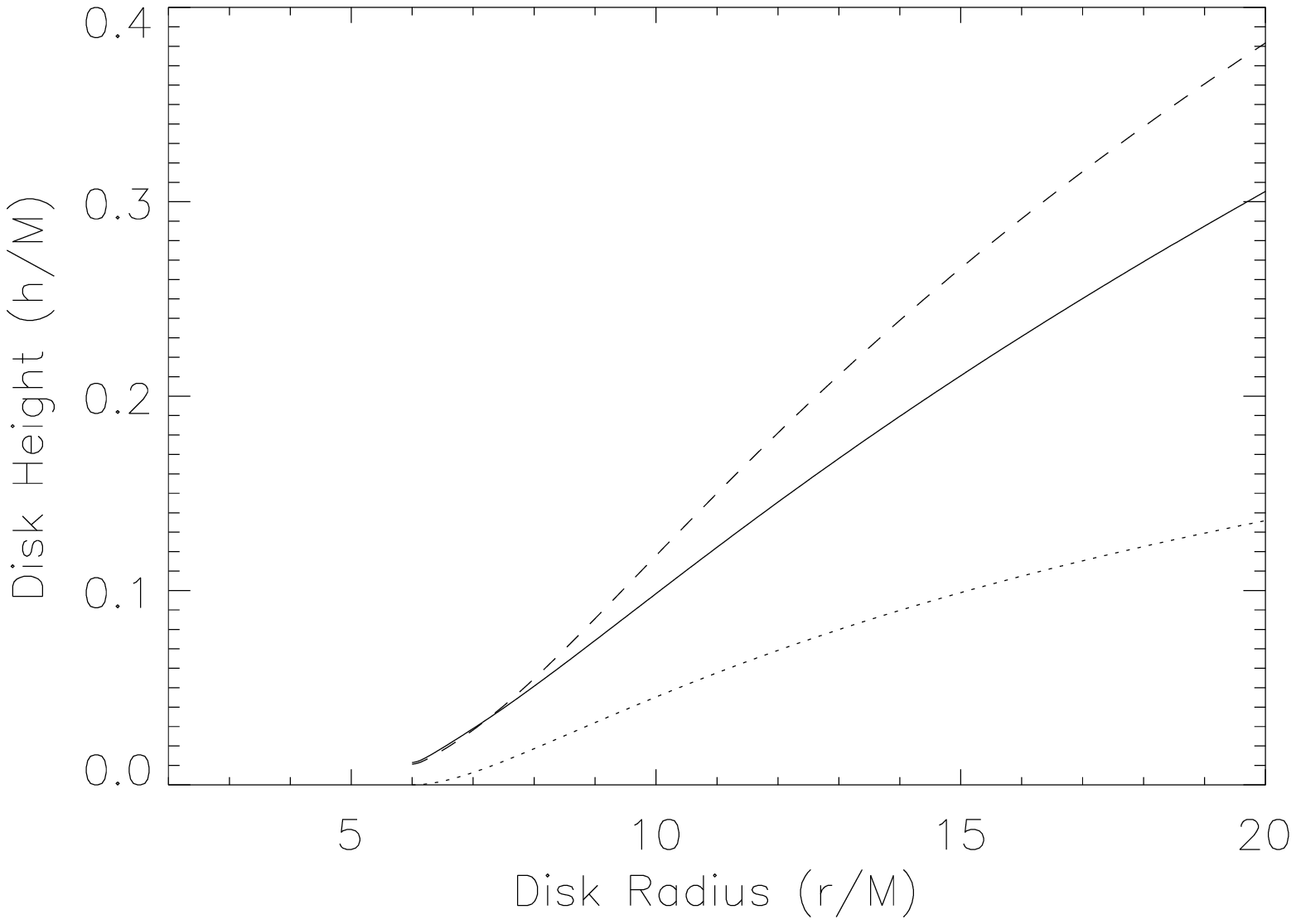}}\\
\scalebox{0.4}{\includegraphics*[40,390][540,700]{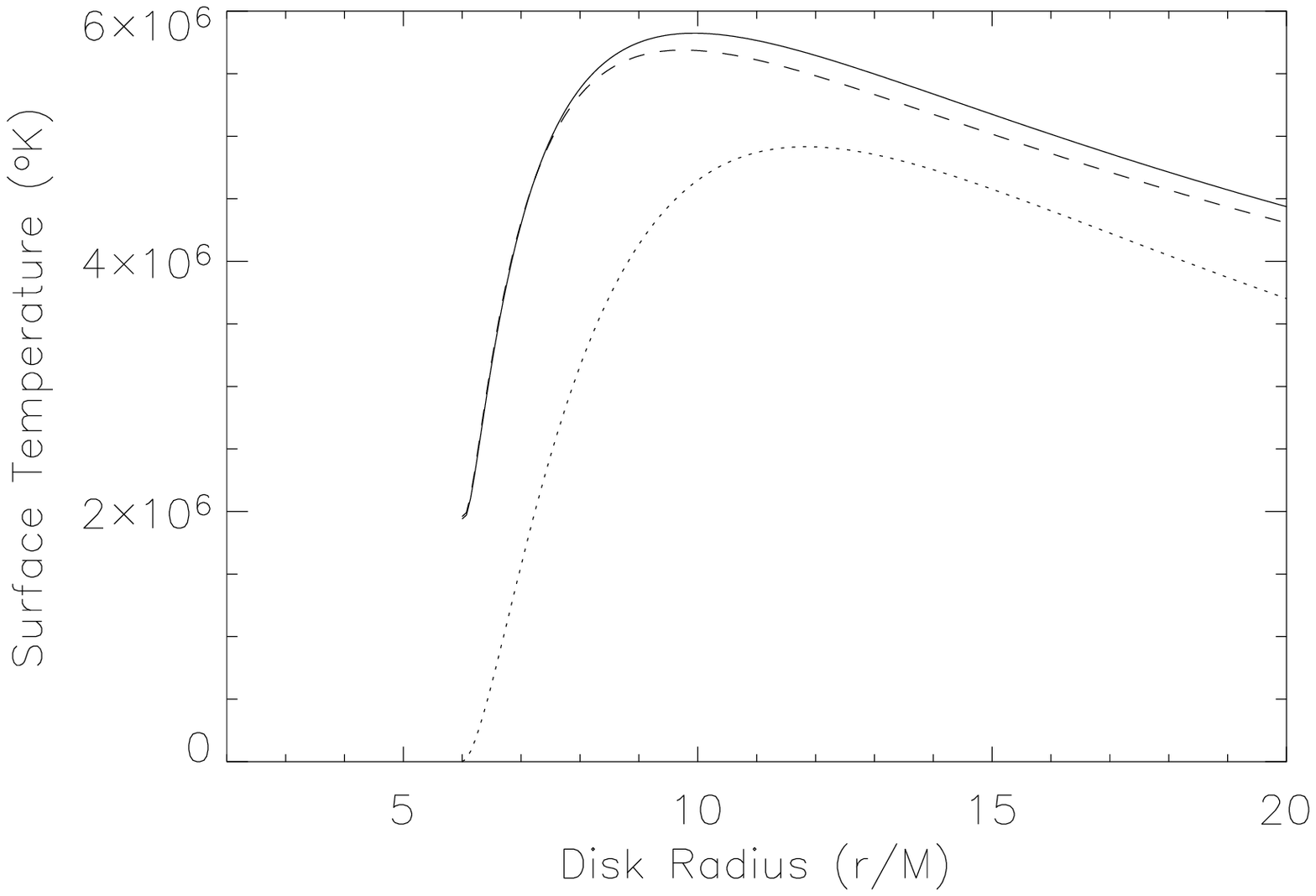}}
\scalebox{0.4}{\includegraphics*[40,390][540,700]{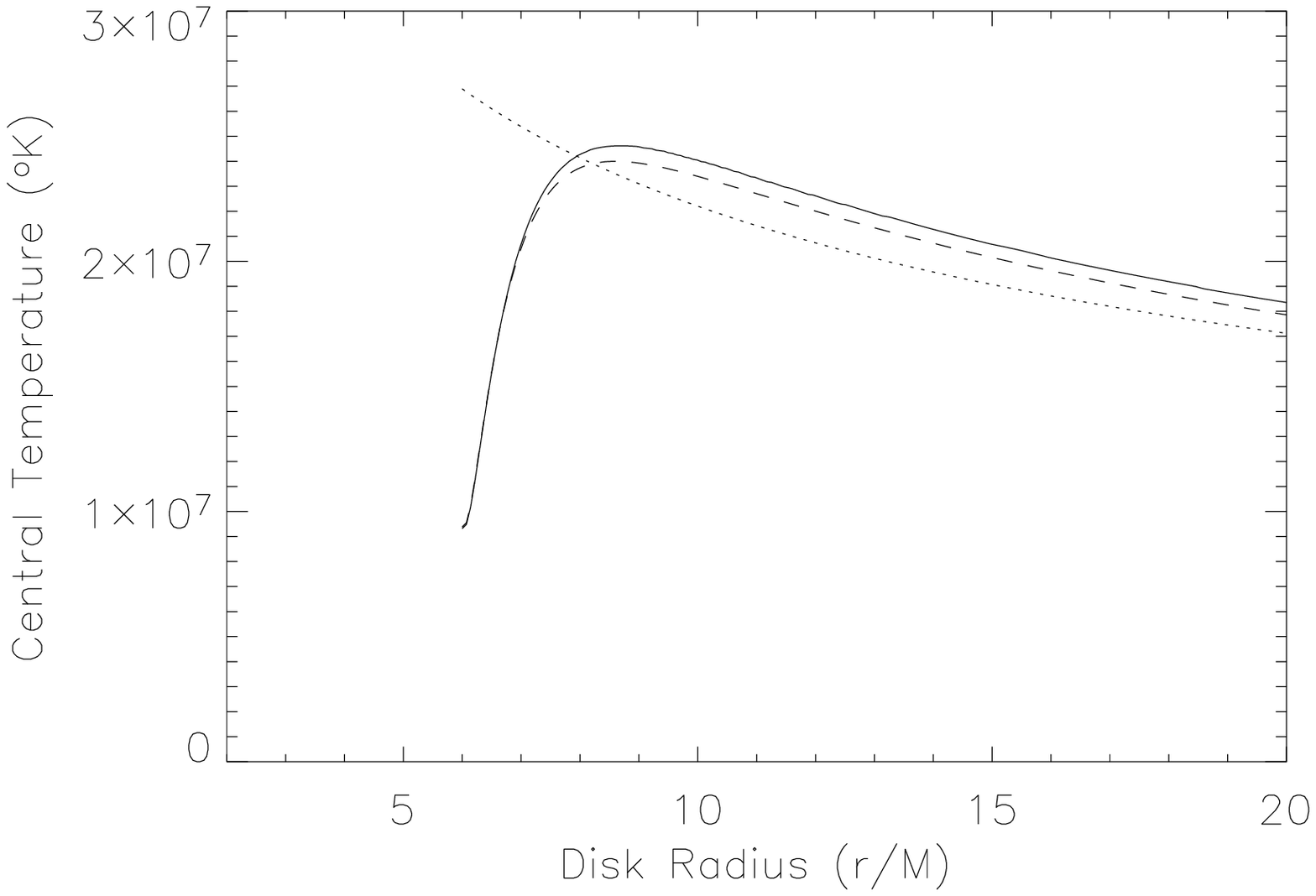}}\\
\scalebox{0.4}{\includegraphics*[40,330][540,700]{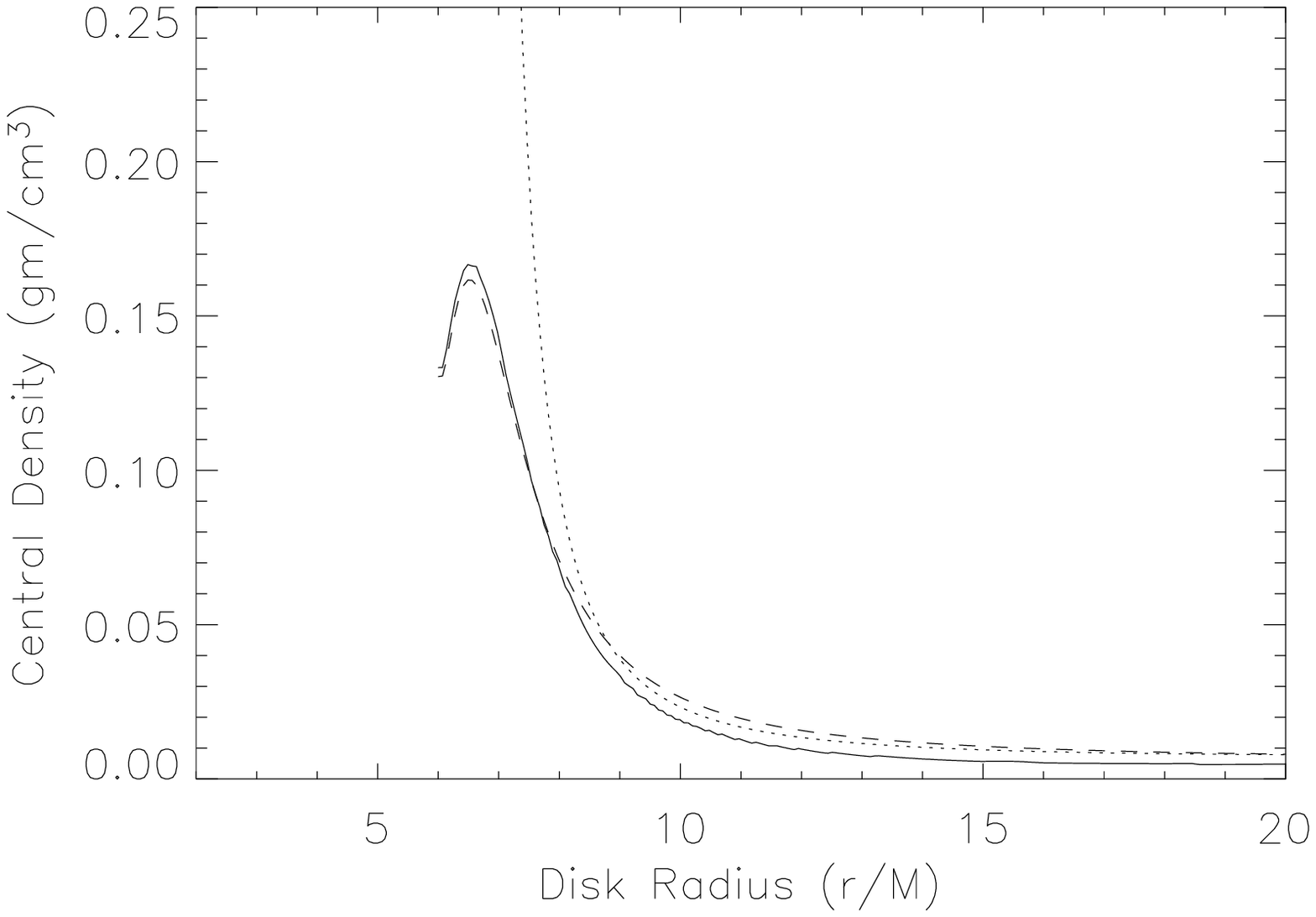}}
\scalebox{0.4}{\includegraphics*[40,330][540,700]{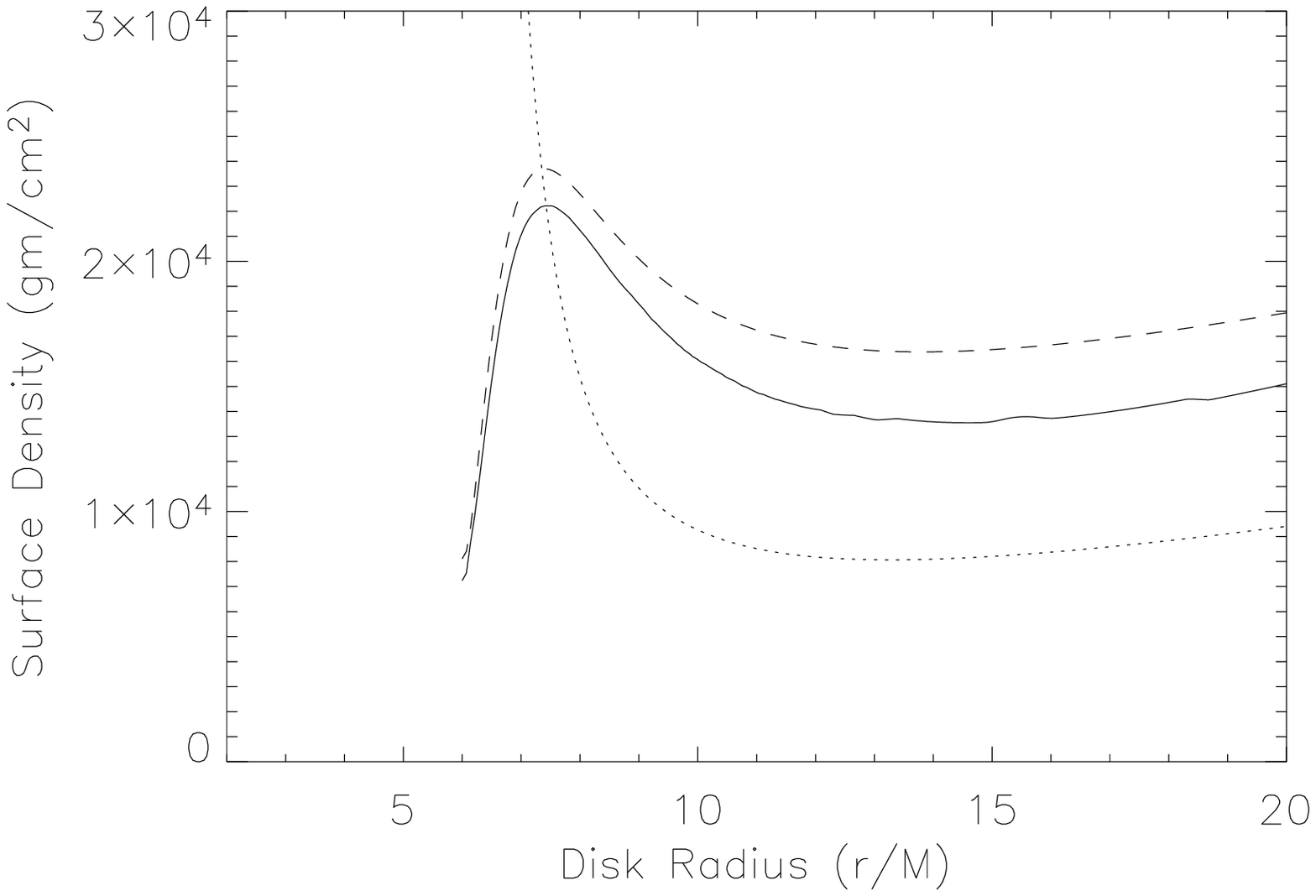}}
\caption[Hydrodynamic fluid variables for an $\alpha$-disk with
$a/M=0$]{\label{nt_compare0} Comparison of hydrodynamic fluid
variables for three versions of the steady-state $\alpha$-disk model:
Full numerical integration of the vertical structure equations (solid
lines); The Novikov-Thorne approximation (dotted line); The revised
analytic approximation derived in the text (dashed line). The black
hole has mass $M=10M_\odot$, spin $a/M=0$, and accretion rate $\dot{M}
= 0.05 \dot{M}_{\rm Edd}$.}
\end{center}
\end{figure}

\begin{figure}
\begin{center}
\scalebox{0.4}{\includegraphics*[40,390][540,700]{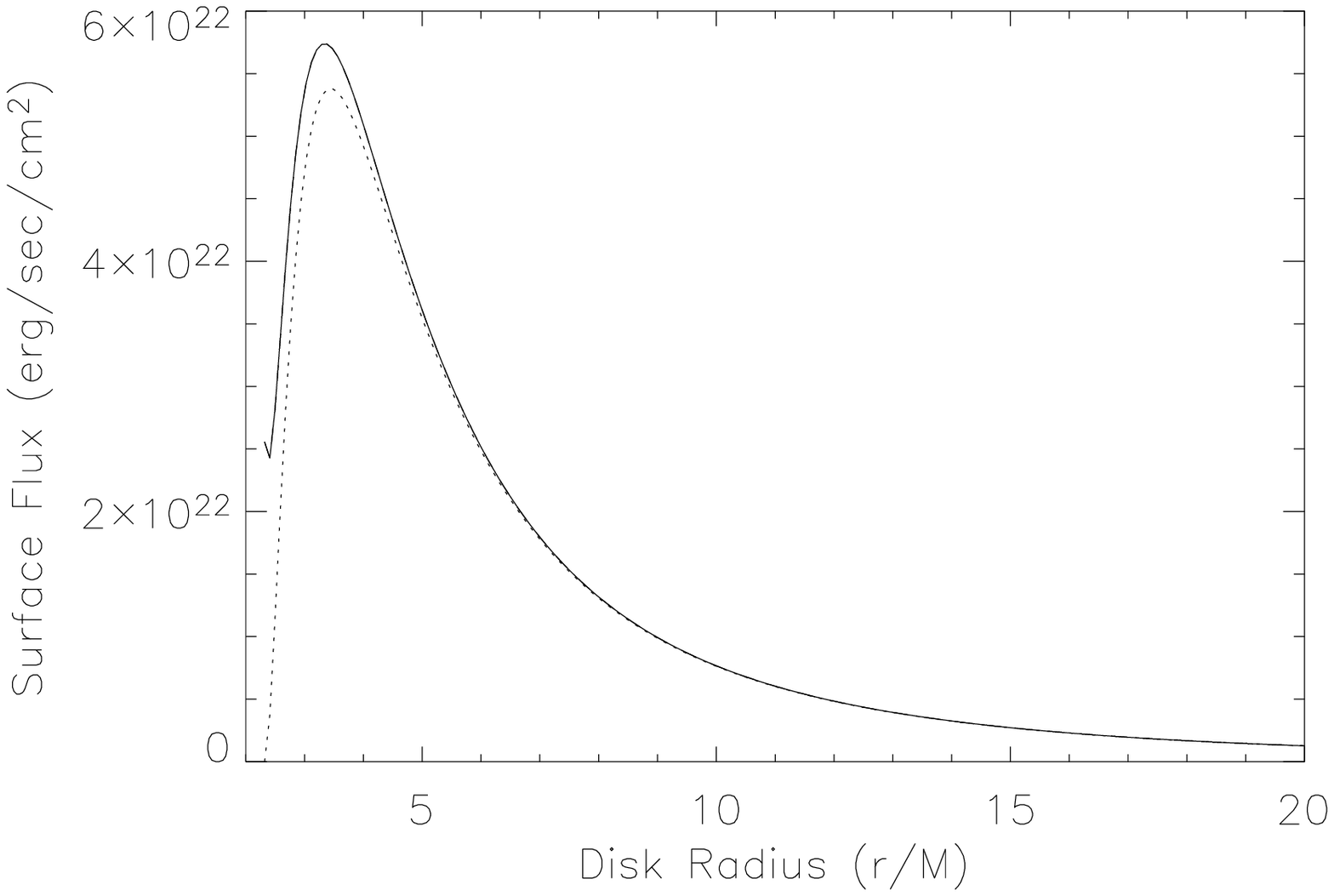}}
\scalebox{0.4}{\includegraphics*[40,390][540,700]{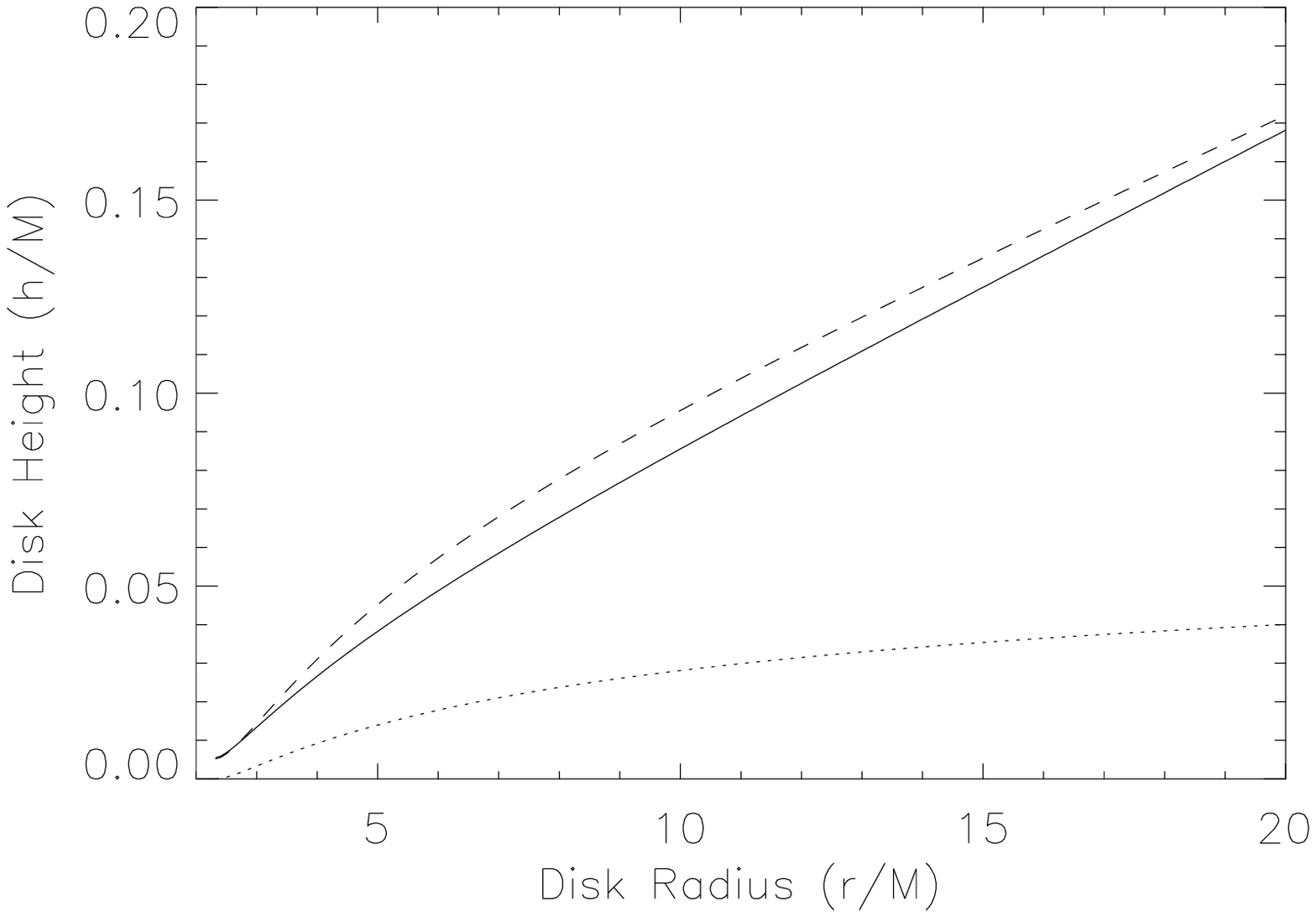}}\\
\scalebox{0.4}{\includegraphics*[40,390][540,700]{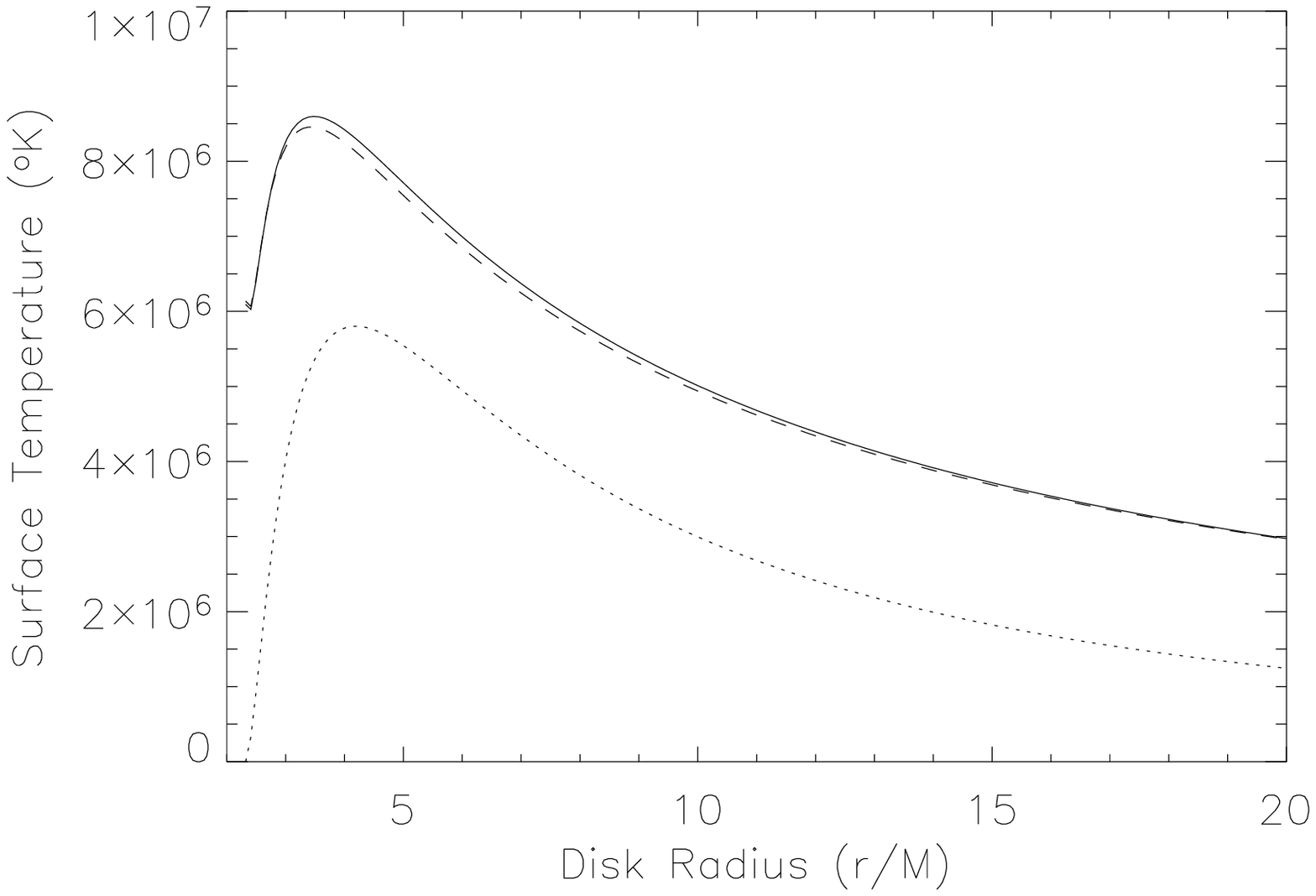}}
\scalebox{0.4}{\includegraphics*[40,390][540,700]{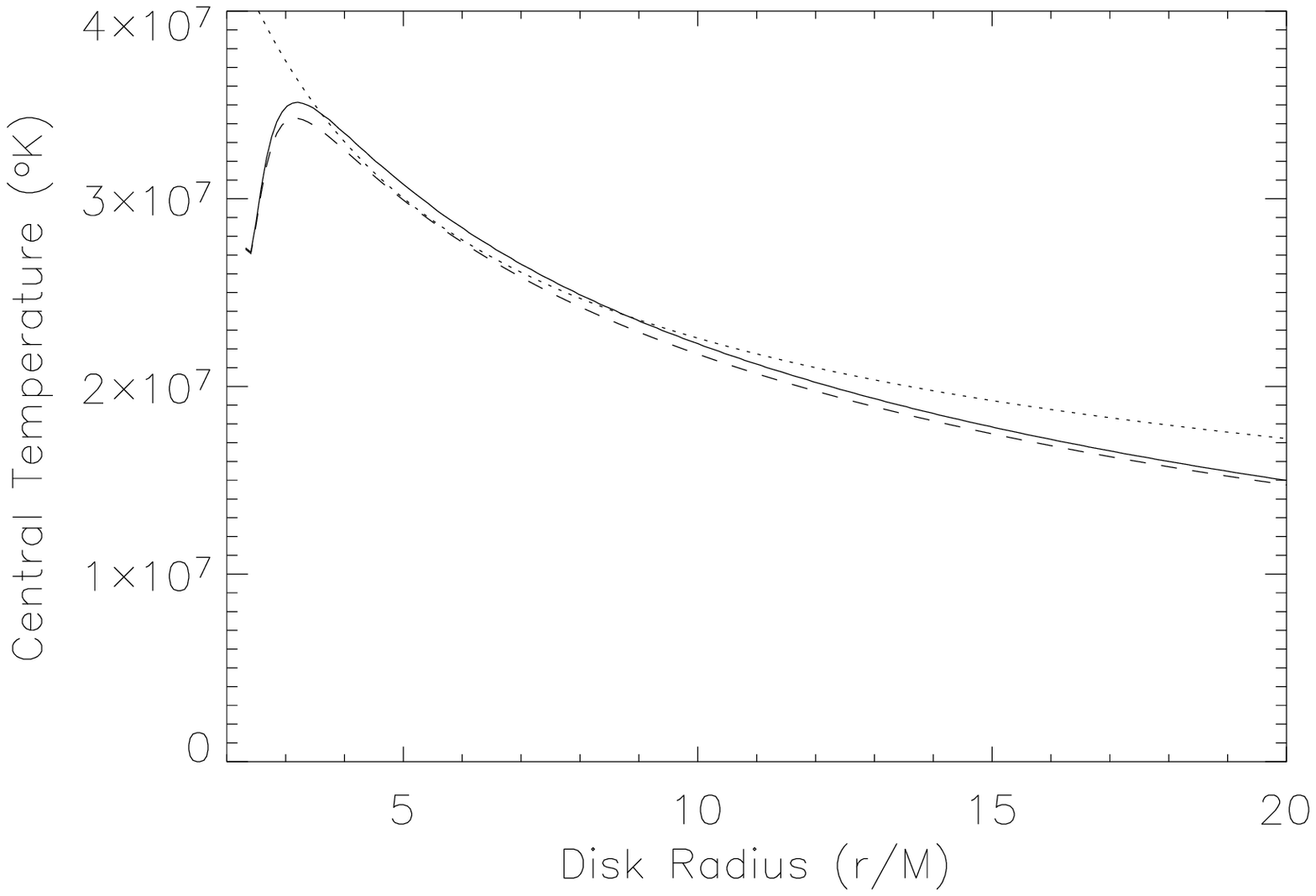}}\\
\scalebox{0.4}{\includegraphics*[40,330][540,700]{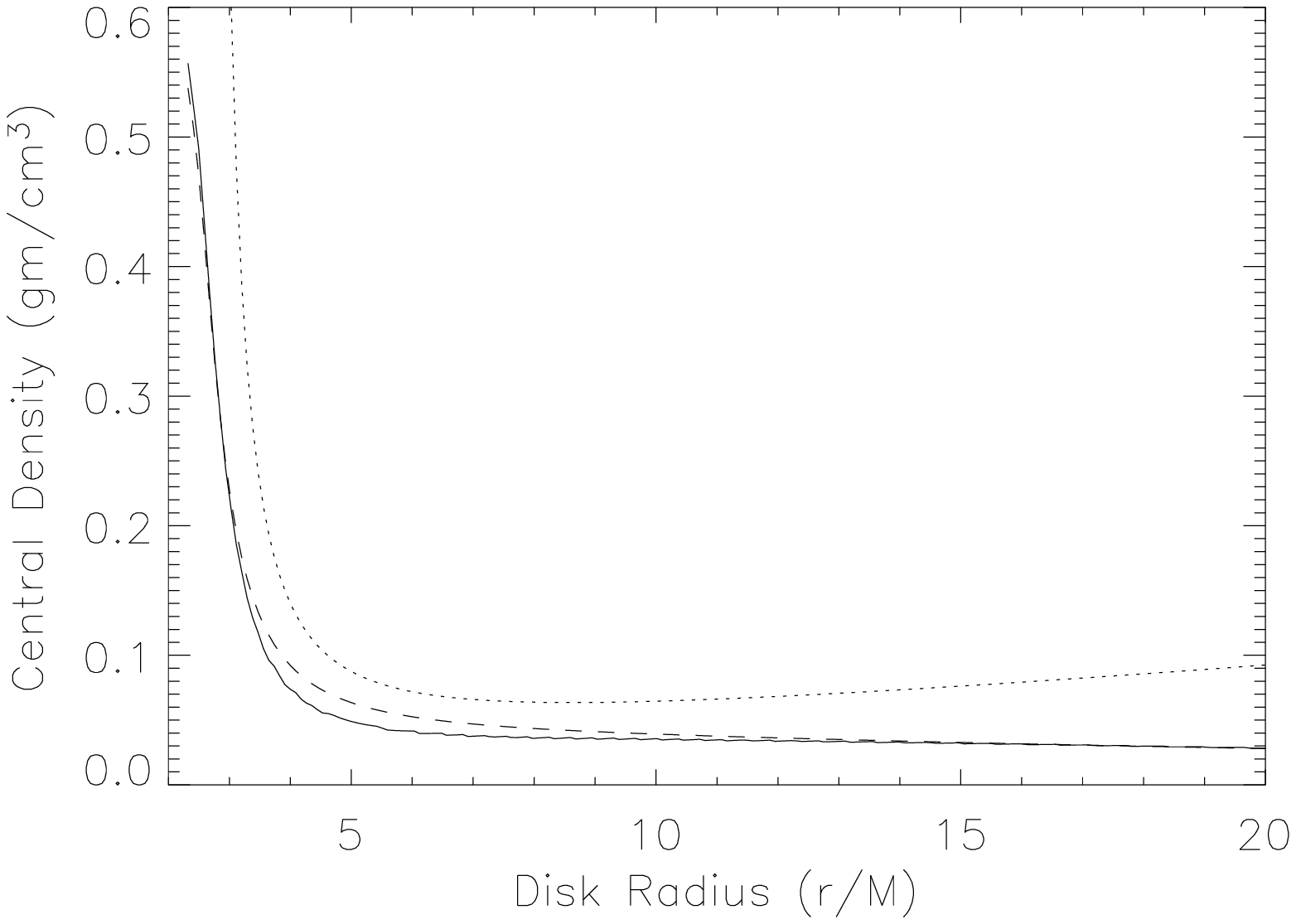}}
\scalebox{0.4}{\includegraphics*[40,330][540,700]{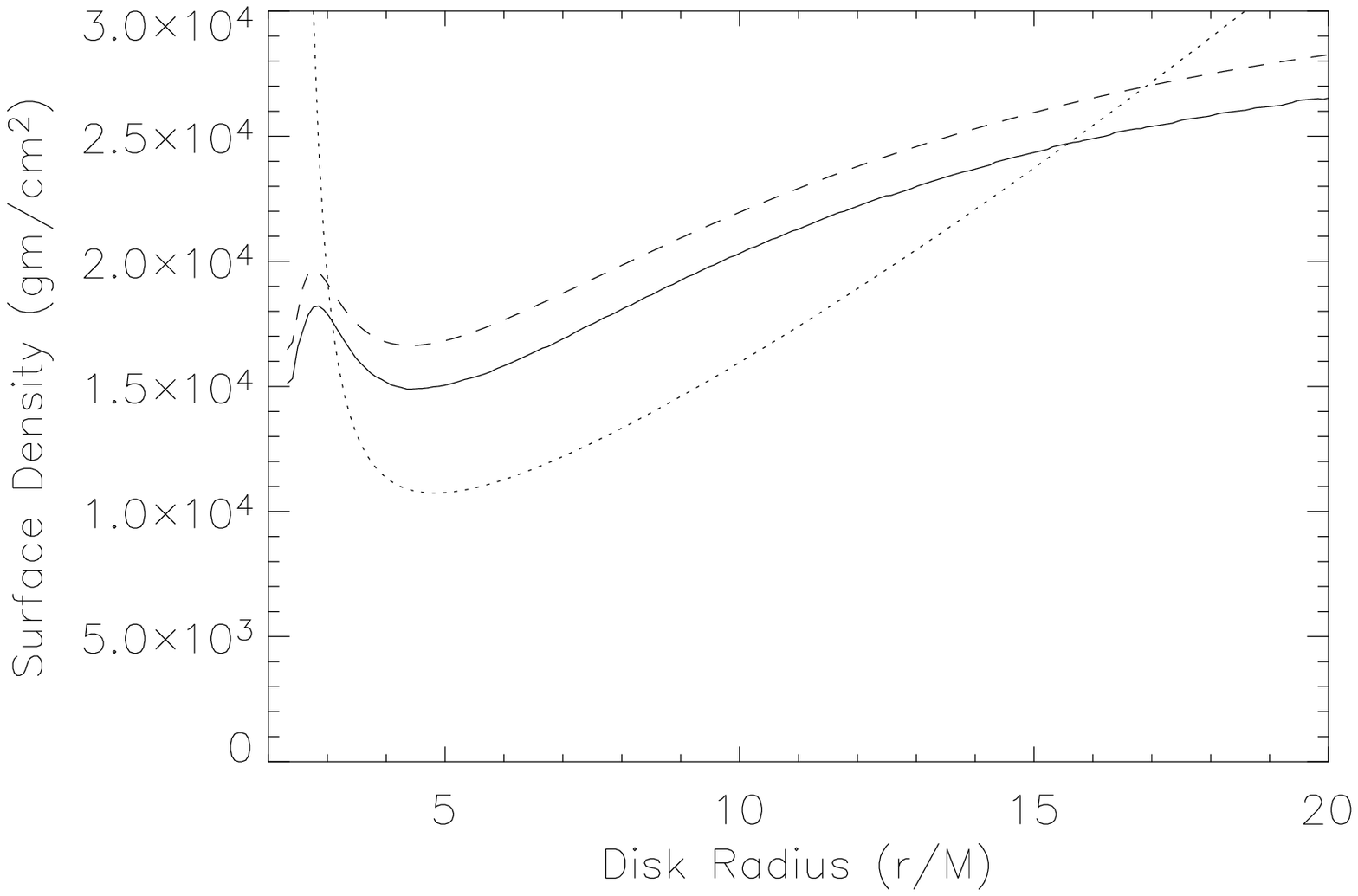}}
\caption[Hydrodynamic fluid variables for an $\alpha$-disk with
$a/M=0.9$]{\label{nt_compare9} Comparison of hydrodynamic fluid
variables for three versions of the steady-state $\alpha$-disk model:
Full numerical integration of the vertical structure equations (solid
lines); The Novikov-Thorne approximation (dotted line); The revised
analytic approximation derived in the text (dashed line). The black
hole has mass $M=10M_\odot$, spin $a/M=0.9$, and accretion rate $\dot{M}
= 0.05 \dot{M}_{\rm Edd}$.}
\end{center}
\end{figure}

In Figure \ref{nt_compare0} we show the comparison of a variety of
fluid variables for three versions of the steady-state
$\alpha$-disk: the numerical integration of the coupled structure
equations (solid line), the NT approximation (dotted
line), and our revised analytic approximation (dashed line). The basic
model parameters used here are $\alpha=0.1$, $M=10M_\odot$, $a/M=0$,
$\dot{M}=0.05 \dot{M}_{\rm Edd}$ ($\dot{M}_{\rm Edd}$ is the mass
accretion rate that gives a total disk flux equal to the Eddington
luminosity, defined below in Section \ref{revised_eddington}). Figure
\ref{nt_compare9} shows the same disk variables for a black hole with
spin $a/M=0.9$ but all other parameters identical.

%For the
%inner-most regions of the disk $(R_{\rm ISCO}<r \lesssim 20M)$, the NT
%solution is:

The surface flux $F(r)$ in our models is determined by integrating
equation (\ref{dZ_dr}) with boundary condition $Z(R_{\rm ISCO})$
representing the net torque on the disk at the ISCO, as described
below in Section \ref{geodesic_plunge}. Despite the fact that our
$F(r)$ is nearly identical to that of NT, the different methods of
solving the vertical structure give very different disk scale
heights. Our model predicts a rather thicker disk and much lower
density atmosphere [NT assume $\rho(h) \approx \rho(0)$]. The
lower-density atmosphere results in a higher surface temperature
through equations (\ref{flux_mod}) and (\ref{flux_mod2}). And since
the NT disk is cut off at the ISCO [$h(R_{\rm ISCO})\to 0$],
the conservation of mass equation (\ref{m_dot}) requires that the
density diverges [$\rho(R_{\rm ISCO}),\Sigma(R_{\rm ISCO}) \to
\infty$]. However, by slightly modifying the inner torque
boundary condition, including gas pressure in the inner disk, and
changing the means of averaging the vertical structure equations, a
very good analytic approximation can be derived for the disk height,
density, and surface temperature. From these fluid variables, we can
produce an accurate multi-colored disk spectrum via the ray-tracing
post-processor (see Section \ref{ntdisk_spectra} below).

\subsection{Revised Eddington Limit}\label{revised_eddington}
We are now in a position to revisit
the traditional Eddington limit on the maximum allowed luminosity for
a gravitationally bound, hydrostatic system. The
gas can be in hydrostatic equilibrium only if the gravitational force on each
proton is greater than the radiation force on each electron. For the
traditional derivation of this limit for a spherically symmetric star,
the force balance equation can be written \citep{hanse94}
\begin{equation}\label{edd_spherical}
\frac{GMm_p}{r^2} > F \frac{\sigma_T}{c} = \frac{L}{4\pi
r^2}\frac{\sigma_T}{c},
\end{equation}
giving an upper limit on the total luminosity as
\begin{equation}
L_{\rm Edd} = \frac{4\pi c GMm_p}{\sigma_T} = 3.5\times 10^4 L_\odot
\left(\frac{M}{M_\odot}\right).
\end{equation}
In our case, we must replace the $1/r^2$ gravitational force with the
local tidal gravity at each radius in the disk:
\begin{equation}\label{edd_planar}
m_p \mathcal{R} z > \frac{F\sigma_T}{c},
\end{equation}
where the classical Thomson cross-section is $\sigma_T=6.65\times
10^{-25}\mbox{ cm}^2$. Taking $z \approx h$, equations
(\ref{h_2}) and (\ref{edd_planar})
give an estimated stability requirement on the central pressure of
\begin{equation}
p_c \lesssim \frac{m_p c \mathcal{R}}{\alpha s_{\hat{\phi}\hat{r}} \sigma_T}.
\end{equation}

For a given value of $r_\star$, the Riemann tensor scales as $\mathcal{R} \sim
M^{-2}$ and the shear scales as $s \sim M^{-1}$, so the maximum
stable pressure is inversely proportional to black hole mass. For
$M=10M_\odot$ and $\alpha = 1$, we find that the maximum stable
luminosity is on the
order of $5\%$ of Eddington. For supermassive black holes with $M >
10^6M_\odot$, no stable solutions exist with these structure
equations, although with a different treatment of the radiation
transport (e.g.\ not diffusion), one might find more
success. Presumably this result would also be sensitive to different
opacities and boundary conditions more appropriate for the typical AGN
disk temperatures of $T_c \sim 10^4-10^5$ K, or the inclusion of
important magnetic pressure and stress \citep{merlo03}. It is also quite
likely that the slim disk geometry simply does not apply to AGN or to
stellar-mass black holes with a high accretion rate. For AGN, a thick,
dusty torus may be more appropriate, and the high-luminosity black
hole binaries may be better described with a quasi-spherical ADAF
geometry (see Section \ref{effect_spectra} below).

\section{Geodesic Plunge Inside the ISCO}\label{geodesic_plunge}
The structure of the innermost region of the disk is based on the
geodesic trajectories 
of plunging particles inside of the ISCO. From the mass continuity
equation (\ref{m_dot}) and integrating the vertical structure
equations to give the surface density $\Sigma$ at the ISCO, the radial velocity
$v^{\hat{r}}$ can be determined, which in turn gives the momentum of a massive
particle at the beginning of its plunge. We assume the gas follows a
geodesic trajectory, conserving both the specific energy and angular
momentum from the ISCO to the horizon. From the Hamiltonian
formulation
\begin{equation}
H(x^\mu,p_\mu) = g^{\mu\nu}p_\mu p_\nu = -1,
\end{equation}
the radial velocity is given by
\begin{equation}
p^r(r) = -\left[-g^{rr}(1+g^{tt}E_0^2-2g^{t\phi}E_0l
+g^{\phi\phi}l^2)\right]^{1/2},
\end{equation}
where the integrals of motion are the specific energy $p_t = -E_0$ and
angular momentum $p_\phi = l$ at the ISCO. Near the plane of the disk,
the metric and inverse-metric (for coordinates $t$, $r$, $z$, and
$\phi$) are given by 
\begin{equation}
g_{\mu \nu} = \left(\begin{array}{cccc}
-\mathcal{DA}^{-1}+(2Ma/r^2)^2\mathcal{A}^{-1} & 0 & 0 & -2Ma/r \\
0 & \mathcal{D}^{-1} & 0 & 0 \\
0 & 0 & 1 & 0 \\
-2Ma/r & 0 & 0 & r^2\mathcal{A} \end{array}\right)
\end{equation}
and
\begin{equation}
g^{\mu \nu} = \left(\begin{array}{cccc}
-\frac{\mathcal{A}}{\mathcal{D}} & 0 & 0 & 
-\frac{2Ma}{r^3\mathcal{D}} \\
0 & \mathcal{D} & 0 & 0 \\
0 & 0 & 1 & 0 \\
-\frac{2Ma}{r^3\mathcal{D}} & 0 & 0 &
\frac{1}{r^2\mathcal{A}}-
\left(\frac{2Ma}{r^3}\right)^2\frac{1}{\mathcal{A}\mathcal{D}} 
\end{array}\right).
\end{equation}
In the coordinate basis, the plunge trajectory 4-velocity inside the
ISCO is
\begin{eqnarray}
p^t &=& -g^{tt}E_0 + g^{t\phi}l, \nonumber\\
p^r &=& -\left[-g^{rr}(1+g^{tt}E_0^2-2g^{t\phi}E_0l
+g^{\phi\phi}l^2)\right]^{1/2}, \nonumber\\
p^z &=& 0, \nonumber\\
p^\phi &=& -g^{t\phi}E_0 + g^{\phi\phi}l.
\end{eqnarray}
In the ZAMO basis,
\begin{eqnarray}
p^{(t)} &=& \sqrt{\frac{\mathcal{D}}{\mathcal{A}}}p^t, \nonumber\\
p^{(r)} &=& \sqrt{\frac{1}{\mathcal{D}}}p^r, \nonumber\\
p^{(z)} &=& 0, \nonumber\\
p^{(\phi)} &=&
-\frac{2Ma}{r^2\mathcal{A}^{1/2}}p^t+r\mathcal{A}^{1/2}p^\phi.
\end{eqnarray}
These components will be necessary for calculating the gravitational
tidal force in the local frame of the plunging gas. The result is
similar to that derived above for circular orbits in
equation (\ref{Riemann}), with one additional term $(p^{(r)})^2
R_{(z)(r)(z)(r)}$ due to the non-zero radial velocity. 

In addition to the modified form of the Riemann tensor, the most
significant characteristic of the plunging region is the expansion of
the gas due to the divergence of nearby trajectories. For a collection
of particles on geodesic orbits in the plane of the disk, an infinitesimal area
$\delta A$ evolves according to the expansion parameter $\theta$, defined as
\begin{equation}\label{expansion}
\theta \equiv \frac{1}{A}\frac{d}{d\tau}\delta A = p^{\alpha}_{;\alpha} =
\frac{\partial}{\partial x^{\alpha}}p^{\alpha} + p^\mu
\Gamma^\alpha_{\mu\alpha},
\end{equation}
where $\tau$ measures the proper time along the trajectory of the gas.
For both Schwarzschild and Kerr black holes, the expansion is given
simply by
\begin{equation}
\theta(r) = \frac{\partial p^r}{\partial r}+2\frac{p^r}{r}.
\end{equation}
Thus the area evolves according to 
\begin{equation}\label{dA_tau}
\delta A = \delta A_0 \exp[\int_0^\tau \theta(\tau')d\tau'] 
= \delta A_0 \exp[\int_{r_{\rm ISCO}}^r dr' \theta(r')/p^r],
\end{equation}
where $\delta A_0$ is the (unit) area of a ``footprint'' of a vertical
column of gas at the ISCO. For smaller values of $p^r(R_{\rm ISCO})$, the
expansion is greater, as the gas falls sharply out of the disk. For
large initial values of $p^r$, the expansion is actually negative and
the gas is compressed as it flows inward through concentric circles of
decreasing $r$, then eventually expands as it approaches the horizon
and is pulled into the black hole on rapidly plunging trajectories.

As we mentioned above, the specific value for $p^r(R_{\rm ISCO})$ is
determined by integrating the vertical structure equations and using
the constant mass accretion relation. In Section \ref{NT_disks}, when
solving the
radial structure of the disk outside of the ISCO, we set the
integrated stress at the ISCO to some small non-zero value. Now we can
determine what this value should be. As explained in \citet{shaku73},
the coefficient of viscosity due to turbulent motion in the gas is 
\begin{equation}
\eta \approx \rho v_{\rm turb} l_{\rm turb},
\end{equation}
where $v_{\rm turb}$ and $l_{\rm turb}$ are the characteristic
velocity and size of a turbulent cell. The turbulent velocity
is limited by the sound speed or else shocks will develop and
dissipate the turbulent energy. MHD simulations suggest these limits
are often nearly equalities. Thus let us set $v_{\rm turb} \approx c_s
\approx \sqrt{p/\rho}$.
The turbulent stress is then \citep{novik73}
\begin{equation}
\alpha p = t_{\hat{\phi}\hat{r}} = \eta |s_{\hat{\phi}\hat{r}}| \approx
\rho c_s l_{\rm turb}|s_{\hat{\phi}\hat{r}}|,
\end{equation}
which combines with equation (\ref{h_1}) to give
\begin{equation}
l_{\rm turb} \approx \frac{\alpha h}{-s_{\hat{\phi}\hat{r}}}
\sqrt{\mathcal{R}/6}
\sim \alpha h.
\end{equation}

This turbulent length scale determines the region over which viscous
torques can act on the gas. Inside of the ISCO, the gas expands
rapidly, thus decreasing the pressure, which in turn is responsible
for creating the viscosity in the gas. Therefore the scale length for
the turbulent cells should be the same as the scale length of the
pressure drop inside of the ISCO. In the NT model, since no stresses
act across the ISCO, there is no means for transporting away angular
momentum and allowing the gas to cross the ISCO. Thus the matter
should start to ``pile up'' at the ISCO, increasing the disk thickness
until the turbulence scale length extends far enough inside the ISCO
to get pulled in by the plunging geodesics. 

A non-zero torque on the
disk at the ISCO increases the overall radiative efficiency of the
disk by effectively removing energy from the accreting matter even
after it has crossed the ISCO. Much of this
energy is then transported outward and radiated at a greater value of
$r$, increasing the temperature of the entire disk
\citep{agol99}. Table \ref{table_eff} shows the efficiency of the
torqued $\alpha$-disk as a function of spin, compared to the
zero-torque NT disk. Also listed are the respective Eddington
accretion rates $\dot{M}_{\rm Edd}$ for a black hole with mass
$10M_\odot$. These accretion rates decrease for increased efficiency,
as it takes less mass to produce the same luminosity.

\begin{table}
\caption[Accretion efficiency for torqued and non-torqued disks]
{\label{table_eff} Accretion efficiency $\eta=L/\dot{M}c^2$ for torqued and
non-torqued (NT) disks. For a given efficiency, $L_{\rm
Edd}=\eta\dot{M}_{\rm Edd}c^2$}
\begin{center}
\begin{tabular}{lcccc}
  BH Spin ($a/M$) & $\eta_{\rm torqued}$ & $\eta_{\rm NT}$ & 
  $\dot{M}_{\rm Edd}$(torqued) & $\dot{M}_{\rm Edd}$(NT) \\
  & & & $\times 10^{19}$ gm/s & $\times 10^{19}$ gm/s \\
  \hline
  0 & 0.058 & 0.056 & 2.29 & 2.37 \\
  0.25 & 0.069 & 0.067 & 1.92 & 1.98 \\
  0.5 & 0.088 & 0.084 & 1.52 & 1.58 \\
  0.75 & 0.128 & 0.118 & 1.04 & 1.12 \\
  0.9 & 0.186 & 0.170 & 0.71 & 0.78 \\
  0.998 & 0.453 & 0.379 & 0.29 & 0.35 \\
\end{tabular}
\end{center}
\end{table}

To estimate the scale length of the pressure drop inside the ISCO,
consider an ideal gas dominated by radiation pressure 
\begin{equation}
p = \frac{a}{3} T^4
\end{equation}
with energy density
\begin{equation}
u = aT^4.
\end{equation}
From the first law of thermodynamics, 
\begin{equation}
d\ln T = -\frac{1}{3}d\ln V
\end{equation}
so
\begin{equation}
p \sim V^{-4/3},
\end{equation}
where $V$ is the volume of the gas, and in our case
$V=h\delta A$. Conservation of mass gives $\rho \sim V^{-1}$, so equation
(\ref{h_1}) gives
\begin{equation}
h^2 \sim \frac{p}{\rho \mathcal{R}} \sim V^{-1/3}\mathcal{R}^{-1} \sim
\delta A^{-1/3}h^{-1/3}\mathcal{R}^{-1} \Rightarrow 
h \sim \left(\frac{1}{\delta A \mathcal{R}^3}\right)^{1/7}.
\end{equation}
Thus the pressure scaling inside of the ISCO can be approximated by
\begin{equation}\label{p_scale}
p(r<r_{\rm ISCO}) \sim \frac{\mathcal{R}^{4/7}}{\delta A^{8/7}},
\end{equation}
where $\mathcal{R}$ and $\delta A$ are given by the geodesic plunge
trajectories. Of course, those trajectories are defined by the initial
inward radial velocity, which is in term determined by the disk
thickness and density at
the ISCO. The consistent disk solution is that in which the pressure
falls off at a length scale of $l_{\rm turb}$:
\begin{equation}\label{p_turb}
p(r_{\rm ISCO}-l_{\rm turb}) \approx \frac{1}{2}p(r_{\rm ISCO}).
\end{equation}

For solar-mass black holes with a given stress parameter $\alpha$,
we are able to find solutions to equation (\ref{p_turb}) for a range
of accretion rates $\dot{M}$ by varying the boundary condition for the
integrated stress at the ISCO $W(R_{\rm ISCO})$. Consider the two
limits: for a small $W$, the flux off the disk surface is small, the
disk is thin, and the surface density $\Sigma$ is small, giving a
large inward velocity $p^r$; however, for a large value of $W(R_{\rm
ISCO})$, the high flux demands a large surface density and thus a
small velocity $p^r$. From the geodesic plunge trajectories, we find
that a small initial inward velocity gives a small scale length for
the plunge (matter falls out over a small range of $r$), while a large
initial velocity gives a longer scale length as the matter ``coasts''
for a while before plunging. These competing factors ensure a solution
to equation (\ref{p_turb}): small $W$ with small $h$ $\Rightarrow$
coasting plunge with large $l_{\rm plunge}$; large $W$ with large $h$
$\Rightarrow$ sharp plunge with small $l_{\rm plunge}$. Thus somewhere
between the two limits a solution for $l_{\rm plunge} = l_{\rm turb}$
exists. 

\section{Numerical Implementation}\label{implicit_scheme}
As we described in Section \ref{NT_disks}, the equations of vertical structure
outside of the ISCO can be integrated in a fashion very similar to
that of the standard stellar structure equations. We have three
coupled first-order differential equations for $p$, $T$, and $q^z$ as
a function of $z$, the vertical height above the accretion disk
midplane. The physical solution is determined by the boundary
conditions on these three equations, setting $q^z(0)=0$, $q^z(h)=F$,
and $T(h)=T_s$, with the surface temperature $T_s$ defined by
equation (\ref{T_s}). 

The actual solution of these structure equations is relatively
straightforward, using a standard fourth-order Runge-Kutta algorithm
with constant step size $dz$. From a Lagrangian mass viewpoint, this
results in finer zone resolution in the outer layers of the disk,
allowing an accurate solution of the atmospheric structure. The
resolution $dz$ is different for each radius in the disk, so that the
thinner inner disk can be divided into roughly the same number of
zones as the thick outer disk. The appropriate step size can be
estimated {\it a priori} from the analytic result (\ref{h_1}) and $dz
\approx h/N_z$ for $N_z$ zones in the disk.

Inside the ISCO, the one-dimensional solution to the vertical
structure equations evolves dynamically in time (proper time of
local free-falling tetrad), so we must replace the coupled ODEs with a
set of hydrodynamic partial differential equations. Following
\citet{bower91}, we adopt an implicit Lagrangian scheme for our
numerical solution to these equations. The Lagrangian approach is
preferred for the one-dimensional problem due to its simplicity and
accuracy in monitoring conserved quantities. 

An implicit scheme is
necessary because of the high sound speed of the radiation
pressure-dominated gas in the inner disk, requiring a very small time
step to satisfy the explicit Courant condition. For typical
resolutions of $N_z \sim 200$, the Courant time step would be
prohibitively small at $dt \sim 10^{-8}$ sec, while the plunge
from the ISCO to the horizon of a $10M_\odot$ black hole could take
$\sim 2-3 \times 10^{-2}$ sec. During this plunge, the gas is
quite ``well-behaved,'' i.e.\ no shock waves or discontinuities in
the state variables, so in the absence of numerical instability, much
larger time steps are appropriate (we typically use $dt \approx 100
\times \Delta t_{\rm Courant}$). 

In the discussion below, we shall use subscripts for spatial indices
and superscripts for temporal indices. The position, velocity,
acceleration, and flux will be defined at the zone boundaries
$k=0,..,N_z$, while the state variables of mass, density, temperature,
pressure, and internal energy will be defined at the zone interiors
$k=1/2,..,N_z-1/2$. A second-order accurate scheme defines the
positions at the whole time steps $n=0,1,...$ and the velocities at
the half-time steps $n=1/2,3/2,...$:
\begin{equation}\label{v_n12}
v_k^{n+1/2} = \frac{z_k^{n+1}-z_k^n}{\Delta t^{n+1/2}}.
\end{equation}
The acceleration of each zone is caused by the tidal gravity and any
pressure gradients in the gas:
\begin{equation}\label{a_n}
a_k^n = \frac{v_k^{n+1/2}-v_k^{n-1/2}}{\Delta t^n} =
\frac{p_{k+1/2}^{n+1}-p_{k-1/2}^{n+1}}{\Delta m_k} 
-R^{\hat{z}}_{\hat{t}\hat{z}\hat{t}}(r) z_k^{n+1},
\end{equation}
where the mass in each zone is given by
\begin{equation}\label{dm1}
\Delta m_{k+1/2}=\rho_{k+1/2}^n(z_{k+1}^n-z_{k}^n)\delta A,
\end{equation}
where we take the footprint of the gas column to have unit
area $\delta A = 1$. The masses on the boundaries are just the
averages of neighboring zones:
\begin{equation}
\Delta m_k = \frac{1}{2}(\Delta m_{k-1/2}+\Delta m_{k+1/2}).
\end{equation}
Note that in equation (\ref{a_n}) the pressure and gravitational
acceleration are defined at the next time step $t^{n+1}$, thus making
this an implicit scheme. 

For our mixture of ionized hydrogen gas and radiation, the internal
specific energy $\varepsilon$ and pressure $p$ are given explicitly as
functions of the fluid density and temperature:
\begin{equation}\label{e_rho_T}
\varepsilon(\rho,T) = \frac{3 k_B T}{m_p} + \frac{aT^4}{\rho}
\end{equation}
and the equation of state given above in equation (\ref{eos}):
\begin{displaymath}
p(\rho,T) = \frac{2k_B T}{m_p}\rho + \frac{a}{3}T^4.
\end{displaymath}
The first law of thermodynamics can be written in the form
\citep{bower91} 
\begin{equation}\label{therm_1}
\left(\frac{\partial \varepsilon}{\partial T}\right)_\rho dT = dQ +
T\left(\frac{\partial p}{\partial T}\right)_\rho \frac{d\rho}{\rho^2},
\end{equation}
where all non-adiabatic contributions (shocks, radiation, etc.) to the
internal energy is included in the term $dQ$. These effects are
included in a separate implicit treatment using the technique of
``operator splitting,'' described below. Thus the purely adiabatic
expansion and compression of the fluid can be described in finite
difference form:
\begin{equation}\label{e_finite_diff}
T_{k+1/2}^{n+1}-T_{k+1/2}^n =
-\frac{T_{k+1/2}^{n+1}(p_{,T})_{k+1/2}^n}{(\varepsilon_{,T})_{k+1/2}^n}
\left(\frac{1}{\rho_{k+1/2}^{n+1}}-\frac{1}{\rho_{k+1/2}^n}\right).
\end{equation}
Here $\varepsilon_{,T}$ and $p_{,T}$ are the partial derivatives of energy
and pressure with respect to temperature. Note that the temperature on
the right hand side of equation (\ref{e_finite_diff}) is also given
implicitly at time $t^{n+1}$. Defining the dimensionless parameter
\begin{equation}\label{gamma_kn}
\Gamma_{k+1/2}^n \equiv
-\frac{(p_{,T})_{k+1/2}^n}{(\varepsilon_{,T})_{k+1/2}^n}
\left(\frac{1}{\rho_{k+1/2}^{n+1}}-\frac{1}{\rho_{k+1/2}^n}\right),
\end{equation}
the temperature at time $t^{n+1}$ is given by
\begin{equation}\label{T_np}
T_{k+1/2}^{n+1} = T_{k+1/2}^n(1+\Gamma_{k+1/2}^n)^{-1}.
\end{equation}
On the right hand side of equation (\ref{gamma_kn}), the density
$\rho_{k+1/2}^{n+1}$ can be estimated \textit{explicitly} to first
order from $v^{n-1/2}_k$ and $v^{n-1/2}_{k+1}$. 

The Lagrangian hydrodynamics conserves the mass in each zone $\Delta
m_{k+1/2}$ between time steps, so the density is given by 
\begin{equation}\label{rho_np}
\rho_{k+1/2}^{n+1} = \frac{\Delta
m_{k+1/2}}{z_{k+1}^{n+1}-z_k^{n+1}}. 
\end{equation}
Equations (\ref{T_np}) and (\ref{rho_np}) and the equation of state
give the pressure at time $t^{n+1}$. This pressure is then used in
equation (\ref{a_n}), which can be combined with (\ref{v_n12}) to give a
set of relations defined on the zone boundaries, expressible as a set
of coupled nonlinear equations
\begin{eqnarray}\label{f_k}
\lefteqn{f_k(z_{k+1}^{n+1},z_k^{n+1},z_{k-1}^{n+1}) = } \nonumber\\ & &
-\frac{1}{\Delta t^n}\left(\frac{z_k^{n+1}-z_k^n}{\Delta t^{n+1/2}}-
\frac{z_k^{n}-z_k^{n-1}}{\Delta t^{n-1/2}}\right) 
+\frac{p_{k+1/2}^{n+1}-p_{k-1/2}^{n+1}}{\Delta m_k} 
-R^{\hat{z}}_{\hat{t}\hat{z}\hat{t}}(r) z_k^{n+1} = 0.
\end{eqnarray}
On the right hand side, the pressure terms $p^{n+1}$ are functions of the
positions $z^{n+1}_{k+1}$ and $z^{n+1}_{k-1}$. The function
$f_k$ is well-defined by the equations above for the
interior zones $(1\le k \le N_z-1)$ and we use linear extrapolation to
give $f_{N_z}$, while planar symmetry requires
$z_{-1}^{n+1}=-z_1^{n+1}$, thus defining $f_0$. The solution to
equation (\ref{f_k}) gives the positions of the zone boundaries $z_k^{n+1}$,
from which all the other hydrodynamic variables can be determined.

\citet{bower91} outline the standard approach to solving this set of
equations using Newton-Raphson iteration and a tridiagonal
solver. Denoting the first order solution to
$f_k(t^{n+1})=0$ by the vector $z_k^i$, equation (\ref{f_k}) can be
written as  
\begin{eqnarray}
\lefteqn{f_k(z_{k+1}^i,z_k^i,z_{k-1}^i)=f_k(z_{k+1}^n,z_k^n,z_{k-1}^n)}
\nonumber\\ & & 
+\left(\frac{\partial f_k}{\partial z_{k+1}^n}\right)^n\Delta z_{k+1}^n 
+\left(\frac{\partial f_k}{\partial z_{k}^n}\right)^n\Delta z_{k}^n
+\left(\frac{\partial f_k}{\partial z_{k-1}^n}\right)^n\Delta
z_{k-1}^n + \mathcal{O}(\Delta z^2).
\end{eqnarray}
Then an approximate solution to (\ref{f_k}) is 
\begin{equation}
z_k^{n+1} = z_k^n + \Delta z_k^n.
\end{equation}
We solve for these $\Delta z_k^n$ iteratively by setting
$f_k(z_{k+1}^i,z_k^i,z_{k-1}^i)=0$ and solving the tridiagonal system 
\begin{equation}
-\left(\frac{\partial f_k}{\partial z_{k+1}^n}\right)^n\Delta z_{k+1}^n 
-\left(\frac{\partial f_k}{\partial z_{k}^n}\right)^n\Delta z_{k}^n
-\left(\frac{\partial f_k}{\partial z_{k-1}^n}\right)^n\Delta
z_{k-1}^n =f_k(z_{k+1}^n,z_k^n,z_{k-1}^n) 
\end{equation}
and then re-evaluating $f_k(z_{k+1}^i,z_k^i,z_{k-1}^i)$ until an
acceptable accuracy is reached for $z_k^{n+1}$ in
equation (\ref{f_k}). This typically take only about seven or eight
iterations to reach machine accuracy, due to the rapid convergence of
the Newton-Raphson root finding algorithm. This accuracy far exceeds
the limiting first-order accuracy of the finite difference equation
for energy (\ref{e_finite_diff}), so we generally use only four or
five iterations in the implicit scheme.

As mentioned above, the technique of operator splitting is employed to
model the energy transfer in the gas via radiation diffusion. Since
the change in heat in a fluid element is given by $dQ$, the rate
of energy flow due to radiation and viscous heating (really turbulent
magnetic stress) can be written:
\begin{equation}\label{dQ_dt}
\frac{dQ}{dt} = \frac{\partial \varepsilon}{\partial T} \frac{dT}{dt}
= -\frac{1}{\rho}\frac{dq^z}{dz}+\frac{\bar{\alpha} p}{\rho},
\end{equation}
where $\bar{\alpha}$ is the compact form of the $\alpha$ parameter
defined in equation (\ref{enrg_gen}), 
and in general we use $\alpha=0.1$.
Converting to the Lagrangian mass coordinate $dm=\rho dz$ and
linearizing $T$, equations (\ref{dQ_dt}) and (\ref{q_z}) give a
single, second-order diffusion equation (with a turbulent heating
source) for the temperature:
\begin{equation}\label{dT_dt}
\frac{dT}{dt} = \left(\frac{\partial \varepsilon}{\partial
T}\right)^{-1} \left[
\frac{4ac}{3\kappa}\frac{d^2T}{dm^2}+\frac{\bar{\alpha} p}{\rho}\right].
\end{equation}
In finite difference form, we have
\begin{eqnarray}\label{T_rad_diff}
\lefteqn{T_{k+1/2}^{n+1} = T_{k+1/2}^n} \nonumber\\ & & 
+ \frac{\Delta t}{(\varepsilon_T)_{k+1/2}^n} 
\left[\frac{4ac}{3\kappa}\frac{(T_{k+1/2}^n)^3}{\Delta m_{k+1/2}^2}
(T_{k+3/2}^{n+1} -2T_{k+1/2}^{n+1}+ T_{k-1/2}^{n+1}) 
+\bar{\alpha}\frac{p_{k+1/2}^n}{\rho_{k+1/2}^n}\right]
\end{eqnarray}
where again the implicit scheme uses the temperatures at the future
time step $t^{n+1}$ on the right hand side. Equation
(\ref{T_rad_diff}) gives another
tridiagonal set of linear equations that solve for
$T_{k+1/2}^{n+1}$. In practice, for each time step, we solve this
system first, then set $T^n=T^{n+1}$ and solve for the new positions
$z_k^{n+1}$ with the adiabatic implicit hydrodynamics described above.  

So far, we have followed a standard one-dimensional approach to the
problem of radiation hydrodynamics. For the relativistic accretion
disk inside the ISCO, we need to include a couple additional GR
effects. First, the gravitational force given by the Riemann tensor in
equation (\ref{a_n}) must be determined in the local frame of a plunging
geodesic particle, as in equation (\ref{Riemann}). The time coordinate
$t^n$ used throughout this Section is then the proper time as measured
in this local tetrad. 

Furthermore, due to the geodesic expansion of the gas inside the ISCO,
the integrated surface density of the disk will fall off rapidly
during the plunge. In other words, the ``footprint'' of a vertical
column of gas at the ISCO will expand in area, while maintaining a
constant mass. We model this expansion by varying the size of the mass
element $\Delta m_{k+1/2}$ for a column of unit area. From equation
(\ref{dA_tau}), we get
\begin{equation}
\Delta m_{k+1/2}(r) = \Delta m_{k+1/2}(r_{\rm ISCO})
\exp[-\int_{r_{\rm ISCO}}^r dr' \theta(r')/p^r],
\end{equation}
thus modifying equation (\ref{dm1}) to give the density
\begin{equation}
\rho_{k+1/2}^n = \frac{\Delta m_{k+1/2}(r^n)}{z_{k+1}^n-z_k^n},
\end{equation}
with $r^n$ being the radial coordinate of the geodesic trajectory at
proper time $t^n$. This ``expanded'' density is that which is actually
used in equations (\ref{a_n}) and (\ref{e_finite_diff}).

Lastly, the viscous stress parameter $\bar{\alpha}$ in equations
(\ref{enrg_gen}, \ref{dQ_dt}, and \ref{dT_dt}) should be modified to
account for the 
geodesic shear $s_{\hat{\phi}\hat{r}}$ inside the ISCO. However, we
find that inside the ISCO, the rapid expansion of the gas and
corresponding drop in the disk's surface density and optical depth
causes radiation diffusion to completely dominate over viscous heating
in the plunge region. Thus we simply set $\bar{\alpha}$ constant at
the value it has at the ISCO. Numerically, we typically use a few
hundred zones in the vertical direction, and achieve reasonable
convergence for time steps of $\Delta t \sim 100 \Delta t_{\rm
Courant}$, thus requiring $\sim 10^3-10^4$ steps to plunge from the
ISCO to the horizon.

\begin{figure}
\begin{center}
%\scalebox{0.45}{\includegraphics*[40,350][545,700]{chap5_f3a.ps}}
%\scalebox{0.45}{\includegraphics*[143,350][550,700]{chap5_f3b.ps}}\\
%\scalebox{0.45}{\includegraphics*[40,350][545,700]{chap5_f3c.ps}}
%\scalebox{0.45}{\includegraphics*[143,350][550,700]{chap5_f3d.ps}}\\
\scalebox{0.8}{\includegraphics{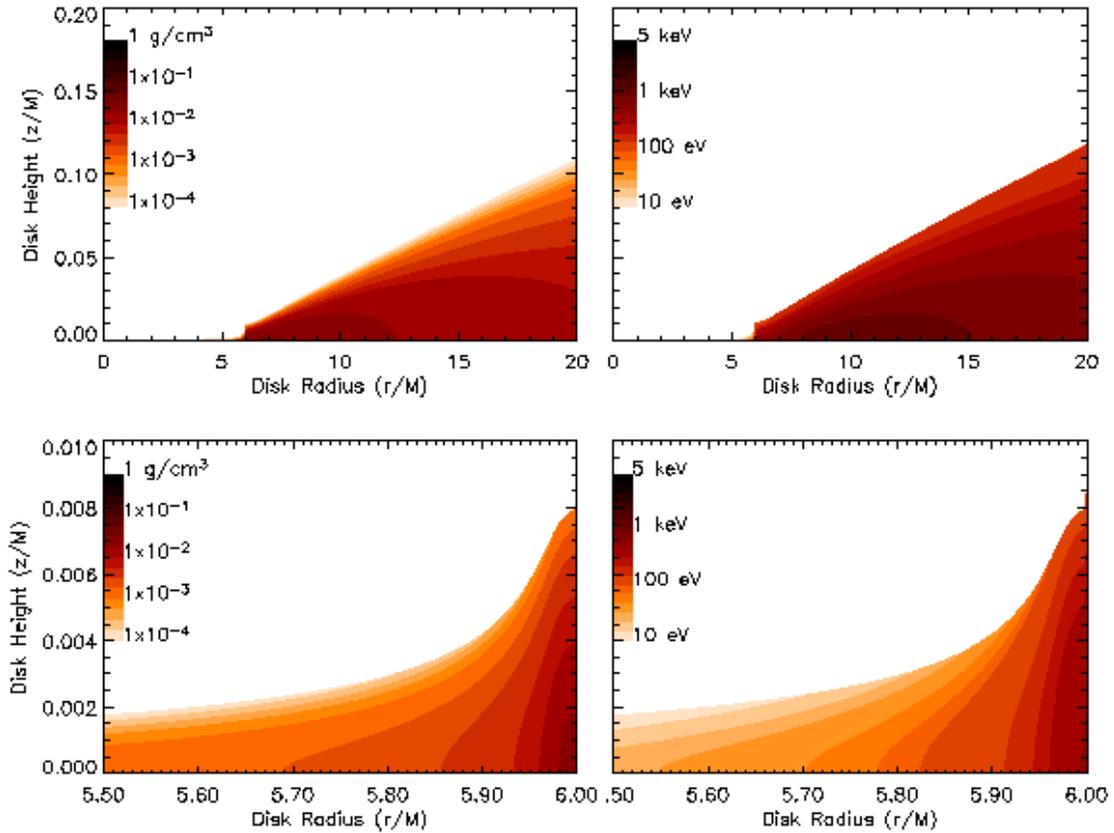}}
\caption[2-D density and temperature profiles of $\alpha$-disk with
$a/M=0$]{\label{nt_profile0} Top panels: density (left) and
temperature (right)
contours in the $r-z $ plane for an axisymmetric $\alpha$-disk around a
Schwarzschild black hole. Bottom panels: plunge region of the disk
immediately inside the ISCO, matching $l_{\rm plunge}\approx l_{\rm
turb} \approx h$. The black hole has mass $10M_\odot$ and accretion
rate $0.02\dot{M}_{\rm Edd}$.}
\end{center}
\end{figure}

Combining all the above results for the disk structure inside and
outside the ISCO, we can now produce a full three-dimensional
(axisymmetric) density and temperature profile for the relativistic
$\alpha$-disk. Figure \ref{nt_profile0} shows the inner disk structure
for a Schwarzschild black hole of mass $10M_\odot$ and accretion rate
$0.02\dot{M}_{\rm Edd}$. Even with a significant stress at the ISCO,
the gas plunges so rapidly inside the ISCO that the density and
temperature fall off quickly in that region. However, while there is
not significant thermal radiation emitted from inside the ISCO, the
total optical depth to electron scattering is still greater than
unity, suggesting that this inner region may still be quite
important as an emitter of fluorescent iron lines (see Section
\ref{transfer_function}). 

\begin{figure}
\begin{center}
%\scalebox{0.45}{\includegraphics*[40,350][545,700]{chap5_f4a.ps}}
%\scalebox{0.45}{\includegraphics*[143,350][550,700]{chap5_f4b.ps}}\\
%\scalebox{0.45}{\includegraphics*[40,350][545,700]{chap5_f4c.ps}}
%\scalebox{0.45}{\includegraphics*[143,350][550,700]{chap5_f4d.ps}}\\
\scalebox{0.8}{\includegraphics{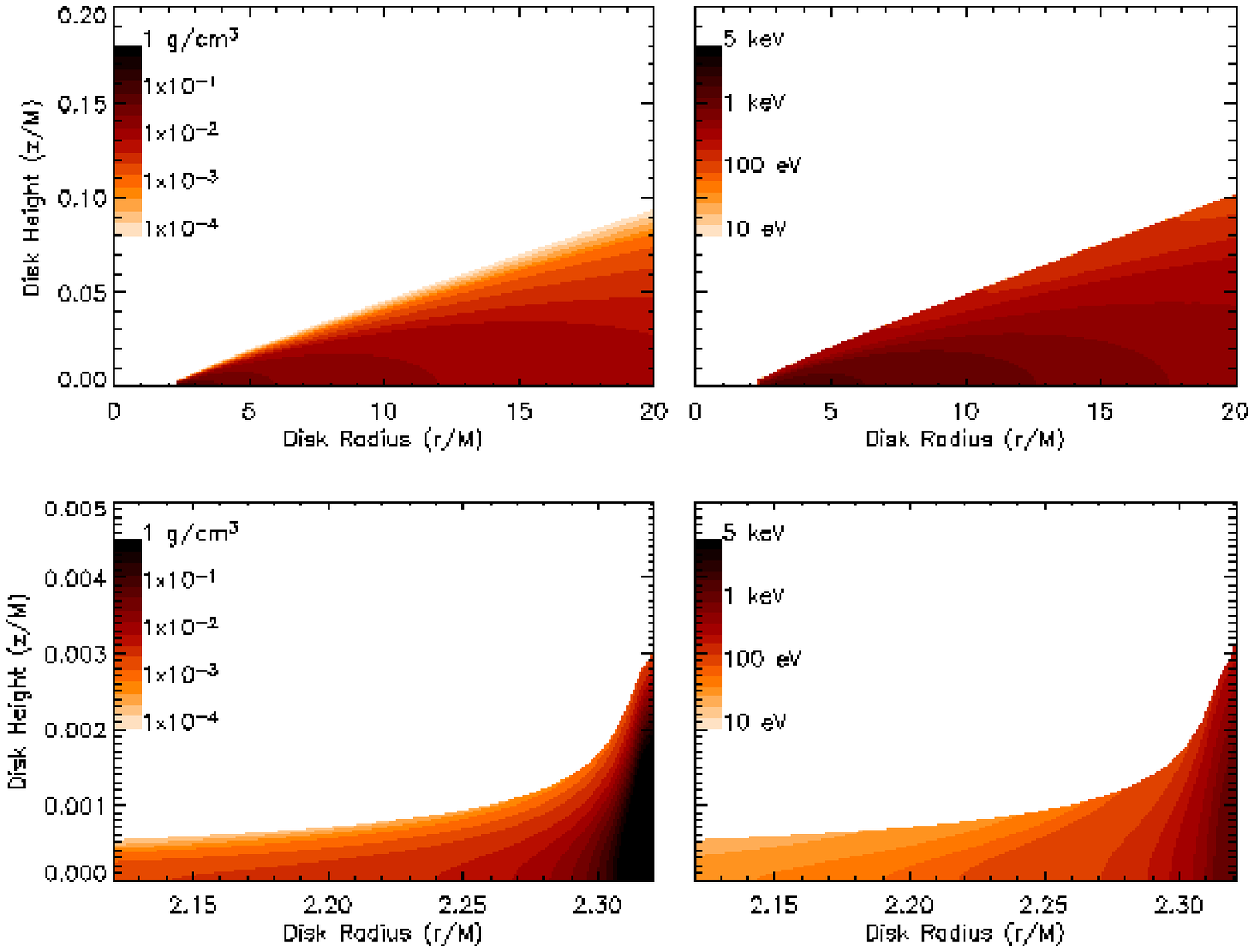}}
\caption[2-D density and temperature profiles of $\alpha$-disk with
$a/M=0.9$]{\label{nt_profile9} Top panels: density (left) and
temperature (right)
contours in the $r-z$ plane for an axisymmetric $\alpha$-disk around a
Kerr black hole with $a/M=0.9$. Bottom panels: plunge region of the disk
immediately inside the ISCO, matching $l_{\rm plunge}\approx l_{\rm
turb} \approx h$. The black hole has mass $10M_\odot$ and accretion
rate $0.02\dot{M}_{\rm Edd}$.}
\end{center}
\end{figure}

Figure \ref{nt_profile9} shows the same density and temperature
profiles, now for a Kerr black hole with spin $a/M=0.9$. For the same
Eddington-normalized accretion rate, the higher spin value leads to a
denser, hotter disk. As the ISCO moves in to $R_{\rm ISCO} \approx
2.3 M$, the tidal gravity in the inner disk becomes stronger,
maintaining hydrostatic equilibrium even for the higher radiation
flux. 

\section{Observed Spectrum of the Disk}\label{ntdisk_spectra}
The net result of the previous three Sections is a complete
three-dimensional description of the density, temperature, and
velocity of the accretion disk everywhere outside the event
horizon. From this tabulated set of data, we can then calculate the
predicted spectral appearance of the accretion system with our
ray-tracing post-processor. For a first
approximation, we will again assume that electron scattering dominates
the opacity for most of the disk and then consider a
two-dimensional ``photosphere'' one optical depth below the surface of
the disk. Given the temperature and density of the disk at the
photosphere, a modified blackbody spectrum for the radiation is
derived by \citet{zeldo69} for an exponential atmosphere and
\citet{shaku72} and \citet{felte72} for a constant density atmosphere. 

While the forms of these modified spectra were presented above in
Section \ref{vertical_structure}, we repeat them here in more
detail. In the approximation of a homogeneous,
isothermal atmosphere with density $\rho_s$ and temperature $T_s$, the
emitted flux is given by 
\begin{equation}\label{F_nu1}
F_\nu = \pi B_\nu \left(\frac{\kappa_\nu^{\rm ff}}{\kappa_{\rm
es}}\right)^{1/2}.
\end{equation}
The term $(\kappa_\nu^{rm ff}/\kappa_{\rm es})^{1/2}$ is due to the
modified path length of a photon that takes a random walk through a
medium dominated by electron scattering [for a detailed derivation,
see e.g.\ \citet{shapi83}].
Here $B_\nu$ is the blackbody brightness with units of [erg sec$^{-1}$
cm$^{-2}$ Hz$^{-1}$ ster$^{-1}$]:
\begin{equation}
B_\nu(T) \equiv \frac{2h\nu^3/c^2}{\exp(h\nu/k_BT)-1} =
\frac{2k_B^3T^3}{h^2c^2} \frac{x^3}{e^x-1}
\end{equation}
and $\kappa_\nu^{\rm ff}$ is the opacity for free-free absorption, for
which we use Kramer's law with \citep{shapi83}
\begin{equation}
\kappa_\nu^{\rm ff} = 1.5\times 10^{25}\rho_s T_s^{-7/2}
x^{-3}(1-e^{-x})\mbox{ cm}^2/\rm{g},
\end{equation}
where we have defined the dimensionless parameter $x\equiv
h\nu/kT$. Now equation (\ref{F_nu1}) becomes 
\begin{equation}
F_\nu = 2.56 \times 10^{-3}\rho_s^{1/2}T_s^{5/4}
\frac{x^{3/2}e^{-x/2}}{(e^x-1)^{1/2}}.
\end{equation}

For the exponential atmosphere with scale height $H$, we have a
slightly different form: 
\begin{equation}
F_\nu = \pi B_\nu \left(\frac{\kappa_\nu^{\rm ff}}{\kappa_{\rm es}^2
\rho H}\right)^{1/3}.
\end{equation}
The scale height can be determined from equation (\ref{expon_atm}) as
\begin{equation}
H = \frac{2k_BT_s}{m_p \mathcal{R} h_{\rm disk}},
\end{equation}
giving the modified spectrum 
\begin{equation}
F_\nu = 3.5\times 10^{-10} (\mathcal{R} h_{\rm disk})^{1/3} T_s^{3/2}
\frac{x^2 e^{-x/3}}{(e^x-1)^{2/3}}. 
\end{equation}

Given the emitted spectrum from each surface element of the disk, we
use the ray-tracing code of Chapter 2 to
calculate the effects of redshift and gravitational lensing to a
distant observer \citep{schni04a}. As described there,
the photon trajectories are traced backwards in time from the observer
with initial energy $E_{\rm obs}=-p_t$. The redshift from the point of
the emitter to the observer is calculated by
\begin{displaymath}
\frac{E_{\rm obs}}{E_{\rm em}} =
\frac{p_\mu(\mathbf{x}_{\rm obs})v^\mu(\mathbf{x}_{\rm obs})}
{p_\mu(\mathbf{x}_{\rm em})v^\mu(\mathbf{x}_{\rm em})},
\end{displaymath}
where for a distant observer at $r\to \infty$, we take
$v^\mu(\mathbf{x}_{\rm obs}) =
[1,0,0,0]$. Lorentz invariance of $I_\nu/\nu^3$ along a photon bundle
gives the observed spectral intensity:
\begin{displaymath}
I_\nu(\rm obs) = I_\nu(\rm em) \frac{\nu^3_{\rm obs}}{\nu^3_{\rm em}}.
\end{displaymath}

Taking $v^\mu(\mathbf{x}_{\rm em})$ as the velocity of a planar
geodesic trajectory and assuming an isotropic emitter $I_\nu({\rm
em}) = F_\nu/\pi$, the observed spectrum is the sum of the redshifted
spectra from each individual path (pixel) ray-traced from the
observer. As we see from Figures \ref{nt_compare0} to
\ref{nt_profile9}, the steady-state $\alpha$-disks for $M=10M_\odot$
are quite thin, with $h/r \lesssim 0.02$ in the inner disk. This
allows us to use the simple transfer function of Section
\ref{transfer_function} to calculate the total disk spectrum. Our job
is made even easier because the spectrum at each radius in the disk is
a function only of the atmospheric temperature and scale height at
that point in the disk. The observed temperature is simply scaled by
the transfer function redshift (much like the cosmological redshift
scales the apparent blackbody temperature of receding stars and
background radiation), allowing us to integrate over the
image plane quite easily. This follows the approach of
\citet{huben00,huben01}, who also include non-LTE transfer in the
atmosphere to model the Lyman-$\alpha$ line in AGN. 

\begin{figure}
\begin{center}
%\scalebox{0.3}{\includegraphics*[40,390][535,700]{chap5_f5_0_00.ps}}
%\scalebox{0.3}{\includegraphics*[100,390][535,700]{chap5_f5_5_00.ps}}
%\scalebox{0.3}{\includegraphics*[100,390][535,700]{chap5_f5_9_00.ps}}\\
%\scalebox{0.3}{\includegraphics*[40,390][535,700]{chap5_f5_0a_45.ps}}
%\scalebox{0.3}{\includegraphics*[100,390][535,700]{chap5_f5_5a_45.ps}}
%\scalebox{0.3}{\includegraphics*[100,390][535,700]{chap5_f5_9a_45.ps}}\\
%\scalebox{0.3}{\includegraphics*[40,390][535,700]{chap5_f5_0b_45.ps}}
%\scalebox{0.3}{\includegraphics*[100,390][535,700]{chap5_f5_5b_45.ps}}
%\scalebox{0.3}{\includegraphics*[100,390][535,700]{chap5_f5_9b_45.ps}}\\
%\scalebox{0.3}{\includegraphics*[40,390][535,700]{chap5_f5_0c_45.ps}}
%\scalebox{0.3}{\includegraphics*[100,390][535,700]{chap5_f5_5c_45.ps}}
%\scalebox{0.3}{\includegraphics*[100,390][535,700]{chap5_f5_9c_45.ps}}\\
%\scalebox{0.3}{\includegraphics*[40,330][535,700]{chap5_f5_0_80.ps}}
%\scalebox{0.3}{\includegraphics*[100,330][535,700]{chap5_f5_5_80.ps}}
%\scalebox{0.3}{\includegraphics*[100,330][535,700]{chap5_f5_9_80.ps}}
\scalebox{0.9}{\includegraphics{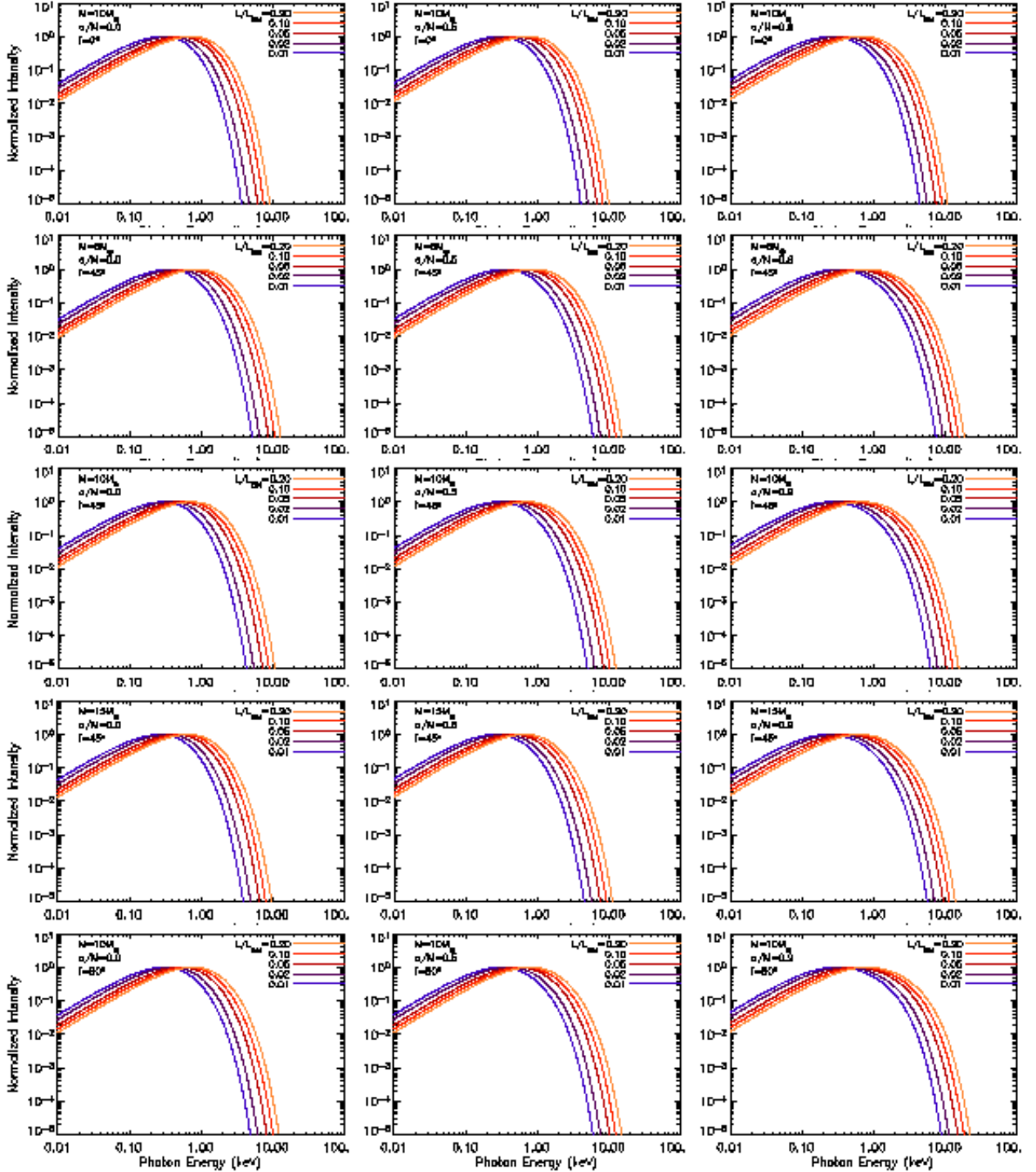}}
\caption[Spectra of thermal steady-state
$\alpha$-disks]{\label{mcd_spectra} Sample of normalized photon energy
spectra
produced by steady-state $\alpha$-disks for a variety of black hole
masses, spins, and inclinations. For each case, a range of Eddington-scaled
luminosities are shown. The peak energy $E_{\rm max}$ scales as in equation
(\ref{Emax_Mdot}) while the ratio $E_{\rm cut}/E_{\rm max}$ is
sensitive to the spin and inclination angle.}
\end{center}
\end{figure}

The resulting modified blackbody spectra are shown in Figure
\ref{mcd_spectra} for a variety of black hole parameters, with
$M=[5,10,15]M_\odot$, $a/M= [0,0.5,0.9]$, and $i =
[0^\circ,45^\circ,80^\circ]$. The disk-integrated spectra are
characterized by a low energy rise with $I_\nu \propto \nu$, followed
by a broad thermal peak around $0.5-2$ keV, and a steep cutoff around
10 keV, consistent with many of the ``Thermal-Dominant'' state spectra
observed with \textit{RXTE} \citep{mccli04}. The specific location of
the peak $E_{\rm max}$ is most sensitive to the black hole's mass and
accretion rate, scaling roughly like the surface temperature given in
NT:
\begin{equation}\label{Emax_Mdot}
E_{\rm max} \sim \alpha^{2/9} \left(\frac{M}{M_\odot}\right)^{-2/9} 
\left(\frac{\dot{M}}{\dot{M}_{\rm Edd}}\right)^{8/9}.
\end{equation}
This relationship may prove very useful in identifying ultra-luminous
X-ray sources as intermediate-mass black holes
\citep{teras04,mille04}, but for a given black hole mass and Eddington
accretion rate, there appears to be relatively little dependence on
spin for this peak energy. 

However, if we are able to measure both the peak location
and the cutoff energy, then it is possible the ratio may be used to
determine the black hole spin. Defining $E_{\rm cut}$ as the point
where the intensity $I_\nu(E_{\rm cut})$ is a factor of $10^5$ smaller than
$I_\nu(E_{\rm max})$, we find that $E_{\rm cut}/E_{\rm max}$ is
relatively insensitive to the black hole mass and accretion rate. Over
a range of masses $5 \le (M/M_\odot) \le 20$ and luminosity $0.01 \le
(L/L_{\rm Edd}) \le 0.2$, this ratio appears to be a function primarily of
black hole spin and disk inclination. Table \ref{table_ratio} shows
the mean values for this ratio for a few different spins and
inclination angles, along with the typical scatter over the
mass/luminosity sample. If the inclination is known from optical
radial velocity curves or broadened iron emission lines, then by
measuring the ratio $E_{\rm cut}/E_{\rm max}$, the spin may be
inferred from Table
\ref{table_ratio}. However, there may be significant observational
challenges to this technique: as we will see in Chapter 6, even in the
Thermal-Dominant spectral state, there is typically still a small
power-law component to the high-energy part of the photon
spectrum. This power-law tail will make it more difficult to
accurately measure the cut off energy $E_{\rm cut}$, but even this
tail may be modeled by the scattering calculations described there,
and possibly might even make this approach more feasible.

\begin{table}
\caption[Ratio of $E_{\rm cut}/E_{\rm max}$ for steady-state
disks]{\label{table_ratio} Ratio of $E_{\rm cut}/E_{\rm max}$ for
relativistic $\alpha$-disks for a range of black hole masses $5 \le
(M/M_\odot) \le 20$ and luminosities $0.01 \le (L/L_{\rm Edd}) \le
0.2$. The ``errors'' quoted are the typical variation of this ratio
over the sample of masses and luminosities.}
\begin{center}
\begin{tabular}{rcccc}
  Inclination ($^\circ$) & \hspace{0.5cm} & & BH Spin ($a/M$) & \\
  & & 0 & 0.5 & 0.9 \\
  \hline
  0 & & 12.3 $\pm$ 0.1 & 13.6 $\pm$ 0.3 & 16.4 $\pm$ 0.8 \\
  45 & & 14.5 $\pm$ 0.1 & 16.9 $\pm$ 0.3 & 22.7 $\pm$ 1.0 \\
  80 & & 15.8 $\pm$ 0.1 & 18.9 $\pm$ 0.2 & 27.9 $\pm$ 0.8 \\
\end{tabular}
\end{center}
\end{table}

Despite the apparent promise of this technique, there are also a
number of systematic errors involved in arriving at these
predicted spectra, not the least of which is the entire premise of an
$\alpha$-based stress/viscosity. Even with our self-consistent
treatment of the ISCO boundary
conditions to apply a non-zero torque to the inner disk, the density
and temperature profiles still do not agree qualitatively with global
MHD simulations. Thus we may be seriously ``under-weighing'' the emission
inside the ISCO. Future work on this subject should attempt to
incorporate the magneto-rotational instability as the driving force
behind angular momentum transport in the disk, especially the inner
regions where the majority of energy is released.

Furthermore, to completely match the X-ray spectra from black holes in
the Thermal-Dominant state, one must also include a small high-energy
power-law component \citep{mccli04}. It is likely that this
high-energy tail is caused by the Compton upscattering of thermal
photons through a hot, low-density electron corona above the
disk. In the next Chapter, we introduce a Monte-Carlo code to
calculate the effect of this scattering on the continuum photon
spectra.

\chapter{Electron Scattering}
\begin{flushright}
{\it
If we knew what it was we were doing, it would not be called research,
would it? \\
\medskip

I am convinced that God does not play dice with the universe.\\
\medskip
}
-Albert Einstein
\end{flushright}
\vspace{1cm}

As we saw in a simplified model in Chapter 4, the scattering of hot
spot photons off of coronal electrons will have a significant effect
on the shape of the observed light curves. As mentioned in the
discussion there, many of the important aspects of the scattering
physics were ignored in the basic model in the interest of deriving an
analytic solution. In this Chapter we develop a more sophisticated
Monte Carlo
model that reproduces many of the same qualitative features of the
simple model, while also introducing a number of new physics
predictions.

\section{Physics of Scattering}
\subsection{Classical Electron Scattering}\label{classical_scattering}
To begin with, we present here a review of the classical scattering of a
plane electromagnetic wave incident on an electron at rest, as derived
in \citet{rybic79}. In the
low-energy limit with $h\nu \ll m_ec^2$, the electric field of the
incoming photon will cause the electron to oscillate with a velocity $v\ll
c$ in the direction of the polarization axis
of the EM wave. The force on the electron will be
\begin{equation}
\mathbf{F} = e \boldsymbol{\epsilon}E_0 \sin \omega_0 t,
\end{equation}
where $\boldsymbol{\epsilon}$ is the direction of the electric field
$E_0$ for a
plane-polarized wave oscillating at angular frequency $\omega_0 =
2\pi\nu$. In response to this incident wave, the electron will produce
a time-varying dipole moment $\mathbf{d}$:
\begin{equation}
\mathbf{d} = e\mathbf{r} = -\left(\frac{e^2 E_0}{m_e\omega_0^2}\right)
\boldsymbol{\epsilon} \sin \omega_0 t.
\end{equation}

\begin{figure}[ht]
\begin{center}
\includegraphics[width=0.7\textwidth]{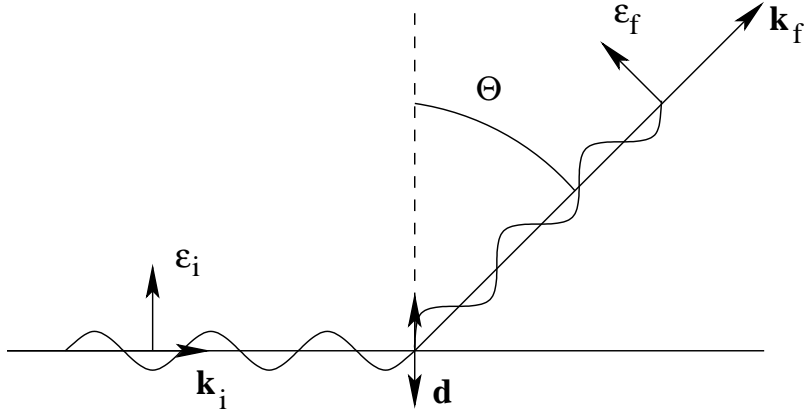}
\caption[Dipole scattering of linearly polarized
radiation]{\label{schem_dipole} Planar geometry of a plane-parallel
wave of electromagnetic radiation $\mathbf{k}_i$ incident on an
electron at rest at the origin. The
incident wave has (vertical) linear polarization in the plane of the
page, denoted by $\boldsymbol{\epsilon}_i$. The wave scatters at an
angle $\Theta$ with respect to the dipole $\mathbf{d}$, which is
parallel to $\boldsymbol{\epsilon}_i$. The scattered wave vector
$\mathbf{k}_f$ is not necessarily in the plane of the page, but must
be in the same plane as the final polarization
$\boldsymbol{\epsilon}_f$ and $\mathbf{d}$.}
\end{center}
\end{figure}

Such a dipole will produce a radiation field with time-averaged power
\begin{equation}
\frac{dP}{d\Omega} = \frac{e^4 E_0^2}{8\pi m_e^2c^3}\sin^2\Theta,
\end{equation}
where $\Theta$ is the polar angle measured with respect to the
polarization (dipole) vector, as shown in Figure \ref{schem_dipole}.
Assuming that the incident flux $\langle S \rangle= E_0^2(c/8\pi)$ is
entirely re-radiated by the dipole, we can define the differential
cross section for polarized radiation:
\begin{equation}
\frac{dP}{d\Omega} = \frac{cE_0^2}{8\pi}
\left(\frac{d\sigma}{d\Omega}\right)_{\rm pol},
\end{equation}
or
\begin{equation}\label{cross_pol}
\left(\frac{d\sigma}{d\Omega}\right)_{\rm pol} = r_0^2\sin^2\Theta.
\end{equation}
Here the classical electron radius $r_0$ is given by
\begin{equation}
r_0 = \frac{e^2}{m_e c^2} = 2.82\times 10^{-13}\mbox{ cm}.
\end{equation}

\begin{figure}[ht]
\begin{center}
\includegraphics[width=0.5\textwidth]{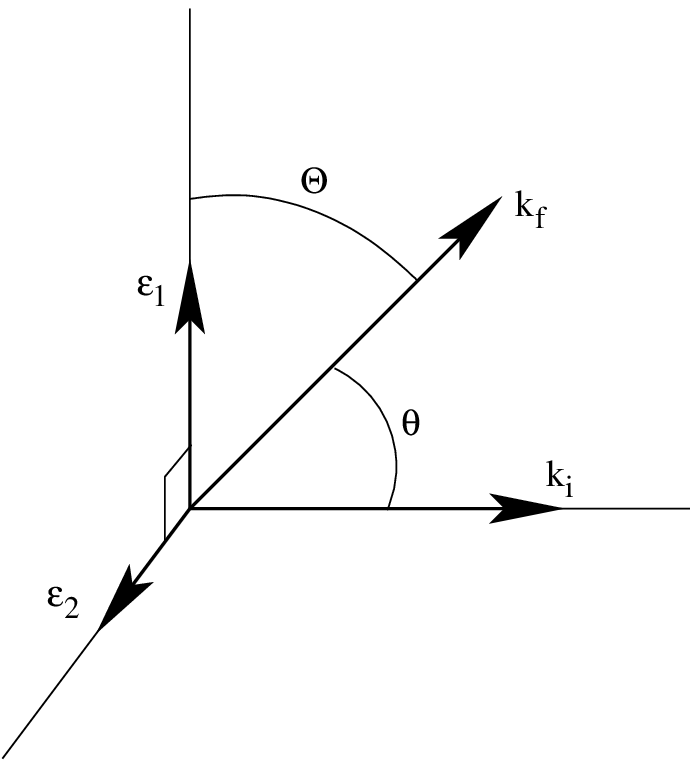}
\caption[Dipole scattering of unpolarized radiation]{\label{schem_unpol}
Scattering geometry for unpolarized radiation incident on an electron
at rest. The incoming radiation is made up of a combination of two
different linear polarizations $\boldsymbol{\epsilon}_1$ and
$\boldsymbol{\epsilon}_2$. The scattered radiation $\mathbf{k}_f$ is
in the plane of $\mathbf{k}_i$ and $\boldsymbol{\epsilon}_1$, with
$\Theta+\theta=90^\circ$.}
\end{center}
\end{figure}

While equation (\ref{cross_pol}) gives the cross section for radiation
polarized in the plane of the page in Figure \ref{schem_dipole}, we
can also use it to calculate
the average cross section for unpolarized radiation. Since unpolarized
radiation is really just the linear combination of two oppositely
polarized waves, we can calculate the unpolarized cross section
by averaging the cross sections for perpendicular polarization
vectors. The geometry for such a system is shown in Figure
\ref{schem_unpol}: an incident wave $\mathbf{k}_i$ is scattered into
$\mathbf{k}_f$, with final polarization in the same plane as
$\mathbf{k}_f$ and the initial polarization
$\boldsymbol{\epsilon}_i$. The initial polarization is either in the
same plane as the two wave vectors ($\boldsymbol{\epsilon}_1$) or
perpendicular to that plane ($\boldsymbol{\epsilon}_2$).

The probability of scattering into $\mathbf{k}_f$ from an initial
polarization of $\boldsymbol{\epsilon}_1$ is the same as in equation
(\ref{cross_pol}), since that is the same geometry as in Figure
\ref{schem_dipole}. If the initial polarization is
$\boldsymbol{\epsilon}_2$, then we can again use equation
(\ref{cross_pol}), evaluated at the perpendicular scattering angle
$\Theta=\pi/2$. The cross section for unpolarized radiation is then
given by the average of the two polarized cross sections:
\begin{eqnarray}\label{cross_unpol}
\left(\frac{d\sigma}{d\Omega}\right)_{\rm unpol} &=& \frac{1}{2}
\left[\left(\frac{d\sigma(\Theta)}{d\Omega}\right)_{\rm pol} +
\left(\frac{d\sigma(\pi/2)}{d\Omega}\right)_{\rm pol}\right]
\nonumber\\
&=& \frac{1}{2}r_0^2(1+\sin^2\Theta) \nonumber\\
&=& \frac{1}{2}r_0^2(1+\cos^2\theta)
\end{eqnarray}
For the rest of the results in this chapter, we will restrict
ourselves to this cross section for nonrelativistic, unpolarized
scattering. In future work, we hope to include a more formal treatment
of the covariant scattering of polarized light
\citep{ports04a,ports04b}. 

It is important to note that when we say ``nonrelativistic,'' this is
a reference to the \textit{photon} energy, not the electron energy. In
the electron rest frame, we require $h\nu \ll m_ec^2$ in order for the
above cross sections to be valid, in which case the scattering is nearly
elastic or \textit{coherent}. For higher energy photons, the
scattering involves quantum effects and requires the ``Klein-Nishina''
cross section [see e.g.\ \citet{heitl54}]. Since we are primarily
interested in the scattering of
photons from a relatively cool thermal accretion disk ($h\nu \sim 1-5$
keV), the classical treatment should suffice.

Even though we treat the scattering as coherent in the electron frame,
in the lab frame energy can be (and often is) transferred from the
electron to the photon. To see this boosting effect, consider a photon
with initial energy $\varepsilon_i$ scattering off an electron with velocity
$\beta$ in the $x$-direction in the ``lab frame'' $K$. In this frame,
the angle between the incoming photon and electron velocity is
$\theta$. In the electron rest frame $K'$, the photon is scattered
at an angle $\theta'$ with respect to the
$x'$-axis. The Doppler shift formula gives (Rybicki \& Lightman,
1979):
\begin{eqnarray}
\varepsilon_i' &=& \varepsilon_i \gamma(1-\beta\cos\theta) \nonumber\\
\varepsilon_f &=& \varepsilon_f' \gamma(1+\beta\cos\theta'),
\end{eqnarray}
where $\gamma = 1/\sqrt{1-\beta^2}$ and 
$\varepsilon_f$ is the post-scattering energy in the lab frame. In the
electron frame, we assume elastic scattering with
$\varepsilon_i'=\varepsilon_f'$, which should be the case for the
typical seed photons from a thermal emitter at $T_{\rm em} \sim 1$ keV.

Averaging over all angles $\theta$
(isotropic) and $\theta'$ (weighted by eqn.\ \ref{cross_unpol}), we
find that the typical scattering event boosts the photon energy by
\begin{equation}\label{ef_ei}
\frac{\varepsilon_f}{\varepsilon_i} \approx \gamma^2.
\end{equation}
For scattering electrons with a thermal velocity distribution around
$T_e$, this factor can be
written
\begin{equation}\label{gamma2_thermal}
\gamma^2 \approx \left(1-\frac{2kT_e}{m_ec^2}\right)^{-1} \approx 1.6
\hspace{1cm}\mbox{for $T_e \approx 100$ keV}.
\end{equation}
To be more precise, we consider a Maxwell-Boltzmann distribution
function in electron momentum $p=\gamma m v$: 
\begin{equation}\label{maxwellian}
f(p)d^3\mathbf{p} \propto 4\pi p^2 \exp\left(-\frac{\sqrt{p^2c^2 +
m_e^2c^4}}{kT_e}\right).
\end{equation} 
For highly relativistic velocities ($\gamma \gg 1$), the distribution
is replaced by the \textit{Juttner distribution}, which involves more
complicated terms including a modified Bessel function of the second
kind \citep{melro99}. For the moderate velocities corresponding to
$T_e \lesssim 100$ keV, equation (\ref{maxwellian}) is accurate enough
for our purposes. 

\subsection{General Relativistic Implementation}
As was emphasized in Chapter 2, many of the classical results of
radiation transport physics can be easily applied to general
relativistic fluids when considered in the appropriate reference
frame. Here too, we can split the problem into two basic pieces: the
ray-tracing of photons in curved space, an inherently general
relativistic process, and the scattering of these low-energy photons
off of hot electrons, a purely classical process in the electron
frame. However, unlike the approach taken in Chapter 2, where the
photons were traced backwards in time from a distant observer to the
emitting region, here it is conceptually easier to trace the photons
\textit{forward} in time from the emitter to the observer, then use
Monte Carlo methods to determine the distribution of scattered
photons. 

This Monte Carlo
approach has its trade-offs: a vastly larger number of photons must be
traced in order to ``observe'' enough at the detector to produce a
reasonable image or spectrum. But all those photons that do not reach
the detector need not be wasted. Consider an enormous spherical shell
detector at large $r$ covering the entire sky surrounding the black
hole. For hot spots on circular obits, any
emitted photon can be mapped in azimuth from its intersection with the
theoretical detector to the ``real'' observer by rotating the photon
back in time. Let the ray-traced photon emitted at spacetime
coordinates $(t_0,\phi_0)$ hit the sphere at $(t_1,\phi_1)$, while the
observer is located at $\phi_{\rm obs}$. If the
emitter has an orbital period of $T_{\rm orb}$ as measured in
coordinate time, then the same identical photon could have come
from the emitter at $(t_0',\phi_0')$, with
\begin{eqnarray}
t_0 - t_0' &=& t_1 - t_{\rm obs}, \nonumber\\
\phi_0 - \phi_0' &=& \phi_1 - \phi_{\rm obs}
\end{eqnarray}
and
\begin{equation}\label{transform_phi}
t_{\rm obs} = t_1 - \frac{\phi_1-\phi_{\rm obs}}{\Omega_\phi} =
t_1 -\frac{T_{\rm orb}}{2\pi}(\phi_1-\phi_{\rm obs}).
\end{equation} 
In this way, every photon is essentially detected at the same location
in azimuth with the appropriate time delay, which in turn can be used
for calculating the light curve and dynamic spectrum of the
emitter. Since the orbital geometry lacks rotational symmetry in the
$\mathbf{e}_\theta$ direction, we cannot produce a similar mapping in
latitude. But all is not lost---by dividing the detector into equally
spaced slices in $\cos\theta$, a single Monte Carlo computation
produces simulated data at all viewer inclinations simultaneously. So
in the end, every ray-traced photon contributes equally to the light
curve and spectrum (except of course those photons that get captured
by the black hole, but even they contribute indirectly by their
absence). 

Unlike the approach in Chapter 2, where the Lorentz invariant
$I_\nu/\nu^3$ was used to appropriately handle relativistic beaming,
when we begin in the emitter's rest frame, the beaming is introduced
automatically by the reference frame transformations. The photons
here should be thought of primarily as particles, and not the continuous
beams we envisioned for the radiative transfer equation in Section
\ref{radiative_transfer}. However, because of the coherent scattering
assumption, the ray-traced path of each photon is energy-independent,
so the photon's final observed energy can be thought of as a fiducial
redshift $E_{\rm obs}/E_{\rm em}$ that can be convolved with the
spectrum in the local emitter frame to produce the total spectrum seen
by the observer [see caveats below following equation
(\ref{energy_compton})].

To get the initial coordinate momentum of each photon, we construct a
tetrad centered on the emitter's rest frame, denoted by tilde
indices $\tilde{\mu}$. Then $\mathbf{e}_{\tilde{t}}$ is parallel to
the emitter's 4-velocity $p^\mu({\rm em})$, $\mathbf{e}_{\tilde{r}}$ and
$\mathbf{e}_{\tilde{\theta}}$ are in the coordinate directions
$\mathbf{e}_r$ and $\mathbf{e}_\theta$, and
$\mathbf{e}_{\tilde{\phi}}$ is given by orthogonality. Recall from
Section \ref{transfer_function} the
expressions for the energy and angular momentum of a particle on a
stable circular orbit around a Kerr black hole:
\begin{equation}
-p_t = \frac{r^2-2Mr\pm a\sqrt{Mr}}{r(r^2-3Mr\pm 2a\sqrt{Mr})^{1/2}}
\end{equation}
and
\begin{equation}
p_\phi = \pm\frac{\sqrt{Mr}(r^2\mp 2a\sqrt{Mr}+a^2)} {r(r^2-3Mr\pm
 2a\sqrt{Mr})^{1/2}}.
\end{equation}
From these we construct the 4-velocity via the inverse metric
$p^\mu({\rm em}) = g^{\mu \nu} p_\nu({\rm em})$, which gives
$\mathbf{e}_{\tilde{t}}$. In Boyer-Lindquist coordinates,
$\mathbf{e}_{\tilde{r}}$ and $\mathbf{e}_{\tilde{\theta}}$ are
trivially normalized as in the ZAMO basis [eqns.\ (\ref{ZAMO_tetradb})
and (\ref{ZAMO_tetradc})]. Writing
\begin{equation}
\mathbf{e}_{\tilde{\phi}} = A\mathbf{e}_t + B\mathbf{e}_\phi,
\end{equation}
the orthonormality conditions are
\begin{subequations}
\begin{eqnarray}
\mathbf{e}_{\tilde{t}}\cdot \mathbf{e}_{\tilde{t}} &=& 
(p^t)^2g_{tt}+2p^t p^\phi g_{t\phi} + (p^\phi)^2 g_{\phi \phi}=-1, \\
\mathbf{e}_{\tilde{t}}\cdot \mathbf{e}_{\tilde{\phi}} &=& 
Ap^tg_{tt}+(Ap^\phi+Bp^t)g_{t\phi} + Bp^\phi g_{\phi \phi}=0, \\
\mathbf{e}_{\tilde{\phi}}\cdot \mathbf{e}_{\tilde{\phi}} &=& 
A^2g_{tt}+2ABg_{t\phi} + B^2 g_{\phi \phi}=1.
\end{eqnarray}
\end{subequations}
\citet{novik73} give analytic expressions for the emitter's tetrad
basis in terms of the functions defined in Section \ref{NT_disks}: 
\begin{subequations}
\begin{eqnarray}
\mathbf{e}_{\tilde{t}} &=& \frac{\mathcal{B}}{\mathcal{C}^{1/2}}
\mathbf{e}_t + \frac{M^{1/2}}{r^{3/2}\mathcal{C}^{1/2}} \mathbf{e}_\phi, \\ 
\mathbf{e}_{\tilde{r}} &=&
\sqrt{\frac{\Delta}{\rho^2}}\, \mathbf{e}_r, \\
\mathbf{e}_{\tilde{\theta}} &=&
\sqrt{\frac{1}{\rho^2}}\, \mathbf{e}_\theta, \\
\mathbf{e}_{\tilde{\phi}} &=&
\frac{\mathcal{F}M^{1/2}}{(r\mathcal{C}\mathcal{D})^{1/2}} \mathbf{e}_t +
\frac{\mathcal{BD}+r^{1/2}\mathcal{AF}M^{1/2}}{r\mathcal{ADC}^{1/2}}
\mathbf{e}_\phi.
\end{eqnarray}
\end{subequations}

In this basis, the initial photon direction is picked randomly from an
isotropic distribution, uniform in spherical coordinates
$\cos\tilde{\theta}=[-1,1]$ and $\tilde{\phi}=[0,2\pi)$. All photons
are given the same initial energy in 
the emitter frame $p_{\tilde{t}} = -E_0$, which is used as a reference
energy for calculating the final redshift with respect to a stationary
observer at infinity. From the basis vectors
$\mathbf{e}_{\tilde{\mu}}$ we construct a transformation matrix
$E_{\tilde{\mu}}^\mu$ as in Section \ref{tetrads} to get the initial
conditions for ray-tracing in the coordinate basis
$p^\mu=E_{\tilde{\mu}}^\mu p^{\tilde{\mu}}$.

\begin{figure}
\begin{center}
\scalebox{0.6}{\includegraphics*[74,330][540,750]{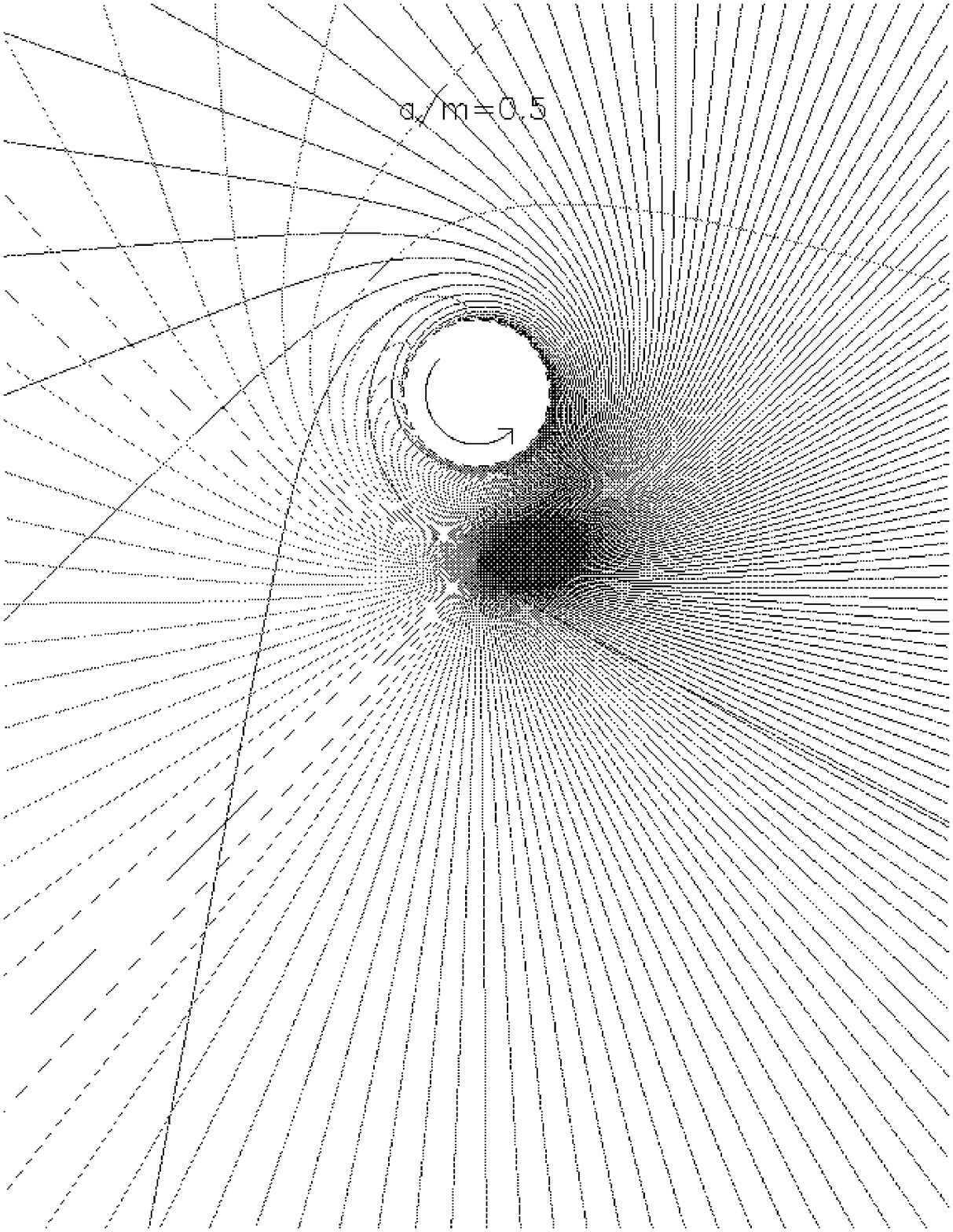}}\\
%\hspace{0.2cm}
\scalebox{0.6}{\includegraphics*[74,330][540,750]{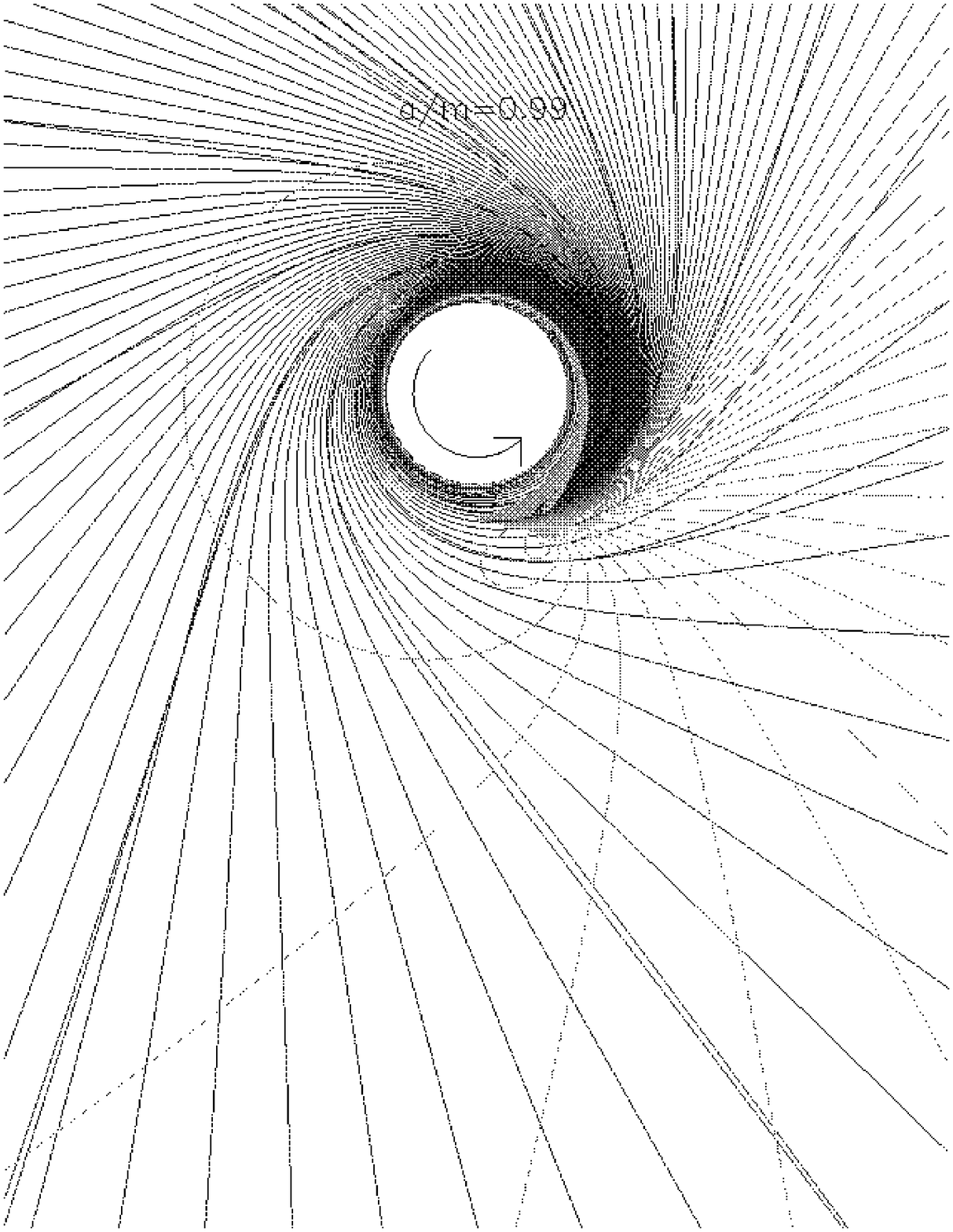}}
\caption[Photons traced from isotropic emitter at
ISCO]{\label{emitter_rays} Planar photons emitted isotropically in the
rest frame of a massive particle on a circular orbit at the ISCO. The
black hole has spin $a/M=0.5$ (top) and $a/M=0.99$ (bottom). The photon
paths are colored according to their red- or blue-shift in energy with
respect to $E_0$ measured in the emitter's frame. For the black hole
with $a/M=0.99$, the ISCO is located inside the ergosphere, so all
photons are forced to move forward in $\phi$, and some are even
created with negative energies ($p_t>0$).}
\end{center}
\end{figure}

Given $p_\mu = g_{\mu\nu}p^\nu$, the photon's geodesic trajectory is
simply integrated using the Hamiltonian formulation according to
equations (\ref{hameq_1}) and (\ref{hameq_2}).
Figure \ref{emitter_rays} shows an ``overhead view'' of photon
trajectories in the plane of the disk, emitted isotropically by a
massive test particle on a circular orbit at the ISCO. Figure
\ref{emitter_rays}a shows a black hole with $a/M=0.5$ and Figure
\ref{emitter_rays}b has $a/M=0.99$. The photons are
colored according to their energy-at-infinity $E_\infty=-p_t$, either
blue- or redshifted with respect to their energy in the emitter frame
$E_0$.  For the black hole
with $a/M=0.99$, the ISCO is located inside the ergosphere, so all
photons are forced to move forward in $\phi$, and some are even
created with negative energies ($p_t>0$). As mentioned above, the
relativistic beaming is done
automatically by the Lorentz boost from the emitter to the coordinate
(or ZAMO) frame, so the blue photons are clearly bunched more tightly
together, as required by the invariance of $I_\nu/\nu^3$. 

As in Chapter 2, the photon's position and momentum are tabulated at
each step along its
path. However, now we have to check at each interval to see if the
photon scatters off an electron. Conveniently, the Runge-Kutta
algorithm takes shorter steps as smaller $r$, where the electron
density tends to be highest, so we can reliably use the differential
formula for the optical depth to electron scattering:
\begin{equation}\label{dtau_es}
d\tau_{\rm es} = \kappa_{\rm es}\rho ds.
\end{equation}
The density $\rho$ is defined in the ZAMO frame and the opacity
$\kappa_{\rm es}$ is given by the classical cross section
derived above in equation (\ref{cross_unpol})
\begin{equation} 
\kappa_{\rm es}= \frac{\sigma_T}{m_p} =
\frac{8\pi}{3}\frac{r_0^2}{m_p} = 0.4 \mbox{ cm}^2/\rm{g}.
\end{equation}
The proper distance $ds$ in equation (\ref{dtau_es}) is calculated
from the path segment $dx^{\hat{\mu}}_i$ as in equation
(\ref{ds2_i}). For relatively small steps, the probability of
scattering after each step is given by $d\tau_{\rm es} \lesssim 0.1$. 

\begin{figure}[ht]
\begin{center}
\includegraphics[width=1.0\textwidth]{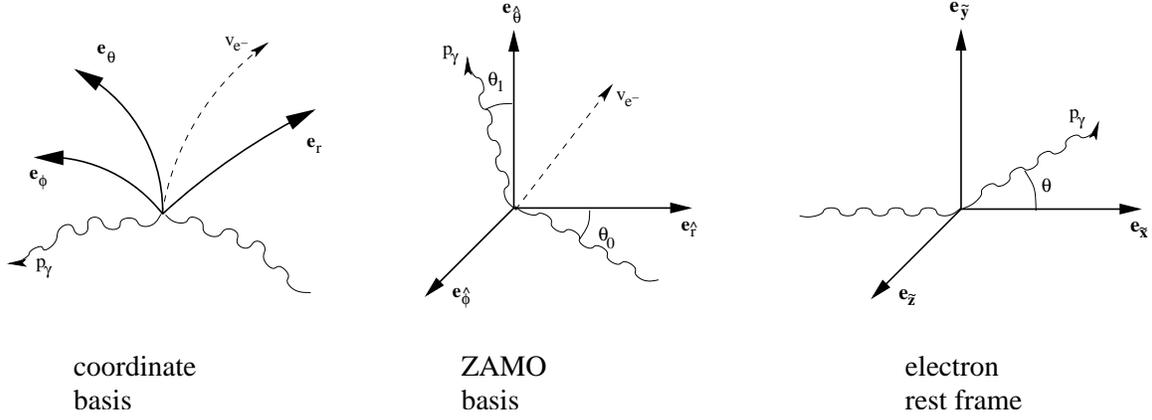}
\caption[Coordinate transformations for electron
scattering]{\label{coordinates} Schematic picture of coordinate
transformations from the coordinate basis $\mathbf{e}_\mu$ in which
the geodesic trajectories are integrated, to the ZAMO basis
$\mathbf{e}_{\hat{\mu}}$ in which the electron density and temperature
are defined, to the electron rest frame basis $\mathbf{e}_{\tilde{\mu}}$
in which the electron scattering angle $\theta$ is given simply by the
Thomson cross section for unpolarized radiation.}
\end{center}
\end{figure}

If the photon does in fact experience a scattering event, we first
transform into the ZAMO basis, where the electron temperature is
defined. Given the electron temperature, we assume an isotropic
distribution of velocities as defined in equation (\ref{maxwellian}),
and pick an electron 4-velocity with random direction in that
basis. Next we must transform to the electron rest frame, in which the
photon scatters according to the Thomson cross section from equation
(\ref{cross_unpol}), reducing a difficult problem in curved
spacetime to a simple classical problem with a single variable---the
scattering angle $\theta$. This set of transformations from
coordinate basis to ZAMO basis to electron rest frame is shown
schematically in Figure \ref{coordinates} (again the ZAMO basis is
denoted by $\hat{\mu}$ subscripts, and the electron frame by
$\tilde{\mu}$, not to be confused with the emitter frame defined
earlier).
  
The transformation from the ZAMO basis to the electron frame is
defined by a Lorentz boost in the direction of the electron 4-velocity
$u^{\hat{\mu}} \to \mathbf{e}_{\tilde{x}}$. Since the scattering
probability is symmetric around 
this axis, the rotational degree of freedom that fixes the other
spatial axes $\mathbf{e}_{\tilde{y}}$ and $\mathbf{e}_{\tilde{z}}$ is
completely arbitrary. One convenient form of the generalized Lorentz
boost is given by \citet{mtw73}:
\begin{eqnarray}
u^\mu &=& [\gamma,\beta n^j] \hspace{1cm} (|n| = 1), \nonumber\\
\Lambda^{t'}_t &=& \gamma, \nonumber\\
\Lambda^{t'}_j &=& \Lambda^{j'}_t = -\beta \gamma n^j, \nonumber\\
\Lambda^{j'}_k &=& \Lambda^{k'}_j = (\gamma-1)n^j n^k+\delta^{jk}.
\end{eqnarray}
The photon momentum in the electron frame is thus given by
$p^{\tilde{\mu}}=\Lambda^{\tilde{\mu}}_{\hat{\mu}}p^{\hat{\mu}}$. 

All that is left to
do is calculate the scattering angle $\theta$. Most portable random
number generators produce a random variable $X$ uniformly distributed
in the range $[0,1]$, and from this we must produce a random variable
$\theta$ with distribution according to equation
(\ref{cross_unpol}). The cross section can be re-written in terms of
the normalized probability distribution function 
\begin{subequations}
\begin{equation}\label{f_theta_scat}
f(\theta)d\theta = \frac{3}{2}\sin\theta(1+\cos^2\theta) d\theta
\end{equation}
or defining $z\equiv \cos\theta$,
\begin{equation}\label{f_x_scat}
f(z)dz = \frac{3}{8}(1+z^2)dz.
\end{equation}
\end{subequations}
Let $g(z)$ be the cumulative distribution function
\begin{equation}
g(z) = \int_{-1}^z f(z')dz'
\end{equation}
so that $g(-1)=0$ and $g(1)=1$. Then given a uniformly distributed
$X$, we can solve for $z=g^{-1}(X)$. Unfortunately, this involves
finding the roots to the cubic equation
\begin{equation}\label{cubic}
z^3+3z+4-8X=0.
\end{equation}
While not trivial, the solution to (\ref{cubic}) is at least unique
[$g(z)$ has no turning points] and can be written in closed
form \citep{zwill96}. After picking a random 3-vector
$p_\perp^{\tilde{j}}$ perpendicular to $p^{\tilde{j}}$ (with the same
magnitude $|p_\perp^{\tilde{j}}|=|p^{\tilde{j}}|=-p^{\tilde{0}}$), we
construct the 4-vector of the scattered photon: 
\begin{equation}
p_f^{\tilde{\mu}} = [p^{\tilde{0}},\cos\theta p^{\tilde{j}}+\sin\theta
p_\perp^{\tilde{j}}].
\end{equation}
$p_f^{\tilde{\mu}}$ is then transformed back to the ZAMO basis with a
boost by $-u^{\hat{\mu}}$, then to the coordinate basis, and then we
continue with integrating the new geodesic trajectory until the next
scattering event or detection by a distant observer.

\section{Effect on Spectra}\label{effect_spectra}

As we showed at the end of the Section \ref{classical_scattering}, the
net effect of this
whole procedure is generally a transfer of energy from the electron to
the photon. One way to quantify this energy transfer is through the
Compton $y$ parameter, defined as the average fractional energy change
per scattering, times the number of scatterings through a finite
medium. For nonrelativistic electrons, \citet{rybic79} show that
the average energy transfer per scattering event is 
\begin{equation}\label{ef_ei2}
\frac{\varepsilon_f-\varepsilon_i}{\varepsilon_i} =
\frac{4kT_e}{m_ec^2}. 
\end{equation}
This is actually slightly higher than the estimate we gave in equation
(\ref{gamma2_thermal}). The reason for this is that the average energy
of a photon in thermal equilibrium with an electron gas is not $kT_e$,
but rather $3kT_e$.

The mean number of scatterings for an optically thin medium is simply
$\tau_{\rm es}$, the total optical depth through the medium. For
optically thick systems, the photons must take a random walk to
escape, so the number of scatterings becomes $\tau^2_{\rm es}$. Thus the
Compton $y$ parameter for a finite medium of nonrelativistic electrons
is 
\begin{equation}\label{compton_y}
y = \frac{4kT_e}{m_ec^2} \mbox{Max}(\tau_{\rm es},\tau_{\rm es}^2).
\end{equation}
For a low-energy soft photon source with multiple scattering events,
the final spectrum
due to inverse-Compton scattering can be calculated using the
\textit{Kompaneets equation}, which is a form of the Fokker-Plank
diffusion equation \citep{kompa57}. For $h\nu \lesssim kT_e$, the
resulting spectrum takes the power-law form
\begin{equation}
I_\nu \sim \nu^{-\alpha},
\end{equation}
with
\begin{equation}\label{powerlaw_alpha}
\alpha = \frac{3}{2}+\sqrt{\frac{9}{4}+\frac{4}{y}}.
\end{equation}
At energies above $kT_e$, the electrons no longer efficiently transfer
energy to the photons, so the spectrum shows a cutoff for $h\nu \gtrsim
kT_e$:
\begin{equation}
I_\nu \sim \nu^3 \exp(-h\nu/kT_e).
\end{equation}

With the assumption of purely elastic scattering, we cannot
actually reproduce this cutoff effect; all photons are scattered
equally, and thus the ratio $\varepsilon_f/\varepsilon_i$ is
independent of energy. Thus equation (\ref{ef_ei2}) would predict
infinite energy boosts until $h\nu \gg m_ec^2$. In reality, higher
energy photons tend to lose
energy in scattering, due to the recoil of the electron. This effect
is relatively easy to calculate from conservation of energy and
momentum in the electron rest frame:
\begin{equation}\label{energy_compton}
\varepsilon_f = \frac{\varepsilon_i}{1+\frac{\varepsilon_i}{m_ec^2}
(1-\cos\theta)}
\end{equation}
To accurately include this effect, we would have to keep track of the
real ``physical'' energy of each photon, instead of the fiducial
redshift method that we currently use to reconstruct the total
spectrum afterwards. Ultimately, this is just a matter of
computational intensity and no real conceptual difficulty. To
first-order, we can treat the thermal photon source as a
monochromatic emitter at $E_0 = 3kT_{\rm em}$, which should give a
reasonable approximation to the true solution.

Before we can actually produce such a spectrum, we must first define
the electron temperature and density profile through which the photons
will scatter. Like relativistic jets, there is still no real consensus
in the literature as to what exactly produces the electron corona
surrounding the black hole and accretion disk. By measuring the
power-law part of the continuum spectrum [see e.g.\
\citet{sunya79,makis86}], it seems clear that the
corona is quite hot ($T_e \gtrsim 50$ keV) and diffuse ($\tau_{\rm
es} \lesssim 5$). Many of the early works on this subject treated the
corona as an isothermal and uniform density sphere with a sharp cutoff
at some radius $R_c$, using the Kompaneets equation to propagate the
seed photons through the corona
\citep{shapi76,sunya80,titar94}. \citet{nobil00} basically follow this
approach, but also allow for two layers in the corona at different
temperatures: an inner hot sphere surrounded by a ``warm'' layer at
slightly lower temperature. More recent Monte Carlo methods
\citep{stern95a,pouta96,yao03,wang04} allow for arbitrary
distributions, but in practice usually only consider similar
isothermal, uniform density profiles.

We have investigated a number of different models, including the
uniform distributions mentioned above as well as isothermal and
polytropic gases in hydrostatic equilibrium (constructed by
integrating the Oppenheimer-Volkoff equation). Both cases require some
arbitrary cut-off radius for the corona in order to have a low enough
optical depth to agree with the observations. Future work will include
a more comprehensive exploration of hydrostatic corona models with
various polytropic equations of state. Perhaps a more physical
option is that of the Advection Dominated Accretion Flow (ADAF) model
proposed by \citet{naray94}, where the self-similar density and
temperature profiles scale as
\begin{subequations}
\begin{eqnarray}
\rho &\propto& r^{-3/2}, \\
T &\propto& r^{-1}
\end{eqnarray}
\end{subequations}
outside of the ISCO. We have ignored the bulk velocity of the inwardly
flowing gas, which will typically have $v_{\rm bulk} \ll v_{\rm
therm}$ in the ADAF model. For larger inflow velocities of $v_{\rm
bulk} \gtrsim 0.1c$, the effect of ``bulk Comptonization'' has been
studied in detail by \citet{psalt01b}.

\begin{figure}[ht]
\begin{center}
\scalebox{0.65}{\includegraphics*[34,320][580,760]{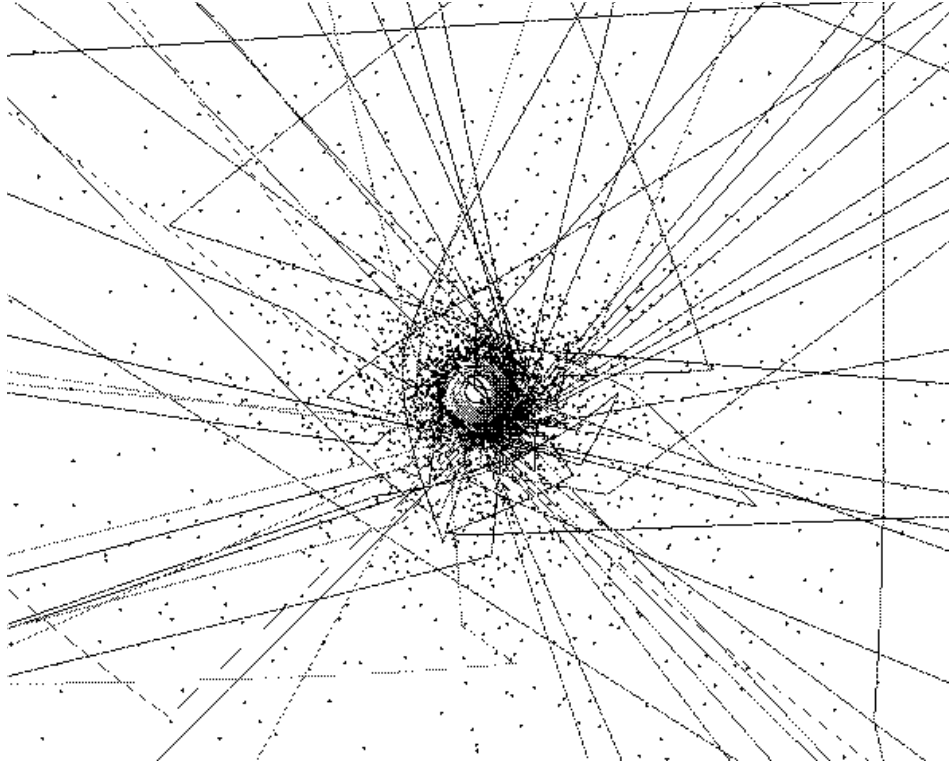}}
\caption[Scattering distribution for an ADAF corona]{\label{ADAF_scatter}
Electron distribution (black dots) for ADAF corona model with
$\tau_{\rm es}\approx 1$. As in Figure
\ref{emitter_rays}a, the photons are emitted isotropically by a
massive particle in a circular orbit at the ISCO of a black hole with
$a/M=0.5$. The photons are colored according to their blue/redshift
with respect to that of the emitter's rest frame. Many of the photons
clearly change color during scattering events, as they are boosted to
higher energies by the hot electrons.}
\end{center}
\end{figure}

For such a profile, most of the scattering events happen at small $r$
and relatively high $T$. Figure \ref{ADAF_scatter} shows roughly what
the electron distribution would look like and where the scattering
takes place. As in Figure \ref{emitter_rays}, the photons are
color-coded according to their blue/redshifted energy
$E_\infty=-p_t$. Upon close inspection, it is clear that this energy
changes slightly during each scattering event via the inverse-Compton
process. 

To create a simulated spectrum for a thermal hot spot or disk, we
follow the steps described above,
starting with isotropic, monochromatic emission in the emitter's rest
frame. To approximate the high-energy cutoff at $h\nu \approx kT_e$,
the initial photon energy is set as $E_0=3kT_{\rm em}$ and equation
(\ref{energy_compton}) is used to limit the runaway energy boosting
from the hot electrons. For decent
spectral and timing resolution, we typically ray-trace $10^7-10^8$
photons, which are either captured by the horizon or detected by a
distant observer with a time and energy label, much like a real
astronomical instrument. For example, if there was no scattering, the
time-averaged ``numerical'' spectrum could be described by the
relativistic transfer function discussed in Section
\ref{transfer_function}, defined over an infinitesimal band in radius
$R_{\rm in} \approx R_{\rm out}=r_{\rm em}$. The
inverse-Compton processes in the corona serve to further broaden this
transfer function, as shown by the curves in \citet{sunya80} and
\citet{titar94}. This transfer function is then normalized to
the rest energy $E_0$ and convolved with the actual emission spectrum
(e.g.\ a thermal blackbody at $kT_{\rm em}$) to give the simulated
observed spectrum. 

\begin{figure}[ht]
\begin{center}
\includegraphics[width=0.8\textwidth]{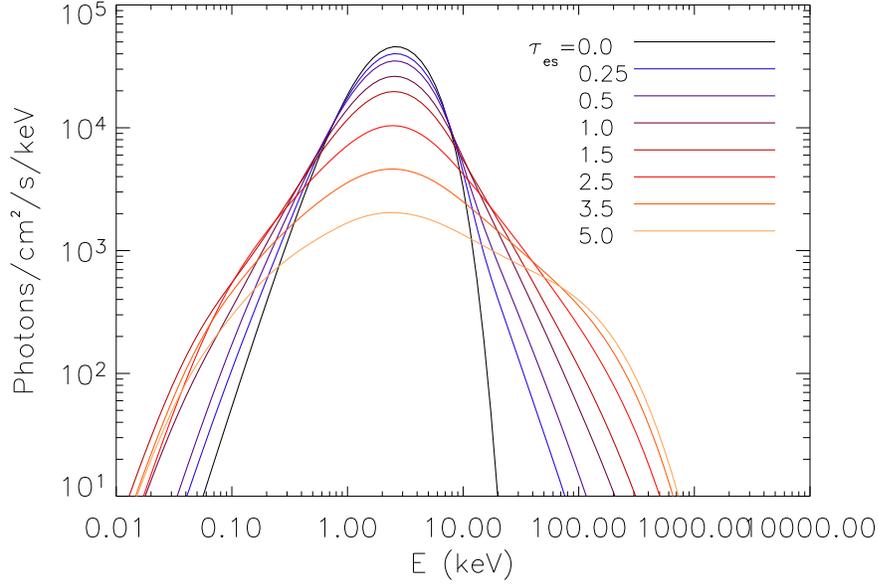}
\caption[Spectra of a thermal source scattering through a hot
corona]{\label{powerlaw_cutoff} Simulated observed spectra of a
thermal hot spot emitter with $T_{\rm em}=1$ keV, on a circular orbit
at the ISCO as in Figure \ref{ADAF_scatter}. The thermal spectrum is
modified by relativistic effects and Compton scattering off a hot
corona with $T_e = (r_{\rm ISCO}/r)100$ keV. The resulting spectrum is
composed of a power-law component with a high-energy cut-off at $h\nu
\approx kT_e$. These two features can be used to infer the corona
temperature and optical depth. Note that the magnitude scale of the
$y$-axis is arbitrary and in general would depend on the luminosity
and distance to the source.}
\end{center}
\end{figure}

Figure \ref{powerlaw_cutoff} shows a set of these simulated spectra
from a hot spot emitter around a black hole with $a/M=0.5$, as in
Figure \ref{ADAF_scatter}. The emission spectrum is thermal in the hot
spot rest frame with $T_{\rm em}=1$ keV. The coronal ADAF model has
$T_e = (r_{\rm ISCO}/r)100$ keV, and electron density $n_e \sim
r^{-3/2}$ for a variety of optical depths $\tau_{\rm es}$. The spectra
are plotted in units of [\#Photons/s/cm$^2$/keV], as is the convention
by many observers, but the actual magnitude of the $y$-axis is
arbitrary, and would normally depend on the distance to the
source. With these units, the power-law section of the
spectrum scales as 
\begin{equation}
N_\nu \sim I_\nu/\nu \sim \nu^{-1-\alpha}
\end{equation}
and the high energy cut-off as
\begin{equation}
N_\nu \sim \nu^2 \exp(-h\nu/kt_e).
\end{equation}
From the slope of the power-law and the location of the cut-off, the
corona temperature and optical depth can be inferred from
observations \citep{pozdn77,sunya79,gilfa94,gierl97,zdzia00}. 

While the spectra in Figure \ref{powerlaw_cutoff} come from a single
hot spot at a single temperature, the total spectrum from a
steady-state disk could be calculated easily by superimposing the
results from many such calculations at different radii. The seed photon
spectrum at each radius would be determined by the results in Chapter
5 [e.g.\ equation (\ref{flux_mod2})]. Currently, many X-ray
observations of black hole binaries fit the
spectrum as a simple superposition of a thermal blackbody peak (either
at a single temperature, or the popular ``multi-color disk'') and a
separate power-law component [see e.g.\
\citet{gierl99,mccli04}]. Our full ray-tracing and Monte Carlo 
scattering approach, while somewhat more computationally intensive,
would give more accurate and physically motivated spectra with which to
compare observations. 
 
\section{Effect on Light Curves}\label{effect_lightcurves}
The spectra in Figure \ref{powerlaw_cutoff} were created by
integrating over the complete hot spot orbital period and over all
observer inclination angles. However, during the Monte Carlo
calculation, it is just as simple to bin all the photons according to
their final values of $\theta$, $t_{\rm obs}$, and energy $-p_t$. With
$10^8$ photons, we achieve decent resolution for $N_\theta = 20$,
$N_t = 100$, and $N_E = 200$. The latitude bins are evenly spaced in
$\cos\theta$ so that a comparable number of photons land in each
zone. The energy bins are spaced logarithmically to include the high
energy tail and also maintain high enough spectral resolution at lower
energies. As described above, the photons at any azimuthal position
can be mapped into the appropriate bin in $t_{\rm obs}$ by equation
(\ref{transform_phi}), assuming a hot spot on a circular periodic
orbit. 

An excellent way to see the effects of scattering on the hot spot
light curves is by plotting the same type of time-dependent
spectrograms we used in the original hot spot model in Chapter
3. However, the spectrograms of unscattered hot spots will appear slightly
different that those produced in Chapter 3, which traced the photons
backwards from the observer to an opaque disk (only passing through
each latitude zone once, thus not producing multiple images --- this
is a feature inherent to the code, not the physical model). The
number of rays
required to resolve the image was then proportional to the solid
angle subtended by the hot spot. When starting at the
emitter, the hot spot can be infinitesimally small and also the
photons can orbit the black hole multiple times, forming the multiple
images inherent in strong gravitational lensing. 

\begin{figure}
\begin{center}
\scalebox{0.45}{\includegraphics*[-24,330][415,720]{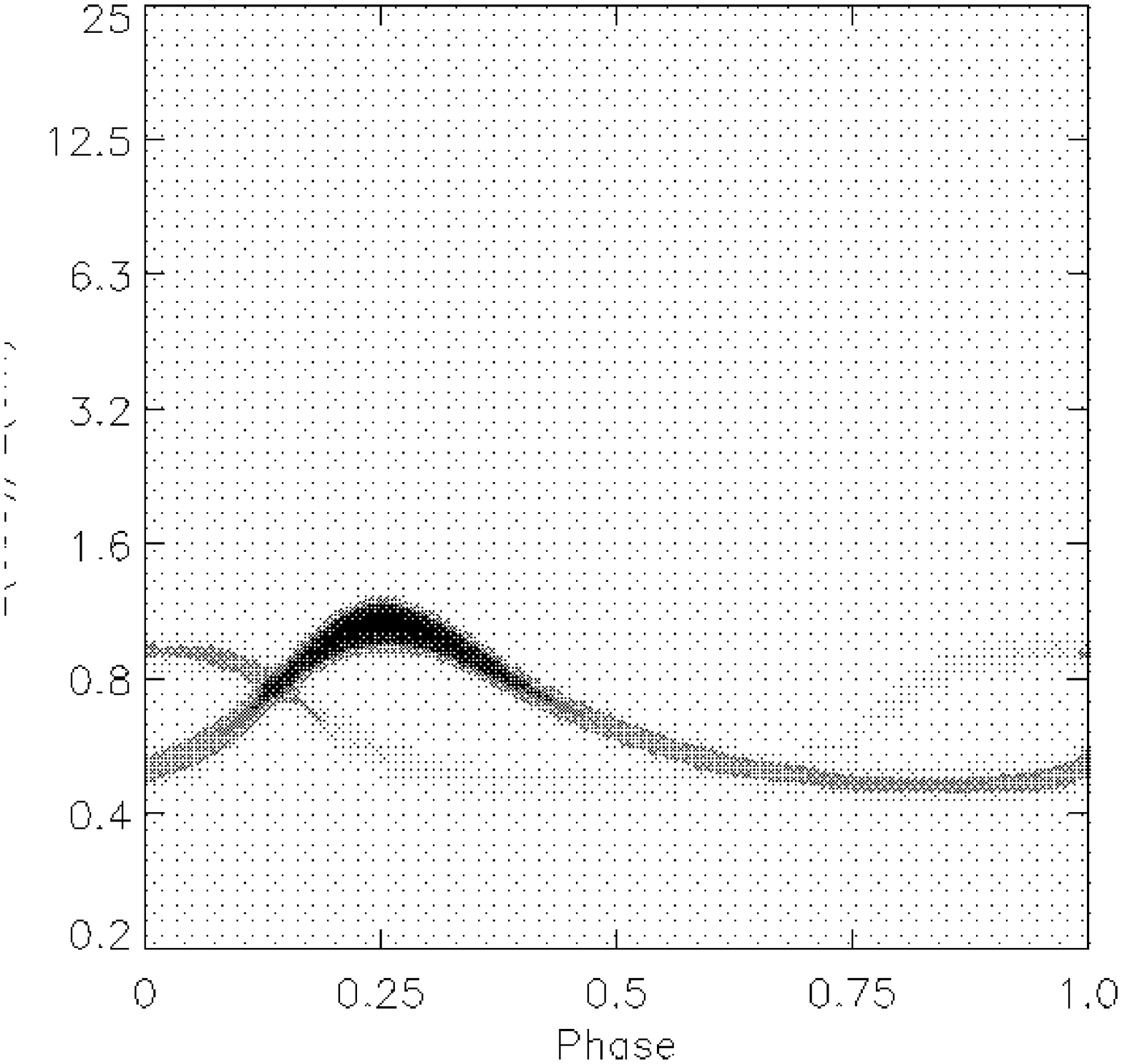}}
\scalebox{0.45}{\includegraphics*[50,330][415,720]{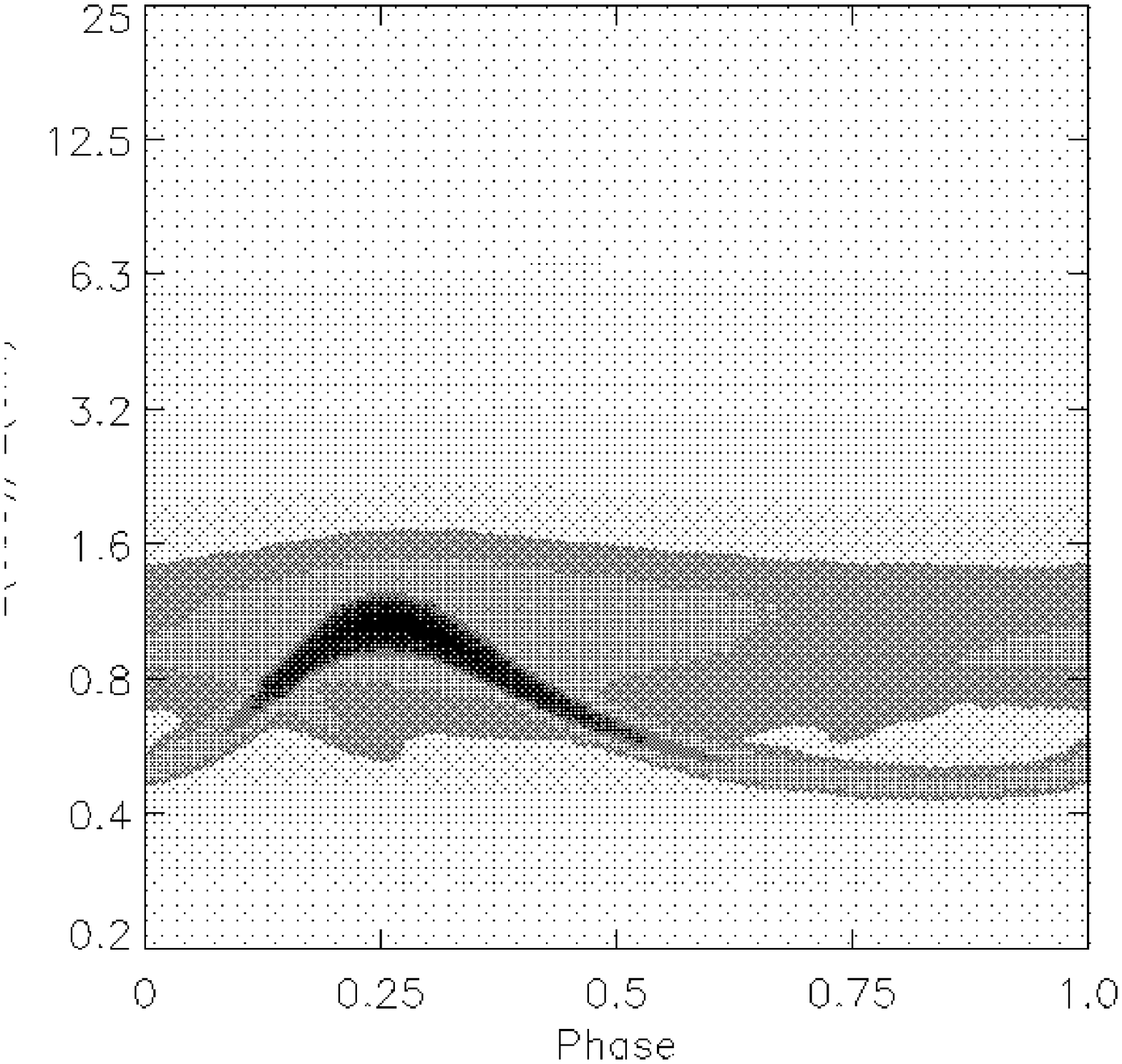}}\\
\scalebox{0.45}{\includegraphics*[-24,300][415,720]{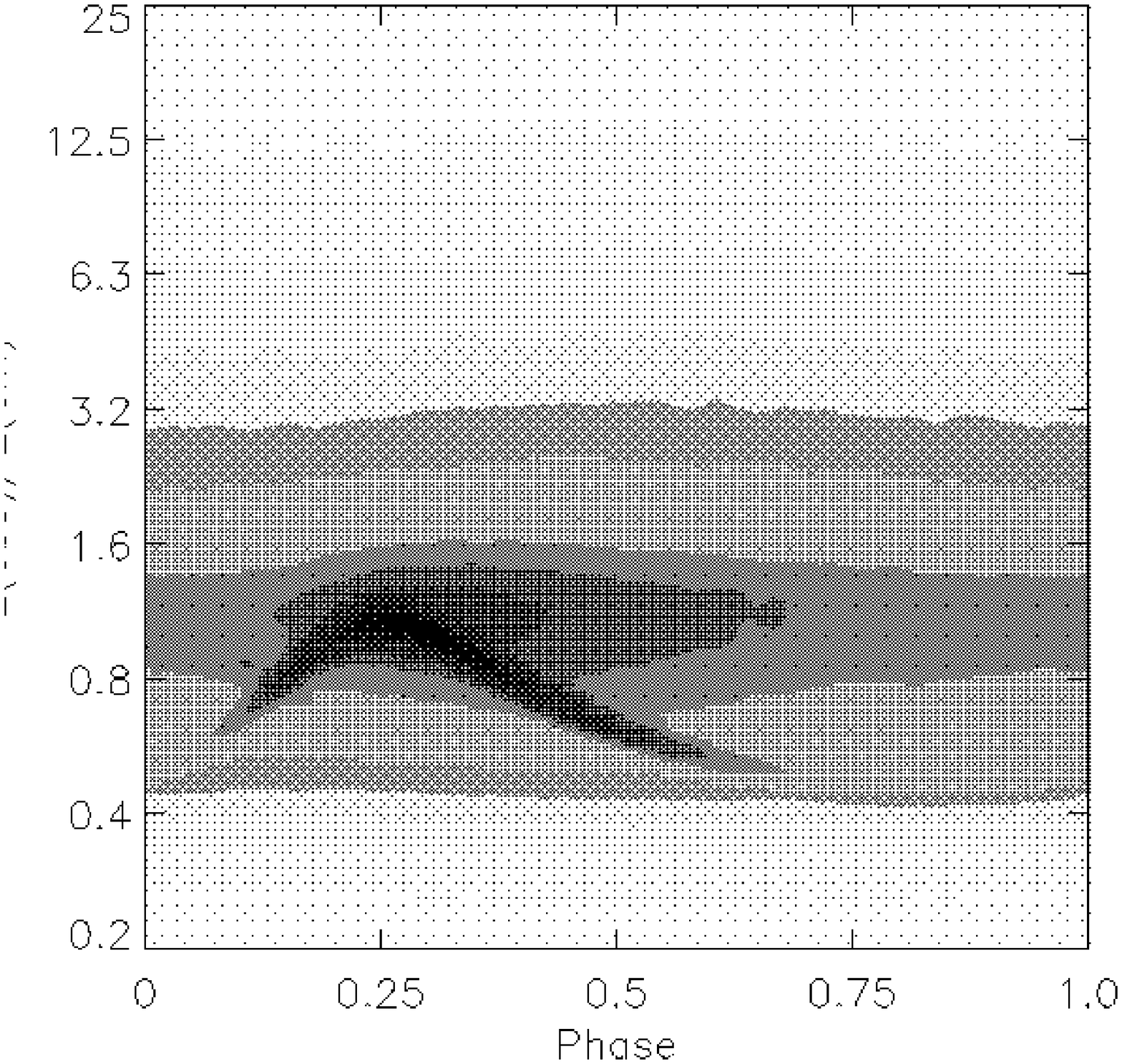}}
\scalebox{0.45}{\includegraphics*[50,300][415,720]{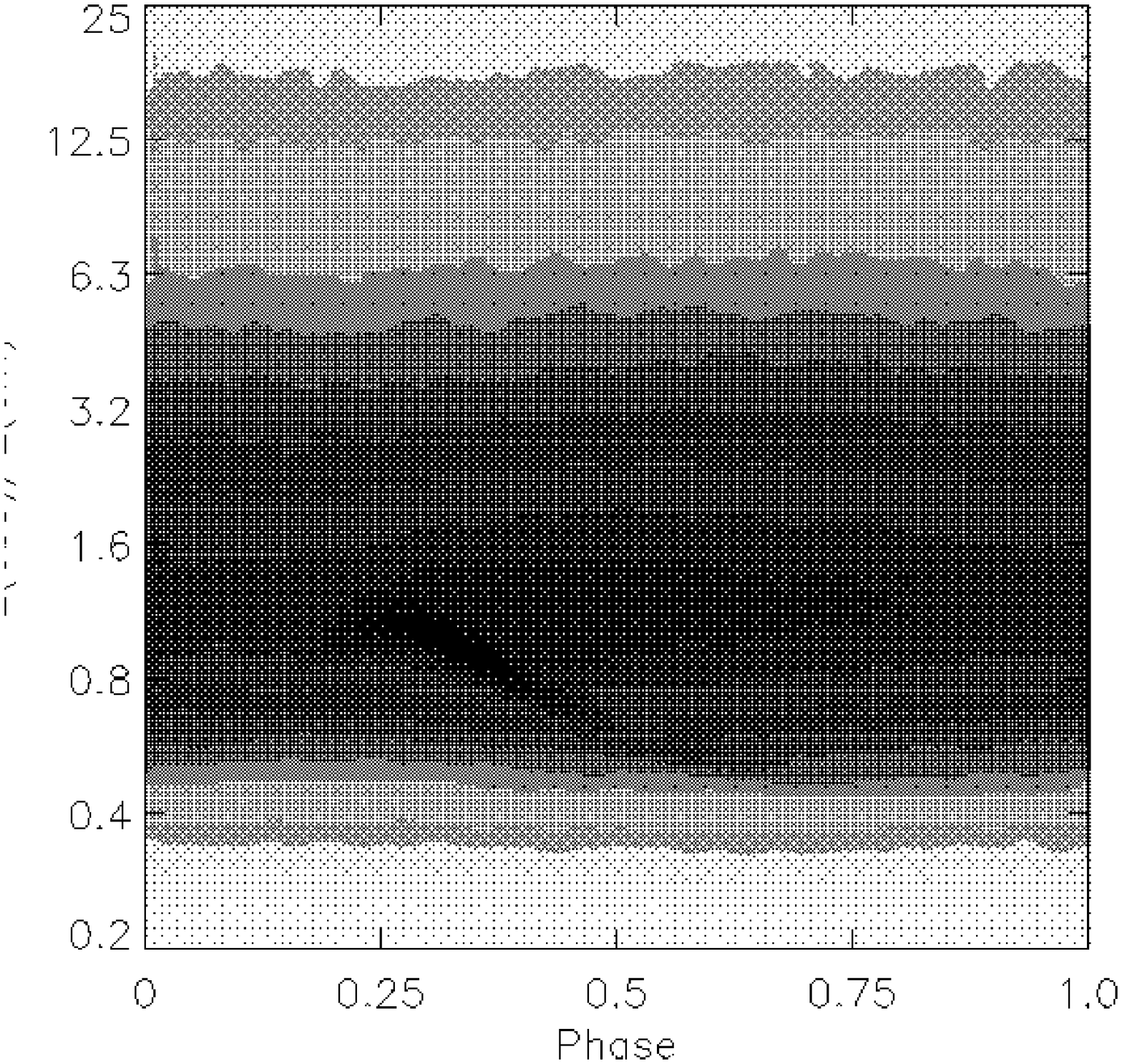}}
\caption[Spectrograms of hot spot emission including
scattering, $i=45^\circ$]{\label{scatter_spectrogram45} Time-dependent
spectra of a monochromatic, isotropic hot spot emitter on an ISCO
orbit with $a/M=0.5$ and inclination angle $i=45^\circ$. The four
panels show spectrograms for systems of
increasing optical depth $\tau_{\rm es}=[0,1,2,4]$. The scattering
clearly spreads out the light curve in phase and photon energy, with a
trend towards greater time delays for higher optical depth, and thus
more scattering events. The logarithmic color scale shows the number
of photons in each time/energy bin, normalized to the peak value for
each panel.}
\end{center}
\end{figure}

These effects are clearly visible in the first panel of Figure
\ref{scatter_spectrogram45}, which shows the time-dependent spectrum of a
monochromatic emitter, as viewed by an observer at $45^\circ$. The
logarithmic color scale shows [\#photons/s/cm$^2$/keV/period],
normalized to the peak intensity in each panel. 
At ``0'' phase, when the emitter is moving away from the observer,
the spectrum shows two distinct lines, one
blueshifted in the forward direction of hot spot motion, and one
redshifted in the backward direction. As the hot spot comes around
towards the observer, the directly beamed blueshifted line dominates,
and then when the phase is $\sim 0.5$ and the hot spot is on the near
side of the black hole, a single line dominates. This is due to the
gravitational demagnification of the secondary images formed by
photons that have to complete a full circle around the black hole to
reach the observer. While these features would most likely be
unresolvable for black hole binaries, they may well be observable in
X-ray flares from Sgr A$^\ast$ as well as other supermassive black
holes [e.g.\ see \citet{bagan01}].

In the subsequent panels, the spectrum is modified by the scattering
of the hot spot photons in the surrounding corona. As in Section
\ref{effect_spectra}, the temperature and density profile of the
corona is given by an ADAF model with $T_e(r_{\rm ISCO})=100$ keV. The
four panels of Figure \ref{scatter_spectrogram45} show increasing
values of $\tau_{\rm es}=[0,1,2,4]$. The effects of scattering on the
spectra are really quite profound. As we described qualitatively in
Chapter 4, the electron corona is like a cloud of fog surrounding a
lighthouse, spreading out the delta-function beam in time and
frequency. Unlike the simple model there, where each photon was
assigned some positive time delay, the Monte Carlo scattering code
shows that some photons actually arrive \textit{earlier} in time by
taking a ``shortcut'' to the observer instead of waiting for the hot
spot to come around and move towards the observer. And of course, the
photons are also spread out in frequency due to the inverse-Compton
effects. As the optical depth increases, the well-defined curve in
Figure \ref{scatter_spectrogram45}a is smeared out into a nearly
constant blur at $\tau_{\rm es} =4$, with a broad spectral peak as in
Figure \ref{powerlaw_cutoff}. Only a slight trace of the original
coherent light curve remains, composed of roughly $1\%$ of the emitted
photons that do not scatter before reaching the observer or get
captured by the event horizon. When $\tau_{\rm es} > 1$, multiple
scattering become more common, so photon shortcuts become rarer,
tending to spread the light curve preferentially to the right (delay in
observer time), as seen in Figures \ref{scatter_spectrogram45}c,d.

\begin{figure}
\begin{center}
\scalebox{0.45}{\includegraphics*[-24,330][415,720]{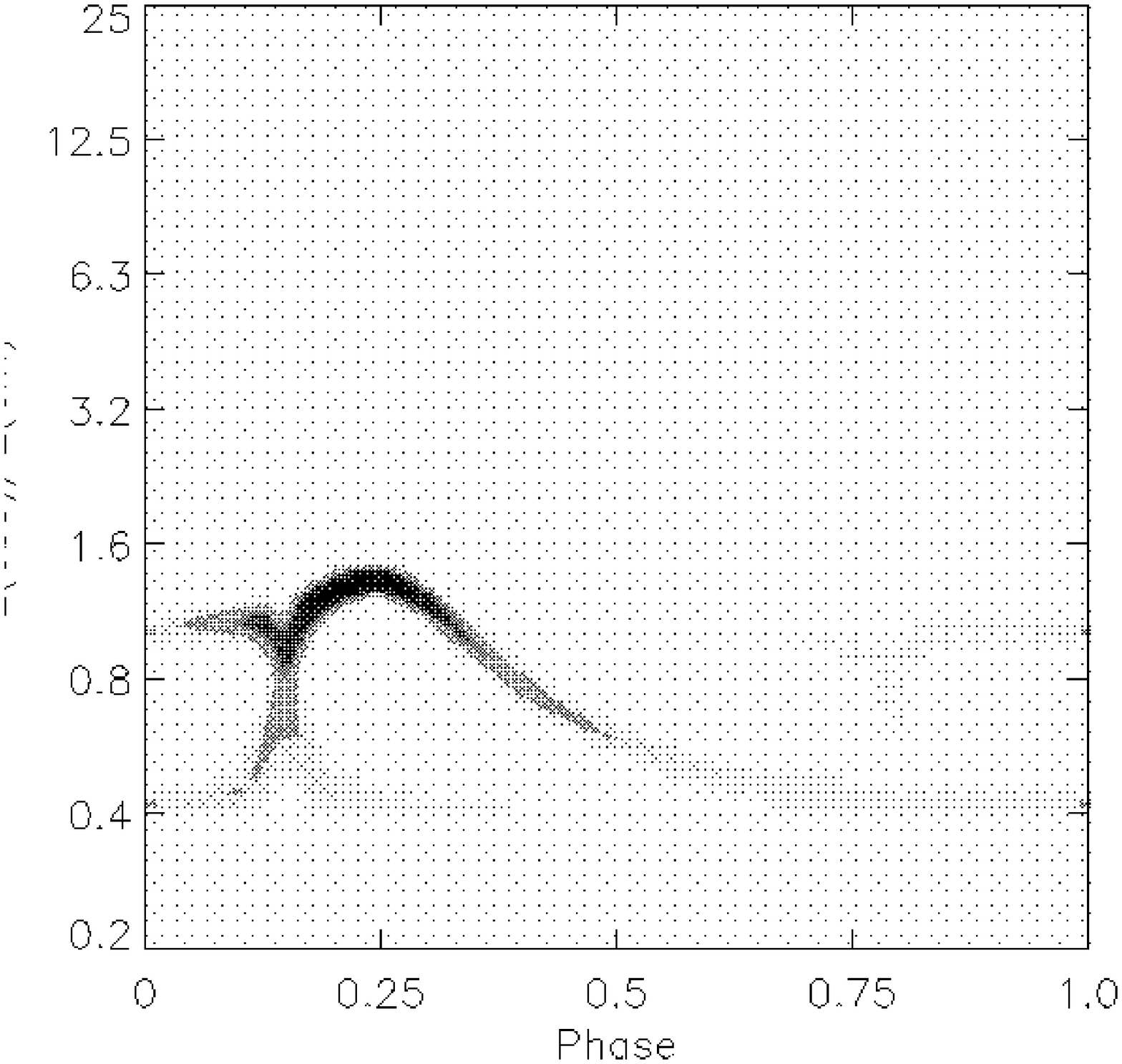}}
\scalebox{0.45}{\includegraphics*[50,330][415,720]{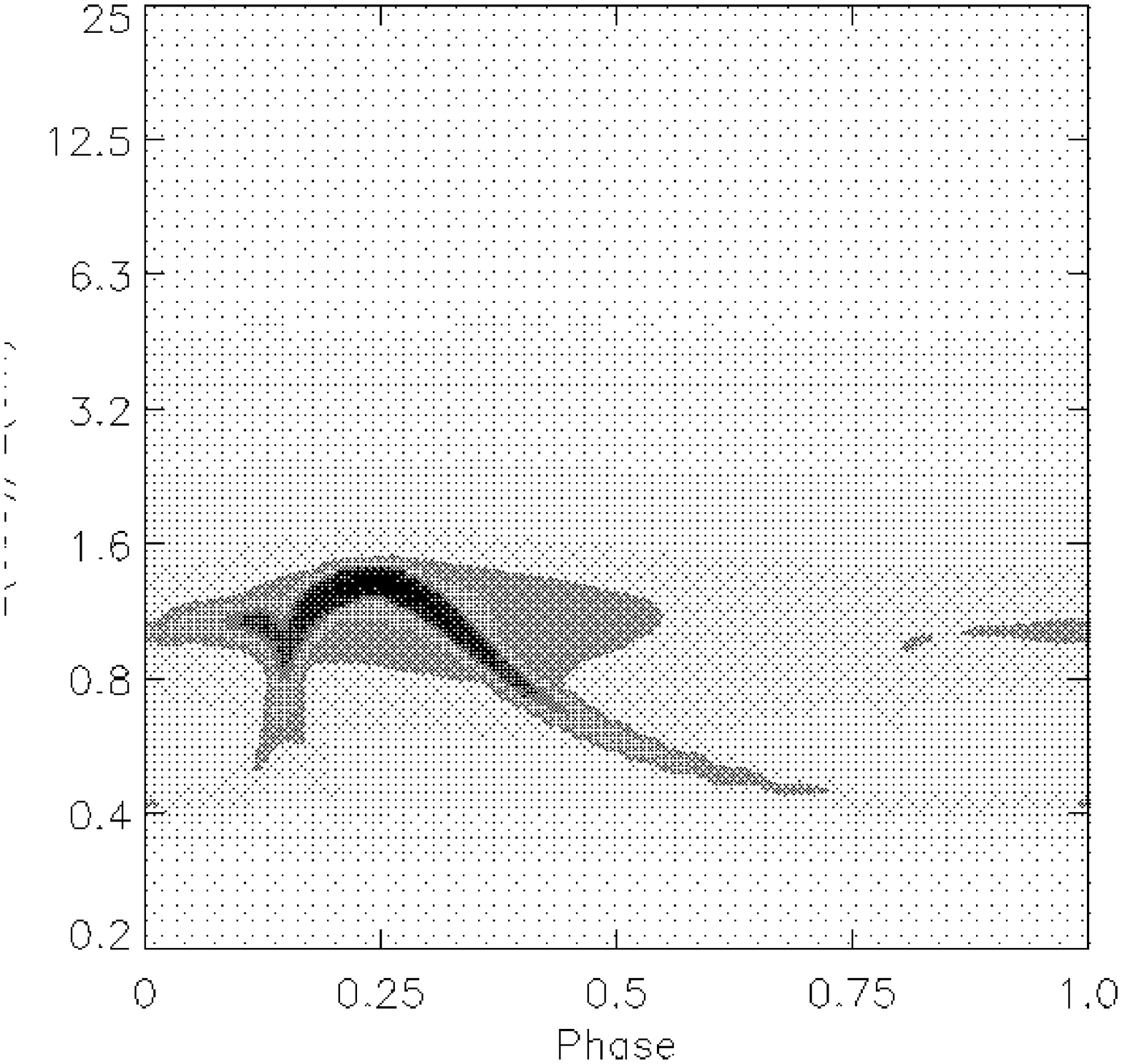}}\\
\scalebox{0.45}{\includegraphics*[-24,300][415,720]{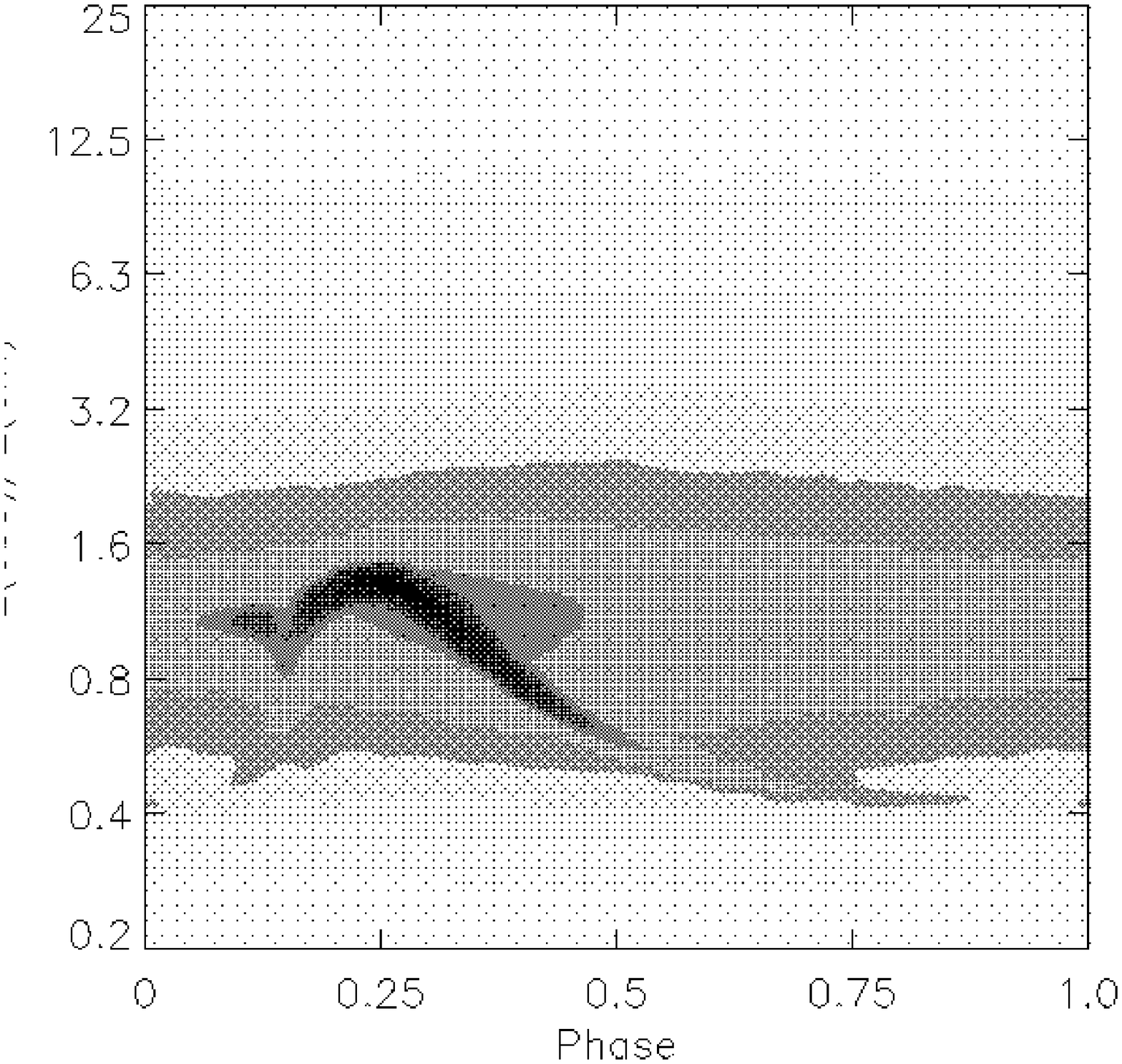}}
\scalebox{0.45}{\includegraphics*[50,300][415,720]{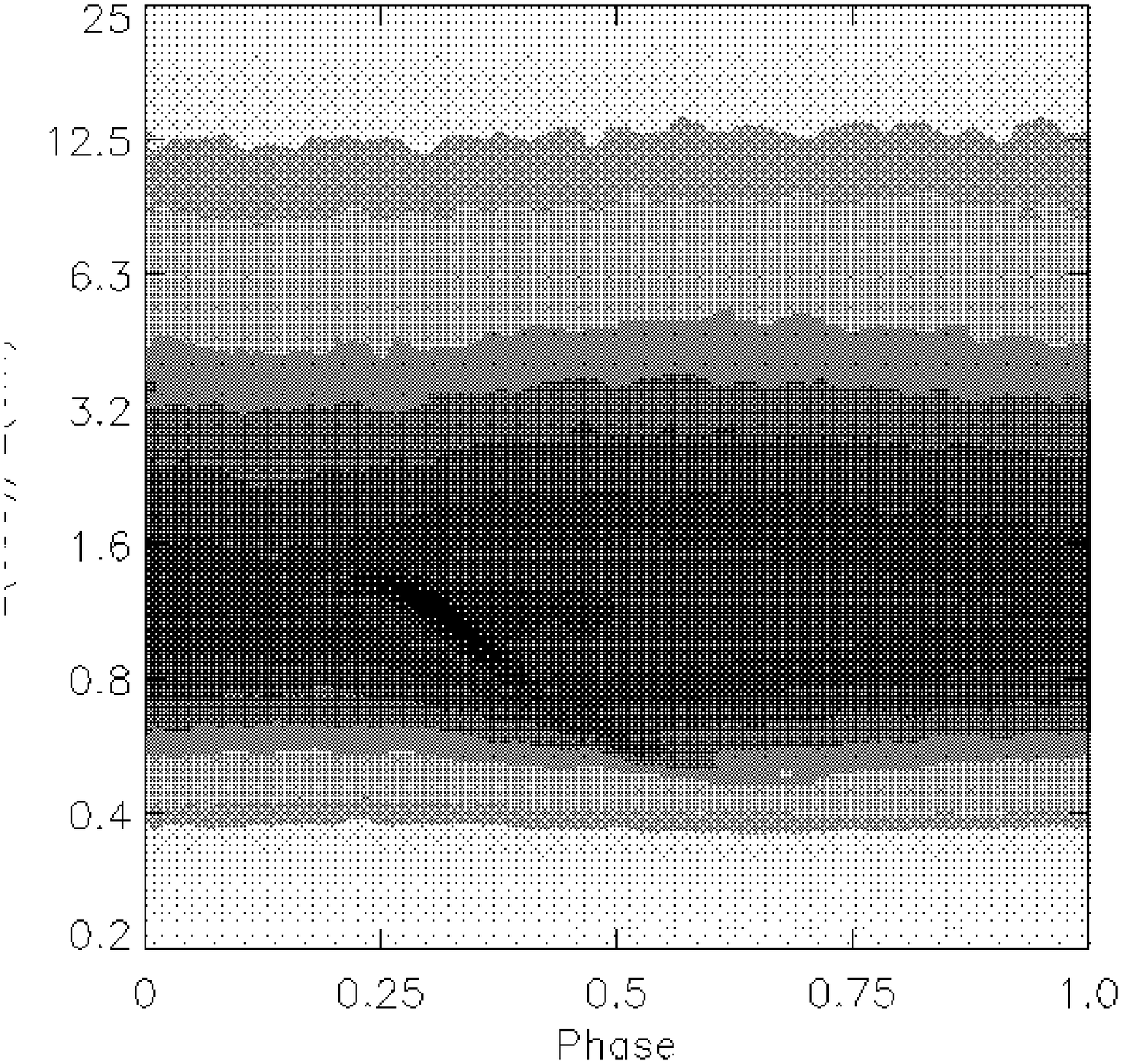}}
\caption[Spectrograms of hot spot emission including
scattering, $i=75^\circ$]{\label{scatter_spectrogram75} Spectrograms
of a monochromatic, isotropic hot spot emitter, as in Figure
\ref{scatter_spectrogram45}, but now with inclination $i=75^\circ$.}
\end{center}
\end{figure}

We show a similar set of spectrograms in Figure
\ref{scatter_spectrogram75}, now with orbital inclination
$i=75^\circ$. The qualitative effects are the same, but the higher
inclination gives stronger relativistic effects and more powerful
beaming. The strongly peaked light curve maintains a somewhat sharper
contrast even for high optical depth. While the detailed spectral and
timing features of Figures \ref{scatter_spectrogram45} and
\ref{scatter_spectrogram75} promise to reveal much about the
underlying physical processes at work, we are still a very long way
from achieving such observational resolution in black hole binary
sources, where the relevant time scales are on the order of
milliseconds. More promising is the application to hot spots from AGN
sources or the X-ray flares around Sgr A$^\ast$ \citep{bagan01}.

\begin{figure}[ht]
\begin{center}
\includegraphics[width=0.8\textwidth]{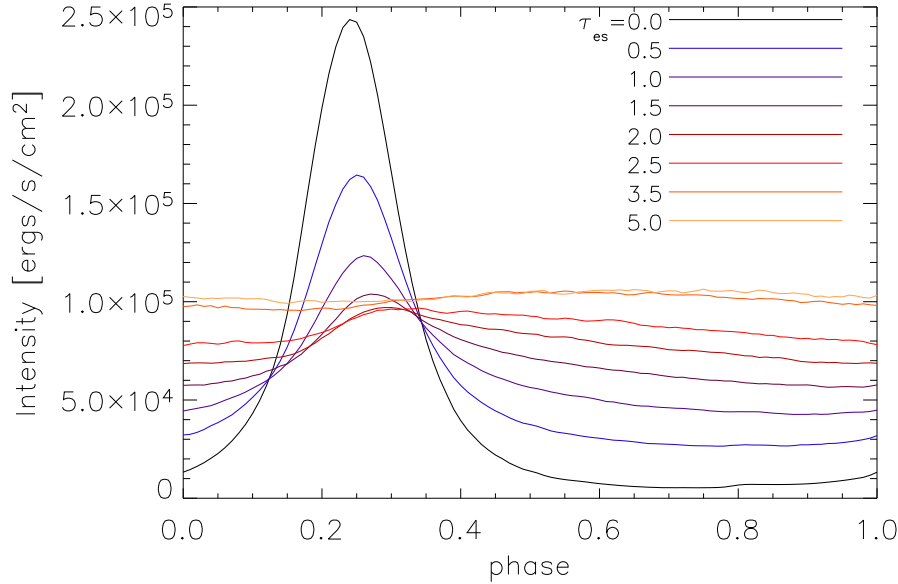}
\caption[Integrated light curves for a range of optical
depths]{\label{lightcurves_tes} Energy-integrated light curves for a
hot spot with orbital parameters as in Figure
\ref{scatter_spectrogram45}. The emitted spectrum is assumed to be
thermal with a hot spot temperature $T_{\rm hs}=1$ keV, integrated
over $0.5-30$ keV in the observer's frame. With increasing optical
depth to scattering, the rms amplitudes decrease significantly,
and their peaks move slightly to the right, due to the time delay from
repeated scattering events.}
\end{center}
\end{figure}

By integrating over broad energy bands such as those typically used in
\textit{RXTE} observations, we can increase our ``signal'' strength
while sacrificing spectral resolution. For millisecond periods, there
will still not be nearly enough photons to provide phase resolution,
but these features may show up statistically in the power spectrum or
bispectrum, as explained in Chapter 4. Figure \ref{lightcurves_tes}
shows a set of integrated light curves for a variety of optical
depths. The black hole and hot spot parameters are as in Figure
\ref{scatter_spectrogram45}, here assuming a thermal emission with hot
spot temperature $T_{\rm hs}=1$ keV. The photons are integrated over
the energy range 0.5-30 keV. 

As the optical depth to electron scattering increases, the rms
amplitude of each light curve decreases as the photons get smoothed
out in time. Similarly, due to the average time delay added to each
photon by the increased path length, the relative location of each
peak is shifted later in time. These amplitudes and phase shifts are
listed in Table \ref{table_tes} for inclinations of $i=45^\circ$ and
$75^\circ$. At even higher optical depths, the
average light curve intensity begins to decrease, as more photons end
up getting captured by the black hole and never reaching the
observer.

\begin{table}[ht]
\caption[Amplitudes and phase shifts of scattered light
curves]{\label{table_tes} Amplitudes and phase shifts of light curve
peaks for hot spot inclinations of $i=45^\circ$ and $75^\circ$ for a
range of optical depths $\tau_{\rm es}$. The amplitude quoted is the
standard deviation $\sigma(I)$ normalized by the mean intensity
$\mu(I)$. The phase shift is where the peak intensity is located in
time, relative to that of the unscattered light curve.}
\begin{center}
\begin{tabular}{lclclclcl}
  & & &$45^\circ$& &\hspace{0.5cm} & &$75^\circ$& \\
  $\tau_{\rm es}$ & & $\sigma/\mu$ & & phase shift & & $\sigma/\mu$ &
  & phase shift \\
  & & & & (periods) & & & & (periods) \\
  \hline
  0 & & 1.14 & & 0 & & 1.51 & & 0 \\
  0.5 & & 0.53 & & 0.01 & & 0.88 & & 0.01 \\
  1.0 & & 0.27 & & 0.02 & & 0.46 & & 0.03 \\
  2.0 & & 0.086 & & 0.06 & & 0.12 & & 0.06 \\
  3.5 & & 0.021 & & 0.29 & & 0.035 & & 0.31 \\
  5.0 & & 0.018 & & 0.35 & & 0.021 & & 0.40 \\
\end{tabular}
\end{center}
\end{table}

\begin{figure}[ht]
\begin{center}
\includegraphics[width=0.45\textwidth]{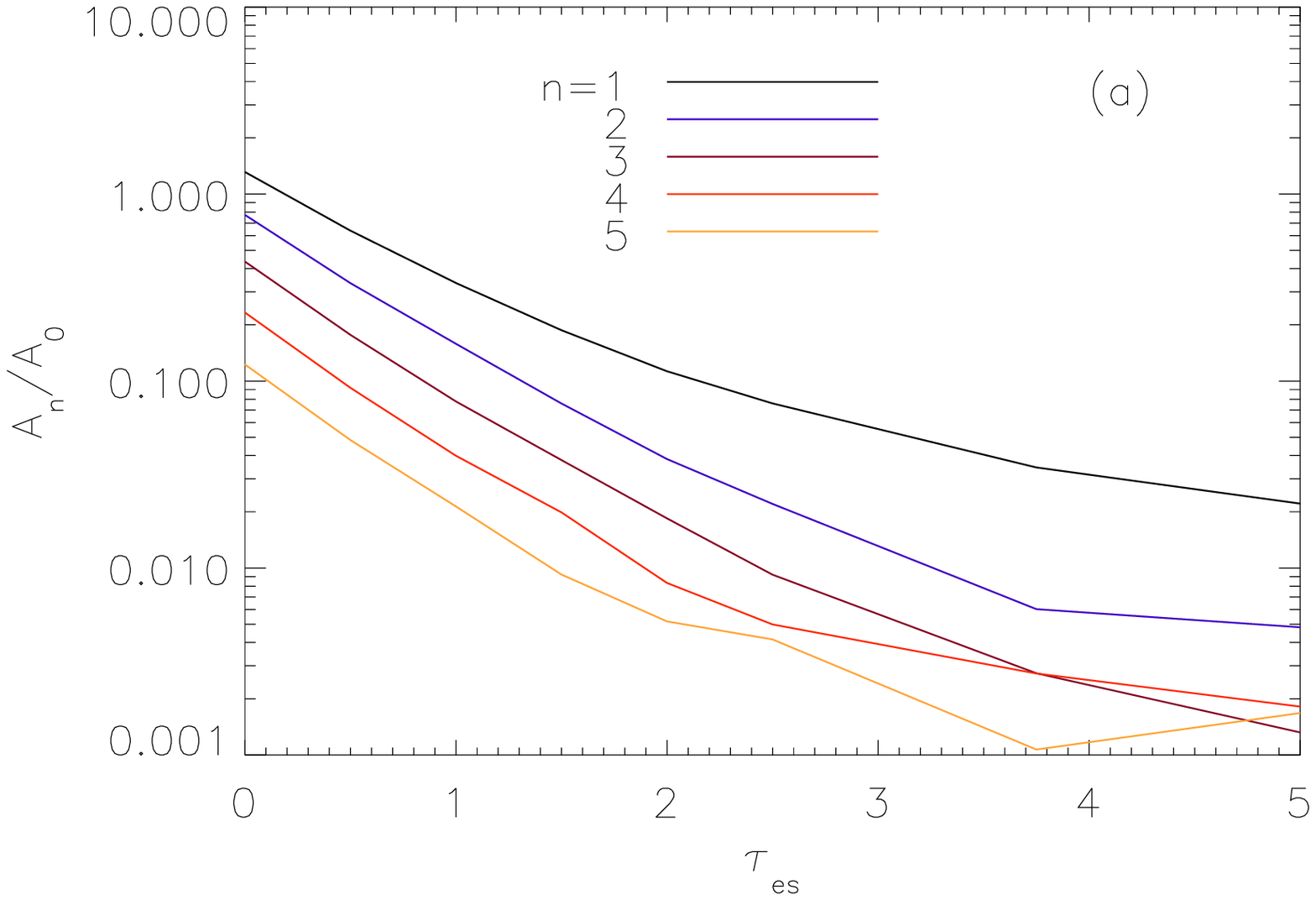}
\includegraphics[width=0.45\textwidth]{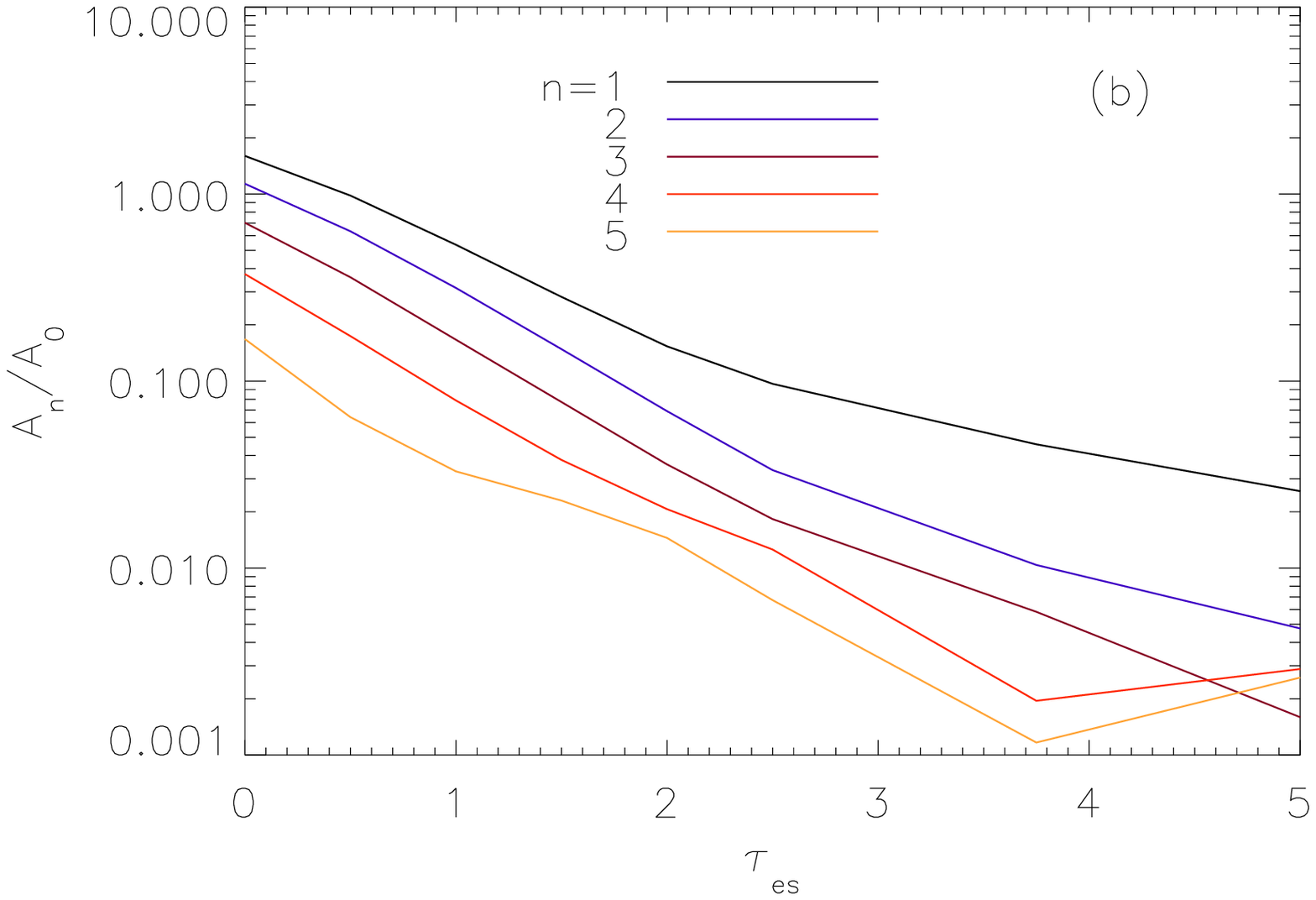}
\caption[Harmonic power as a function of optical
depth]{\label{harmonics_tes} Normalized Fourier amplitudes $A_n/A_0$
for hot spot light curves as in Figure \ref{lightcurves_tes}, for
inclinations of (a) $45^\circ$ and (b) $75^\circ$. As the optical
depth to electron scattering increases, the modulation amplitudes of
the light curves decreases. The higher-order harmonics $n>1$ are
damped even more with respect to the fundamental mode at $n=1$.}
\end{center}
\end{figure}

In all likelihood, the relative phase shifts would be nearly
impossible to detect, regardless of the instrument sensitivity, since
to do so would require measuring the light curve from a single
coherent hot spot at two different optical depths. It is difficult to
imagine a scenario where the coronal properties could change on such
short time scales [yet it is possible that a fixed hot spot on the
surface of an X-ray pulsar might actually be used for this technique;
see \citet{ford00} and \citet{gierl02}]. However, the higher harmonic
peaks of the different
light curves may in fact be measurable with the next-generation X-ray
timing mission, or under extremely favorable conditions, even with
\textit{RXTE}. In Figure \ref{harmonics_tes} we show the damping of
the Fourier modes $A_n/A_0$ with increasing optical depth. Not only does
the overall amplitude of modulation decrease with increased
scattering, but also the relative amplitudes of the higher harmonics
$(n>1)$ decreases relative to the fundamental $(n=1)$. 

\begin{figure}[ht]
\begin{center}
\includegraphics[width=0.45\textwidth]{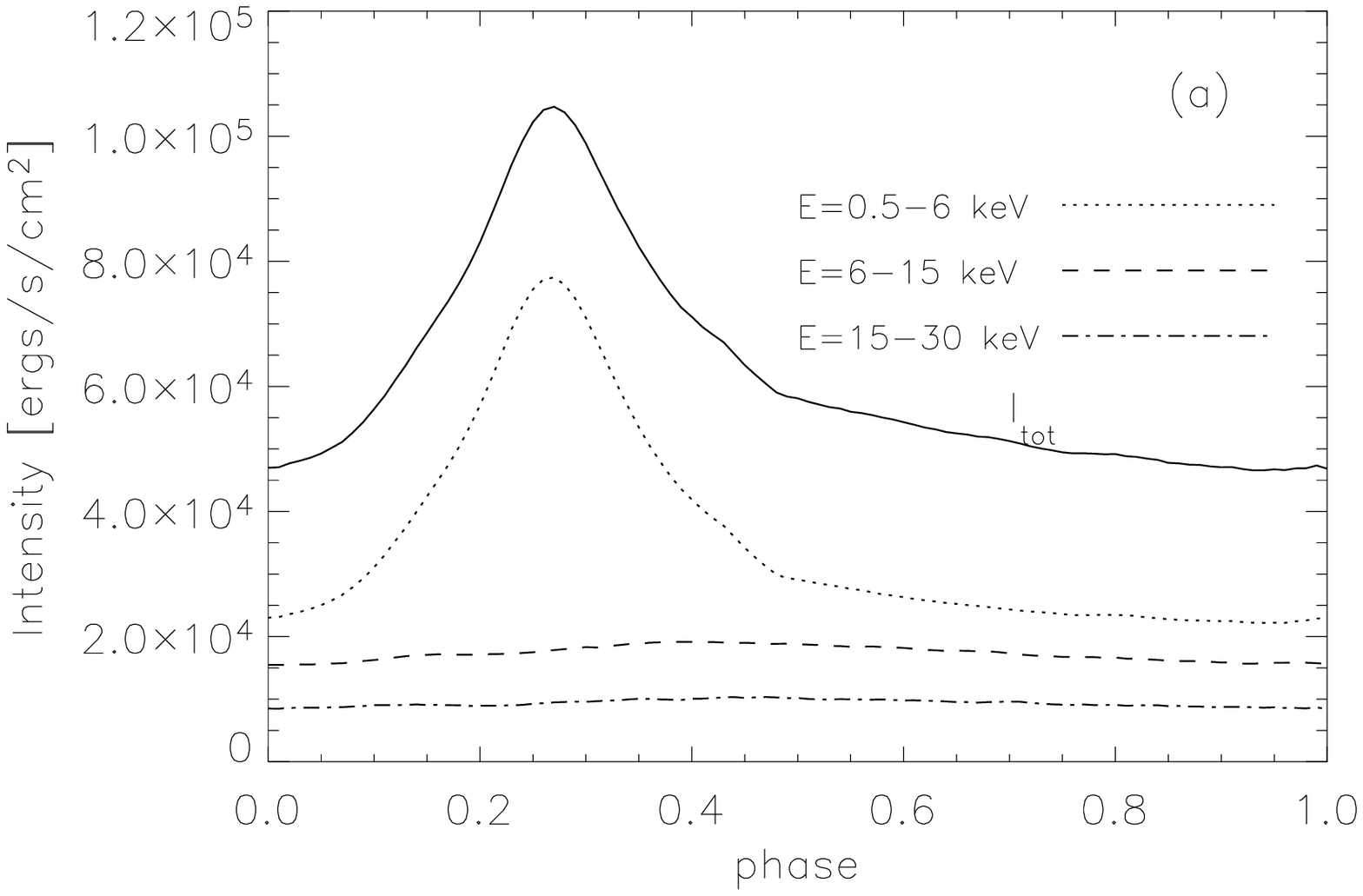}
\includegraphics[width=0.45\textwidth]{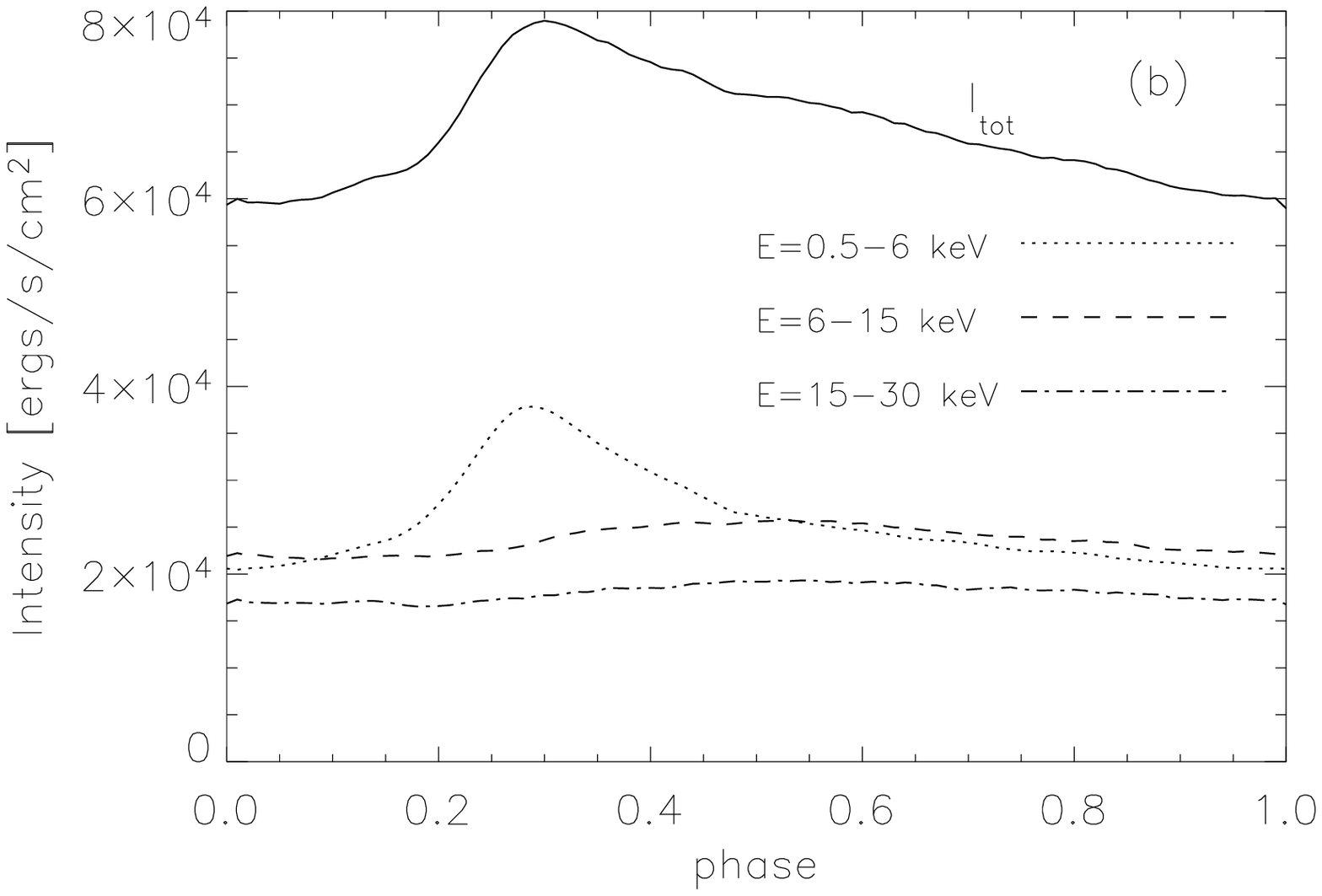}
\caption[Light curves in different \textit{RXTE} energy
bands]{\label{lightcurves_eband} Hot spot light curves in a few
different \textit{RXTE} energy bands (we have expanded the lowest
energy band down to 0.5 keV to include the thermal emission of a hot
spot at $T_{\rm hs}=1$ keV). The hot spot inclination is $75^\circ$
and the coronal properties are as in Figure
\ref{scatter_spectrogram75}. The optical depth to scattering is
$\tau_{\rm es}=1.5$ in (a) and $2.5$ in (b). The higher energy light
curves are made from photons that have experienced more scattering
events, boosting their energy and delaying their arrival time.}
\end{center}
\end{figure}

While the absolute peak shifts for hot spot light curves at different
optical depths would probably not be detectable, the relative shifts of
simultaneous light curves in different energy bands may be observable,
at least on a statistical level with a cross-correlation
analysis. Since the average scattering event boosts photons to higher
energy bands and also causes a net time delay due to the added
geometric path, the light curves in higher energy bands should be
delayed with respect to the lower energy light curves. A few of the
typical energy bands used for \textit{RXTE} observations are $2-6$,
$6-15$, and $15-30$ keV. To fully cover the peak emission from a
thermal hot spot at 1 keV, we expand the lowest energy band in our
calculations to cover $0.5-6$ keV. The light curves in these three
bands are plotted in Figure \ref{lightcurves_eband} for
$i=75^\circ$. The low energy band resembles the unscattered light
curve plus a roughly flat background, while the higher energy light
curves show a much smaller modulation with a significant phase shift
($\sim 0.3$ periods) due to the additional photon path lengths. 

\section{Implications for QPO Models}\label{implications_QPO} 

The original motivation for the application of scattering to the hot
spot model was to answer a few important questions raised by
\textit{RXTE} observations:
\begin{itemize}
\item The distinct lack of power in higher harmonics at
integer multiples of the peak frequencies.
\item The larger significance of high frequency QPO detections in the
higher energy bands (6-30 keV) relative to the signal in the lower
energy band (2-6 keV).
\item The trend for these HFQPOs to exist
predominantly in the ``Steep Power Law'' (SPL) spectral state of the
black hole.
\end{itemize}

Beginning with the final point, it appears to be quite reasonable that
the physical mechanism causing the power law spectra is the
inverse-Compton scattering of cool, thermal photons off of hot coronal
electrons. The \textit{steep} power law suggests a small-to-moderate
value for the Compton $y$ parameter, inferred from equation
(\ref{powerlaw_alpha}), in the range $0.5 \lesssim y \lesssim 10$. From
equation (\ref{compton_y}), this suggests either a small optical depth
or a small electron temperature. To gain insight into which of these
two options is more likely, we need to address the other two
observational clues.

In Chapter 4, we first proposed the scattering model as an explanation
for harmonic damping. There, the model simply assigned a random time
delay to each photon detected by the observer, but the photons
still followed their unobstructed geodesic paths. Thus, photons beamed
towards the observer at the peak of the light curve were still beamed
towards the observer, but delayed slightly in phase. With the more
careful treatment in this chapter, we include not only
the temporal, but also the spatial effects of electron scattering. The
photons originally beamed toward the observer are now scattered in the
opposite direction, while the photons emitted away from the observer
can now be scattered back to him. This smoothes out the light curve in
time more effectively than the localized convolution functions used in
Chapter 4. At the same time, the scattering is not completely
isotropic [see eqn.\ \ref{cross_unpol}], so some
modulation remains. This modulation is probably also supported in the
Kerr geometry because of the bulk rotation of the corona gas (treated
``at rest'' in the ZAMO frame): photons emitted with angular momentum
parallel to the black hole spin get swept along by the electrons
moving in the same direction, maintaining a slight level of beamed
modulation. Thus, to maintain a significant modulation in the light
curve, we require a relatively small optical depth, reducing the
smoothing effects of the scattering.

The fact that most HFQPOs appear more significantly in higher energy
bands also points towards Compton scattering off hot
electrons. However, as the calculations above show (see Fig.\
\ref{lightcurves_eband}), with the basic thermal disk/hot spot model,
the light curves actually have \textit{smaller} amplitude fluctuations
in the higher
energy bands, as these scattered photons get smoothed out more in
time. Furthermore, while the higher harmonic modes are successfully
damped in the scattering geometry, so is the fundamental peak. Thus,
in order to agree with observations, the hot spot overbrightness would
need to be much higher than the values quoted in Chapters 3 and 4. 

Based on these arguments alone, we find it unlikely that the
HFQPOs are coming from a cool, thermal hot spot getting upscattered by
a hot corona. From the photon continuum spectra of the SPL state,
there appears to be a hot corona with Compton $y \sim 1$, but as
Figure \ref{lightcurves_eband}a shows, the lowest energy ($<6$ keV) band
has by far the greatest amplitude modulations. It is possible that the
relative modulation would appear smaller due to the added flux from
the rest of the cool, thermal disk, but much of this steady-state
emission should
also get scattered to higher energies, further damping the modulations
in the 6-30 keV bands. 

The high luminosity of the SPL state (also called the Very
High state) suggests that the thermal, slim disk geometry may not be
appropriate here. Perhaps it is more likely that these cases
correspond to an ADAF model, traditionally associated with very low or
very high accretion rates. Since the ADAF model cannot radiate energy
efficiently, the gas in the innermost regions (where causality
arguments focus our attention) will be much hotter than in the
$\alpha$-disk paradigm. Thus \textit{hot} hot spots with $T_{\rm hs}
\gtrsim 5$ keV could be forming inside a
small ADAF coronal region, providing seed photons that are already in
the higher energy bands, and are only moderately upscattered by the
surrounding corona. In this model, the harmonic damping would be
primarily caused by the formation of arc hot spots as in Chapter 4. 

Of course, this schematic description still leaves many of the original
QPO questions unanswered, like why should the hot spots form at
special radii with commensurate frequencies. If a global oscillation
[such as the pressure-dominated torus of \citet{rezzo03a}] is
producing the seed photons, scattering will also damp out the light
curve modulations, although probably not as much, since relativistic
beaming plays a smaller role in that axisymmetric
geometry. Ultimately, the aim of
this research is to construct a formalism in which these different
models can be accurately evaluated and compared directly to the
observed data.

\chapter{Conclusions and Future Work}
\begin{flushright}
{\it
Science is a wonderful thing if one does not have to earn one's living
at it.\\
\medskip

I never think of the future. It comes soon enough.\\
\medskip
}
-Albert Einstein
\end{flushright}
\vspace{1cm}

\section{Summary of Results}
In this concluding Chapter, we will briefly summarize the results of
the Thesis, their relevance to X-ray observations, and directions for
future work. 
\subsection{Ray-tracing}
In Chapter 2, we developed the foundation of a fully relativistic
ray-tracing code that can be used as a ``post-processor'' to analyze
the output data from three-dimensional hydrodynamics simulations. This
code is based on a Hamiltonian formulation of the equations of motion
in Boyer-Lindquist coordinates. In Section \ref{Hamiltonian_eom} we
showed how certain symmetries in $t$ and $\phi$ can be used to reduce
the dimensionality of the ray-tracing problem. This results in a system of five
coupled first-order differential equations for
$[r,\theta,\phi,p_r,p_\theta]$, which are solved numerically using an
adaptive step fifth-order Runge-Kutta integrator (Section
\ref{num_methods}). The fourth
integral of motion, Carter's constant $\mathcal{Q}$, is used as an
independent check of the accuracy of the numerics. 

While our code uses Boyer-Lindquist coordinates to perform the global
ray-tracing calculation, we also discuss the advantages of the Doran
coordinate system in Section \ref{doran_coord}, which is particularly
useful for modeling processes inside and near the horizon. These
coordinates are based on observers free-falling from infinity, so they
experience no coordinate singularity at the horizon, and are thus
especially convenient for producing movies of exotic black hole
processes such as passing through inner and outer horizons into
parallel universes.

After the geodesic ray-tracing is performed, we have a tabulated set
of positions and momenta along each photon path. Coupled with the
density and temperature profiles from some dynamic disk model, we
can integrate the radiative transfer equation along the path length to
produce time-varying images and spectra of the disk. To successfully
merge the ray-tracing and radiative transfer pieces of the code, in
Section \ref{tetrads} we
introduced a tetrad formalism that defines a locally flat, orthonormal
basis at each point on the photon path. 

In Section \ref{transfer_function} we applied the ray-tracing code to
a simple disk model composed of test particles on planar circular
orbits normal to the black hole rotation axis. Assuming each particle
is emitting monochromatic, isotropic radiation in its rest frame, we
calculated the \textit{transfer function} that maps redshifts from the
plane of the disk to the observer's image plane. This transfer
function can be used to model the broad iron emission lines observed
in black hole binaries and AGN. For disks that extend inside the ISCO, we
showed that the emission line profile is not sensitive to the black
hole spin, and thus at this point is not an especially promising
method for measuring $a/M$.

\subsection{The Hot Spot Model}
In order to apply the post-processor code developed in Chapter 2 to
\textit{RXTE} timing observations, we introduced a simple
geodesic hot spot model in Chapter 3, and then expanded it in Chapter
4 to fit data from XTE J1550--564. The basic hot spot model is
composed of a small region of overbrightness moving on a geodesic orbit
around the black hole, which in turn produces a periodic modulation in
the X-ray light curve. In Section \ref{overbrightness_amp} we showed
how this modulated emission can be added to a steady-state disk flux
and used to infer the size and overbrightness of the hot spot. 

While the actual light curve is not resolvable over such short periods
as $3-10$ msec, the harmonic structure of the Fourier modes can be used
to infer the orbital inclination and hot spot arc length, and to a
lesser degree, the
eccentricity and black hole spin (Sections \ref{harm_inc_spin} and
\ref{noncircular_orbits}). We showed how the radial coordinate
frequency can form beat modes with the fundamental orbital frequency,
giving Fourier power at $\nu_\phi \pm \nu_r$. For closed orbits with
$\nu_\phi=3\nu_r$, these modes have commensurate frequencies $2:3:4$,
as observed in a growing number of black hole binaries (with the
$\nu_\phi+\nu_r$ mode damped due to the shearing of the hot spot into
an arc).

In Section \ref{nonplanar_orbits}, we described the effects of
Lense-Thirring precession on non-planar orbits and examined the
possibility that this precession is responsible for producing the low
frequency QPOs often seen concurrently with the HFQPOs. If these three QPO
peaks (one LF and two HF) are in fact linked, then the black hole mass
\textit{and} spin could be determined independently by matching the
peak frequencies to the geodesic frequencies $\nu_{\rm LT}$,
$\nu_\phi-\nu_r$, and $\nu_\phi$. If the LFQPO is in fact not related
to the hot spot emitter, then the location of the two HFQPOs alone
could be used to determine the spin for a given black hole mass.

In Chapter 4, we expanded the hot spot model to account for the peak
broadening in the QPO power spectrum, deriving two analytic models for
the superposition of multiple hot spots with finite lifetimes and a
range of coordinate frequencies. For a collection of hot spots with
random phases, the finite lifetime of each hot spot causes every
delta-function peak of the periodic power spectrum to be broadened
into a Lorentzian with exactly the same width. Assuming the
hot spots are formed on commensurate orbits with some finite resonance
width in radius (and thus a range of coordinate frequencies), the
different QPO peaks will all be broadened by a different
amount. In Section \ref{freq} we showed how this differential peak
broadening is sensitive to the black hole spin and may be used to map
out the spacetime in the innermost regions of the accretion disk.

In Section \ref{scatter}, we derived a simple electron scattering
model for a low-density corona surrounding the hot spot emitter. The
primary effect of this coronal scattering was to assign each photon
a randomized time delay due to the added path length to the detector,
smoothing out the light curve in time. While this does not affect the
width of the QPO peaks, it does cause a significant damping of the
higher harmonic modes. This scattering model and the formulae for the
peak widths can be combined into a single
analytic expression for the power spectrum, allowing us to fit the
data by minimizing $\chi^2$ over some small set of model
parameters. In Section \ref{data} we applied these models to the
\textit{RXTE} data for XTE J1550--564, and were able to explain the
power spectra of type A and type B QPOs with different hot spot arc
lengths and lifetimes. 

Finally, in Section \ref{bispectrum}, we introduced the use of higher
order statistics such as the bispectrum and bicoherence as a means for
distinguishing between various QPO models. In particular, we showed
how the random phase broadening and the coordinate frequency
broadening would have distinctly different signatures in the
bispectrum contours. For a next-generation X-ray timing mission, these
signatures could be used to map out the spacetime around accreting
black holes and serve to further constrain or rule out the hot spot
model.

%\subsection{QPO Power Spectrum}
\subsection{Steady-state Disks}
As a test-bed for the ray-tracing post-processor and to gain more
insight into the X-ray spectrum of the steady-state disk, in Chapter 5
we developed a relativistic $\alpha$-disk model, based on the work of
\citet{shaku73} and \citet{novik73}. While their treatment can be
thought of as a one- or two-zone model (uniform density; temperature
defined at disk mid-plane and surface), we actually integrate the
complete set of vertical structure equations for density, temperature,
pressure, and energy flux. Coupled with the Novikov-Thorne equations
for radial structure, with the appropriate selection of boundary
conditions for the disk atmosphere, the vertical structure equations
have a unique solution at each radius.

To self-consistently model the torque on the inner edge of the disk,
in Section \ref{geodesic_plunge} we showed how the accreting gas expands
and plunges along geodesic trajectories inside of the ISCO. Following
a one-dimensional column of gas in the frame of the plunging particle,
we can model the time-dependent vertical structure of the innermost
disk with a Lagrangian approach to the partial differential equations
of hydrodynamics. We found that the plunging disk pressure falls off at
a scale length of $l_{\rm plunge}$ inside the ISCO. Since the angular
momentum transport (torque) acts over a turbulent length scale of
$l_{\rm turb} \approx h$, we can solve for the integrated stress (and
thus the surface density $\Sigma$ and radial velocity $p^r$) at
the ISCO by setting $l_{\rm plunge} = l_{\rm turb}$. 

In Section \ref{implicit_scheme} we outlined the numerical methods
used to solve for the disk structure inside and outside of the
ISCO. The Lagrangian hydrodynamics is based on an implicit scheme
described in \citet{bower91}. We found that the density and
temperature profiles of the disk fall off rapidly inside the ISCO, but
that even a small torque can significantly affect the temperature and
total efficiency of the $\alpha$-disk outside of the ISCO. Given the
temperature and scale
height of the disk atmosphere, in Section \ref{ntdisk_spectra} we
used the relativistic ray-tracing code (see Chapter 2) to calculate
the modified ``multi-colored'' spectrum of the disk. The peak of the
spectrum occurs at $E_{\rm max}$, which is a function of the black
hole mass and Eddington-scaled accretion rate. Defining a high-energy
cutoff such that $I(E_{\rm cut})=10^{-5}I(E_{\rm max})$, we showed
that the ratio $E_{\rm cut}/E_{\rm max}$ may be used to determine the
inclination and/or spin of the black hole system. 

\subsection{Electron Scattering}

Motivated by the fact that most black hole HFQPOs are seen in the
``Very High'' or ``Steep Power-Law'' spectral state \citep{mccli04},
we replaced the simple coronal scattering model from Section
\ref{scatter} with a more detailed Monte Carlo treatment,
including angular
dependence and energy transfer via the inverse-Compton
effect. Reversing the ray-tracing paradigm of Chapter 2, in Chapter 6
we explained how thermal photons are traced from isotropic hot spot
emitters through an ADAF-type corona, and then detected by a distant
observer. Like the radiative transfer equation in Section
\ref{radiative_transfer}, the electron scattering can be treated
classically in the appropriate reference frame, and then the post-scattered
photon is transformed back to the coordinate basis and proceeds along
its new geodesic trajectory.

The electron scattering has two major observable effects: it modifies
the photon spectrum, adding a power-law tail at high energies (Section
\ref{effect_spectra}), and it smoothes out the light curve in time as
the photons get re-directed into new time bins (Section
\ref{effect_lightcurves}). We showed how the slope of this power-law
component can be used to determine the density (optical depth) and
temperature of the corona, assuming a self-similar ADAF
profile. We also predicted significant phase shifts in the
light curves observed in different \textit{RXTE} energy channels,
since photons that experience more scattering events tend to get a
larger energy boost and longer time delay before reaching the
detector. This effect may ultimately be observable with a
next-generation timing mission and the careful application of higher
order statistical analysis.

In Section \ref{implications_QPO} we summarized some of the major results of
QPO observations and compared them to the predictions of the
Monte Carlo hot spot scattering calculations. We concluded that the
seed photons causing the QPO modulations in higher energy channels
most likely are coming from an intrinsically hot source, and \textit{not}
getting upscattered by coronal electrons. For an ADAF corona, the
steep power law component of the X-ray spectrum suggests an optical
depth of $\tau_{\rm es}
\approx 1$ and electron temperature $T_e \approx 100$ keV. The arc
shearing described in Chapters 3 and 4 is still a likely candidate for
damping higher harmonic modes in the power spectrum. It is also
quite possible that a more ``global'' oscillation in the inner disk
(e.g.\ one that does not rely
on hot spot beaming)is producing the seed photons,
which then propagate like sound waves through an optically thick
corona, boosting them in energy while damping out the higher
harmonics.

\section{Caveats}\label{caveats}
In the interest of full disclosure, we include here a number of
caveats and qualifications for the models and methods used throughout
the Thesis. Some are more significant than others, but all should be kept
in mind when evaluating the results presented above. 

First and foremost, the geodesic hot spot model still lacks
convincing physical explanations for the following questions: \\

\noindent
(1) How are the hot spots formed? \\
(2) How long should the hot spots survive and what causes their
destruction? \\
(3) What is a reasonable size and overbrightness for the
typical hot spot? \\
and perhaps most importantly, \\
(4) \textit{Why should the hot
spots have special orbits with commensurate coordinate frequencies?}\\

The first three questions may eventually be answered empirically by
global MHD simulations,
but even the most sophisticated 3-D global GR codes still lack important
radiation physics, an essential ingredient in any accretion disk
model. As far as the fourth question, an answer may lie in the
heuristic resonance models of Abramowicz \& Kluzniak, but these still
have a long ways to go before producing a convincing argument for the
formation of hot spots along special commensurate orbits.

When fitting the relative amplitudes of the QPO peaks in Section
\ref{data}, we found that the hot spot arc length was a
well-constrained parameter of the model for both type A and type B
QPOs (although the two types gave two different values for $\Delta
\phi$). For the type
B power spectrum in particular, it was especially important to have a
very long arc to damp out the higher frequency peaks at $\nu_\phi$ and
$\nu_\phi+\nu_r$,
while amplifying the power at $\nu_\phi-\nu_r$. Is it reasonable to
think that over observations of thousands of seconds, the
random hot spot arc lengths could be so consistent and well constrained?
Ideally, any model that is proposed to explain commensurate QPOs
should be quite robust and applicable to a range of black hole masses
and spins. At this point, the geodesic hot spot model still requires a
little too much ``fine tuning'' to satisfy this robustness
criterion, but is nonetheless a powerful tool as a building
block for more physical models.

As mentioned in Chapter 5, the steady-state $\alpha$-disks can be
quite useful for estimating accretion efficiency and temperatures for
multi-color disk models, but do not necessarily give an accurate
treatment of the innermost regions of the disk, particularly inside
the ISCO. While our solution for the turbulence length gives a
reasonable first-order estimate for the torque at the inner edge, the
geodesic plunge still does not agree well with MHD simulations, which
show little if any change in the local temperature and density of the
disk at the ISCO. Furthermore, to quote \citet{novik73}, ``Almost all of the
uncertainties and complications of the model are lumped into the
vertical structure. Ten years hence one will have a much improved
theory of the vertical structure, whereas the equations of (averaged,
steady-state) radial structure will presumably be unchanged.'' Well,
more than thirty years later, we still do not have a complete understanding of
the vertical structure. And as we saw in Section
\ref{vertical_structure}, the vertical structure (particularly that of
the atmosphere) will have a significant effect on the emitted
spectrum. There is growing consensus that the $\alpha$
viscosity/turbulence model is not correct (certainly for a constant
$\alpha$), and the most likely candidate for angular momentum transfer
seems to be the magneto-rotational instability of
\citet{balbu91}. Despite the progress made with MHD simulations,
there is not yet an elegant way of incorporating this process into an
analytic steady-state model like the $\alpha$-disk. 

One of the important assumptions of Chapter 5 was that the accretion
disks are thin, with $h/r \ll 1$. While this appears to be reasonable
for low accretion rates, the hydrostatic structure equations will
break down in the limit of high luminosity (an important limit for the
Very High state associated with HFQPOs). Our revised derivation of the
Eddington
luminosity may help quantify these limits, but is still in a very
early stage of development and has not yet been studied rigorously for
a broad range of black hole masses and accretion rates. Similarly,
while we argued in Section \ref{geodesic_plunge} that a
self-consistent solution for the torque exists at the ISCO, this line
of reasoning was based on the assumption that with increasing mass
accretion rates, the disk just gets thicker and thicker until $l_{\rm
plunge} \approx l_{\rm turb}$. It is now clear that this limit may not
exist for equilibrium slim disks solutions without violating the
modified Eddington limit of Section \ref{revised_eddington}.

Lastly, in Chapter 6 we assumed a static ADAF model for the corona
geometry. While the photon energy spectrum seems to be dependent only
on the Compton $y$ parameter, and not the detailed density and
temperature profiles of the corona, the dependence of the light curves
on these parameters has not been examined comprehensively. Also, while
the corona is treated as static in the ZAMO frame, the underlying hot
spot is moving relativistically through this medium on circular planar
orbits. A consistent model is needed to explain how these two very
different pieces of the accretion flow could exist
simultaneously. This ultimately brings us back to a fundamental
question raised throughout the Thesis: What exactly does the accretion
geometry look like in the Very High/SPL state? Until we can arrive at
a decent answer to this question, none of the various theoretical
models will be acceptable as providing an unambiguous explanation for
the source of high frequency QPOs.

\section{New Applications of Current Code}
The methods and results presented in this Thesis provide a number of
different directions for proceeding with future work. They can be
roughly divided into three categories, in order of increasing labor
requirements and potential scientific reward: (1) Applications of the
current ray-tracing code to
answer new questions and analyze new simulations and observations; (2)
Adding new physics modules to the basic structure of the existing
code; and (3) Developing entirely new models to explain black hole
QPOs and observations of X-ray spectra. In the next three Sections we
will outline a few ideas for each of these categories and try to
evaluate their relative promise for producing important astrophysical
results. 

The first obvious application of the current code is simply to apply
it to a much wider range of model parameters in order to
carry out a comprehensive study of the features of the QPO power
spectrum. In particular, we would like to understand better the
harmonic dependence on orbital eccentricity and hot spot shape (not
just the Gaussian arcs considered in Chapters 3 and 4), as well as
exploring the upper limits to the allowable eccentricity. The results
in Chapter 5 also only scratch the surface of what might be learned
from the thermal disk model. The ISCO boundary condition should be
studied more carefully to see if a self-consistent solution really
exists for a wider range of black hole and accretion parameters. The
scattering model in Chapter 6 must also be examined to better understand
the dependence (if any) on black hole mass and spin, as well as the
corona parameters. 

The results of Chapters 5 and 6 may easily be
combined to produce a complete integrated spectrum of the
multi-colored disk, modified by scattering both in the disk atmosphere
and surrounding hot corona. With this integrated approach we hope to
explain the observed spectra over a broad range of luminosity states
such as those enumerated by \citet{esin97}. This approach could also
lead to a more detailed understanding of radiation transport from one
radius of the disk to another. In particular, by modeling the photons
that get inverse-Compton scattered from the corona back to the disk,
we should be able to produce a more accurate model for the fluorescent
iron emission profile.

As was mentioned in Section \ref{history_observ}, \citet{mille05} have
recently discovered a correlation between the phase of the low
frequency QPO and the instantaneous shape of the iron emission line in
the black
hole source GRS 1915+105. Our ray-tracing code is ideally suited for
investigating the possibility of whether the emission may be coming
from an inclined ring or torus precessing at the Lense-Thirring
frequency. As this ring precesses around the black hole spin axis, the
solid angle seen by the distant observer should oscillate
periodically, thus modulating the X-ray light curve. At the same time,
the transfer function from the ring to the image plane will also vary
periodically, changing the shape of any relativistic emission
lines. Our code should be able to fit the data with a small number of
model parameters, which may even be used to constrain the black hole
spin. Furthermore, the higher-order statistical methods of Section
\ref{bispectrum} might be applied to these LFQPOs to further constrain
the emission model.

The ultimate purpose of the ray-tracing code has always been to
use it as a totally general post-processor analysis tool for any 3-D
(magneto)hydrodynamic accretion simulation. We believe that we are now
in a position to begin applying it as such and producing quantitative
predictions with which to compare observations. It has recently been
used to produce preliminary light curves from the oscillating torus model of
Rezzolla \& Zanotti. These light curves are found to be nearly sinusoidal and
almost entirely limited to a single fundamental frequency mode. We
plan on investigating other initial conditions and emission mechanisms
to see whether the torus may in fact be able to produce commensurate
QPOs with 3:2 ratios. Another exciting collaboration that should
develop in the near future is with John Hawley's group at the
University of Virginia. Even if we cannot identify clear QPOs, by
post-processing their global MHD simulations, we hope to understand
more about the continuum photon energy distribution and power spectra
for turbulent disks.

\section{New Features for Code}
In addition to applying the existing code to new problems, we also
plan to add new physics capabilities to the code to make it more
accurate and useful for a larger number of applications. For the hot
spot model, we would like to be able to include more complicated
temporal evolution for the hot spot emissivity. Instead of simply
turning ``on'' and ``off'' instantaneously, the emission should be
able to evolve continuously like turbulent modes that grow, saturate,
and eventually decay. While the analytic methods derived in Chapter 4
can model this type of evolution reasonably well, it is always nice to
be able to simulate it directly with actual hot spots being viewed by
the ray-tracing code.

The Monte Carlo calculations of Chapter 6 could be expanded
significantly to include a more careful treatment of the relativistic
scattering (i.e.\ the Klein-Nishina cross section), the geometry and
dynamics of the electron corona [bulk velocity and angular momentum;
see \citet{psalt01b}], and the addition of photon polarization. For
increased computational efficiency, the Monte Carlo scattering may be
replaced by a more elegant density matrix formalism for the radiation
field \citep{ports04a,
ports04b}. While we expect the revised cross sections for polarized
light to have only a small effect on the integrated light curves
(multiply-scattered photons will be preferentially scattered at
different angles than for unpolarized light), it is possible that the
detailed calculations may give interesting predictions for a
next-generation observatory that is sensitive to X-ray polarization
\citep{sunya85,laor91,dovci04}. Also, new models for the corona should
be investigated, e.g.\ ``clumpy'' coronae made up of optically thick
clouds that might affect the light curves and spectra of the scattered
photons \citep{fuers04}.

After integrating the electron scattering calculations with the
$\alpha$-disk modified thermal emission, we hope to ``streamline'' the
process in order to produce many different disk spectra in an
efficient manner. In this way, one could fit spectral data with an
accurate model consisting of only a few key parameters such as the
black hole mass, spin, accretion rate, inclination, and the coronal
density/temperate profile. Ultimately, we would like to make the code
publicly available, much like the popular CMBFAST code
\citep{selja96}, so that many different users can analyze stellar-mass,
intermediate-mass, and supermassive black hole spectra, 

Finally, it is possible that with slight modifications, the
ray-tracing code could be used to model even more exotic physical
processes around black holes. In recent years, there have been a
number of very exciting
observations of Sgr A$^\ast$, home to the supermassive black
hole at the center of our galaxy [see, e.g.\ \citet{bagan01,genze03}].
\citet{tsuch04} and \citet{aharo04} have both reported significant
detection of extremely
high energy $\gamma$-ray emission from the vicinity of Sgr
A$^\ast$. One speculative but extremely exciting possible explanation
for this TeV emission could be the annihilation of dark matter
particles in the central cusp of the galactic halo [\citet{bergs04}
and references therein]. The very same code used to trace hot spot
photons could also calculate the redshift and energy
distribution of the annihilation $\gamma$-rays, assuming a simple model
for their production. Eventually, with the proper theoretical
foundation, these observations may also be used to map out the spacetime
around black holes, and even understand the fundamental particle
physics of dark matter.

\section{Development of New Models}
While it is definitely a useful achievement to develop a
post-processor ray-tracing code capable of analyzing any general accretion disk
model, what would be really exciting is to develop a new physical
model that could predict the existence of HFQPOs from first
principles. The very fact that there are currently so many alternative (and
certainly not completely convincing) models in the literature was what
motivated us to focus on the post-processor approach in the first
place. But as they often say, ``the more the merrier,'' so we propose
to add a couple more possibilities to this growing list.

Perhaps the simplest to analyze would be a set of spiral density waves
forming in a relatively cold, thin disk of test particles on geodesic
orbits. Much like the sweeping spiral arms that make up many galaxies
\citep{toomr64,toomr72}, these accretion disk density waves would be
produced around regions in phase
space where the epicyclic orbits overlap and form caustic sheets
\citep{gottl02}. While the individual particles in the disk will be
orbiting at the same geodesic frequencies as the hot spot model, the
spiral density arms may appear to be moving at quite a different
velocity, perhaps even explaining the low frequency QPOs. Also, the
formation of these waves may be closely related to the resonant
interaction between azimuthal and radial coordinate frequencies, just
as in the forced resonance model for hot spot formation. 

As we mentioned at the end of Section \ref{implications_QPO}, the QPOs
may be coming from a more global oscillation in the inner regions of
the corona. We have recently begun to investigate the possibility of
forming radiation ``eigenmodes'' that can grow in a type of resonance
cavity around the black hole. For example, consider a ring of hot gas
in a planar circular orbit, with a non-axisymmetric $m=2$ temperature
perturbation in $\phi$. The two opposite points of maximum temperature, and
thus emission, will ``see'' each other greatly magnified by the
gravitational lensing of the central black hole. These points will
thus absorb more radiation than the rest of the ring, growing even
hotter, thus amplifying any small initial perturbations. A global
ray-tracing calculation could be used to find any special radii where
such radiation eigenmodes would form and evolve. This model seems
particularly appropriate for the hot, low-density, quasi-spherical
ADAF geometries that might form at high luminosities.

Despite the optimistic language of \citet{novik73} quoted above in
Section \ref{caveats}, we still do not have a real physical model for
the transport of angular momentum and energy for a thin accretion
disk, particularly inside the ISCO. Such a model would be critical for
understanding the shapes of
broad iron emission lines, which almost certainly originate in the
inner disk, and thus for measuring black hole spin. A successful model
would most likely incorporate the magneto-rotational instability as
the primary source for turbulent viscosity and angular momentum
transport, which may possibly be done
analytically with the heuristic treatment of \citet{gammi04}. At the
same time, the model must also include a means for treating the
radiation diffusion through the
disk, which will likely result in thinner, cooler disks than those
predicted by the current generation of global MHD simulations. If we
could derive such an elegant, analytic model, it could also be used
to form the basis for detailed perturbation analysis, returning us to
the central problem of giving an explanation for high frequency QPO
emission.

\appendix
\chapter{Formulae for Hamiltonian Equations of Motion}
The equations of motion for the reduced Hamiltonian $H_1$ in
Boyer-Lindquist coordinates, as given in Chapter 2, are repeated here:
\begin{displaymath}
H_1(r,\theta,\phi,p_r,p_\theta,p_\phi;t) = -p_t = \omega p_\phi
+\alpha\left(\frac{\Delta}{\rho^2}p_r^2 
+\frac{1}{\rho^2}p_\theta^2 +\frac{1}{\varpi^2} p_\phi^2
+m^2\right)^{1/2},
\end{displaymath}
and according to classical theory:
\begin{subequations}
\begin{eqnarray}
\frac{dx^i}{dt} &=& \frac{\partial H_1}{\partial p_i}\\
\frac{dp_i}{dt} &=& -\frac{\partial H_1}{\partial x^i}.
\end{eqnarray}
\end{subequations}
For convenience of notation, we define the quantity $D^2$ as
\begin{equation}\label{def_X}
D^2(r,\theta,\phi,p_r,p_\theta,p_\phi) =
\frac{\Delta}{\rho^2}p_r^2 
+\frac{1}{\rho^2}p_\theta^2 +\frac{1}{\varpi^2} p_\phi^2
+m^2.
\end{equation}
Then for an arbitrary variable $y \in (x^i,p_i)$, the partial derivative
of $H_1$ can be written
\begin{equation}\label{dH_dy}
\frac{\partial H_1}{\partial y} = \frac{\partial}{\partial y} 
(\omega p_\phi) + \frac{\partial \alpha}{\partial y}D -
\frac{1}{2}\frac{\alpha^2}{p_t+\omega p_\phi} 
\frac{\partial D^2}{\partial y}.
\end{equation}
The first set of Hamiltonian's equations are straightforward to
produce: 
\begin{subequations}
\begin{eqnarray}\label{dxi_dt}
\frac{dr}{dt} &=& \frac{\partial H_1}{\partial p_r} =
-\frac{p_r}{p_t+\omega p_\phi}\frac{\alpha^2 \Delta}{\rho^2}, \\
\frac{d\theta}{dt} &=& \frac{\partial H_1}{\partial p_\theta} =
-\frac{p_\theta}{p_t+\omega p_\phi}\frac{\alpha^2}{\rho^2}, \\
\frac{d\phi}{dt} &=& \frac{\partial H_1}{\partial p_\phi} = \omega
-\frac{p_\phi}{p_t+\omega p_\phi}\frac{\alpha^2}{\varpi^2}.
\end{eqnarray}
\end{subequations}
The momentum equations are a bit more involved, but there are only two
of them (for $p_r$ and $p_\theta$; $p_\phi$ is conserved):
\begin{eqnarray}\label{dpi_dt}
\frac{dp_i}{dt} = -\frac{\partial H_1}{\partial x^i} &=&
-\frac{\partial \omega}{\partial x^i}p_\phi + \frac{p_t + \omega
 p_\phi}{\alpha} \frac{\partial \alpha}{\partial x^i} + \nonumber \\
& & \frac{\alpha^2}{2(p_t + \omega p_\phi)}
 \left[\frac{\partial}{\partial x^i}
\left(\frac{\Delta}{\rho^2}p_r^2 +\frac{1}{\rho^2}p_\theta^2 
+\frac{1}{\varpi^2} p_\phi^2 \right)\right].
\end{eqnarray}
The relevant spatial derivatives are as follows:
\begin{subequations}
\begin{eqnarray}
\frac{\partial\omega}{\partial r} &=& -\frac{\omega^2}{2Ma}
\left[3r^2+a^2(1+\cos^2\theta)-\frac{a^4}{r^2}\cos^2\theta\right] \\
\frac{\partial\omega}{\partial \theta} &=& -\frac{\omega^2}{2Ma}
\left[\left(2Ma^2-a^2r-\frac{a^4}{r}\right)\sin\theta\cos\theta \right] \\
\frac{\partial\alpha^2}{\partial r} &=& -\alpha^4
\left(\frac{2M}{\Delta\rho^2}\right)
\left(\frac{a^4-r^4}{\Delta}-\frac{2r^2a^2\sin^2\theta}{\rho^2}\right) \\
\frac{\partial\alpha^2}{\partial \theta} &=& -\alpha^4
\left[\frac{4Ma^2r\sin\theta\cos\theta(a^2+r^2)}{\Delta\rho^2}\right] \\
\frac{\partial}{\partial r}\left(\frac{1}{\varpi^2}\right) &=& 
-\frac{2}{\varpi^4}\left[\sin^2\theta\left(r+\frac{2Ma^2\sin^2\theta
(a^2\cos^2\theta-r^2)}{\rho^4}\right)\right] \\
\frac{\partial}{\partial \theta}\left(\frac{1}{\varpi^2}\right) &=& 
-\frac{4\sin\theta\cos\theta}{\varpi^4}\left[2Ma^2\sin^2\theta
\left(\frac{r^2+a^2}{\rho^4}+\frac{1}{\rho^2}\right)+(r^2+a^2)\right] \\
\frac{\partial}{\partial r}\left(\frac{\Delta}{\rho^2}\right) &=& 
\frac{2}{\rho^2}\left(r-M-\frac{r\Delta}{\rho^2}\right) \\
\frac{\partial}{\partial \theta}\left(\frac{\Delta}{\rho^2}\right) &=& 
\frac{2}{\rho^4}a^2\Delta\sin\theta\cos\theta \\
\frac{\partial}{\partial r}\left(\frac{1}{\rho^2}\right) &=& 
-\frac{2r}{\rho^4} \\
\frac{\partial}{\partial \theta}\left(\frac{1}{\rho^2}\right) &=& 
\frac{2}{\rho^4}a^2\sin\theta\cos\theta.
\end{eqnarray}
\end{subequations}
\newpage

\chapter{Summing Periodic Functions with Random Phases}
In this Appendix we derive the shape of a QPO peak in Fourier space
broadened by
the summation of multiple periodic functions combined with random
phases. There are many different accepted conventions for discrete and
continuous Fourier transforms \citep{press97}, so we begin by defining the
forward- and reverse-transforms between the time and frequency domains ($t$
and $\nu$). For a Fourier pair $f(t)$ and $F(\nu)$,
\begin{equation}
F_j = \frac{1}{N_s} \sum_{k=0}^{N_s-1} f_k e^{-2\pi i j k/N_s} \to
F(\nu) = \frac{1}{T_f}\int_0^{T_f} f(t) e^{-2\pi i \nu t} dt
\end{equation}
and
\begin{equation}
f_k = \sum_{j=0}^{N_s-1} F_j e^{2\pi i j k/N_s} \to
f(t) = T_f\int_{-\nu_N}^{\nu_N} F(\nu) e^{2\pi i \nu t} d\nu,
\end{equation}
where $f(t)$ is defined on the time interval $[0,T_f]$ and $\nu_N=1/(2\Delta
t)$ is the Nyquist frequency for a sampling rate $\Delta
t=T_f/N_s$. With this convention, $f(t)$ and $F(\nu)$ conveniently have the
same units and Parseval's theorem takes the form
\begin{equation}\label{spectral}
\int_0^{T_f} f^2(t)dt = T_f^2\int_{-\nu_N}^{\nu_N}F^2(\nu)d\nu.
\end{equation}
For such a time series $f(t)$, the power spectrum is defined as
$F^2(\nu)$, the squared amplitude of the Fourier transform.

Consider a purely sinusoidal function 
\begin{equation}
f(t) = A\sin(2\pi\nu_0 t+\phi),
\end{equation}
where $\phi$ is some constant phase. If there are an integer number of
complete oscillations within the time $T_f$, or 
in the limit of $T_f\to \infty$, the Fourier transform of $f(t)$ will be 
\begin{equation}
F(\nu) = \left\{ \begin{array}{cl}
\frac{A}{2}e^{i(\phi-\pi/2)} & \nu = \nu_0 \\
\frac{A}{2}e^{-i(\phi-\pi/2)} & \nu = -\nu_0 \\
0 & {\rm otherwise} \end{array} \right. .
\end{equation}
If we then truncate the function $f(t)$ by multiplying it with a boxcar
window function $w(t)$ of length $\Delta T$, the convolution theorem gives
the transform of the resulting function $g(t)$:
\begin{eqnarray}
g(t) = f(t)w(t) \Leftrightarrow G(\nu)=(F\star W)(\nu).
\end{eqnarray} 
In the case where the window function is longer than a single period and
short compared to the total sampling time ($1/\nu_0 < \Delta T \ll T_f$), the
convolved power $G^2(\nu)$ can be well approximated by 
\begin{equation}\label{sin1}
G^2(\nu) \approx \frac{A^2}{4T_f^2}
\frac{\sin^2[\pi(\nu \pm \nu_0)\Delta T]}{\pi^2(\nu \pm \nu_0)^2}.
\end{equation}

For $f(t)$ real, $F(-\nu)=F^\ast(\nu)$ and since we are primarily concerned
with the power spectrum $F^2(-\nu)=F^2(\nu)$, we will generally
consider only positive frequencies (unless explicitly stated otherwise).
Of course, when calculating the actual observable power in a
signal, both positive and negative frequencies must be included.

All the information about the phase $\phi$ of $f(t)$ and
the location in time of the window function is contained in the complex
phase of the function $G(\nu)$. This phase information is important
when considering the total power contributed by a collection of
signals, each with a different time window and random phase. 
When summing a series of complex functions with random
phase, the total amplitude adds in quadrature as in a two-dimensional
random walk. Therefore combining $N$ different segments of $f(t)$, each of
length $\Delta T$ and random $\phi$, gives a Fourier transform with amplitude
$\sqrt{N}|G(\nu)|$, and thus the net power spectrum is $NG^2(\nu)$.

The result in equation (\ref{sin1}) is valid only if every segment of
$f(t)$ has the exact same sampling length $\Delta T$ and frequency
$\nu_0$. Motivated by the physical processes of radioactive decay, we
assume here an exponential distribution for the lifetime of each
segment. For a characteristic lifetime of $T_l$, the differential
probability distribution of lifetimes $T$ for coherent segments is
\begin{equation}
P(T)dT = \frac{dT}{T_l}e^{-T/T_l}.
\end{equation}
Over a sample time $T_f \gg T_l$, the number of segments with a
lifetime between $T$ and $T+dT$ is given by
\begin{equation}\label{distribution}
dN(T) = \frac{T_f}{T_l^2}e^{-T/T_l}dT.
\end{equation}

Assuming for the time being that each coherent section of the signal is
given by the sinusoidal function $f(t)$ used above, we can sum all the
individual segments to give the total light curve $I(t)$ with
corresponding power spectrum
\begin{eqnarray}\label{lorentz1}
\tilde{I}^2(\nu)&=&\int_0^\infty G^2(\nu,T) dN(T) \nonumber\\
&=& \left(\frac{A}{2\pi T_f}\right)^2 \int_0^\infty
\frac{\sin^2[\pi(\nu-\nu_0)T]}{\pi^2(\nu-\nu_0)^2}
\frac{T_f}{T_l^2}e^{-T/T_l} dT \nonumber\\
&=& 2A^2\frac{T_l}{T_f} \frac{1}{1+4\pi^2 T_l^2 (\nu-\nu_0)^2}.
\end{eqnarray}
Hence we find the shape of the resulting power spectrum is a Lorentzian
peaked around $\nu_0$ with characteristic width 
\begin{equation}
\Delta \nu = \frac{1}{2\pi T_l}.
\end{equation}

Since the boxcar
window represents an instantaneous formation and subsequent destruction
mechanism, the resulting power spectrum contains significant power
at high frequencies, a general property of discontinuous functions. 
A smoother, Gaussian window function in time gives a Gaussian profile
in frequency space: 
\begin{equation}\label{gauss}
w(t)=\exp\left(\frac{-t^2}{2T^2}\right) \Leftrightarrow 
W(\nu) = \sqrt{2\pi}\frac{T}{T_f}
\exp\left(\frac{-\nu^2}{2\Delta \nu^2}\right)
\end{equation}
where again the characteristic width is given by $\Delta \nu = 1/(2\pi
T)$. After integrating over the same distribution of lifetimes
$dN(T)$ as above, we get the power spectrum
\begin{equation}\label{erfc}
\tilde{I}^2(\nu) = 4\pi N_{\rm spot}A^2\frac{T_l}{T_f}z^3
\left[\sqrt{\pi}(1+2z^2){\rm erfc}(z)e^{z^2}-2z \right],
\end{equation}
where we have defined 
\begin{equation}
z \equiv \frac{1}{4\pi T_l(\nu-\nu_0)}.
\end{equation}
For large $z$ (near the peak at $\nu=\nu_0$), equation (\ref{erfc})
can be approximated by the narrow Lorentzian
\begin{equation}
\tilde{I}^2(\nu) \approx 4\pi N_{\rm spot}A^2\frac{T_l}{T_f}
\frac{1}{1+48\pi^2 T_l^2(\nu-\nu_0)^2}.
\end{equation}

As with the boxcar window, the exponential lifetime distribution has the
effect of narrowing the peak of the net power spectrum compared with
that of a single Gaussian segment of the light 
curve with length $T_l$. These results
are in fact easily generalized. For any set of localized, self-similar
window functions $w(t,T)=w(t/T)$, the corresponding power spectra
$W^2(\nu;T)$ can be approximated near $\nu=0$ as a Lorentzian:
\begin{equation}
W^2(\nu;T) \approx \frac{T^2}{T_f^2}
\frac{1}{1+\beta^2T^2\nu^2},
\end{equation}
with $\beta$ a dimensionless constant over the set of $w(t,T)$. The
characteristic
width of $W^2(\nu,T)$ is thus defined as $1/(\beta T)$. Integrating
over the lifetime distribution $dN(T)$ from equation
(\ref{distribution}), the net power function is given by 
\begin{equation}
\tilde{I}^2(\nu) \approx \tilde{I}^2(\nu_0)
\frac{1}{1+12\beta^2T_l^2(\nu-\nu_0)^2}.
\end{equation}
We see now that the general effect of an exponential distribution of
sampling lifetimes is to decrease the peak width, and thus increase
the coherency, by a factor of $\sqrt{12}$. 

\newpage

%% This defines the bibliography file (thesis.bib) and the bibliography style.
%% If you want to create a bibliography file by hand, change the contents of
%% this file to a `thebibliography' environment.  For more information 
%% see section 4.3 of the LaTeX manual.

%\bibliographystyle{natbib}
%\bibliography{thesis}

\end{document}